\documentclass[nonacm, acmtog]{acmart}

\usepackage{booktabs}

\usepackage{multirow}
\usepackage{xpatch}
\usepackage{placeins}
\usepackage{amsmath}
\usepackage{makecell}

\usepackage{xcolor}
\definecolor{greenish}{RGB}{0,109,44}
\definecolor{RED}{RGB}{255,0,0}
\definecolor{lightGreen}{RGB}{173,223,138}
\definecolor{lightBlue}{RGB}{166,206,227}
\definecolor{middleBlue}{RGB}{0,128,255}
\definecolor{darkerBlue}{RGB}{0,102,204}
\definecolor{lightGray}{RGB}{206,206,206}
\definecolor{lightYellow}{RGB}{255,238,170}
\definecolor{lightOrange}{RGB}{255,204,170}
\definecolor{lightRed}{RGB}{223,173,138}

\usepackage[normalem]{ulem}

\usepackage[ruled]{algorithm2e} 

\SetAlFnt{\small}
\SetAlCapFnt{\small}
\SetAlCapNameFnt{\small}
\SetAlCapHSkip{0pt}

\let\oldnl\nl
\newcommand{\nonl}{\renewcommand{\nl}{\let\nl\oldnl}}

\newlength\inlen
\newcommand\myinput[1]{%
  \nonl\settowidth\inlen{\KwIn{}}%
  \setlength\hangindent{\inlen}%
  \hspace*{\inlen}#1\\}

\newlength\outlen

\usepackage{mathtools}
\DeclarePairedDelimiter{\norm}{\lVert}{\rVert}

\def \bezier {B{\'e}zier}
\newcommand{\etal}{~et al{.}}

\begin{document}
\title{Robust Containment Queries over Collections of Trimmed NURBS Surfaces via Generalized Winding Numbers}

\author{Jacob Spainhour}
\orcid{0000-0001-8219-4360}
\affiliation{
  \institution{Lawrence Livermore National Laboratory}
  \city{Livermore}
  \country{USA}}
\email{jspainhour@llnl.gov}

\author{Kenneth Weiss}
\orcid{0000-0001-6649-8022}
\affiliation{%
  \institution{Lawrence Livermore National Laboratory}
  \city{Livermore}
  \country{USA}}
\email{kweiss@llnl.gov}

\renewcommand\shortauthors{J. Spainhour and K. Weiss}

\begin{abstract}
We propose a containment query that is robust to the watertightness of regions bound by trimmed NURBS surfaces, as this property is difficult to guarantee for in-the-wild CAD models.
Containment is determined through the generalized winding number (GWN), a mathematical construction that is indifferent to the arrangement of surfaces in the shape. 
Applying contemporary techniques for the 3D GWN to trimmed NURBS surfaces requires some form of geometric discretization, introducing computational inefficiency to the algorithm and even risking containment misclassifications near the surface.
In contrast, our proposed method leverages properties of the 3D solid angle to solve the relevant surface integral using a boundary formulation with rapidly converging adaptive quadrature.
Batches of queries are further accelerated by \textit{memoizing} (i.e.\ caching and reusing) quadrature node positions and tangents as they are evaluated.
We demonstrate that our GWN method is robust to complex trimming geometry in a CAD model, and is accurate up to arbitrary precision at arbitrary distances from the surface.
The derived containment query is therefore robust to model non-watertightness while respecting all curved features of the input shape.

\end{abstract}

%
%

\ccsdesc[500]{Computing methodologies~Shape analysis}
\ccsdesc[500]{Computing methodologies~Parametric curve and surface models}
%
%

\keywords{Winding number, point containment query, trimmed CAD models, robust geometry processing, trimmed NURBS}

\begin{teaserfigure}
  \includegraphics[width=\linewidth]{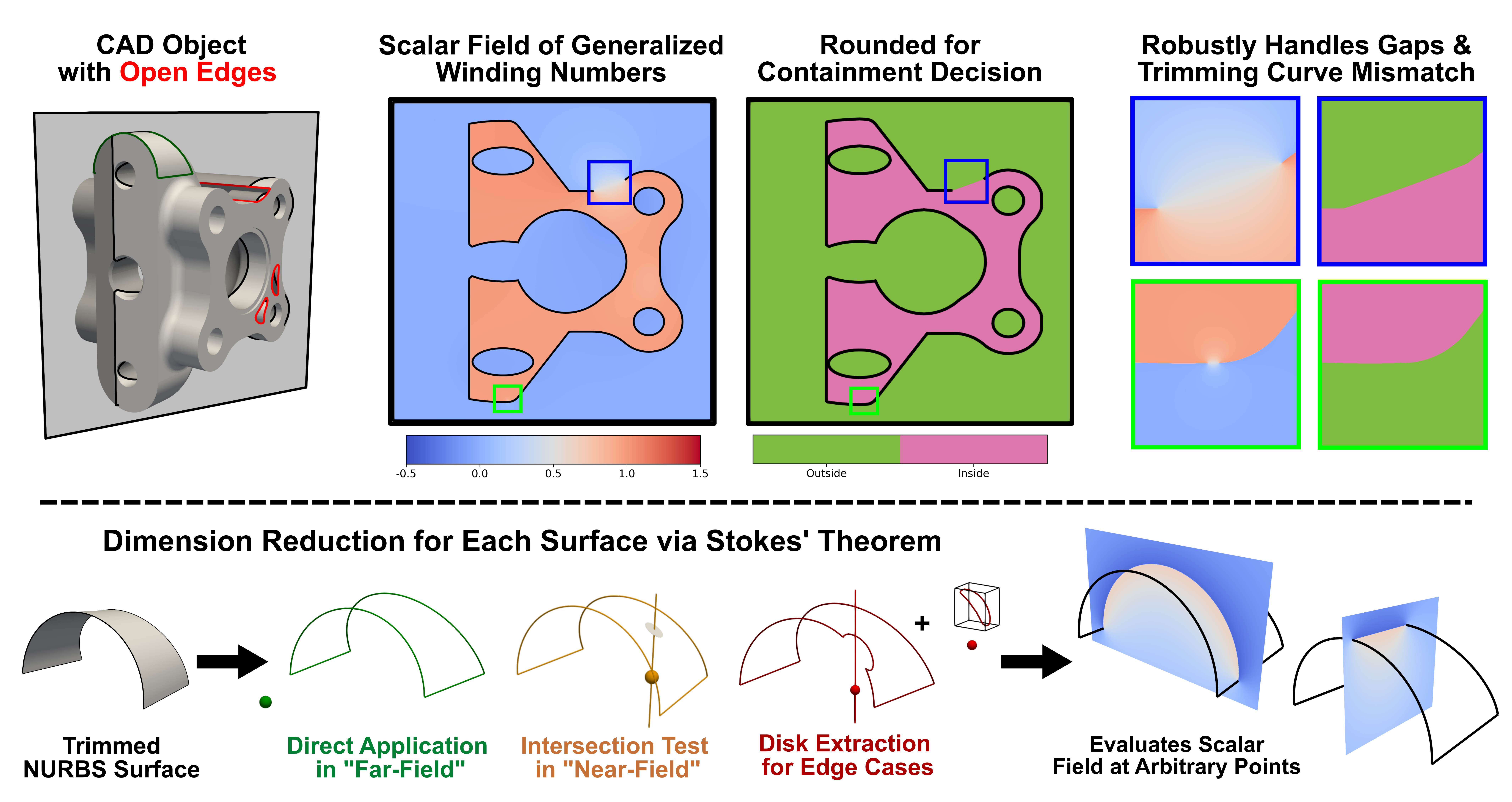}
  \caption{CAD models composed of trimmed NURBS surfaces are ubiquitous in engineering and design, but robustly defining an interior for such objects is challenging in the presence of gaps or overlaps between adjacent patches.
  We present a method to evaluate the field of \textit{generalized winding numbers} for such surfaces, providing a containment query that is indifferent to model watertightness. 
  Our algorithm uses Stokes' theorem to reformulate the problem as a 1D line integral over the boundary of each patch, solved with an adaptive quadrature technique that is accurate to user tolerance. 
  When Stokes' theorem is theoretically unfounded (i.e.\ near the surface), we perform a line-surface intersection test to recover the computational advantages of our boundary formulation.
  Edge cases near surface boundaries are handled accurately and robustly by extracting a parametric disk from the surface to be processed separately.
	}
  \Description[Graphical Abstract]{In the top half of the figure a machine part CAD shape with open edges is shown alongside a 2D slice through the middle showing the GWN field. The scalar field is rounded to the nearest integer to produce a plot of the containment decision. Two inset images are shown, one around a small open edge and one against a large open edge. In both cases, the containment decision shows a reasonable partition between inside and outside. In the bottom half of the figure the dimension-reduction properties of Stokes' theorem are shown as divided into the three cases: ``Far-Field'', ``Near-Field'', and ``Edge-Cases''.}
	\label{fig:graphical_abstract}
\end{teaserfigure}

\maketitle

\section{Introduction}\label{sec:introduction}

\begin{figure}[tb]
    \centering
    \includegraphics[width=\linewidth]{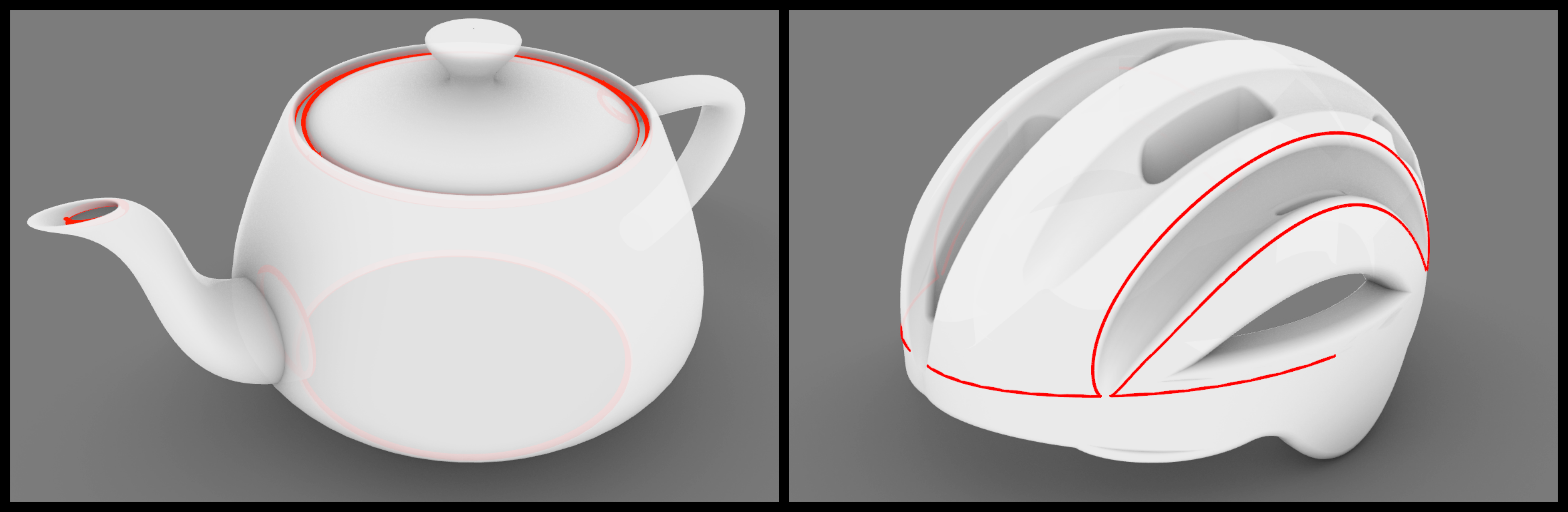}
    \caption{In the exaggerated example (left) we see open and non-manifold edges, highlighted in red, that prohibit a strict definition of the shape's interior volume. Analogous issues are present in the in-the-wild example (right) which are numerically significant, but otherwise unnoticeable to a human observer and therefore difficult to predict or correct.
	}
  \Description{Exaggerated and in-the-wild example of non-watertight CAD objects.}
	\label{fig:bad_mesh_example}
\end{figure}

In computer graphics and adjacent fields, 3D objects are often defined in terms of a boundary representation (B-Rep) of the surface.
While B-reps are very useful for flexibly designing 3D objects within a CAD system, they often fail to be \textit{watertight}, in the sense that they do not define a legitimate interior volume (see Figure~\ref{fig:bad_mesh_example}).
For example, disconnected edges between component surfaces may be an intentional design decision, but are just as likely to result from human error during modeling, or differing tolerances between applications~\cite{kasik-2005-cad-challenges,mcadams-11-elasticity} (see Figure~\ref{fig:bad_mesh_example}).
The use of Boolean operations during model construction can also introduce problematic non-manifold edges.
Issues of watertightness are greatly exacerbated in the context of trimmed NURBS (non-uniform rational B-spline) surfaces, where such gaps are nearly mathematically inevitable due to the complexities of projecting 3D surface-surface intersection curves into 2D parametric trimming curves~\cite{sederberg-08-watertightnurbs, zou-24-cadsurfacemeshfem}.
Nevertheless, we are often interested in determining whether a given point is contained within the shape in a way that conforms to the provided boundaries, but is robust enough to provide reasonable decisions in the vicinity of such geometric errors.

A potential solution is to restore model watertightness directly via surface repair, e.g.\ closing gaps and/or clipping overlaps, but manual correction is often time-consuming and error-prone.
Automated systems for surface repair in the context of CAD models with arbitrary curved and trimmed surfaces is an active area of research, and is typically applied globally during export or mesh-generation~\cite{smith-10-watertightcadrepair,marchandise-12-cadrepair,xiao-2021-nurbsrepair}.
However, the focus of such work is often the creation of watertight models composed of linear elements (e.g.\ triangles), which necessarily degrade the geometric fidelity of the model.

Nevertheless, discretization of individual trimmed surfaces into a linear mesh permits application of not only more traditional surface repair techniques, but also the growing body of work that defines containment queries for non-watertight shapes~\cite{Jacobson-13-winding,Barill-18-soupcloud, feng-24-heatsigneddistance}.
However, there are many contexts in which this type of approximation through discretization would either be too costly to capture complex geometry, or has the potential to introduce unforeseen and otherwise unacceptable errors in downstream applications.

We take multiphysics simulation as an important class of examples, and point to the work of Sevilla\etal\ for a comprehensive review of complications that arise when piecewise linear boundaries are used to approximate curved geometry in multiphysics contexts~\cite{sevilla-08-nurbsfem,sevilla-11-nurbsfem3d,sevilla-18-adaptnefem}.
This has led to the development of isogeometric analysis techniques that operate directly on the NURBS representation defined by CAD geometry~\cite{hughes-05-isogeometric, marussig-17-isogeometricreview}.

Similarly, methods for immersogeometric analysis model fluid-structure interactions by placing a complicated CAD model on a simpler computational grid which records the numerical solution~\cite{kamensky-15-immersogeometric,fromm-2023-trimmedimmersogeometric}.
In such methods, it is necessary to know whether points on this grid are inside or outside the CAD model, which can be particularly challenging for points located arbitrarily close to the B-Rep.

In this work, we seek to extend the capabilities of these and other techniques by introducing the first direct evaluation method for containment queries on non-watertight B-Reps composed of trimmed NURBS surfaces (see Figure~\ref{fig:graphical_abstract}).
We accomplish this through the \textit{generalized winding number} (GWN)~\cite{Jacobson-13-winding}, a mathematical construction that is useful not just in defining such a query, but in a variety of other geometry processing routines, including Boolean operations~\cite{trettner-22-ember} and surface reconstruction~\cite{Jacobson-13-winding}. 
Specifically, we present in this work the following principal contributions:
\begin{itemize}
  \item We extend the theory of generalized winding numbers to the context of trimmed NURBS patches.
  With this, we define a containment query for B-Reps composed of unstructured collections of such shapes that is robust to (and indeed indifferent to) watertightness.
  \item We describe a novel algorithmic framework for evaluating the GWN with respect to a trimmed NURBS surface, greatly improving efficiency and stability by reformulating the relevant surface integral as a line integral along the surface boundary.
  We evaluate these integrals with an adaptive quadrature scheme to achieve results up to arbitrary accuracy, and further improve performance with a \textit{memoization} strategy that caches and reuses geometric calculations for each surface. 
  Within our algorithm, we robustly handle \textit{trim tests} (visibility queries over trimming curves in the 2D parameter space of each surface) with the 2D GWN.
  \item We accurately evaluate the 3D GWN field for points arbitrarily close and even coincident with the surface by directly identifying and compensating for the jump discontinuity in the GWN field across surfaces.
  This permits evaluation of the GWN both in the ``far-field'' and ``near-field'' of the NURBS surface without any discretization.
\end{itemize}

Practical applications of the GWN typically use a spatial index to accelerate queries (e.g.\ in the works of~\citet{Jacobson-13-winding} and~\citet{Barill-18-soupcloud}), which we intend to pursue in follow-up work.
Instead, our focus in this work is on exact and efficient evaluation of the GWN with respect to individual trimmed NURBS surface patches, the cost of which scales nearly linearly with the number of trimming curves.
Our algorithm has been implemented in Axom, a BSD-licensed open-source library of Computer Science infrastructure for HPC applications~\cite{Axom_CS_infrastructure}.

\section{Background and Related Work}\label{sec:background}
\subsection{Containment Queries in 3D CAD Applications}\label{chapter_winding_3d:sec:containment_queries}
There are many strategies to evaluate containment queries for CAD applications.
In the typical ray casting method, an arbitrary ray is extended out from the point of interest and the number of intersections with bounding geometry are recorded. 
Assuming a watertight B-Rep, each such intersection increments a crossing number whose parity determines containment.

There is considerable overlap with the adjacent task of ray tracing for the purposes of rendering, which typically considers a specific, pre-selected ray direction, and only requires calculation of the first intersection of a ray with the B-Rep.
Much of the research focus in ray tracing relates to triangulated surfaces, as many hardware platforms are capable of calculating ray-triangle intersections directly, exactly, and massively in parallel~\cite[Chapter 3]{haines-19-raytracinggems}.
However, there is also interest in applying such techniques directly to trimmed NURBS surfaces~\cite{schollmeyer-09-trimming,schollmeyer-19-renderingnurbs, xiong-23-eter, zhu-2025-projectionbsp}.
Of particular note are recent methods for ray tracing bilinear patches, as, similar to triangles, they admit direct geometric formulae for computing their intersections with rays~\cite{reshetov-19-coolpatches}.

On more general curved surfaces, surface-line intersection routines are more complex and iterative~\cite{nishita-90-raytracing, martin-00-newtonraytracing,efremov-05-robustbezierclipping, shen-16-matrixraytracing,marschner-21-sosraytracing}.
Although the direction of the cast ray is arbitrary in principle, containment strategies for watertight shapes which utilize ray casting are highly sensitive to the specific configuration of surface and cast ray.
This issue is compounded in the non-watertight case, as a cast ray passing through a numerical gap between surfaces results will result in an unexpectedly incorrect containment result.

\begin{figure}
    \includegraphics[width=\linewidth]{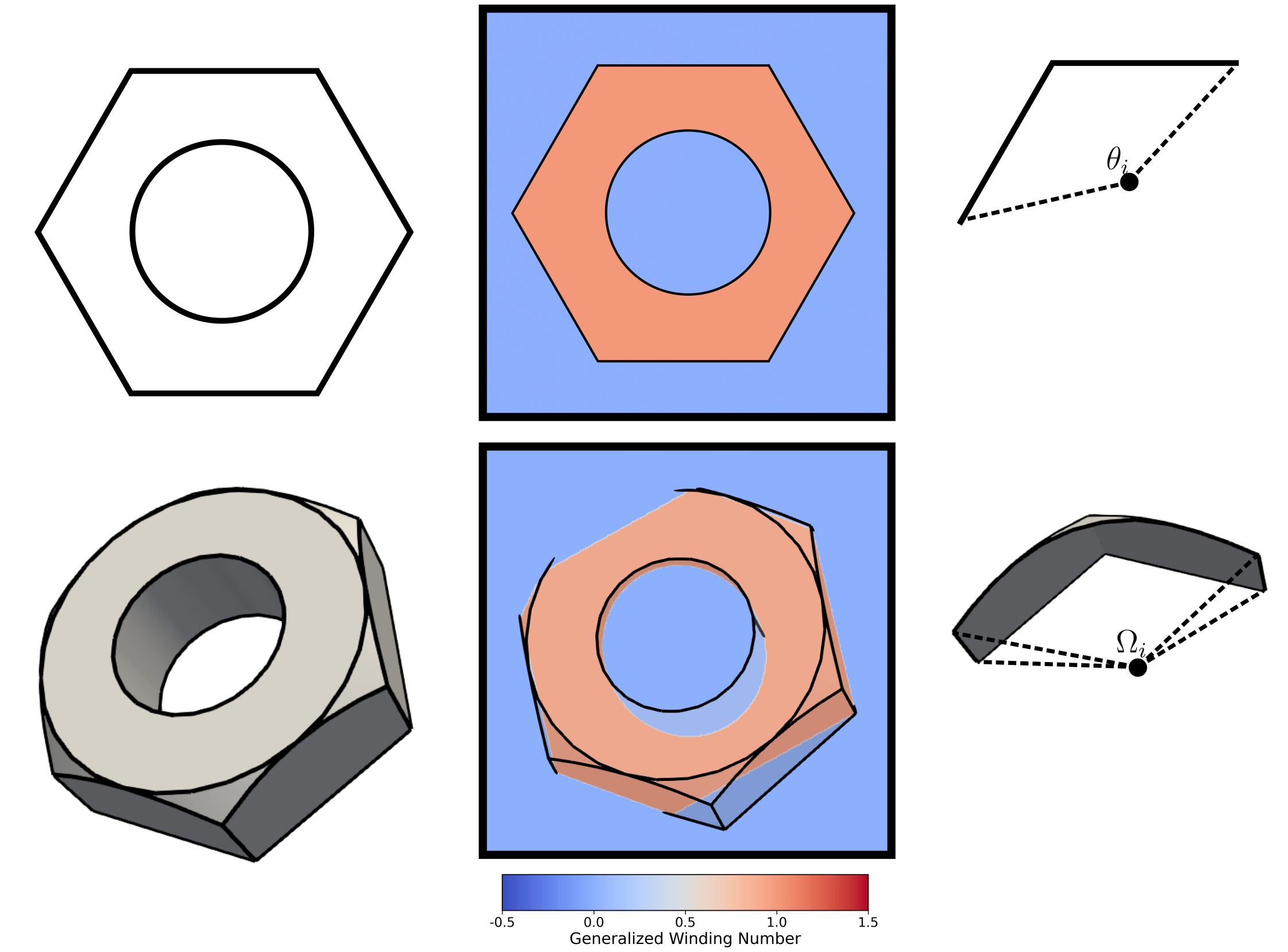}
    \caption{In 2D (top) and 3D (bottom), the GWN of watertight shapes (left) is integer valued (middle), 
			while the fractional GWN of a non-watertight shape (right) is equal to the signed angle subtended by its boundary. 
			To more easily visualize the 3D GWN as a scalar field, we consider a 2D slice of the field as it intersects with the surface.}
\Description[Simple comparison of 2D and 3D GWN]{The GWN field is shown for two representations of a mechanical nut shape. On the top is a 2D representation, the 2D GWN in a bounding box of the shape, and an image depicting the GWN as a subtended angle. On the bottom is a 3D representation, a 2D slice of the 3D GWN field, and an image depicting the GWN as a subtended solid angle.}
			\label{fig:simple_gwn_example}
\end{figure}

Alternatively, winding number methods count the number of times the shape encloses the point, thereby implicitly handling configurations which are difficult for ray casting, even for watertight shapes~\cite{Carvalho-95-pointinpolyhedron,hormann-2001-pointinpolygon}.
These methods can then be \textit{generalized} by summing the signed angle subtended by individual boundary components so that the resulting containment decision is indifferent to watertightness (see Figure~\ref{fig:simple_gwn_example}).

The generalized winding number (GWN) is a real-valued scalar field that degrades smoothly, yet rapidly away from integer values near gaps and/or overlaps between boundary components.
This field can be interpreted as defining containment, for example, by rounding its values to the nearest integer and applying either a non-zero or even-odd rule.
Although other uses of the GWN field are possible, we are primarily concerned with its evaluation for the purposes of defining this robust containment query.

\subsection{GWN Evaluation for Discretized Surfaces}\label{sec:generalized_winding_numbers}

The generalized winding number was first conceptualized to provide containment queries on messy STL geometry, doing so via a hierarchical spatial index and divide-and-conquer algorithm~\cite{Jacobson-13-winding}.
This work was later extended to entirely disconnected triangle soups by~\citet{Barill-18-soupcloud}, improving computational performance with an error-controlled approximation of the GWN with respect to groups of triangles that are far from the query point.

Evaluation methods for the GWN field defined by triangle meshes and soups are particularly computationally efficient, as the solid angle which defines the 3D GWN can be directly and exactly computed for polygonal shapes.
However, if the input surfaces are curved, a na{\"i}ve application of these methods must necessarily use some geometric approximation of the surface to obtain the necessary input type, which in turn causes the GWN field of the approximation to diverge from the ground truth.
At the same time, even generating a tessellation of a trimmed NURBS surface is a non-trivial task, and high-resolution meshes can be prohibitively expensive to generate and store for complex models~\cite{xiong-23-eter} 

The work of~\citet{Barill-18-soupcloud} defined the GWN of oriented point cloud data, which could be sampled directly from a CAD-derived surface.
Indeed, \citet{balu-23-immersogeometric} do so in the context of multiphysics simulation, describing a method which uses a point cloud randomly sampled from a CAD model to define an interior volume for the otherwise messy or complex shape.
However, even if the weights for each point are accurately computed from the ground-truth surface (which may not be the case depending on implementation), such a discretization results in significant errors in the evaluated GWN field, particularly for points near the surface. 

In comparing our objectives to these existing methods (or na{\"i}ve adaptations of existing methods to curved geometry), we find it useful to distinguish between two types of errors in the computed GWN field.
The first is \textit{inaccuracy} in the GWN calculation, where the error in the GWN field causes the derived containment query to result in misclassifications.
This is almost exclusively a concern for points near the surface, particularly when the GWN is computed using a low-quality intermediate discretization of the surface.
In such cases, the query point can be identified as on one side of the discretized surface while being on the opposite side of the true surface, resulting in an off-by-one error in the GWN field.

\begin{figure}
    \includegraphics[width=\linewidth]{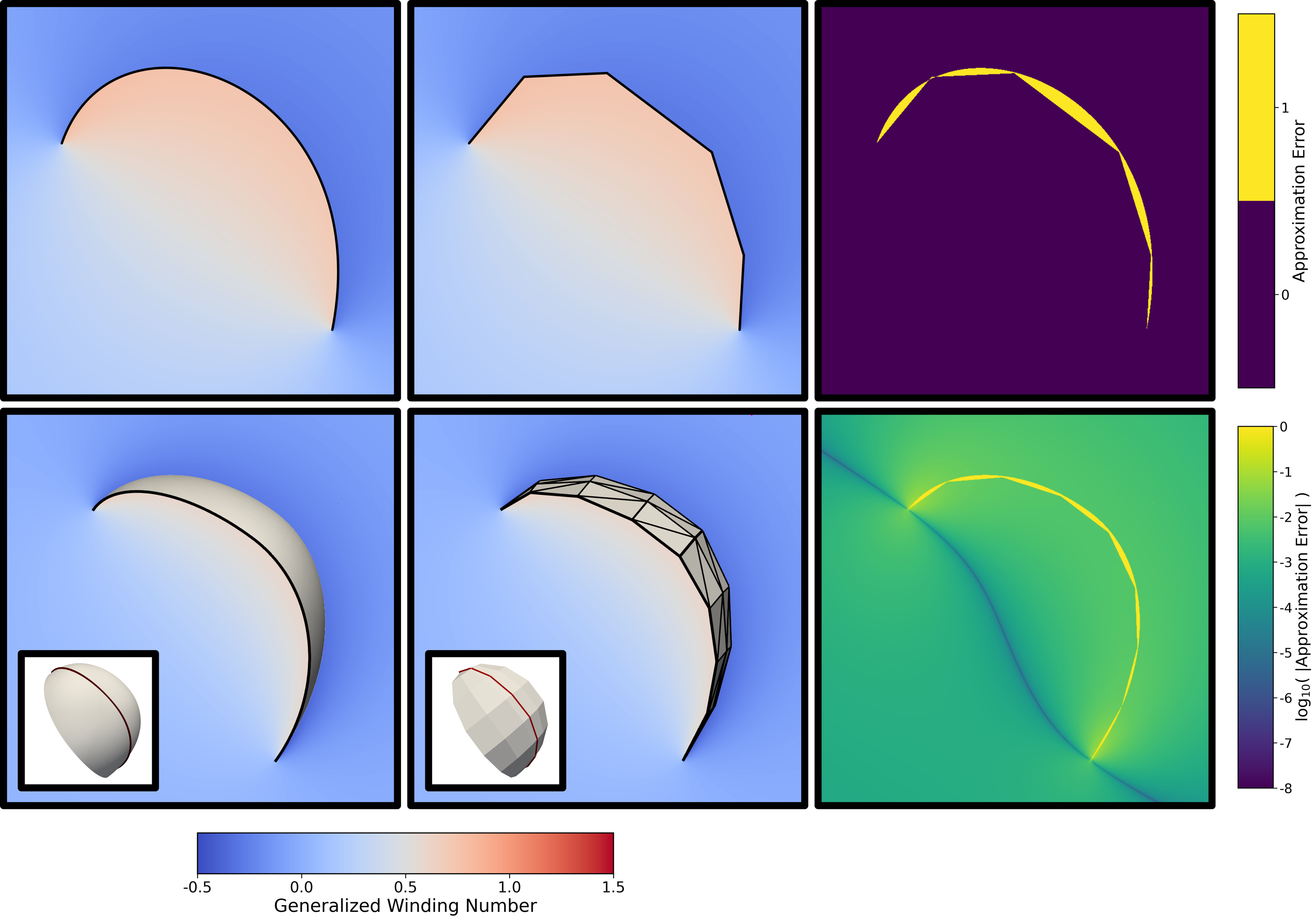}
    \caption{Comparing the difference (right, log scale) between the GWN field of the original curved geometry (left) and its linearly discretized approximation (middle).
    While the induced approximation error in 2D (top) is integer-valued, the analogous distribution of errors in 3D (bottom, displayed on a 2D slice of the 3D surface shown in the inset) is non-zero almost everywhere.
    }
    \Description[Comparison between the GWN of a curved/discretized object in 2D/3D]{On the top is shown the GWN field for a 2D curve, a polyline which approximates the curve, and a plot of the absolute error between them. It is shown that the error is either 0 or 1 everywhere. On the bottom is the GWN for a 3D surface, a triangulation which approximates the surface, and a plot of the absolute error between them. It is shown that the error is nonzero everywhere.}
	\label{fig:discretization_comparison}
\end{figure}

The second is numerical \textit{imprecision} in the computed fractional component of the GWN field.
While it is necessarily true in 2D that, for a fixed set of input queries, there exists some linear discretization of the input curve that produces the correct GWN field at each query, this is no longer the case in 3D.
Even if a query point is correctly placed on the proper side of the surface, the mismatch between the boundaries of the original surface and \textit{any} triangulation introduces imprecision in the computed field that is difficult to predict from the triangulation alone (see~Figure~\ref{fig:discretization_comparison}).
While these errors decrease with increasing mesh resolution, they never vanish completely.

As we shall see, it is common for 3D GWN evaluation methods to control for each type of error independently.
In general, we find that methods which discretize the surface geometry for the purposes of evaluating the GWN field precisely for far-away query points often do so at the cost of accuracy nearer to the surface.
Instead, our work introduces a GWN method that is accurate at arbitrary points relative to the input surface, and maintains precision through error-controlled numerical methods, all in service of avoiding misclassifications in the derived containment query. 
\subsection{GWN Evaluation for General Surfaces}\label{sec:gwn_for_curved_geometry}

In this section we describe contemporary methods for evaluating the GWN field of curved surfaces in 3D, and more broadly explain the fundamental differences from the 2D analog.

The 2D GWN $w_C(q)$ at a query point $q$ with respect to a curve $C$ is defined as the total signed angle subtended by the curve, translated to place $q$ at the origin, and normalized by a factor of $2\pi$.
This results in the following integral of the (signed) differential angle~$d\theta$,
\begin{align}\label{eqn:winding_number_angle}
    w_C(q) := \frac{1}{2\pi}\int_{C - q}d\theta,
\end{align}
where $C-q$ denotes translating the curve to make $q$ the origin.

Evaluating this line integral directly via numerical quadrature is difficult, as the near-singular behavior of the integrand means that for points close to the curve, conventional quadrature methods are very unstable.
For linear elements with (translated) endpoints $a$ and $b$, however, this subtended angle has a simple and direct formula,
 \begin{align}
    w_{L - q}(0) := \frac{1}{2\pi}\text{arctan} \left( \frac{\norm{a\times b}}{a\cdot b} \right).
 \end{align}
This, along with the property that closed curves have integer-valued winding numbers, has been leveraged to compute the GWN of an \textit{arbitrary} curve in 2D without the use of any quadrature scheme by subtracting the GWN of the curve's closure from that of the closed shape~\cite{spainhour_24_robustcontainment2d}.
Indeed, for far away points, this continues to be the most efficient way of processing any individual curve for any single query point~\cite{liu-2025-closedform-wn}.

Theoretically, this framework can be applied directly to the 3D problem.
The 3D GWN $w_S(q)$ at query point $q$ with respect to a surface $S$ is defined as the total solid angle subtended by the surface, again translated by $q$, and normalized by a factor of $4\pi$:
\begin{align}\label{eqn:winding_number_solid_angle}
    w_S(q) := \frac{1}{4\pi}\int_{S - q}d\Omega,
\end{align}
which can be written in Cartesian coordinates as: 
\begin{align}\label{eqn:cartesian_surface}
    w_S(q) &= \frac{1}{4\pi}\iint_{S - q} \frac{x\cdot \hat n}{\norm{x}^3}\,dS.
\end{align}

There exist a number of quadrature techniques in the literature that could be immediately applied to Equation~\ref{eqn:cartesian_surface}, many of which can in principle be used even in the presence of complicated trimming geometry~\cite{gunderman-20-cad, chin-21-cubature, gunderman-21-trimmednurbsintegration}.
However, the integrand's fundamental near-singularity makes direct numerical evaluation difficult for query points near the surface. 

The hope, then, is that this too can be avoided through an exact formula for the analogous linear element.
Indeed, the GWN for a 3D triangle defined by (translated) vertices $a$, $b$, and $c$ is given by
\begin{align}\label{eqn:winding_number_triangle}
    w_{\triangle - q}(0) := \frac{1}{2\pi}\text{arctan} \left( \frac{\scriptstyle \norm{a\cdot (b \times c)}}{\scriptstyle \norm{a}\norm{b}\norm{c} + (a\cdot b)\norm{c} + (b\cdot c)\norm{a} + (c\cdot a)\norm{b}} \right).
\end{align}

But while any 2D curve can be closed with a straight line, only a trivial subset of 3D surfaces can be exactly closed by a collection of triangles.
For more general 3D surfaces, the closing surface must also be a high-order surface~\cite{sederberg-08-watertightnurbs}, and so computing the GWN of the closing surface (for the purposes of subtracting the value from an integer winding number of the closed surface) is typically just as difficult as computing the GWN of the original.

To find a more direct and consistent solution method, we utilize a known property of the 3D solid angle which states that in certain contexts, it and the derived GWN field can be defined entirely from the boundary of the surface~\cite{Jacobson-13-winding, binysh-2018-maxwell, gillespie-2024-raytracingharmonic}.
In this spirit, one could imagine adapting the method of~\citet{Jacobson-13-winding} to a 3D curved surface $S$ by constructing a linear approximation of the surface boundary $\partial S'$ (with $\partial S'~\approx~\partial S$), along with a coarse triangulated closure $\overline{S'}$ whose GWN field can be computed exactly.
For query points far from the original surface, this approach is somewhat reasonable---although some precision is lost in the discretization of $\partial S'$, the GWN field of the approximated closure is related to the GWN field of the original surface for far-away points through the simple formula $w_S~=~-w_{\overline{S}}~\approx~-w_{\overline{S'}}$.

However, the simplicity and practicality of this boundary-focused approach necessarily breaks down for near-surface points.
To reiterate, deriving $w_S$ from $w_{\overline{S}}$ in general requires calculating $w_{S\cup\overline{S}}$, i.e.\ the integer winding number of the query within the closed surface, which requires consideration of the original, undiscretized surface to preserve accuracy close to the surface.
In this case, using $\overline{S'}$ as a computational proxy for $\overline{S}$ complicates things further, as $w_{S\cup\overline{S'}}$ is no longer integer valued, since, by construction, $\overline{S}$ and $\overline{S'}$ do not share the same boundary.

In this way, accurate calculation of the 3D GWN requires consideration of the \textit{entire} undiscretized surface, including its boundary.
To our knowledge, only the method of~\citet{sawhney-20-montecarlogeometry} handles curved input directly without discretization, solving certain classes of PDEs---including that which implicitly defines the GWN---by means of random walks and closest-point projections.
In the vernacular of the previous section, this method is accurate near the surface interior by respecting the geometric fidelity of curved surfaces, but risks a noisy and inconsistent containment query near surface boundaries, and loses precision for individual queries due to its stochastic nature.

Overall, our proposed GWN method compensates for the complexities of 3D space by replacing all forms of discretization error, as visualized in Figure~\ref{fig:discretization_comparison}, with a rapidly converging adaptive quadrature method whose error can be tightly controlled.

\subsection{GWN Evaluation for Trimmed NURBS Surfaces}

An advantage of working with trimmed NURBS is that they provide direct access to the boundary curves of each surface as a collection of 2D trimming curves defined in the surface's parameter space.
Each trimmed NURBS surface within a model can have several complicated and disconnected boundary components,
and it is critical that we assume no structure between the component surfaces in the input B-Rep, as this is what makes any GWN algorithm robust to the shape's watertightness.  

To explore the implications of this problem context, we consider the recent ``One-Shot'' approach of~\citet{martens-2025-oneshot} which, like the idealized adaptation of~\citet{Jacobson-13-winding} to curved surfaces in the previous section, evaluates the GWN of parametric surfaces with open edges by discretizing \textit{only} the open edges within connected components of the shape. 
These polyline edges are projected onto a unit sphere around the query point, partitioning it into a graph of spherical polygons. 
After mapping this collection to integer crossing numbers determined through a single ray-surface intersection test, the total GWN is evaluated as the sum of each integer, weighted by the approximate surface area of the associated polygon.
The quality of the boundary discretization naturally influences the precision of the area calculation, but also the accuracy of the subsequent containment decision.

While the suggested ray-surface intersection routine does take into consideration the original (undiscretized) surface, the \textit{mapping} of crossing numbers to spherical polygons may still introduce off-by-one errors depending on the quality of the boundary discretization. 
Although careful selection of the ray direction may reduce the likelihood of misclassification, this risk represents a fundamental limitation of using unconsidered discretizations of the input shape.

Although the One-Shot approach is aimed at computing the GWN for parametric surfaces, we find its high-level approach unsuitable for our problem context.
For example, a stated limitation of the One-Shot method is that it is most effective for surfaces with simple boundaries, as is typical for the preprocessed curve networks used as examples in that work.
On the other hand, a trimmed NURBS surface can have an arbitrarily complicated boundary, and so partitioning the unit sphere projection of these boundaries becomes combinatorially intractable in practical contexts.
Importantly, while the cost of the line-surface intersection procedure is reduced for query points which are far from the surface (and could theoretically be eliminated with minor algorithmic adjustments), the cost of this partitioning procedure remains relatively constant across space, and even demands a higher-resolution boundary discretization as query points become further away and partitioned regions become smaller.
Even so, as we demonstrate in Section~\ref{sec:results_integration}, our method is more performant even when applied to simple shapes.

Additionally, working with large collections of input surfaces means that the vast majority of query points will be far from all but a few surfaces.
As we will see, our approach efficiently handles this scenario by operating directly on the provided boundary curves without the need for any auxiliary data structure or intersection routine.
Although both the proposed algorithm and the One-Shot method must be run on each component surface individually for a given query point, our approach effectively amortizes the increased cost of handling the few nearby surfaces.

\section{Method}\label{sec:methods}
In this work, we present a generalized winding number algorithm that can be accurately evaluated at arbitrary points in space relative to the CAD model, with precision up to an arbitrary user tolerance.
We assume that each component surface is an oriented trimmed tensor-product NURBS patch, with trimming curves that are explicitly defined in the parameter space of the knot spans of the patches by closed loops of 2D NURBS curves.
This is the default representation for trimming curves in the STEP and IGES interchange formats and a common assumption in the literature~\cite{piegl-1997-nurbs,sheng-1992-triangulation, hamann-1996-tessellation,piegl-1995-tessellating,sloup-2021-optimizing,krishnamurthy-2009-optimized}.
The orientation of the patch is determined by the direction of each parameter-space knot span.
Since we use the 2D GWN method of~\citet{spainhour_24_robustcontainment2d} to robustly determine containment in the parameter space of each patch, we treat the collection of trimming curves as unstructured, significantly simplifying their processing.
For each query point, we evaluate our algorithm for each component surface in the model and sum the results, making the algorithm indifferent to watertightness by the linearity of the GWN.

\subsection{Algorithm Overview}\label{sec:methods_overview}
Our algorithm is based on a direct application of Stokes' theorem, which reduces the dimension of Equation~\ref{eqn:cartesian_surface} to an integral over the boundary of each surface.
For notational simplicity, we assume that the surface $S$ is translated so that the query point $q$ is at the origin.

For each translated surface $S$ of the model, we evaluate the GWN at query point $q = 0$ by evaluating
\begin{align}
    w_S &= \frac{1}{4\pi}\oint_{\partial S}\left\langle \frac{yz}{(x^2+y^2)\,\norm{\vec{x}}}, \frac{-xz}{(x^2+y^2)\,\norm{\vec{x}}}, 0\right\rangle\cdot d\vec{\Gamma}\label{eqn:adjusted_stokes}\\
        &+ \sum_{\left\{ \substack{\text{intersections of}\\S\text{ and }z\text{-axis}} \right\} }
						 \begin{cases}
							   0.5  & \parbox[c]{.5\columnwidth}{if $q$ is on the \textit{positive} side of $S$,} \\
							   0.0  & \parbox[c]{.5\columnwidth}{if $q$ is \textit{on} $S$,} \\
							  -0.5  & \parbox[c]{.5\columnwidth}{if $q$ is on the \textit{negative} side of $S$,} 
						 \end{cases}
						\nonumber
\end{align}
where $q$ is on the positive side of $S$ if for intersection $(0, 0, z_0)$, $z_0 > 0$.

The first term is the Stokes'-derived boundary formulation, which we evaluate using an error-controlled adaptive Gaussian quadrature scheme.
This evaluation is accelerated across batches of queries through a memoization strategy which caches and reuses geometric data for each surface, such as surface points and tangent vectors at quadrature nodes (detailed in Section~\ref{sec:adaptive_quadrature}).
The second term is a correction term that accounts for any jump discontinuities in the GWN field over the surface.

The specific evaluation strategy used for a given point, as well as the associated computational costs, varies according to its relative position to the surface $S$ and its boundary edges $\partial S$, as shown in Figure~\ref{fig:graphical_abstract}.
These cases can be categorized as:
\begin{itemize}
    \item \textbf{Far-field}: For points sufficiently far from the surface, i.e.\ outside a bounding box encompassing the surface, the number of intersections is known to be zero.
	As such, we evaluate the GWN directly with our unadjusted boundary formulation. 
    \item \textbf{Near-field}: For points closer to the surface, we use the same boundary formulation, but must also perform a line-surface intersection test to identify an appropriate correction term.
    \item \textbf{Edge-cases}: Some points, such as those near a boundary edge of $S$ require special treatment. In these cases, we split $S$ along new trimming curves to extract the problematic portion and reprocess the resulting surfaces.
\end{itemize}

The specific form of Equation~\ref{eqn:adjusted_stokes} assumes that the correction term is determined by the $z$-axis. 
However, the intersection test can be performed with an \textit{arbitrary} oriented line by subsequently rotating the surface so that the positive direction of the line is parallel to the positive $z-$axis.
As intersections with lines that are near-tangent to the surface are the most difficult to identify and resolve, we heuristically select a line with a direction equal to an average surface normal, which we compute for each surface as preprocessing.
In cases where the symmetry of the surface results in an average normal with near-zero magnitude, we instead select the line randomly.

We present the full algorithm to compute the GWN of an arbitrary surface in Algorithm~\ref{alg:generalized_winding_number}.
In the following subsections, we justify our method mathematically by providing an overview of Stokes' theorem, and then describe how its application to this problem naturally delineates each of the above cases.
We then discuss our strategy for evaluating the GWN field at points which are coincident to the surface as well as our accelerated adaptive quadrature.

\subsection{Reformulation with Stokes' Theorem}\label{sec:stokes_method}

The connection between the 3D solid angle and an associated line integral formulation has been independently explored in several application contexts, including physical optics~\cite{asvestas-1985-solidangle}, soft matter physics and electrostatics~\cite{binysh-2018-maxwell}, and perhaps most directly in differential geometry~\cite{ranikci-2017-solidangle}.
For example, there are many contexts in which the solution to specific line integrals over closed 3D contours can be more efficiently evaluated by means of computing the (often approximated) solid angle of some interior surface~\cite{binysh-2018-maxwell}.
Common among these works is either the selection or presupposition of a particular line $L$ which specifies the particular form of the boundary formulation, as is the case in the presented work.
Indeed, the work of~\citet{asvestas-1985-solidangle} presents a solution strategy which most closely mirrors our own, where signed intersections between the line $L$ and a planar surface are distinguished according to the location of the intersection, with the numerical result further adjusted through an analytically derived correction term. 

Directly applying such methods to the present graphics context is restricted by the fact that the targeted geometry is more complex and cannot be \textit{a priori} discretized (see Section~\ref{sec:generalized_winding_numbers}).
However, we have the advantage of being able to select the line $L$ which is most computationally advantageous, and further manipulate the boundary curves of the surface to ensure our algorithm's robustness.  
As such, we present our own derivation of these principles in differential geometry from the ground up, as doing so leads to our proposed computational treatment of general trimmed NURBS surfaces. 

In its most general form, Stokes' theorem states that if a vector field $F(x)$ is defined with continuous first derivatives on a region in $\mathbb{R}^3$ containing the surface $S$, then 
\begin{align}\label{eqn:stokes_theorem}
    \int_S \left(\nabla \times F\right) \cdot d\vec{S} = \oint_{\partial S} F\cdot d\vec{\Gamma},
\end{align}
where $\partial S$ is the total boundary of the surface, which may contain multiple disconnected curves.

To apply Stokes' theorem to the integral in Equation~\ref{eqn:cartesian_surface}, we take 
\begin{align}\label{eqn:surface_integrand}
    \nabla \times F = \frac{1}{4\pi}\cdot\frac{\vec{x}}{\norm{\vec{x}}^3}.
\end{align}
Recall that we have made the assumption that the query point $q$ is located at the origin, and so this integrand is continuous on the surface whenever $\vec{x} \neq 0$, i.e.\ the query point does not coincide with the surface (we treat this case explicitly in Section~\ref{sec:coincident_winding_number}).

There are many distinct vector fields $F$ that satisfy Equation~\ref{eqn:surface_integrand}, and so one could evaluate the GWN for an arbitrary surface with the following 1D line integral, 
\begin{align}\label{eqn:stokes_theorem_equation}
    w_S &= \frac{1}{4\pi}\oint_{\partial S}\left\langle \frac{yz}{(x^2+y^2)\,\norm{\vec{x}}}, \frac{-xz}{(x^2+y^2)\,\norm{\vec{x}}}, 0\right\rangle\cdot d\vec{\Gamma},
\end{align}
assuming that the theoretical conditions of Stokes' theorem are met for the surface $S$.

This reformulation of the problem statement comes with a number of computational advantages.
First, evaluating such an integral via numerical quadrature along the lower-dimensional space naturally improves the efficiency of the total method.
This is a well-established advantage of Stokes' theorem, and is the basis of a method proposed by \citet{gunderman-21-trimmednurbsintegration} for evaluating arbitrary integrals on trimmed NURBS surfaces.
In this problem context, we can evaluate the ``antiderivative'' $F$ exactly for the relevant $\nabla \times F$, further reducing the number of quadrature nodes in a scheme of fixed order $n$ from $O(n^2)$ to $O(n)$.

Second, such a boundary reformulation fundamentally improves the stability of the ensuing quadrature scheme, as query points which are near to the surface but \textit{not} near the boundary are not subject to the same near-singular behavior in the corresponding integrand.
The resulting line integrals are additionally more amenable to a geometrically adaptive quadrature method, the computational burden of which is dramatically reduced by considering subdivision of 3D space curves instead of entire trimmed surfaces.

However, not every surface can be treated directly by the formulation in Equation~\ref{eqn:stokes_theorem_equation}.
To meet the conditions of Stokes' theorem, both the original vector field and its antiderivative must be continuous---not just on its domain of integration (the boundary of the surface) but also on the \textit{original} domain of integration (the surface itself).
Notably, the relevant vector field $F$ is discontinuous whenever $x^2 + y^2 = 0$ on the surface, i.e.\ whenever the surface intersects a line extending in both $z$ directions from the query point.
In such cases, we cannot mathematically justify the direct use of Stokes' theorem, even if the resulting line integral is itself continuous.

We note that the specific antiderivative in Equation~\ref{eqn:stokes_theorem_equation}, characterized by a line of discontinuities on the $z$-axis, was chosen arbitrarily.
Indeed, one can analogously define a \textit{family} of antiderivatives $F$, each characterized by a \textit{different} line of discontinuities.
\begin{figure}[tb]
    \centering
    \includegraphics[width=\linewidth]{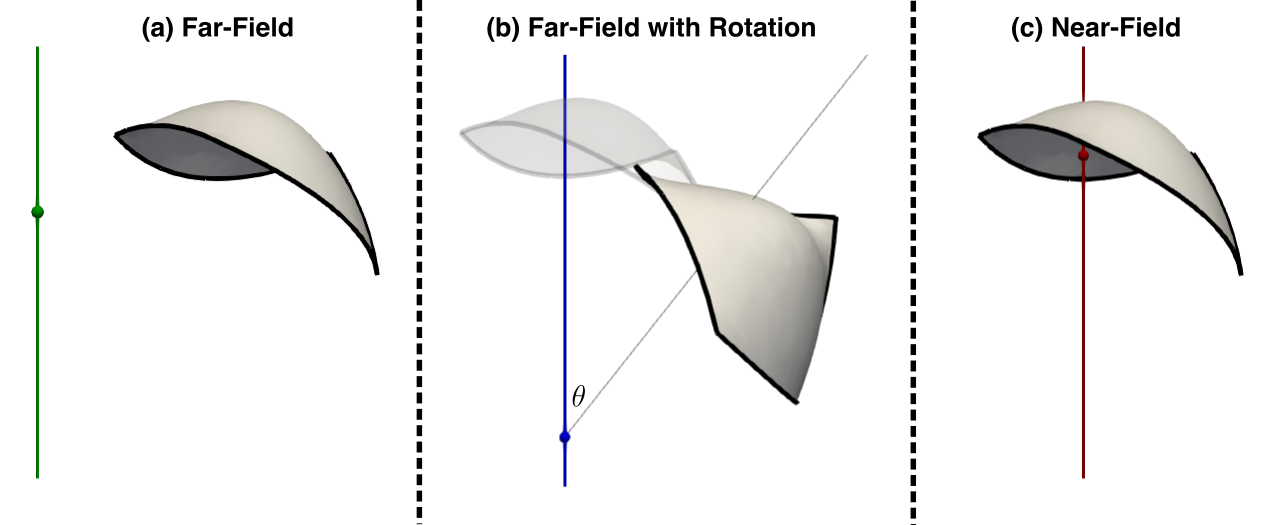}
    \caption[Example configurations for surface, query point, and line of discontinuity.]{
			Far- and near-field points are classified according to whether the line of discontinuities for antiderivative $F$
			does not intersect (a), 
			can be rotated to avoid intersecting (b), or 
			unavoidably intersects (c) the surface.}
            \Description{Example configurations for surface, query point, and line of discontinuity.}
		\label{fig:rotation_cases}
\end{figure}

In Figure~\ref{fig:rotation_cases}, we see examples of different configurations of the surface and the line.
As shown in Figure~\ref{fig:rotation_cases} (b), it is practical to view different choices for $F$ not as explicit vector fields, but as different rotations of the NURBS patch via its control points so that the desired line of discontinuities is aligned to the $z$-axis.
In this way, applying Stokes' theorem requires choosing a line which does not intersect the surface, and rotating the surface accordingly.
In practice, one cannot always identify such a line for query points which are nearer to the surface, nor does one necessarily exist (see Figure~\ref{fig:rotation_cases} (c)).

The remainder of this subsection details our three cases.
\subsubsection{\textbf{Far-field}: Direct Application of Stokes' Theorem}\label{sec:direct_stokes}

If an axis-aligned bounding box of the shifted surface does not contain the origin (i.e.\ the query point is exterior to the bounding box of the original surface), then it must be the case that at least one coordinate axis does not intersect the surface, trivially providing a line $L$ to apply in our Stokes' theorem boundary formulation.
Indeed, if the $z$-axis does not intersect the bounding box, then we can directly apply the formulation in Equation~\ref{eqn:stokes_theorem_equation} to evaluate the GWN.
In the event that surface bounding box does not contain the origin, but \textit{does} intersect the $z$-axis, then we can instead consider a form of the integrand that corresponds to the $x$- or $y$-axis, one of which necessarily does not intersect the surface.
In these cases, we use the corresponding analogs of antiderivative $F$ within Equation~\ref{eqn:stokes_theorem_equation}:
\begin{align}
    w_{S, x\text{-axis}} &= \frac{1}{4\pi}\oint_{\partial S}\left\langle 0, \frac{zx}{(y^2+z^2)\,\norm{\vec{x}}}, \frac{-yx}{(y^2+z^2)\,\norm{\vec{x}}}\right\rangle\cdot d\vec{\Gamma}
\end{align}
and
\begin{align}
    w_{S, y\text{-axis}} &= \frac{1}{4\pi}\oint_{\partial S}\left\langle \frac{xy}{(x^2+z^2)\,\norm{\vec{x}}}, 0, \frac{-zy}{(x^2+z^2)\,\norm{\vec{x}}}\right\rangle\cdot d\vec{\Gamma}.
\end{align}

While our implementation first checks an axis-aligned bounding box to identify far-field scenarios, other convex shapes containing the shifted surface can also be used to guarantee a non-intersecting line $L$, again followed by the appropriate rotation of the surface.
For example, our implementation also checks a precomputed oriented bounding box for each NURBS surface, allowing more points to be classified as far-field.
In any case, these are the most commonly evaluated forms of the integral, as an arbitrary query point will be considered far from all but a few surfaces in a typical CAD model.

\begin{figure}[tb]
    \centering
    \includegraphics[width=0.8\linewidth]{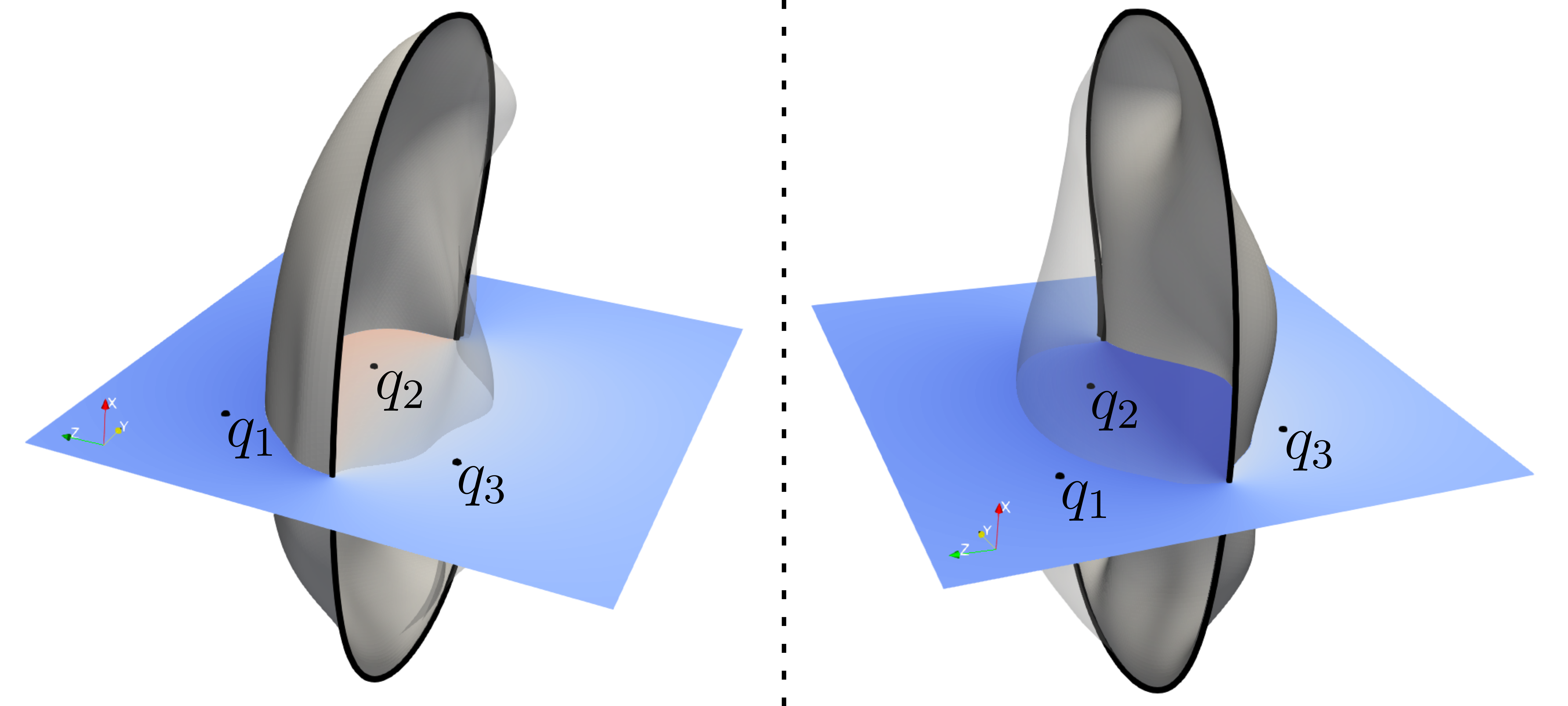}
        \begin{tabular}{cccc}
        Query & Left  & Right &  \\ 
        Point & Shape GWN & Shape GWN & Difference \\ \cmidrule{1-4}
        $q_1$ & -0.20139 & -0.20139 & 0.0\\
        $q_2$ &  0.62912 & -0.37088 & 1.0\\
        $q_3$ &  0.27805 &  0.27805 & 0.0\\
    \end{tabular}
    \caption[Intuition for 3D GWN algorithm.]{We illustrate a property of the GWN for two surfaces which share a boundary, viewed from different angles. In general, any two surfaces which share a boundary generate the same GWN field, modulo some integer which depends on the orientation of the query point relative to the surface.}
		\label{fig:intuition_3d_extra}
    \Description[Illustrative example of 3D GWN algorithm]{The GWN for 2 surfaces which share a boundary is shown on a 2D slice. The two surfaces generate an identical GWN field except in the region between them.}
\end{figure}

The validity of this far-field formulation can be made more intuitive by making concrete a property of the GWN introduced in Section~\ref{sec:gwn_for_curved_geometry} -- namely, that the GWN field of a surface can, for far away points, be defined entirely by its boundary.
More specifically, any two surfaces $S$ and $S'$ which share a boundary $\partial S = \partial S'$ generate an identical GWN field at all points outside a volume implied by $S\cup S'$.
Furthermore, if one were to continuously deform $S$ into $S'$, then the value of the GWN at a fixed query point $q$ inside this volume would remain constant up until the surface passes through $q$, at which point the GWN value would jump by one.
We depict this scenario explicitly in Figure~\ref{fig:intuition_3d_extra}, which further illustrates the specific form of Equation~\ref{eqn:adjusted_stokes}.
Although the GWN field is determined almost entirely by boundary data (and completely so for far-away points), we must still account for the influence of the internal surface; as we will show, it affects the GWN value only by a constant.

\subsubsection{\textbf{Near-field}: Analytically Adjusted Stokes' Theorem}\label{sec:adjusted_stokes}

If the surface bounding box does contain the origin (i.e.\ the query point is inside the surface bounding box), we can no longer guarantee that there exists a line $L$ which does not intersect the surface.
Instead, we must explicitly test for signed intersection between some oriented line $L$ and the surface.
We consider cases to be ``near-field'' when all intersections between $L$ and the surface occur away from the surface edges and other degeneracies.
When these conditions are not met, the query point is handled as an ``edge-case''.

Naturally, distinguishing the near-field and edge cases requires first identifying the location of all intersections between the surface and the line in the parameter space of the surface.
As noted in Section~\ref{sec:background}, there are many numerical ray casting algorithms one could apply to find signed intersections between this line and the surface, and in principle, any can be used for the purposes of evaluating the GWN.
We describe our own geometric subdivision approach in Section~\ref{sec:line_surface_intersections}, which in contrast to many algebraic alternatives, takes advantage of the broader GWN algorithm's high tolerance for imprecision in the intersection routine.

Because any intersection between $L$ and the surface violates the conditions of Stokes' theorem, direct application of the boundary formulation is theoretically unfounded and, in general, evaluates to an incorrect GWN.
Our proposed correction term in Equation~\ref{eqn:adjusted_stokes} permits evaluation of the GWN using the same boundary integral formulation as Equation~\ref{eqn:stokes_theorem_equation}, provided that for each intersection (if any), the value $0.5$ is added or subtracted according to the orientation of the surface and line at the point of intersection to compensate for the jump discontinuity present in the GWN field along the surface.

\begin{figure}[tb]
    \centering
    \includegraphics[width=\linewidth]{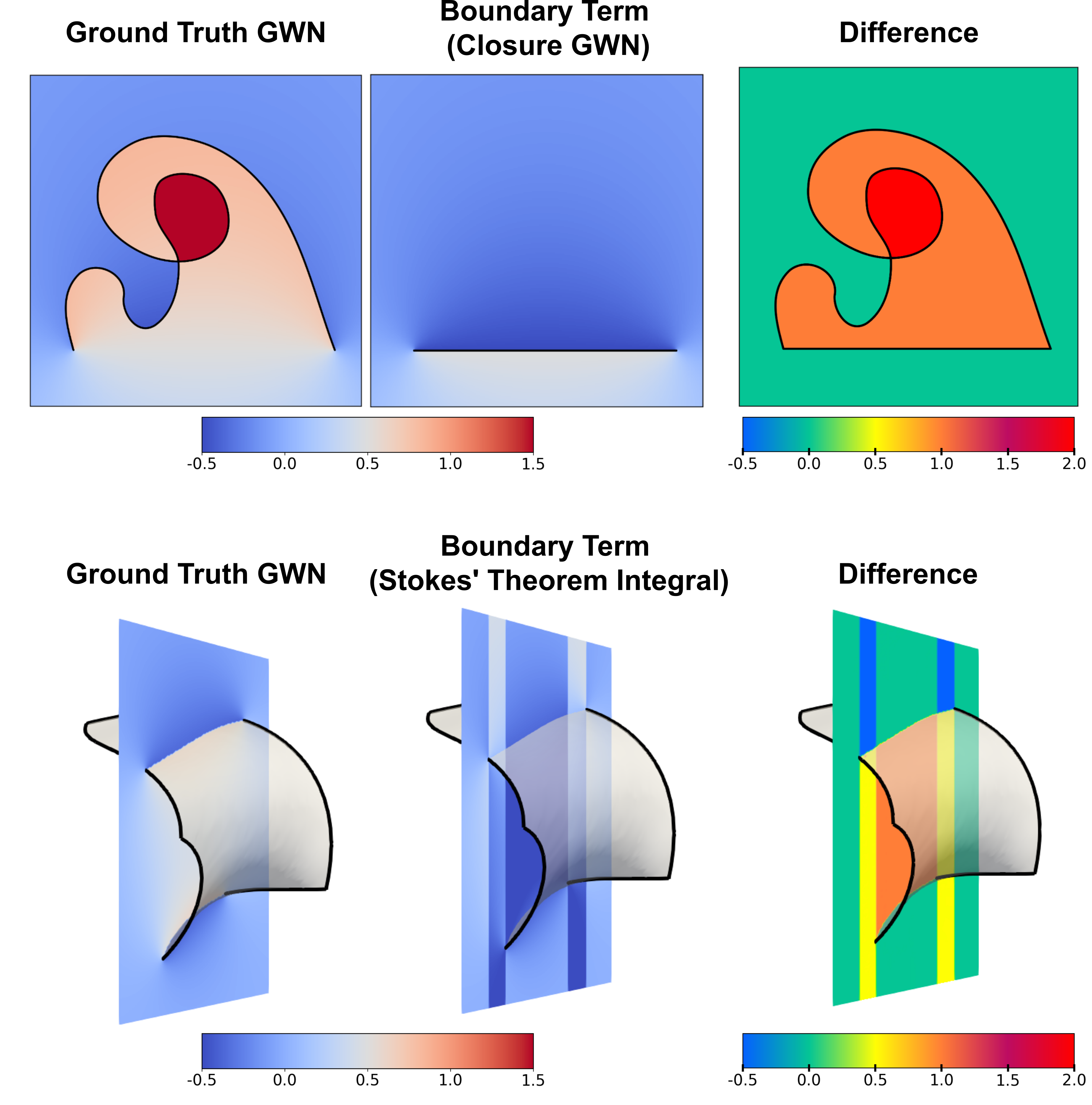}
    \caption[Intuition for Boundary Formulation.]{We illustrate a property of the GWN in both 2D (top) and 3D (bottom). The difference between the GWN for a curved surface (left) and that of a particular boundary term (middle) is always a constant value. In 2D, when this boundary term is the GWN of the curve's linear closure, the difference is integer valued. In 3D, when this boundary term is the integral component of Equation~\ref{eqn:adjusted_stokes}, the difference is always \textit{half}-integer valued. This suggests a method for GWN evaluation by evaluating the boundary term, and adjusting it by identifying the proper constant.}
    \Description[Illustrative example of boundary formulation]{A 2D and 3D example of the GWN for a curved shape, the scalar field generated by the associated boundary formulation, and the difference between them. In both cases, the difference is a piecewise-constant field.}
		\label{fig:intuition_total}
\end{figure}

This is illustrated in Figure~\ref{fig:intuition_total} (bottom), showing that the difference between the ground truth GWN and the specific boundary formulation in Equation~\ref{eqn:stokes_theorem_equation} is always, perhaps unexpectedly, half-integer valued.
As a result, the true GWN can always be \textit{recovered} from the boundary formulation by identifying the proper half-integer value.

In some sense, the existence of an analytic fix that depends only on the orientation of the query point relative to the surface is a natural consequence of the double-layer potential which, when integrated along the surface, defines the GWN scalar field~\cite{Barill-18-soupcloud,bang-23-multipole}.
The use of this specific correction term is made more intuitive by means of analogy in two spatial dimensions (with a direct proof provided in Appendix~\ref{sec:appendix}).
As summarized in Section~\ref{sec:gwn_for_curved_geometry} and demonstrated in Figure~\ref{fig:intuition_total}, the GWN for an arbitrary curve in 2D (left) can always be computed as the difference between an integer winding number for the curve closed by a straight line (right) and a generalized winding number defined by that straight line closure (middle).
Crucially, the GWN of the closure is uniquely defined by the two boundary points of the original curve.
From the lens of our 3D GWN framework, the GWN of the closure can be thought of as a ``boundary term'' which must be ``corrected'' by the integer winding number of the closed shape.

To make the comparison explicit, we see that in both 2D and 3D, the value of the GWN is the sum of a fractional component which does not depend on any interior geometry, and an integer component which does.
Although the fractional component in 2D reduces to a subtended angle, while the 3D equivalent is not the GWN of any particular shape, both integer components can be thought of as determined by the result of a line intersection test.
Most importantly, points which are known to be far from the shape do not require any correction at all.
Returning our focus to 3D, we emphasize that although  our line-surface intersection test does come at additional cost, it is necessary only for the surfaces nearest to the query point.

\subsubsection{\textbf{Edge Cases}: Disk Extraction via Trimming Curves}\label{sec:edge_cases}

\begin{figure*}[tb]
    \centering
    \includegraphics[width=0.66\linewidth]{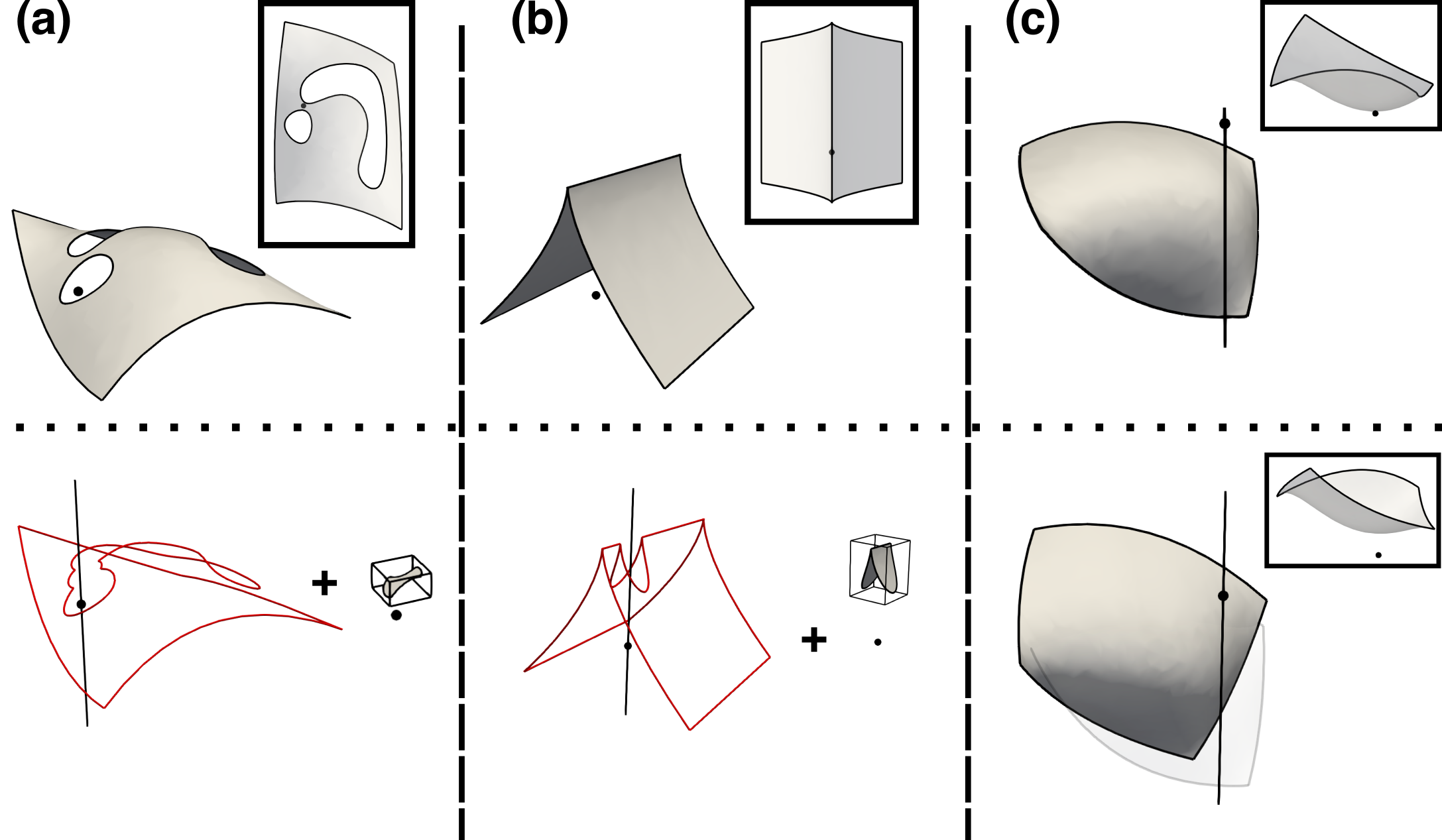}
    \caption[Edge-Case Evaluation for 3D GWN.]{Edge-Case Evaluation: For query points within a surface bounding box, we consider a line containing the point. 
	If the line intersects the surface near a boundary curve (a) or 
	at a point for which a normal cannot be defined such as a cusp (b), we remove the intersection by trimming the surface along a small disk in parameter space 
  and feed the remaining disk back into our algorithm for reprocessing.
	If the intersection occurs at a tangent point between the line and the surface (c), we apply a random rotation to the surface that does not change the GWN evaluation, but is more likely to result in a case that is easier to handle.}
    \Description[Edge-Case Evaluation for 3D GWN]{Several examples of edge-case handling are shown, including the line of discontinuities intersecting (a) the boundary of the surface, (2) a cusp on the surface, (c) at a point of tangency with the surface. In the first two cases, a parameter-space disk is removed from the surface and processed separately. In the third, the surface is randomly rotated and processed again.}
		\label{fig:edge_cases}
\end{figure*}

We now consider the remaining edge cases for which the query point is inside the surface bounding box, but the correction term in Equation~\ref{eqn:adjusted_stokes} cannot be applied.
As shown in Figure~\ref{fig:edge_cases}, this occurs when the line of discontinuities intersects the surface at a boundary, at a cusp, or is tangent to the surface.
We treat all edge cases by either finding a new line (in the case of tangencies) or by removing from the surface a disk of radius $r$ in parameter space around the point of intersection.
This guarantees that the line does not intersect the surface at the otherwise problematic intersection point, and permits evaluation via Equation~\ref{eqn:adjusted_stokes}.
The extracted disk is then fed back into the original algorithm for reprocessing.
Since the produced disk has a much smaller bounding box than the original surface, the computational cost of the reprocessing is significantly reduced.

While the choice of disk radius $r$ is mathematically arbitrary, it can have important consequences on computational performance and the selection of other numerical tolerances.
Through observation and numerical experiments (presented in full in Section~\ref{sec:adaptive_quadrature_performance}), we have found that fixing the radius $r$ to be 1\% of the bounding box diagonal offers a reasonable tradeoff between these considerations.

Furthermore, the proposed strategy for handling edge cases is such that it could be used to accurately compute the GWN of \textit{any} query point (at increased computational cost).
Because most edge cases can be identified by a zero normal vector at the point of intersection, we can conservatively treat any intersection with a \textit{near}-zero normal vector as an edge case without significantly impacting the aggregate performance of the method.

Similarly, we treat any intersection within a distance $r$ of the surface boundary as an edge case.
Because this should also include intersections that occur \textit{outside} the trimmed surface, during preprocessing, we linearly extend the parameter space of each untrimmed NURBS patch by $r$~\cite{wolters-2000-extensions}.
Accounting for ``near-misses'' in this way also improves the robustness of the GWN algorithm to the precision of the line-surface intersection routine.

The only edge case which is not well-handled by this strategy is a line that is tangent (or near-tangent) to the surface, as the surface remaining after the disk is removed may still intersect the line.
In these cases, we instead rotate the surface (i.e.\ consider a different line) and feed the rotated surface into the original algorithm (see Figure~\ref{fig:edge_cases} (c)).
While this is the most computationally expensive type of edge case, it is also the least common due to our choice of the cast ray to be an approximate surface normal (See Section~\ref{sec:line_surface_intersections}).

\begin{algorithm}[tb]
    \caption{\texttt{TrimmedSurfaceGWN} 
				Evaluate the generalized winding number for a trimmed NURBS surface.
		}\label{alg:generalized_winding_number}
    \DontPrintSemicolon
    \KwIn{$S$: Trimmed NURBS surface}
    \myinput{$q$: Query point}
    \myinput{$\epsilon_q$: Tolerance for adaptive quadrature}
    \myinput{$\epsilon_{ls}$: Tolerance for surface-line intersection}
    \KwOut{$w_{S}$: The GWN evaluated at ${q}$}
    \nonl\;
    $S \gets S - q$ \tcp*{Translate patch so q is at the origin}
    $w_S = 0$ \tcp*{Initialize the GWN}
    \nonl\;
    $r \gets 0.01 * \texttt{UVBoundingBoxDiagonal}(S)$\;
    $S \gets \texttt{ExtendUntrimmedPatch}(S, r)$ \tcp*{Per~\cite{wolters-2000-extensions}}
    \nonl\;
    \eIf{$q \notin \texttt{BoundingBox}(S)$}{
        \tcc{Rotate patch to not intersect $z$-axis}
        $L \gets \texttt{Non-intersecting Line}$\;
        $S \gets \texttt{RotateToZ}(S, L)$\;
    }{
        \tcc{Compute $z$-axis intersections with $S$ (Alg.~\ref{alg:line_patch_intersect})}
        $z, u, v \gets \texttt{LinePatchIntersections}(S, \text{z-axis}, \epsilon_{ls})$\;
        \ForEach{intersection point $\{(u_0, v_0), z_0\}$}{
            $\vec{n} \gets S_u(u_0, v_0) \times S_v( u_0, v_0 )$\;
            \uIf{ $S(u_0, v_0)$ is an interior point of $S$ with normal $\vec{n} \neq 0$}{
                \tcc{Add constant according to orientation}
                \uIf{ $\vec{n}_z z_0 > 0$ }{
                    $w_S = w_S - 0.5$
                }\uElseIf{ $\vec{n}_zz_0 < 0$ }{
                    $w_S = w_S + 0.5$
                }
            
            }\uElseIf{$S$ is tangent to the $z$-axis at $S(u_0, v_0)$}{
                \tcc{Rotate the surface and try again}
                \Return $\texttt{TrimmedSurfaceGWN}(\texttt{RandomRotation}(S), 0)$\;
            }\uElse{
                \tcc{Extract a parameter space disk S\_disk, and update the trimming curves of S}
                $S, S_{\text{disk}} \gets \texttt{ExtractParameterDisk}(S, r)$\;
                $w_S = w_S + \texttt{TrimmedSurfaceGWN}(S_{\text{disk}}, 0)$\;
            }
        }

    }
    
    \ForEach{1D Curve in $\texttt{BoundaryCurves}(S)$}{
        $w_S = w_S + \texttt{EvaluateLineIntegral}(C, \epsilon_q)$ \tcp*{Alg.~\ref{alg:adaptive_quadrature_recursive}}
    }
    
    \Return $w_S$
\end{algorithm}

\subsection{GWN for Points Coincident with the Surface}\label{sec:coincident_winding_number}

Strictly speaking, the generalized winding number of query points coincident with the surface is undefined, as the fundamental mathematical definition of the GWN (Equation~\ref{eqn:cartesian_surface}) is discontinuous.
This is a problem unique to curved shapes, as the GWN of a point coincident with a line or triangle is exactly equal to zero.
However, the GWN algorithm should still return reasonable values for points coincident to curved shapes (alongside the relevant error flag) to make the method more amenable to downstream applications.

In all cases, coincident or otherwise, it is important that the GWN algorithm produce consistent results for nearby points.
This prohibits the common solution of perturbing the query points slightly to avoid coincidence, as this may move truly interior or exterior points across a boundary.

Analogous to the 2D case explored in~\citet{spainhour_24_robustcontainment2d}, we define the winding number of coincident points to be the average of the two values of the GWN computed across the jump discontinuity of the scalar field.
Fortunately, we know by the results of Section~\ref{sec:adjusted_stokes} that in most cases, the correction term contributes a value of $0.5$ if the query point is on one side of the surface, and $-0.5$ if it is on the other. 
This means that if the query point is \text{on} the surface, these two values effectively cancel out, and no correction term needs to be added to the boundary formulation for the intersection, as shown in Equation~\ref{eqn:adjusted_stokes}. 

The only unaddressed case is when the query point is coincident with a boundary curve of the surface, i.e.\ a coincident edge-case.
This case is uniquely difficult to solve via our Stokes' theorem reformulation, as the evaluated \textit{antiderivative} is also non-trivially discontinuous along its domain. 
We handle this in much the same way as the other edge-cases in Section~\ref{sec:edge_cases}, where we remove a disk from the surface around the query point.
The GWN calculation for the larger subdivided surface is performed as normal, but we consider the GWN of the removed disk to be zero, an assumption that only holds \textit{because} the query point is coincident to the disk.
To strengthen this assumption, we remove a much smaller disk than in the general case.
Indeed, a similar approach is used by \citet{liu-2025-closedform-wn} in 2D to remove a segment of the curve and treat its contribution to the GWN as zero.

\subsection{Accelerated Adaptive Quadrature}\label{sec:adaptive_quadrature}

Although the 1D line integral in Equation~\ref{eqn:adjusted_stokes} is more numerically stable than the 2D surface integral equivalent in Equation~\ref{eqn:cartesian_surface}, its robust evaluation requires a numerical integration scheme that is guaranteed to be precise for the near-singular integrand. 

In our implementation, we utilize a recursive subdivision approach, applying a fixed-order Gauss-Legendre quadrature rule at each level of recursion.
We stop the recursion when the difference between the evaluated integral on the entire curve and the sum of the integrals on the two halves is less than a user-defined quadrature tolerance $\epsilon_q$.
Thus, in most cases, $\epsilon_q$ determines how many digits of precision are achieved in the computed GWN field, although it is possible for more lenient values of $\epsilon_q$ that the two terms will agree at an incorrect value.
Using a Gaussian rule at each recursion level is particularly useful as it is capable of maintaining high precision even without specific structure in the integrand.
We specifically chose this recursive subdivision strategy because it effectively concentrates quadrature nodes near unstable regions of the integrand, i.e.\ when the query point is close to the 3D space curve which defines the domain of integration.
Elsewhere, where the integrand is stable, we maintain the typically high precision of Gaussian quadrature without additional computational expense.

To further improve computational performance, we cache the quadrature nodes and surface tangent vectors for each trimming curve at each level of recursion, and reuse these values across multiple queries.
Such a memoization strategy also compensates for the inability of Gauss-Legendre quadrature nodes to be reused across refinement levels (as would be the case for a nested quadrature rule such as Gauss-Kronrod), allowing the method to fully utilize the precision of the original rule.
On the other hand, this strategy can only be used if the trimming curves of a patch remain unchanged throughout the evaluation of the GWN at a single query point. 
In particular, since our edge-case resolution involves modifying trimming curves within the extraction disk of radius $r$, our memoization strategy cannot be applied to the modified curves in these cases.
We present the full algorithm in Algorithm~\ref{alg:adaptive_quadrature_recursive} and explore its performance more fully in Section~\ref{sec:adaptive_quadrature_performance}, particularly as the parameter $\epsilon_q$ relates to disk radius $r$.

\section{Numerical Experiments and Results}
\subsection{Numerical Integration of Boundary Formulation}\label{sec:results_integration}

\begin{figure}[t]
    \centering
    \includegraphics[width=\linewidth]{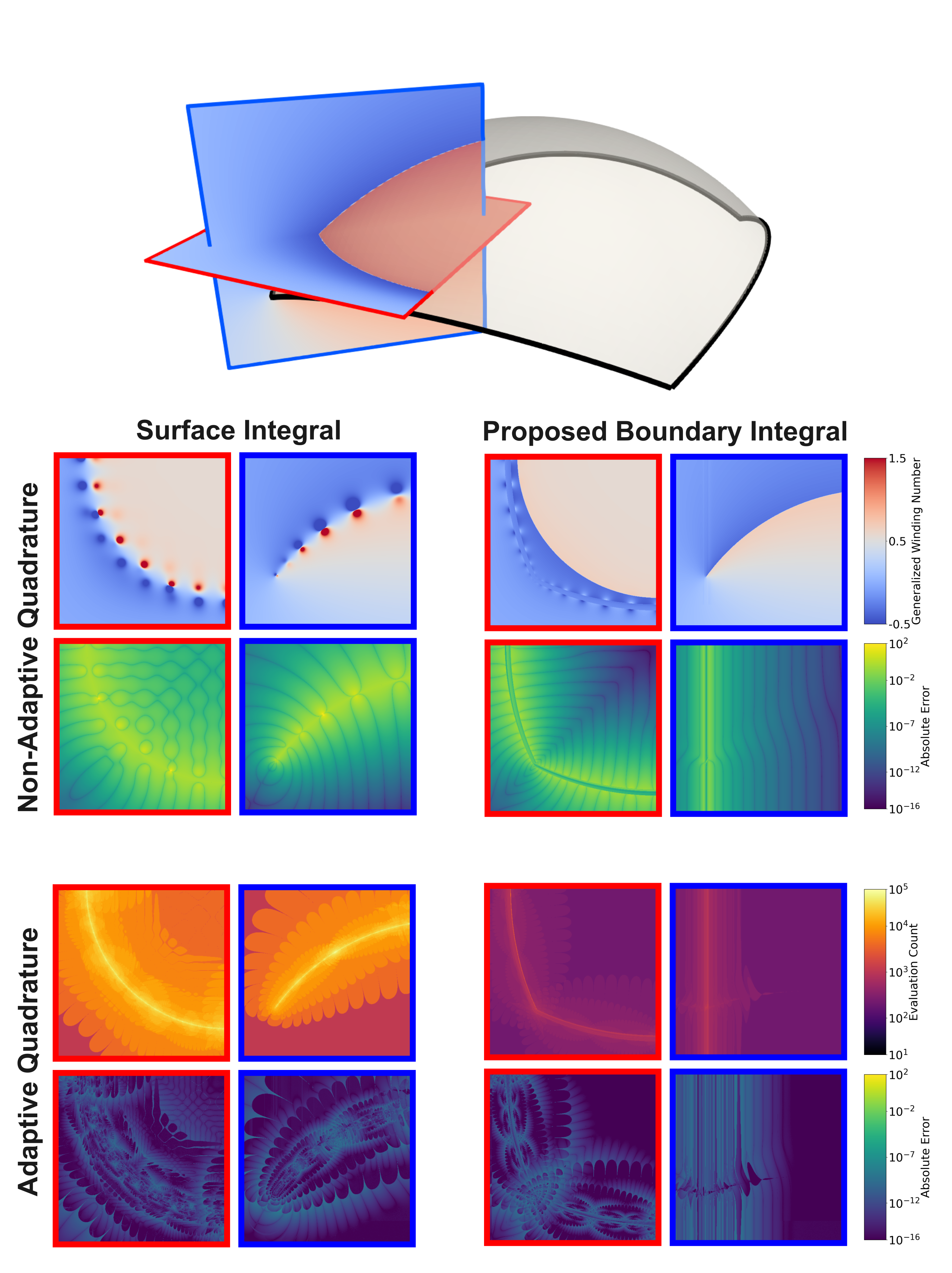}
    \caption[Comparison of numerical integration schemes for the 3D GWN.]{
      Comparing surface-based 2D tensor product Gaussian quadrature schemes (left) to boundary-based 1D Gaussian quadrature schemes (right). 
      For each, we consider a fixed (``non-adaptive'') 39$^{\text{th}}$-order rule which requires $20^2$ evaluation points for the surface integral and $4\cdot 20$ evaluation points for the boundary integral (top) 
      and an adaptive quadrature scheme which uses recursive bisection to ensure an absolute error bound of $10^{-6}$ (bottom). 
      The boundary formulation produces less error when evaluated as a non-adaptive scheme, and reduces evaluation cost as an adaptive scheme.
      Evaluation counts and absolute errors are presented in log-scale.
    }
    \Description[Comparison of numerical integration schemes for the 3D GWN.]{Comparisons between integration with a boundary formulation vs. tensor product rule, and fixed-order vs adaptive rules, valuated on slices of the top portion of a sphere. The plots for the fixed-order rules shows different spatial distributions of error. The plots for the adaptive rules show how both result in the same error, but with higher cost for the surface rule.}
    \label{fig:quadrature_slices}
\end{figure}

We first justify our choice of a boundary formulation in our solution method by comparing our 1D numerical integration scheme that solves the boundary form in Equation~\ref{eqn:adjusted_stokes} to a 2D numerical integration scheme that solves the surface integral of Equation~\ref{eqn:cartesian_surface}.
We perform this comparison on one of six biquartic patches representing a sphere~\cite{cobb-1988-tiling} where we evaluate the GWN field along two slices using a 20-point Gaussian quadrature rule.
The derived 1D method involves applying this 20-point rule to each boundary curve for a total of $4\cdot 20$ evaluations of the NURBS surface,
while the 2D scheme involves a tensor product of 20-point rules for a total of $20^2$ evaluations of the NURBS surface.

We show the results of this comparison in the ``Non-Adaptive Quadrature'' row of Figure~\ref{fig:quadrature_slices}, where errors in the 1D scheme are only present near the surface boundary, while errors in the 2D scheme are present for all points near the surface.
The accuracy of this 2D scheme, and in some sense even its applicability, would be further compromised by the presence of trimming curves, which are far more naturally handled in our boundary scheme.

The error in both methods can be mitigated through the use of adaptive quadrature, describing a ``Boundary-adaptive'' and ``Surface-adaptive'' scheme respectively. 
The performance of each method is shown in the ``Adaptive Quadrature'' row of Figure~\ref{fig:quadrature_slices}, showing that the surface-adaptive scheme would require thousands or even tens of thousands of times as many surface evaluations to achieve the same level of accuracy as our boundary-adaptive strategy, a cost which is further exacerbated when subdividing trimmed surfaces.

\begin{figure}
    \centering
    \begin{tabular}{ccc}
        \includegraphics[width=0.3\linewidth]{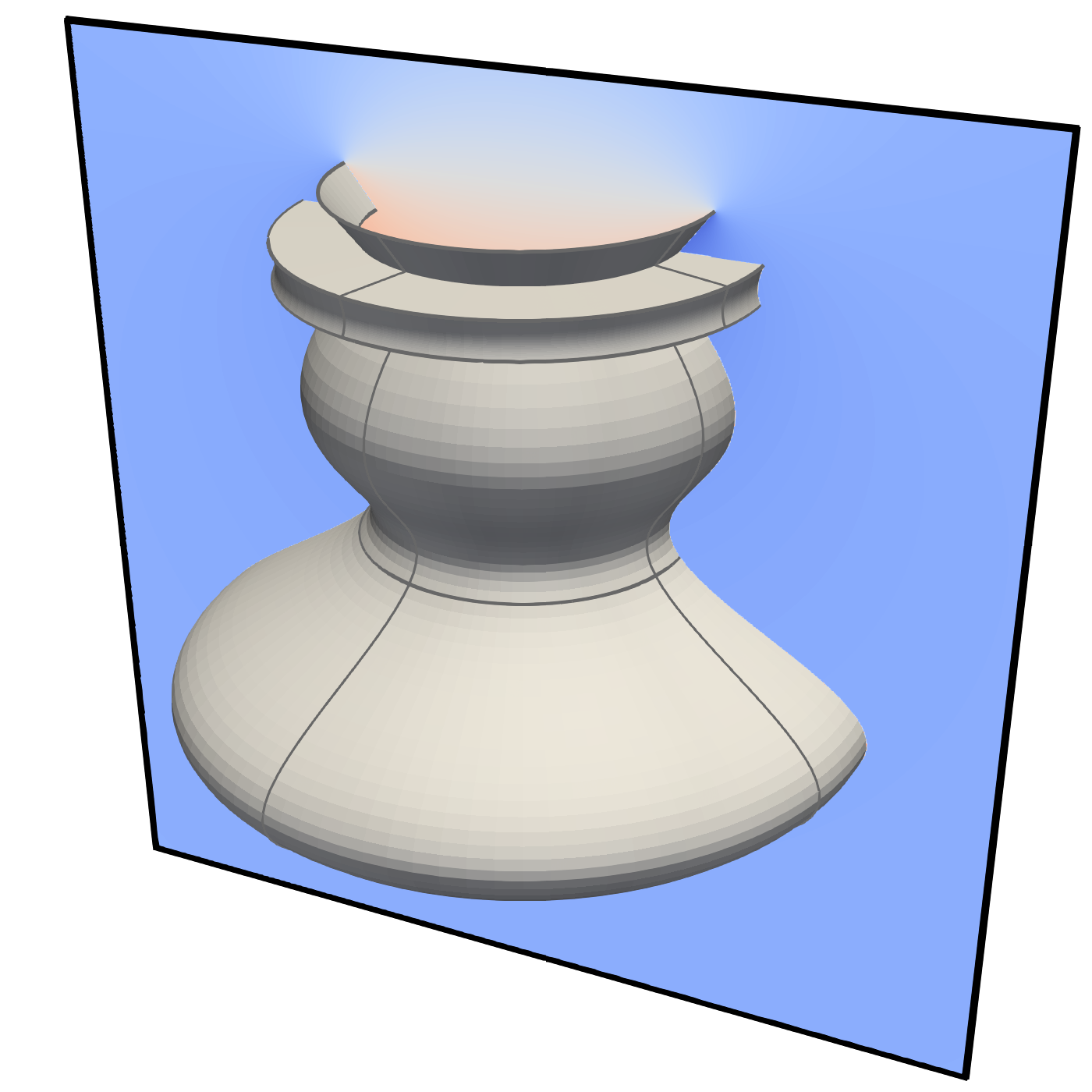} & \includegraphics[width=0.3\linewidth]{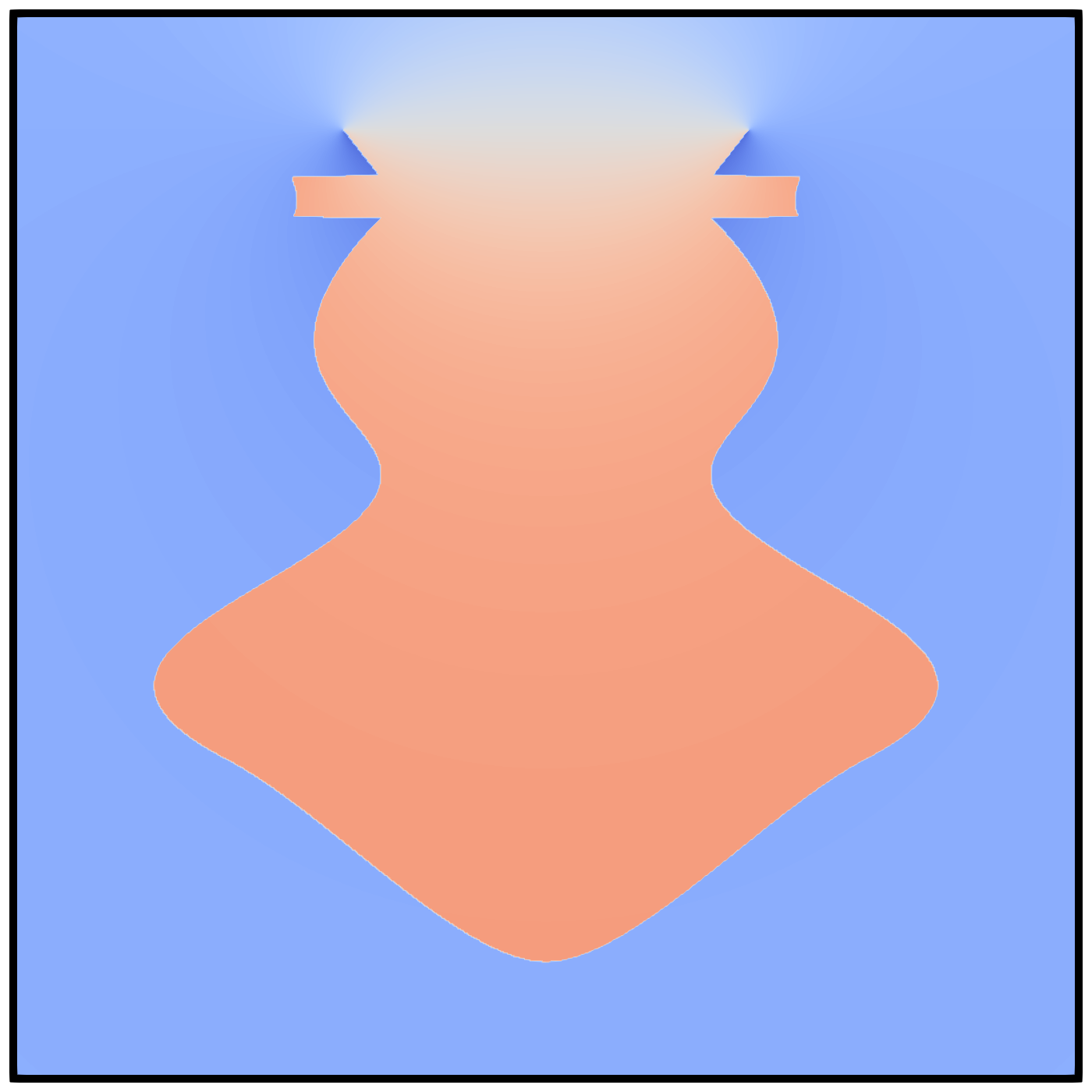} &
        \includegraphics[width=0.3\linewidth]{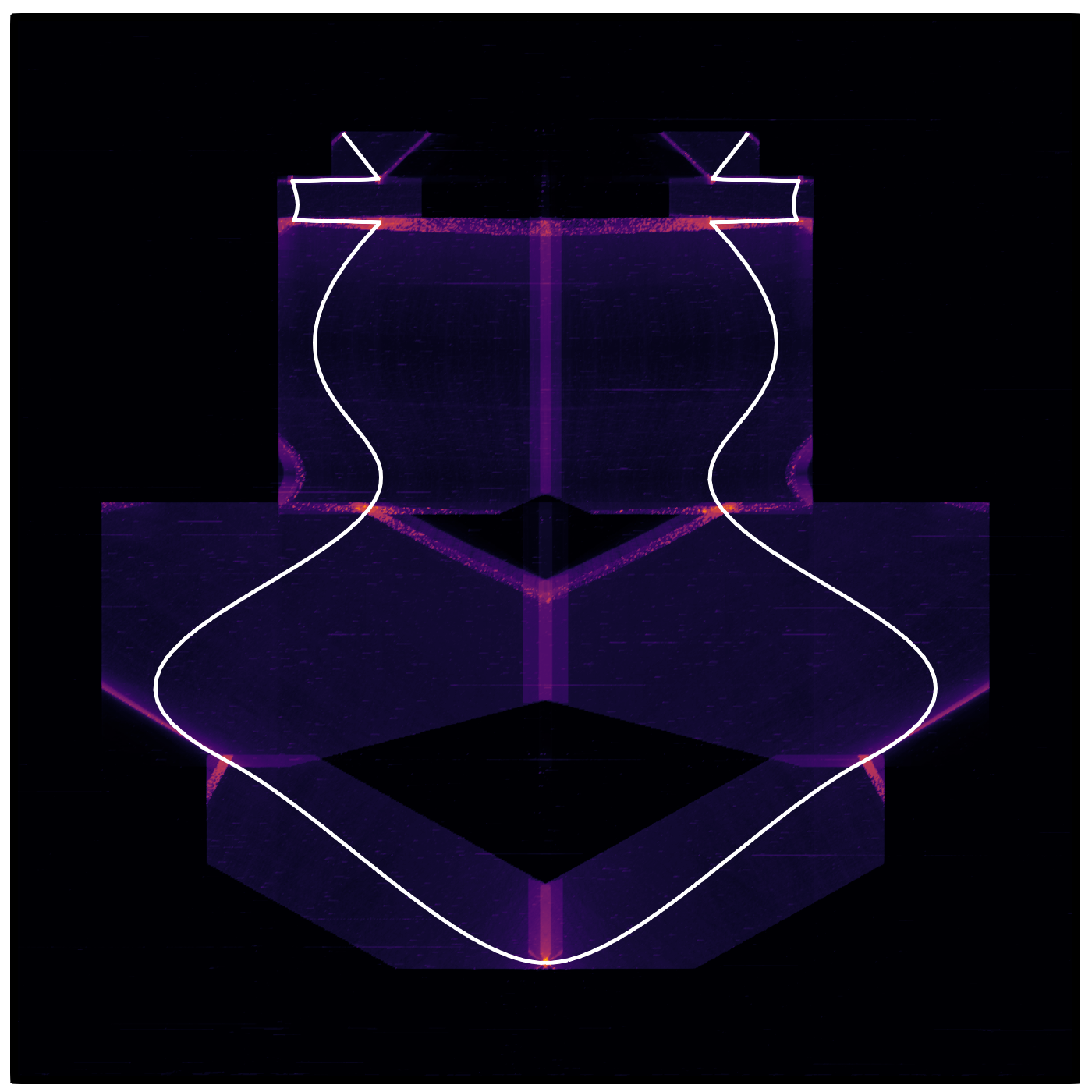}\\
        \includegraphics[width=0.3\linewidth]{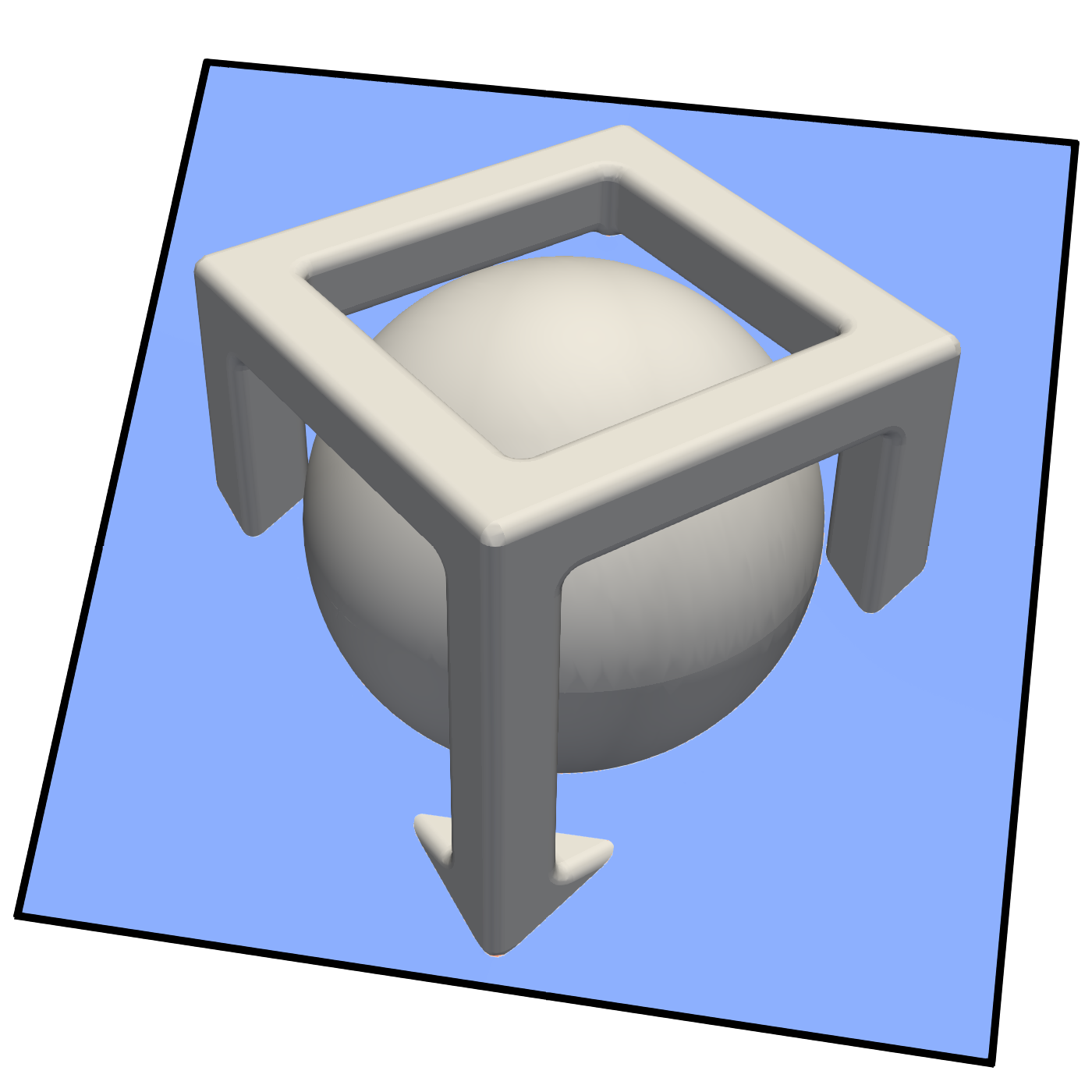} & \includegraphics[width=0.3\linewidth]{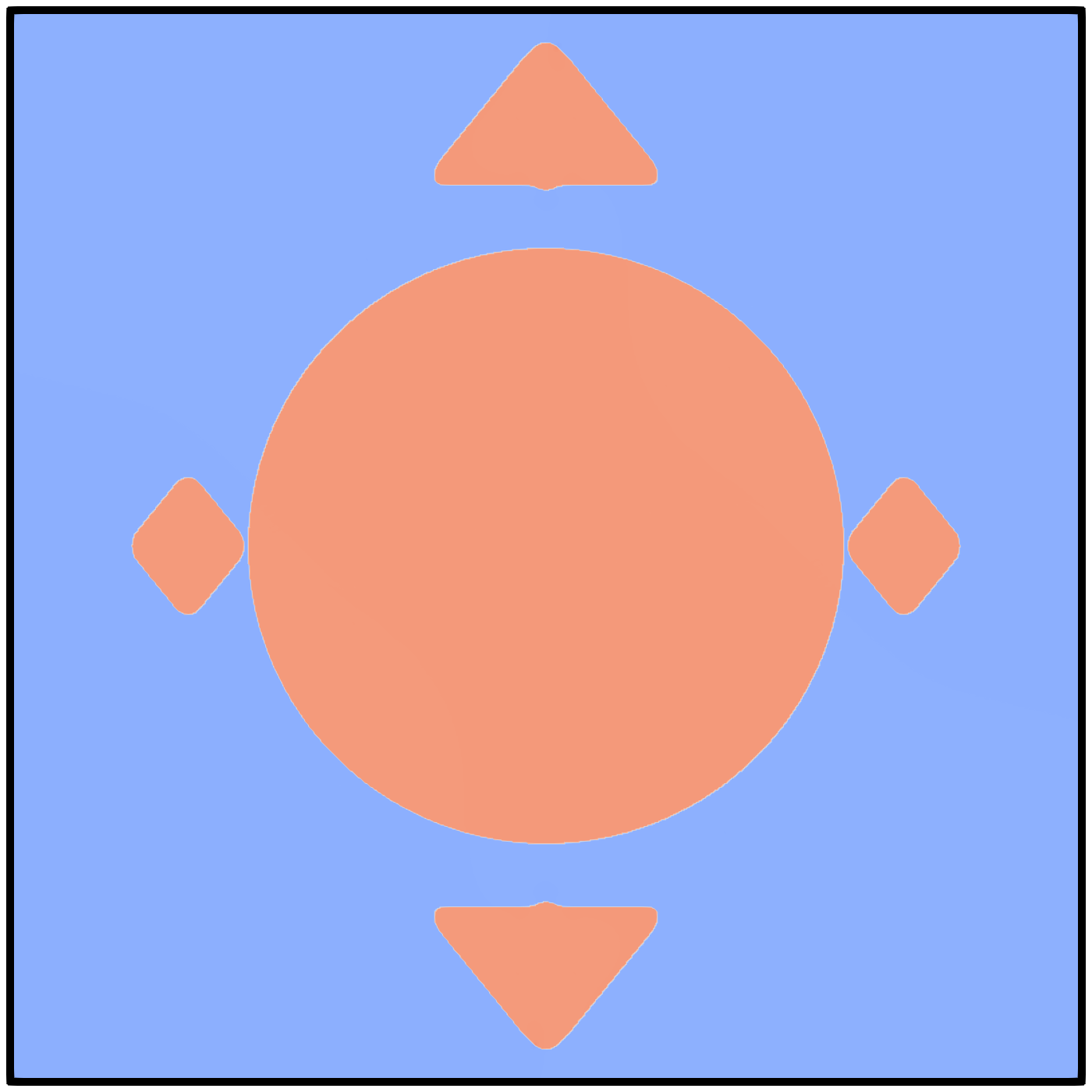} &
        \includegraphics[width=0.3\linewidth]{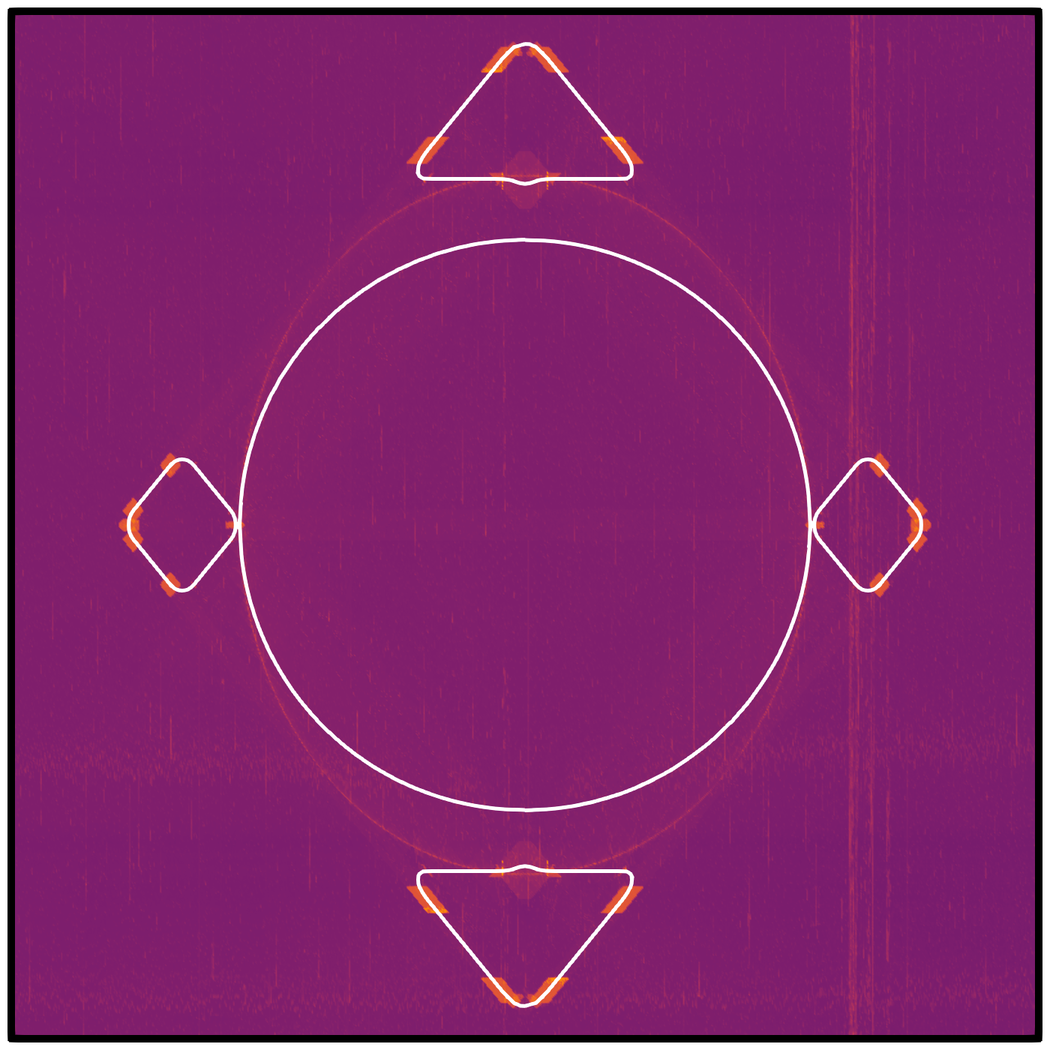}\\
      & \includegraphics[width=0.3\linewidth]{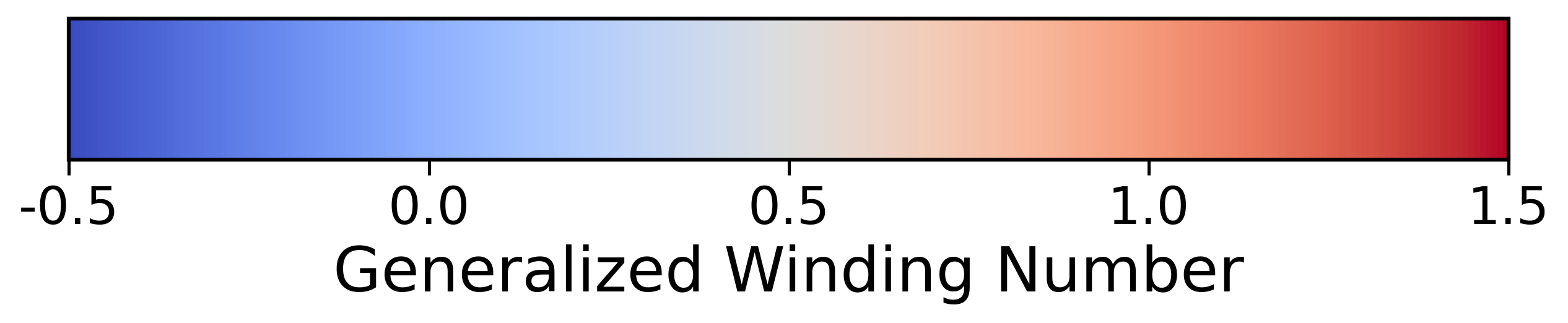} &
        \includegraphics[width=0.3\linewidth]{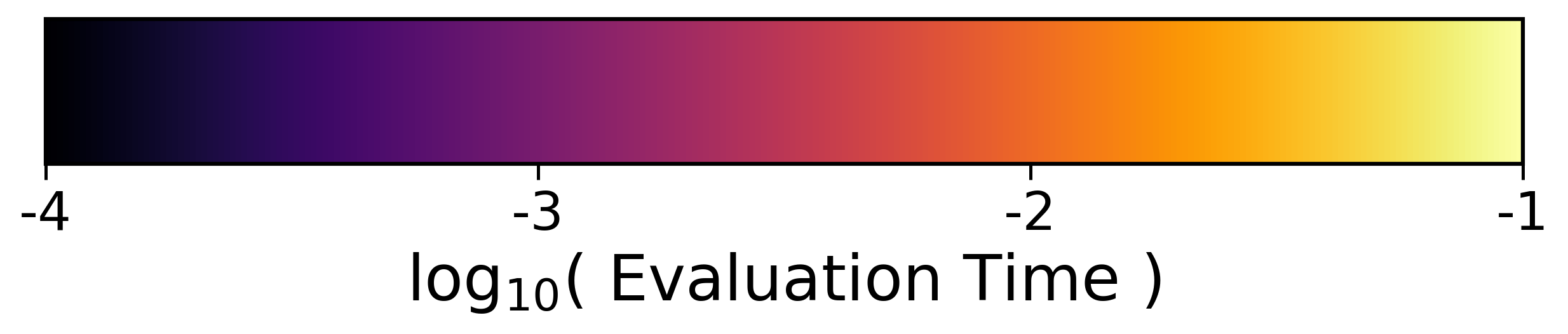}\\
    \end{tabular}
    \caption{GWN fields and evaluation times along an $800\times 800$ slice of grid points. 
    The 28-patch ``Vase'' surface of revolution (top) is adapted from an example in~\citet{martens-2025-oneshot}, which reports a threaded processing time of 484 seconds,
    while our approach takes 97.7 seconds serially (approximately 0.15ms per query point). 
    Our approach takes 829.9 seconds (approximately 1.29ms per query point) on the more complicated 192-patch filleted ``Box-Sphere'' shape (bottom).
		}
    \Description[GWN fields and per-query evaluation times for a simple and complex shape]{The GWN field and per-query point evaluation times are shown for a simple shape and one with many more surfaces. In the plot of evaluation times, the most expensive points are near the surface boundary. The shape with more surfaces has increased evaluation time across the domain.}
    \label{fig:vase_comparisons}
\end{figure}

In~\cite{martens-2025-oneshot}, this surface-adaptive scheme is referred to simply as ``adaptive quadrature'', which largely explains their characterization that quadrature is inapplicable for complex CAD surfaces.
In contrast, as can be seen in Figure~\ref{fig:vase_comparisons}, a boundary-adaptive strategy outperforms the reported metrics of that work on similar shapes, and is capable of doing so even without taking advantage of per-query parallelism, the planar characteristics of the input boundary curves, or any structure in the input set of query points.
Figure~\ref{fig:vase_comparisons} also demonstrates on practical examples how algorithm performance is influenced by the position of the query point. 
As expected, ``near'' points close to surface interiors have negligible overhead compared to ``far'' points, 
while only those close to a surface \textit{boundary} are more expensive since they must often be treated as edge-cases.

\subsection{Accuracy Evaluation}\label{sec:accuracy_results}

\begin{figure*}
    \centering
    \begin{tabular}{ccccc}
        \includegraphics[width=0.18\linewidth]{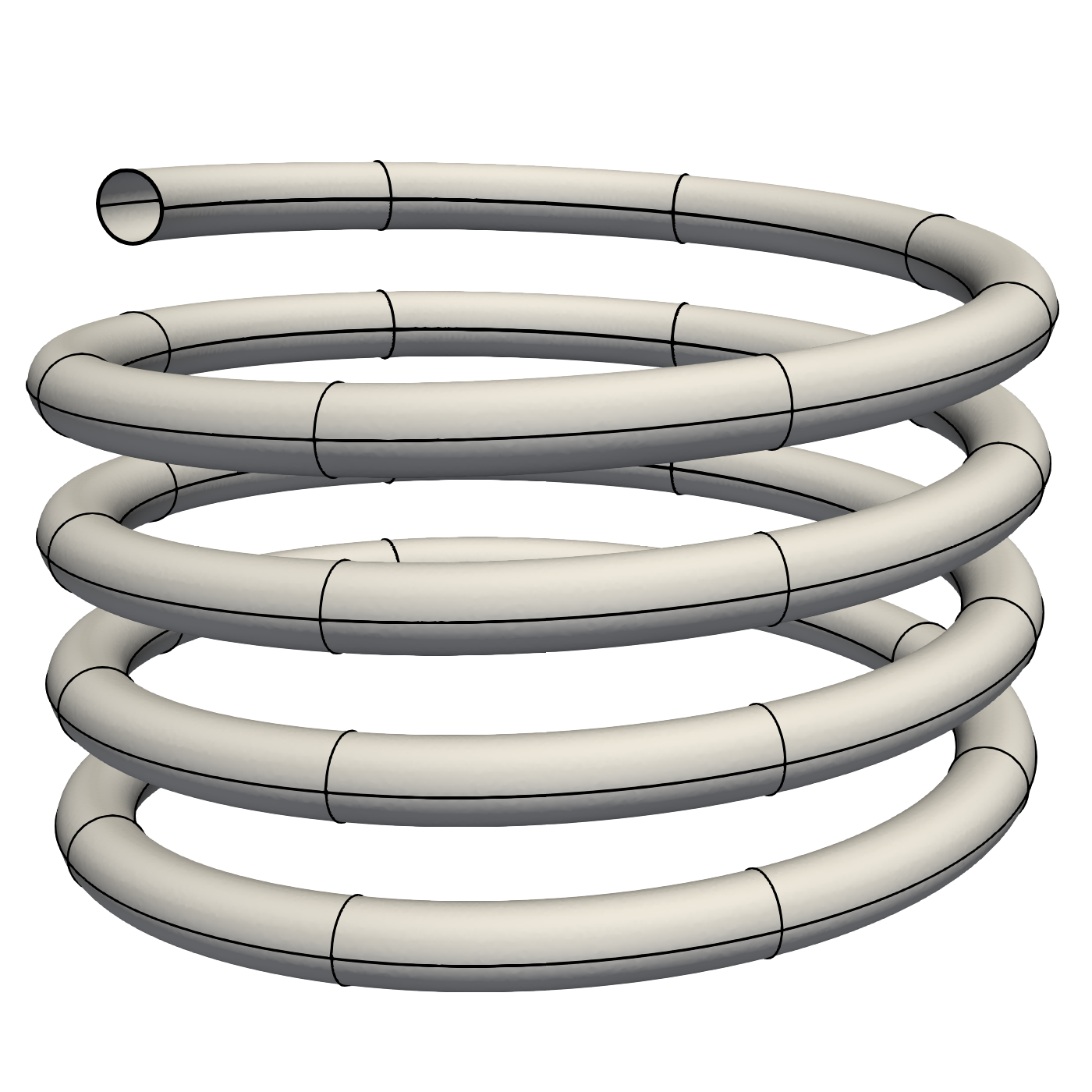} &
        \includegraphics[width=0.18\linewidth]{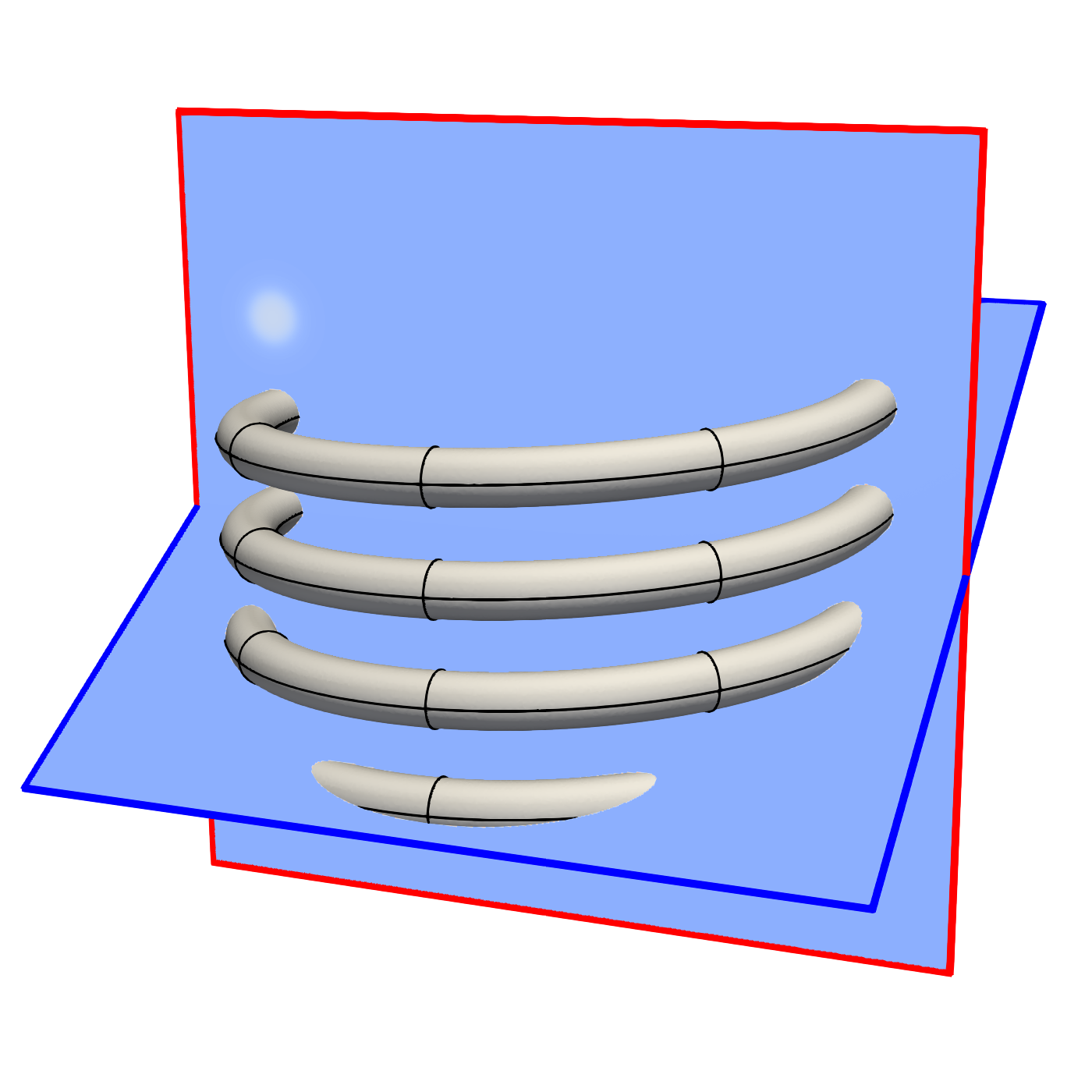} & 
        \includegraphics[width=0.18\linewidth]{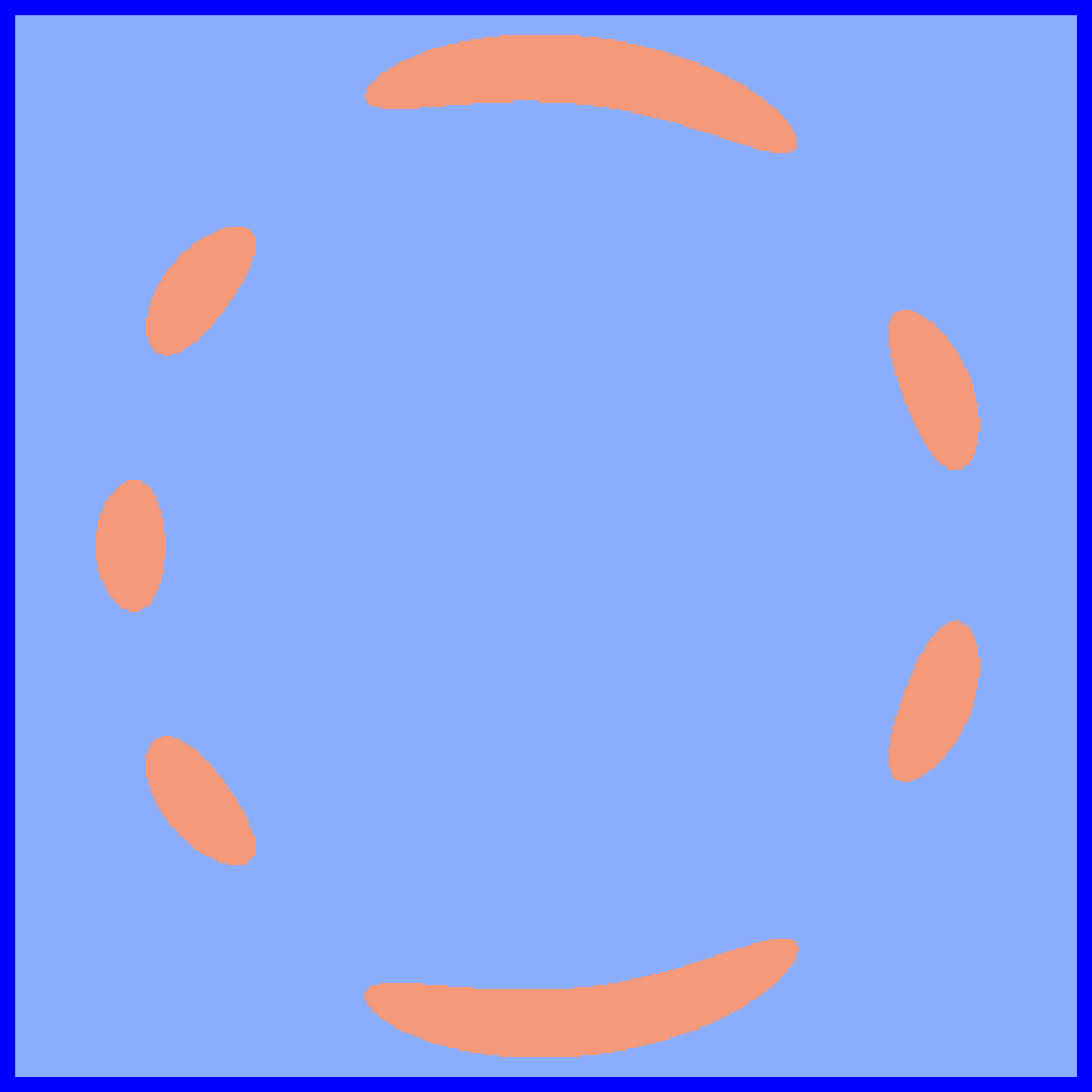} & 
        \includegraphics[width=0.18\linewidth]{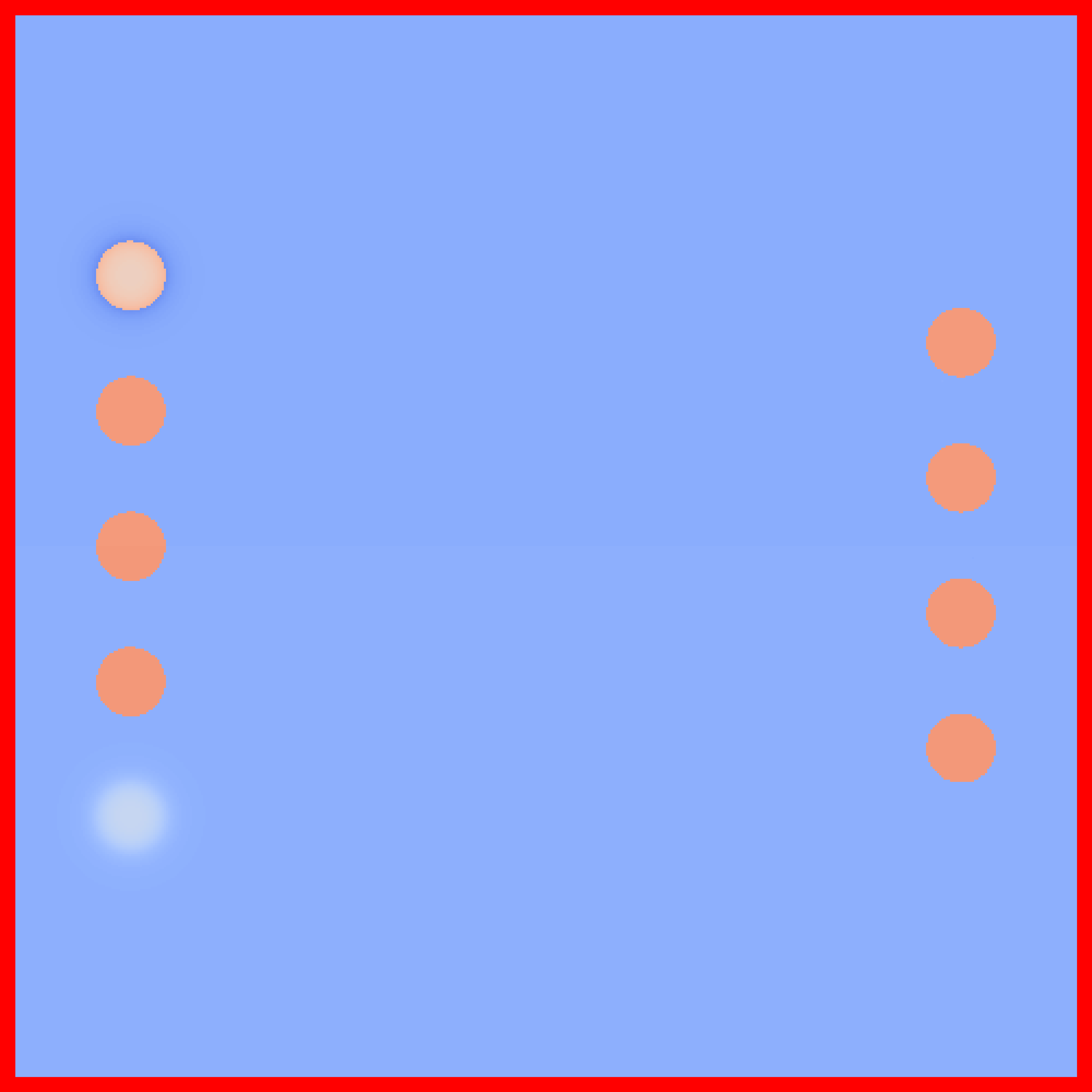} & 
        \includegraphics[width=0.18\linewidth]{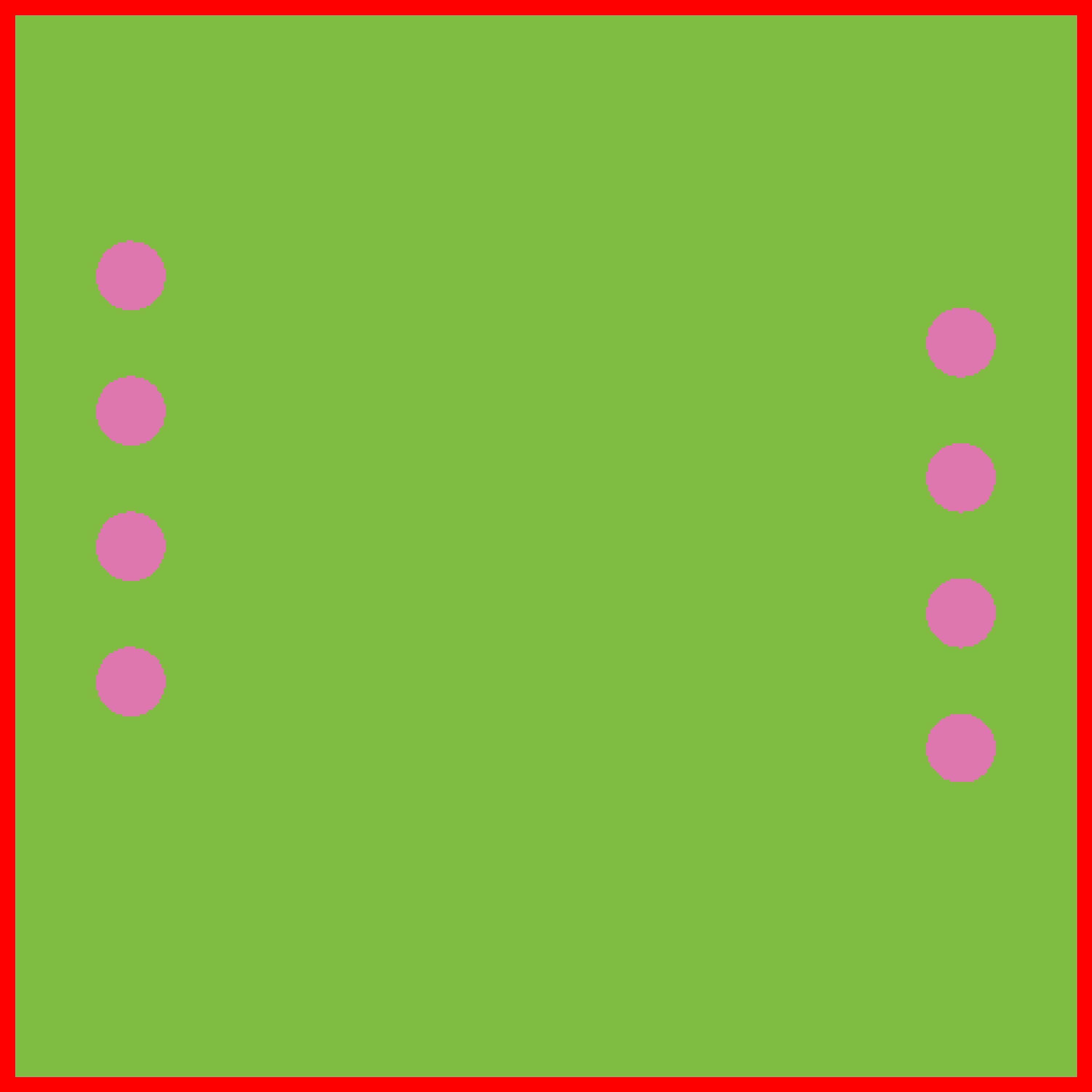} \\
        \includegraphics[width=0.18\linewidth]{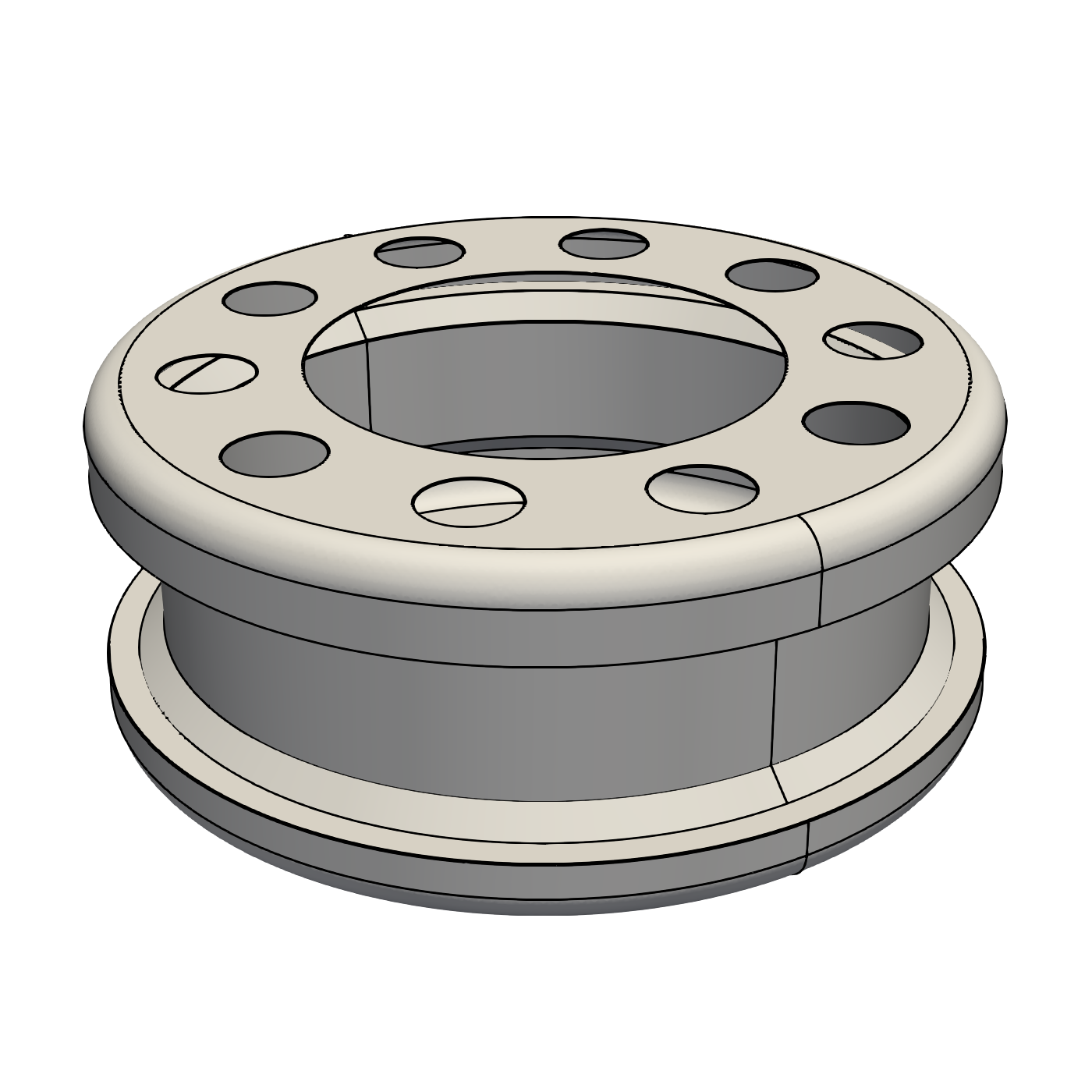} & 
        \includegraphics[width=0.18\linewidth]{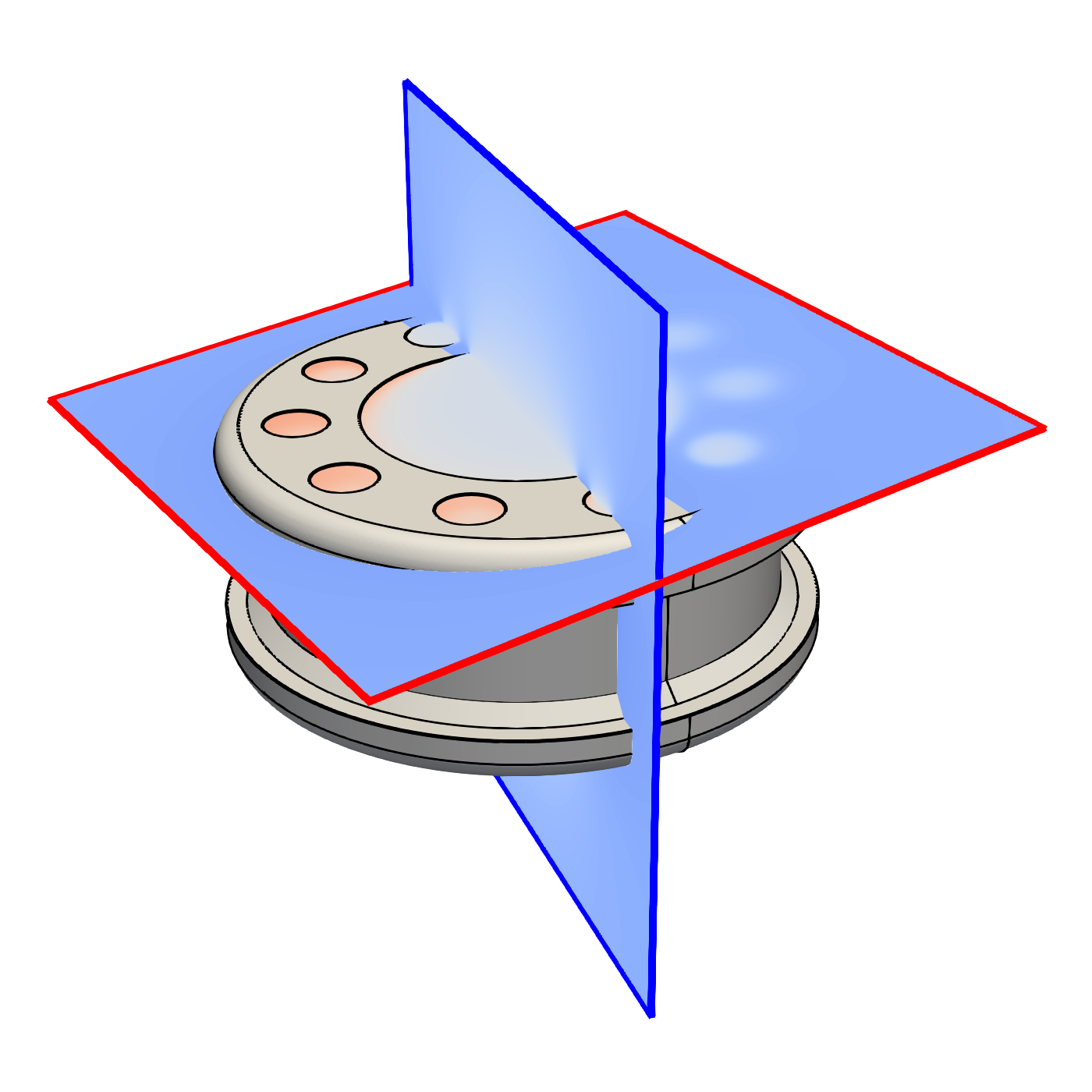} & 
        \includegraphics[width=0.18\linewidth]{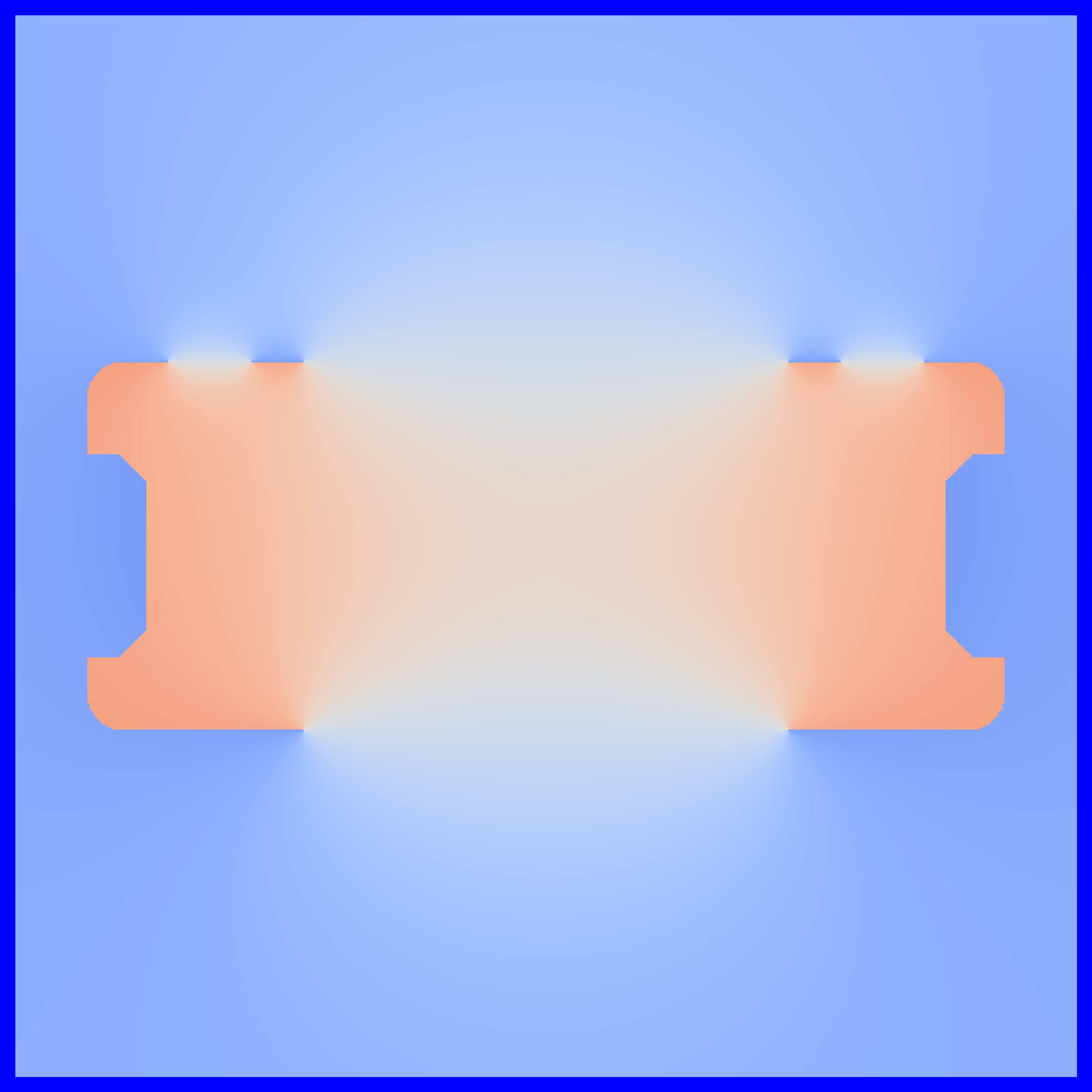} & 
        \includegraphics[width=0.18\linewidth]{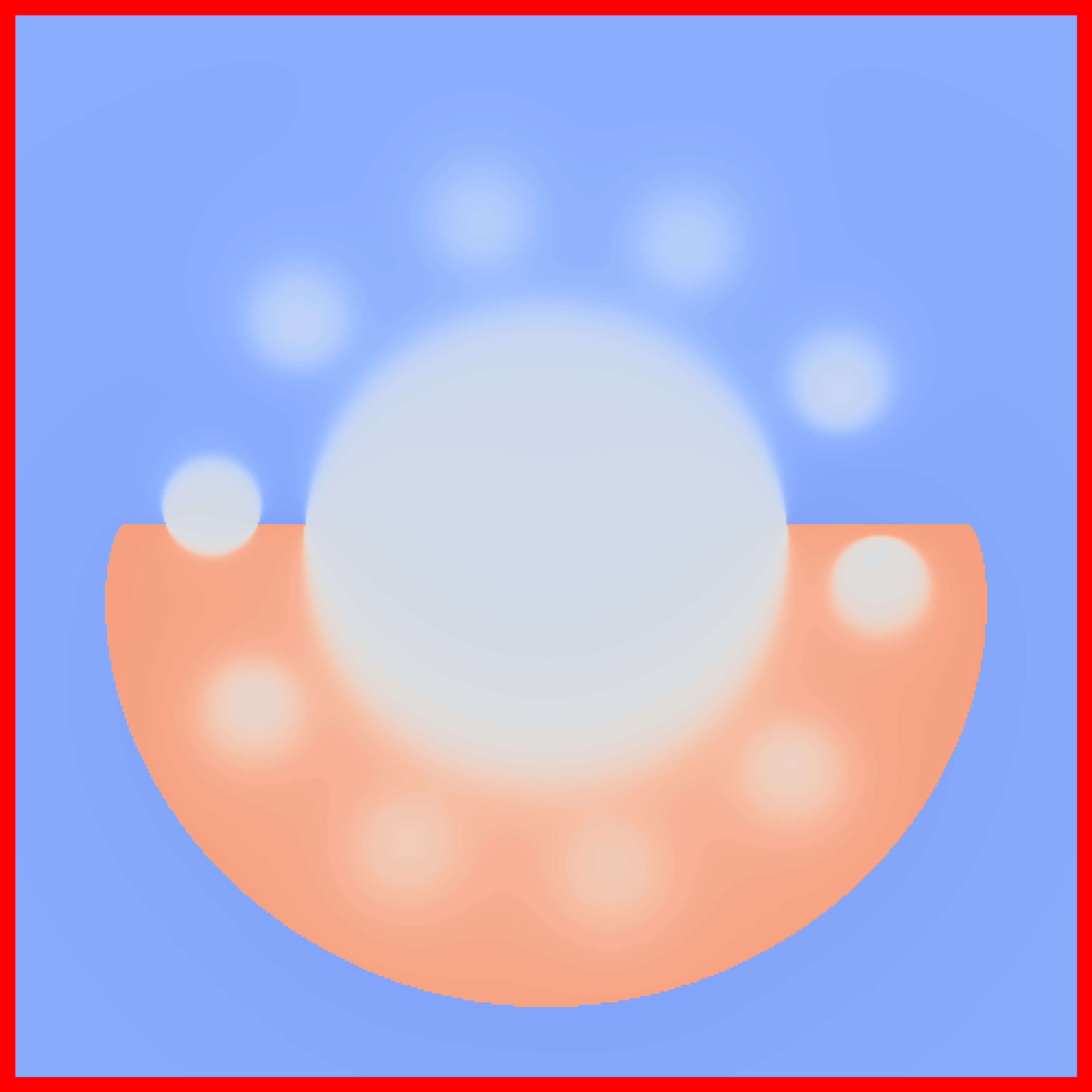}  & 
        \includegraphics[width=0.18\linewidth]{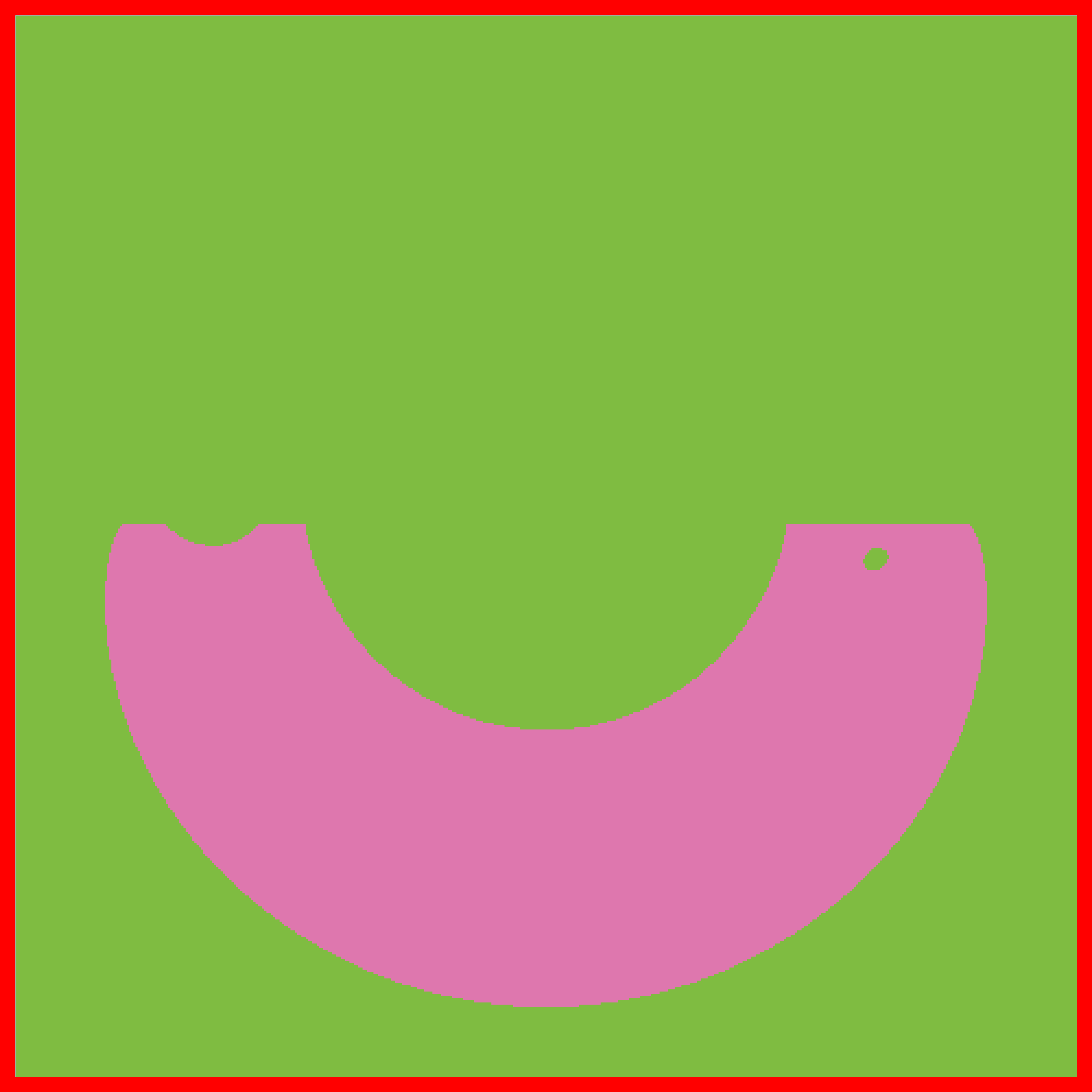} \\
        \includegraphics[width=0.18\linewidth]{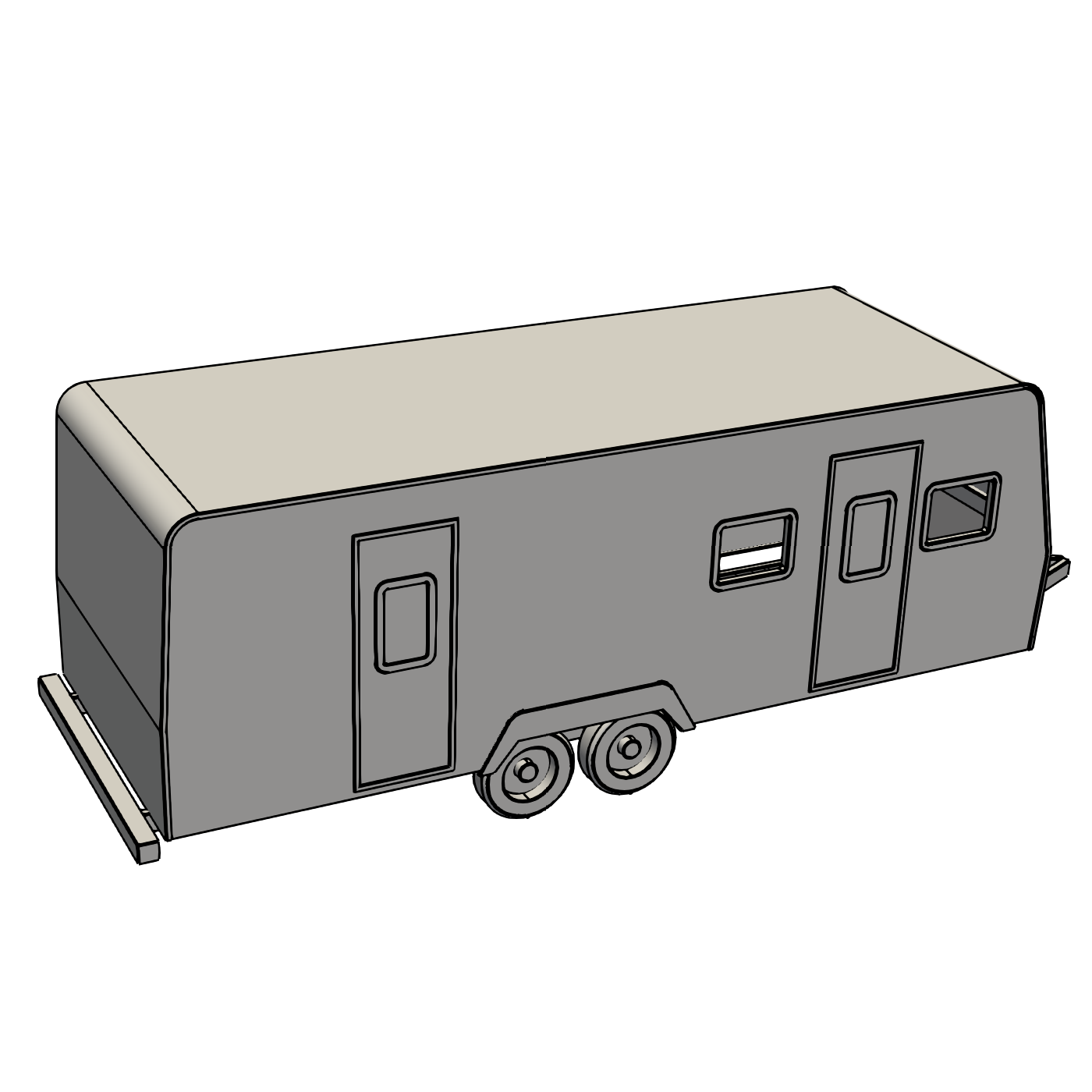} & 
        \includegraphics[width=0.18\linewidth]{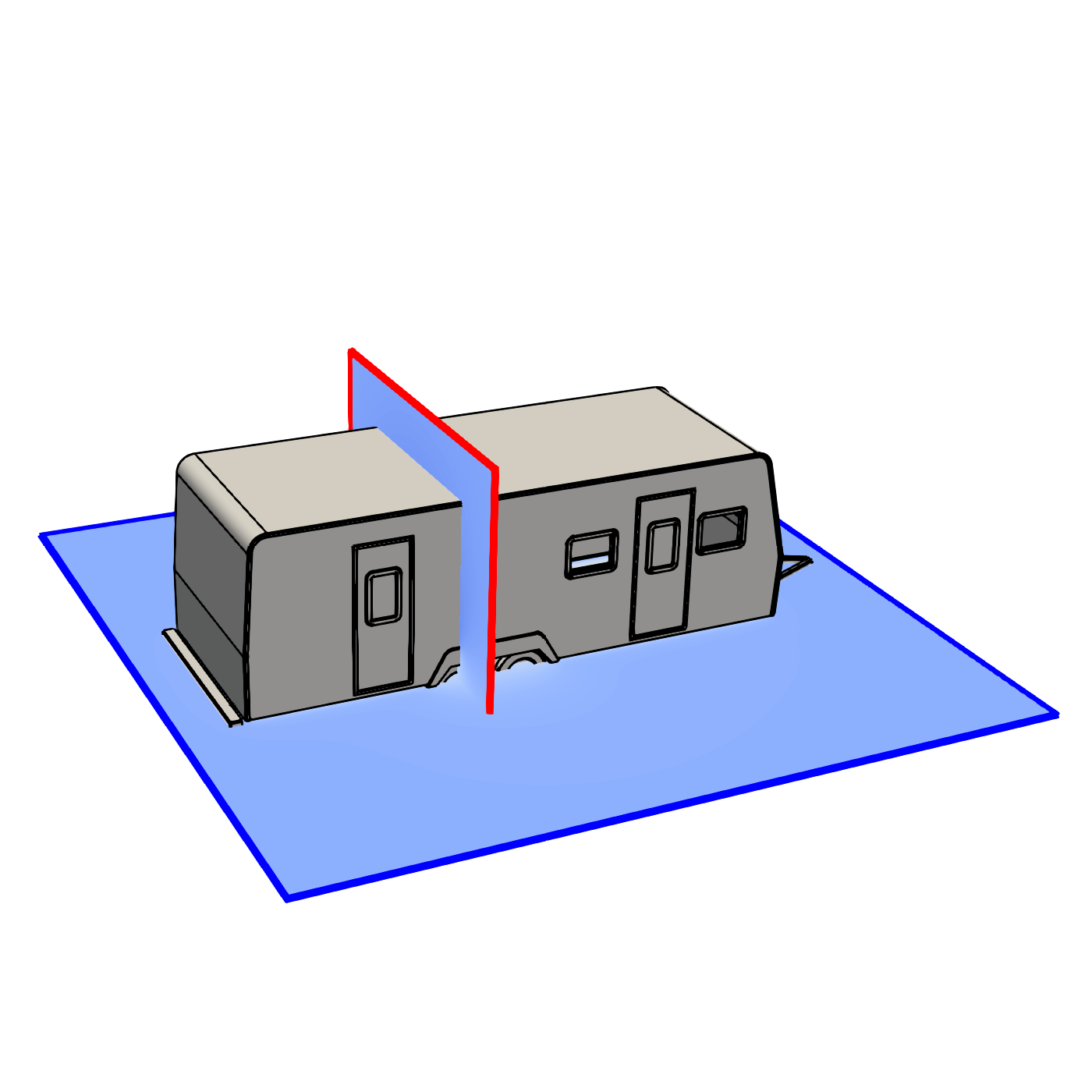} & 
        \includegraphics[width=0.18\linewidth]{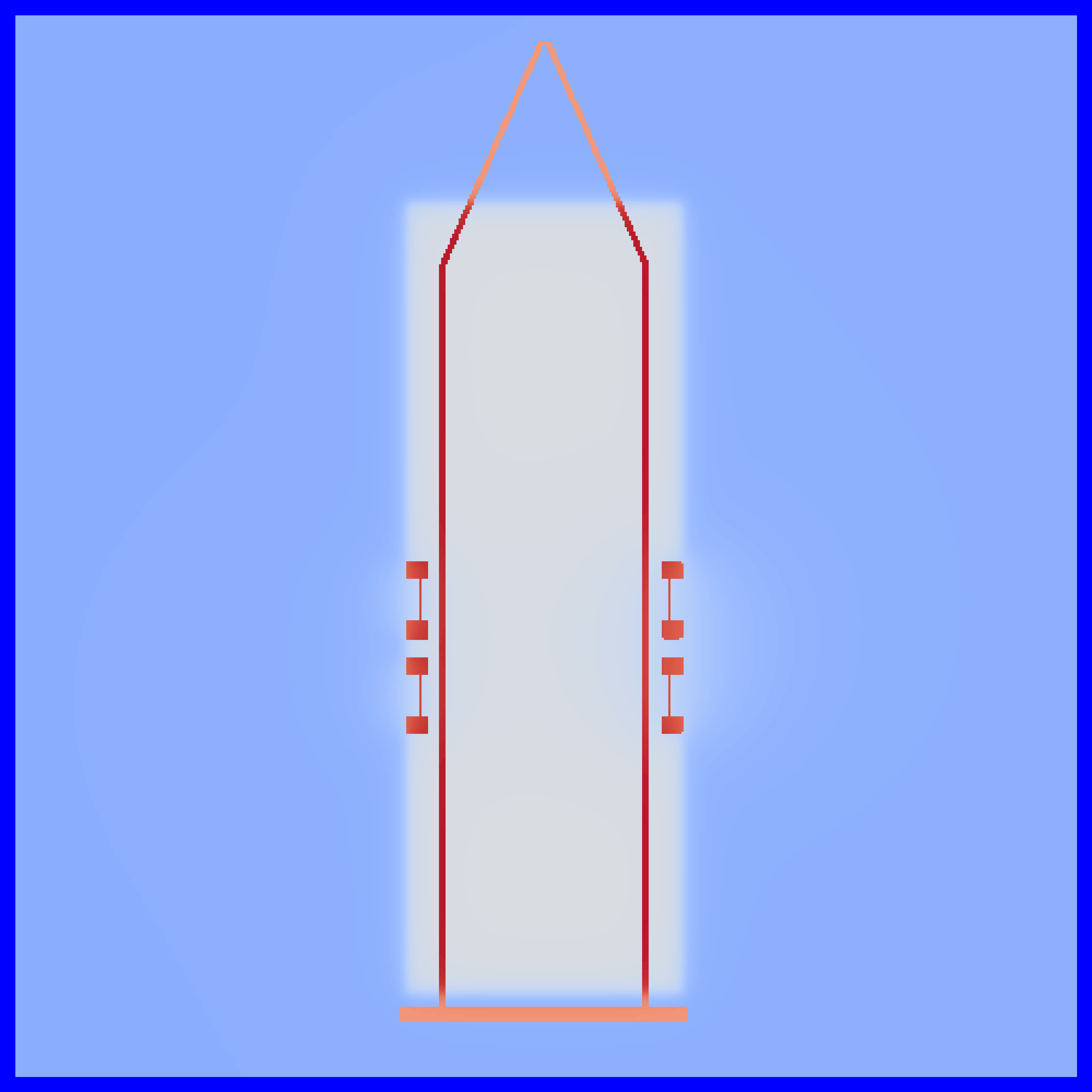} &
        \includegraphics[width=0.18\linewidth]{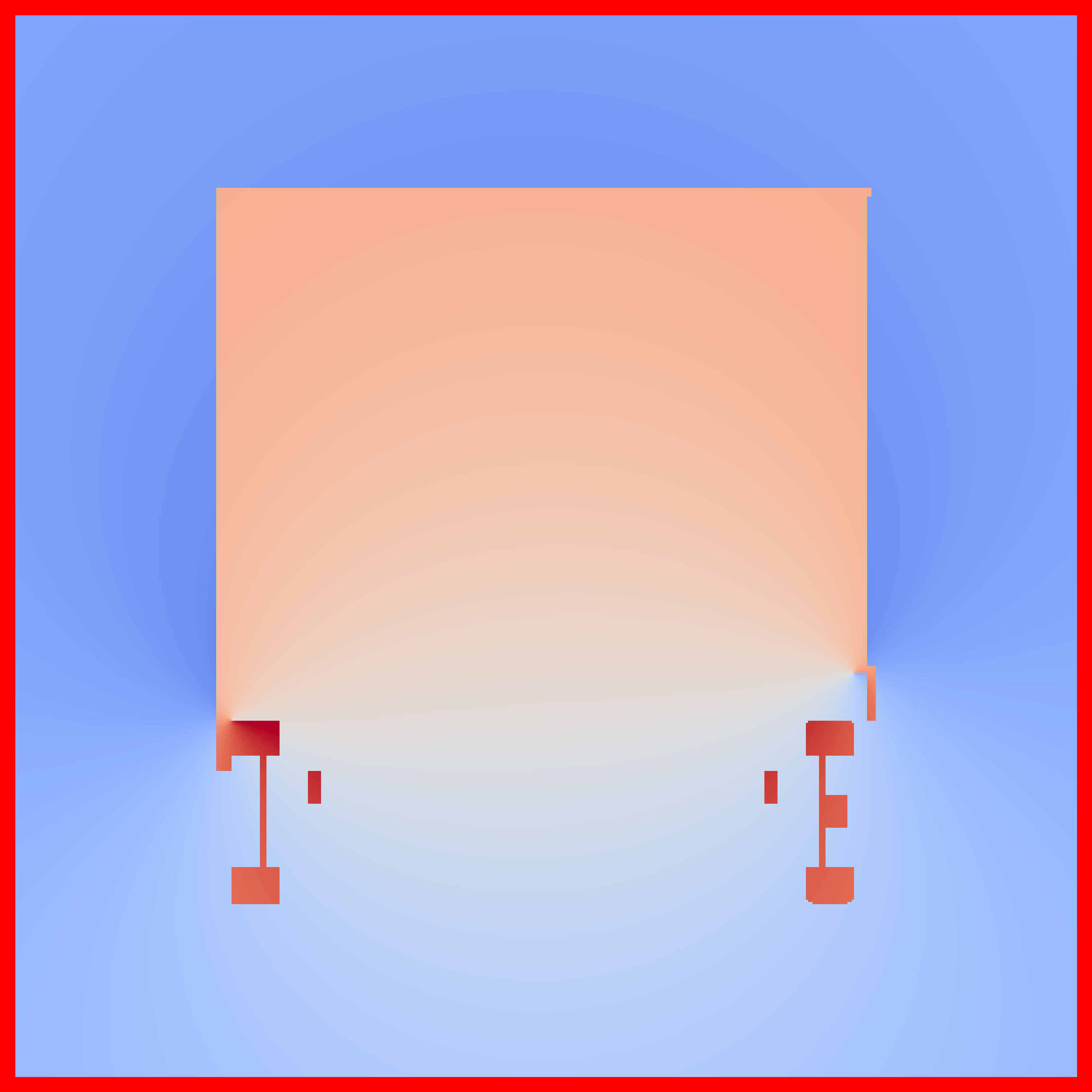} & 
        \includegraphics[width=0.18\linewidth]{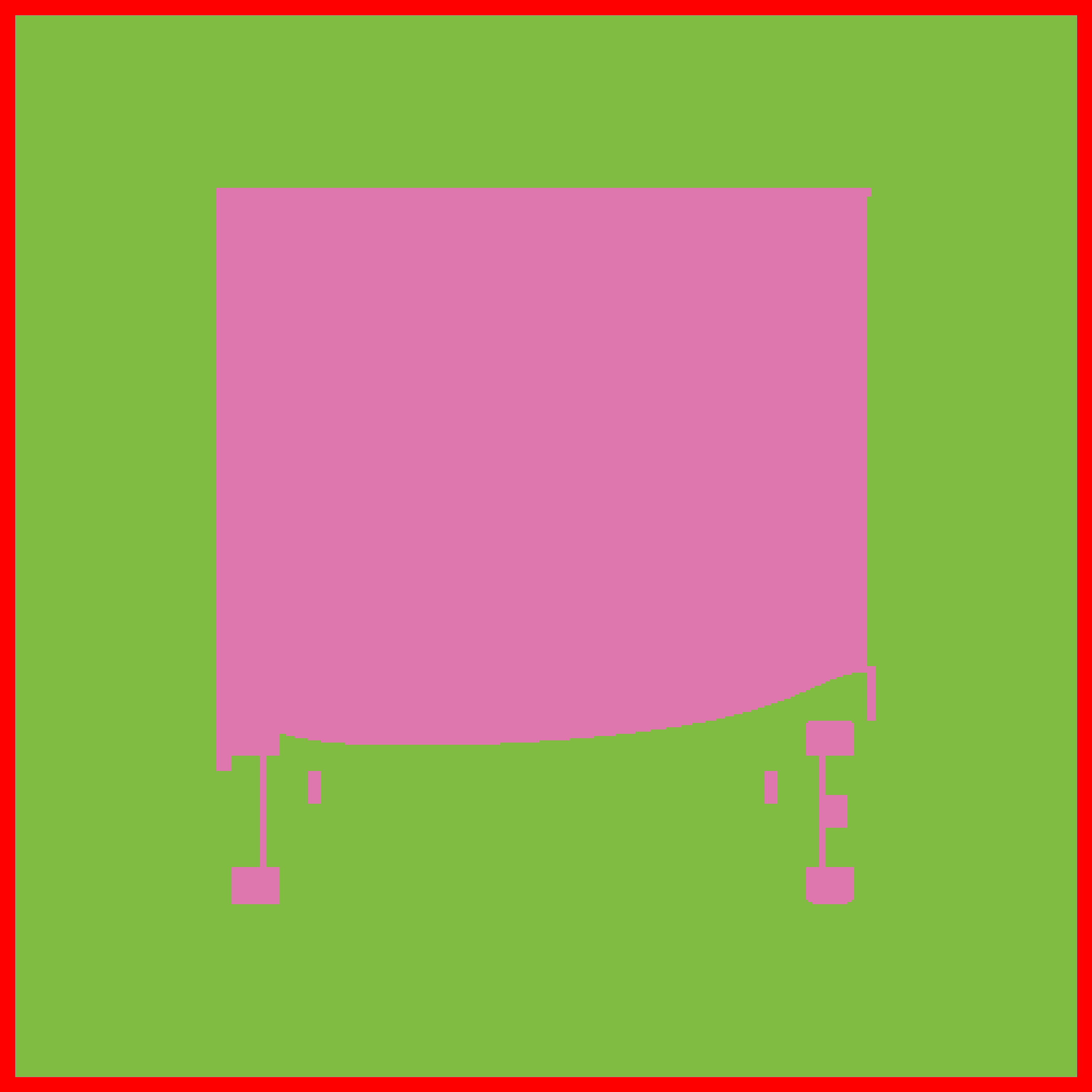} \\
        \includegraphics[width=0.18\linewidth]{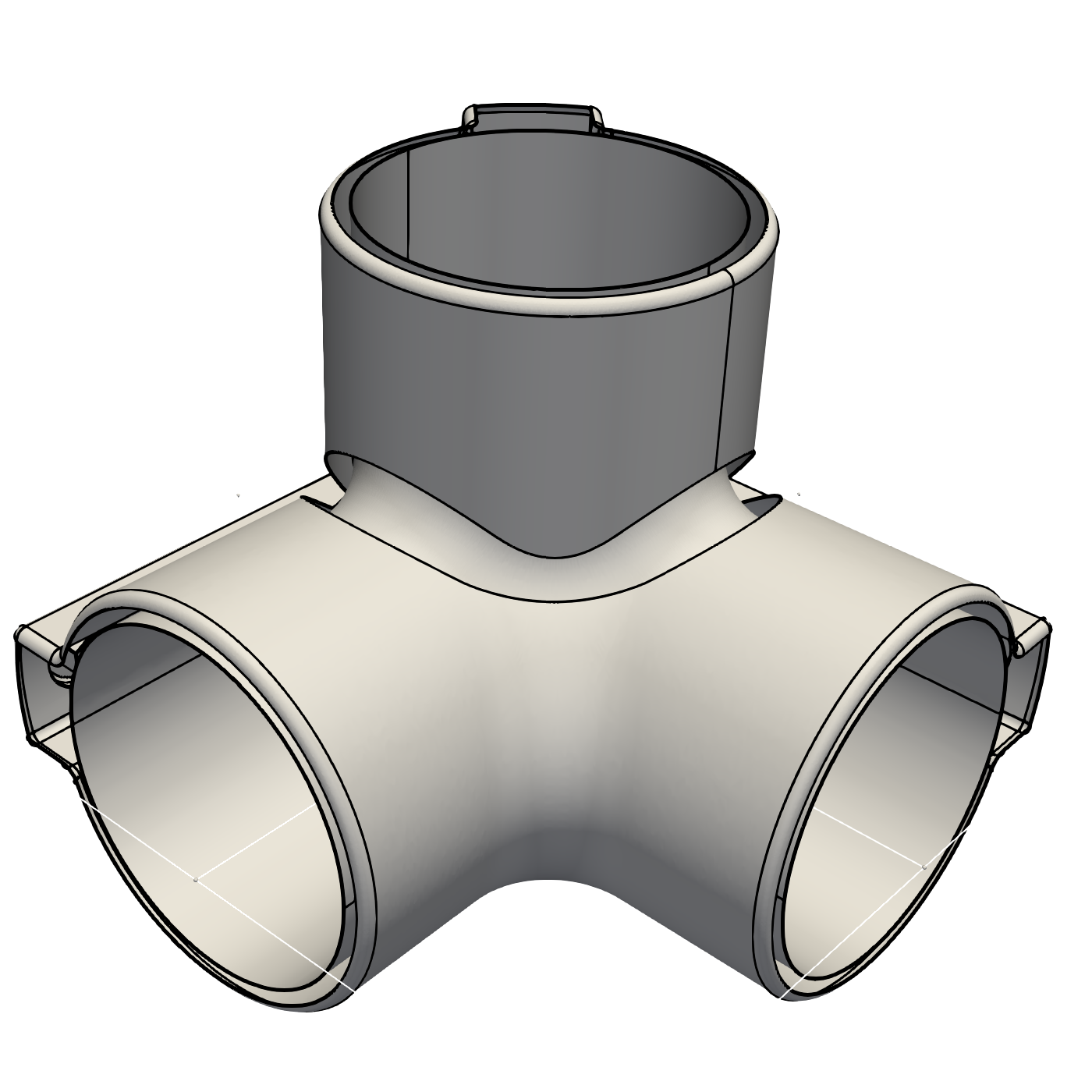} & 
        \includegraphics[width=0.18\linewidth]{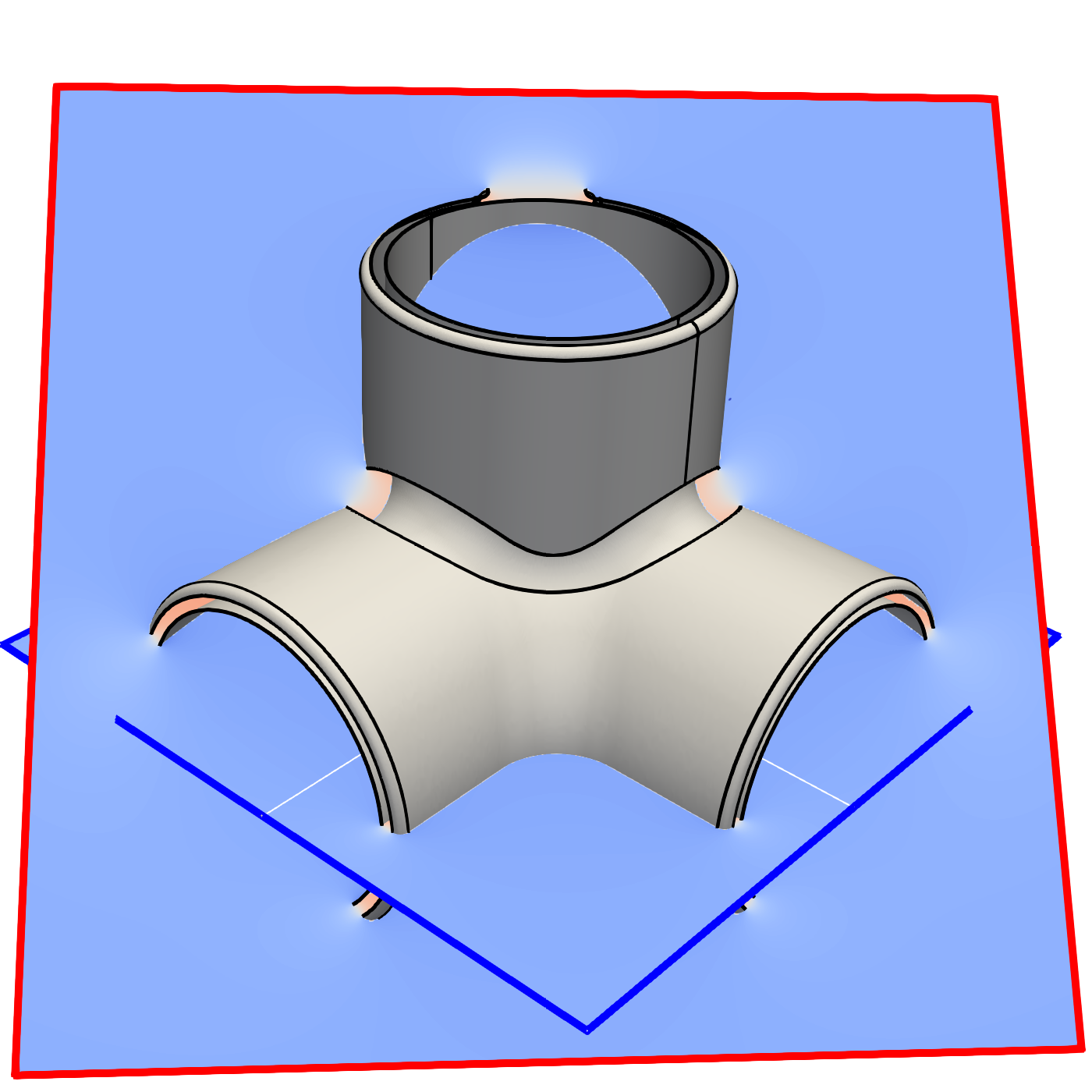} & 
        \includegraphics[width=0.18\linewidth]{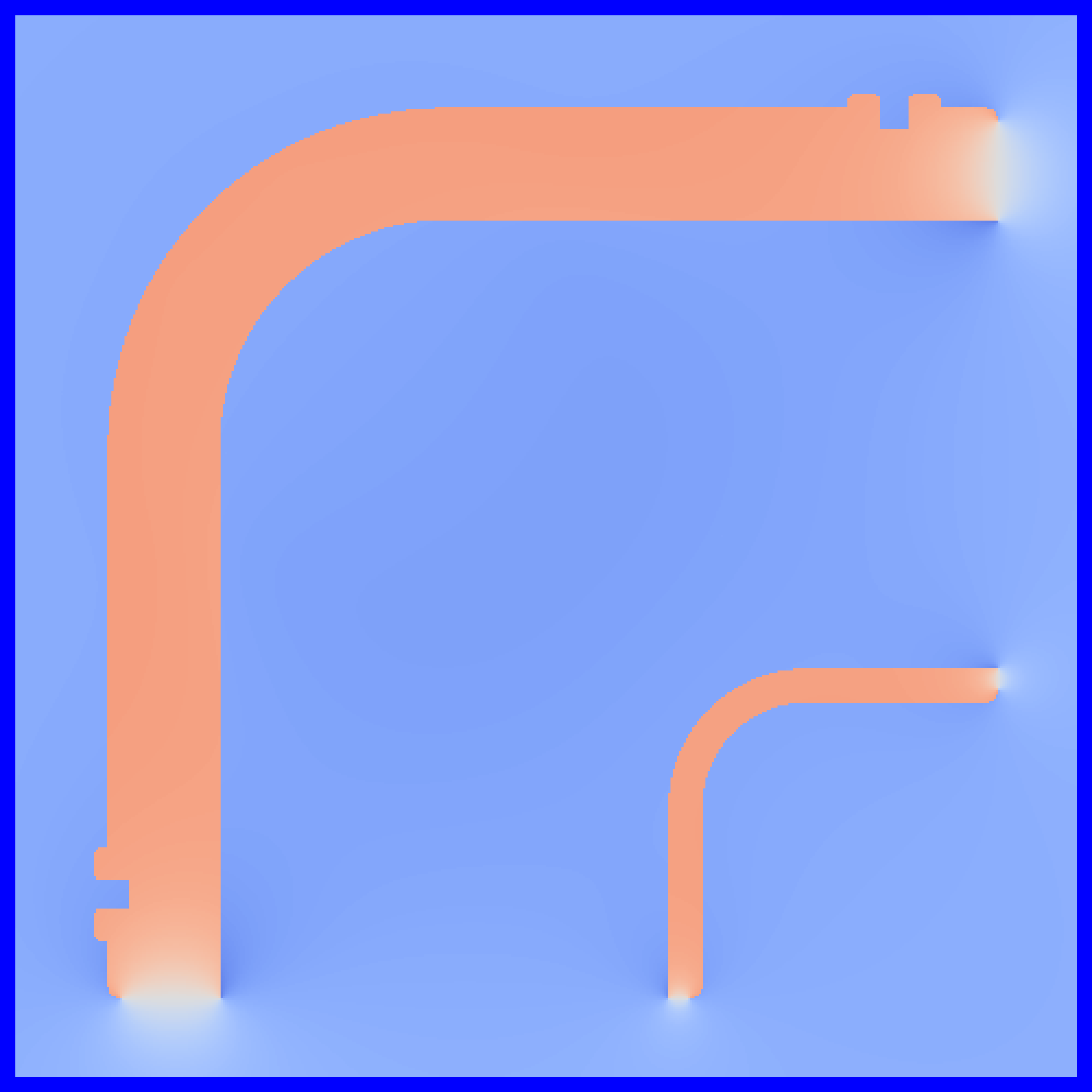} & 
        \includegraphics[width=0.18\linewidth]{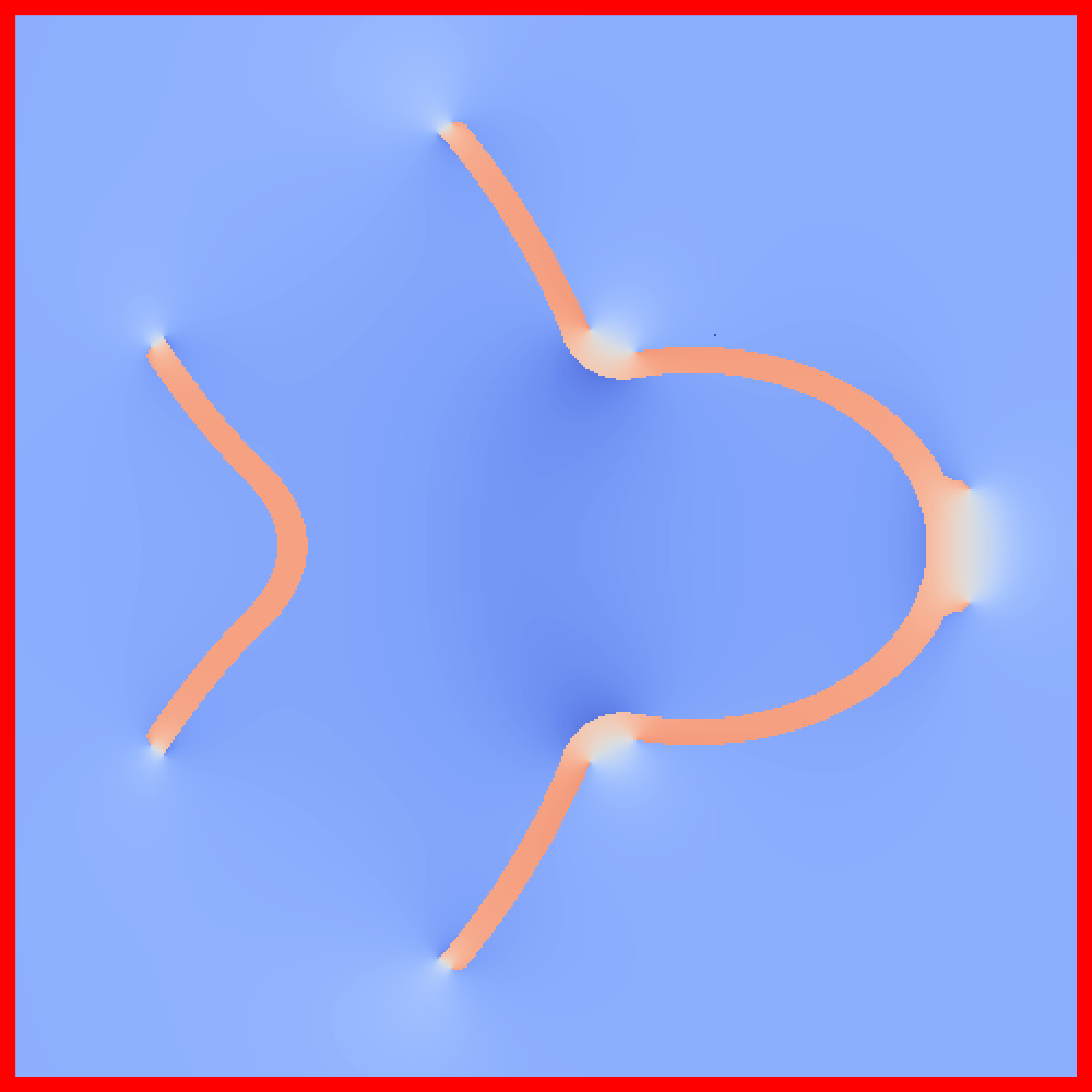} & 
        \includegraphics[width=0.18\linewidth]{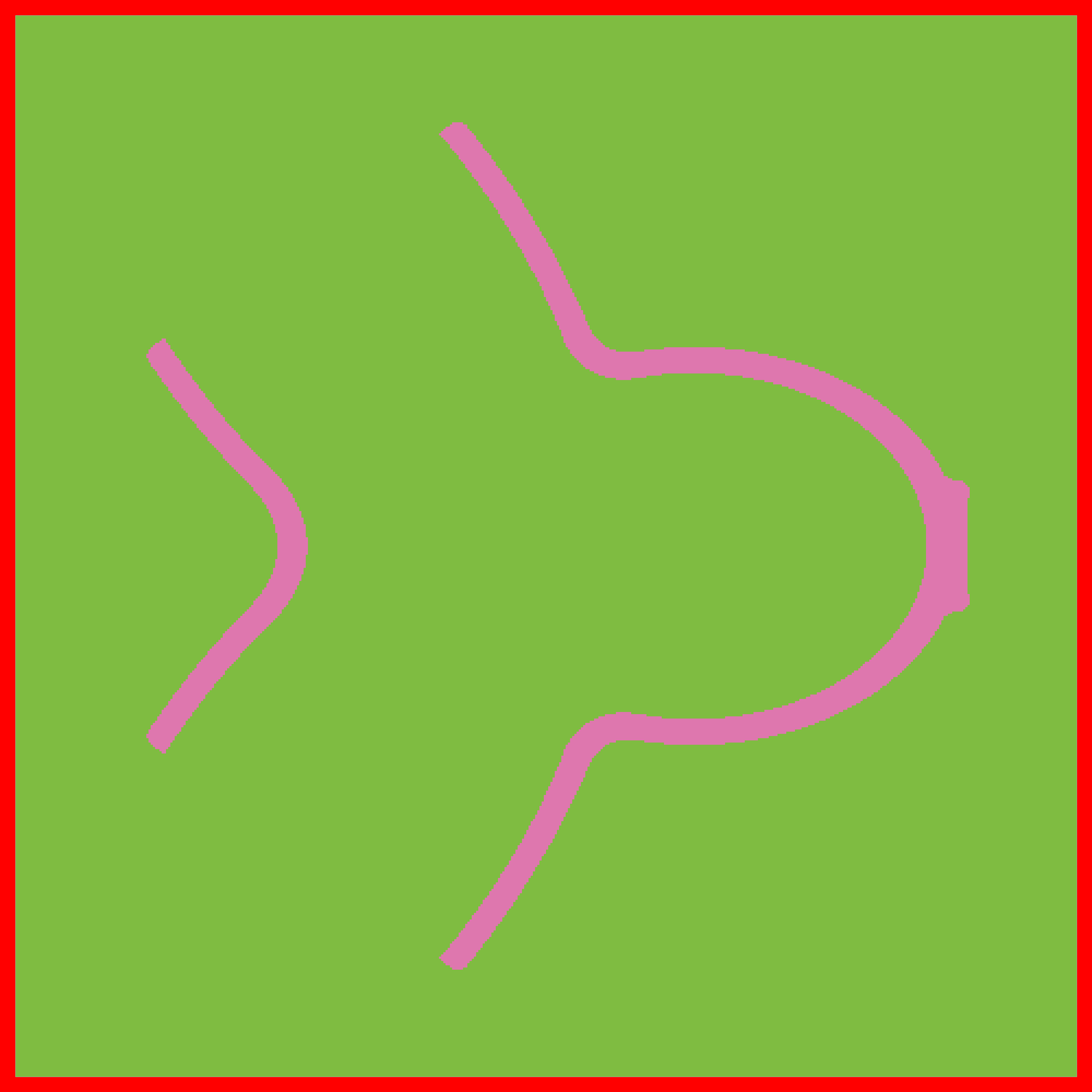} \\
        & & \includegraphics[width=0.18\linewidth]{figures/colorbars/gwn_colorbar_15.png} & & \includegraphics[width=0.18\linewidth]{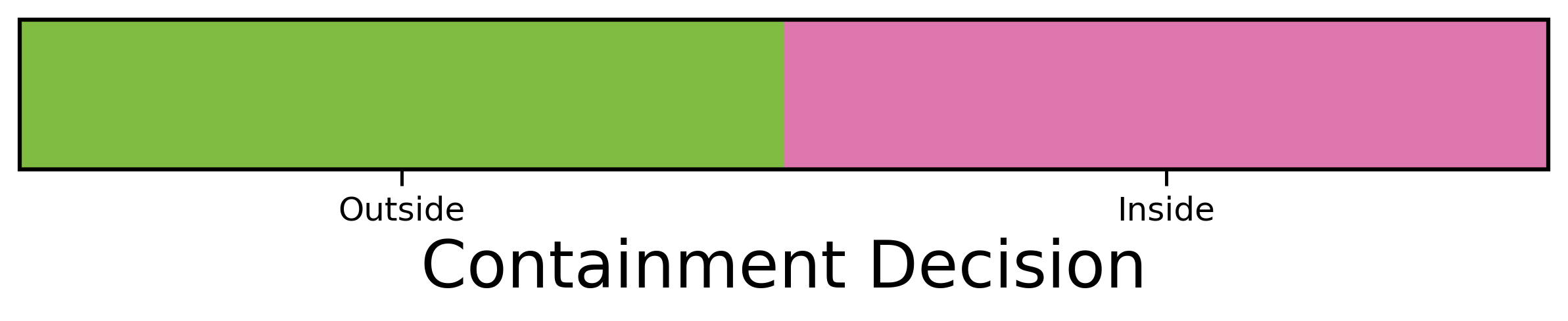}
    \end{tabular}
    \caption[
        Demonstration of 3D GWN field on CAD shapes.]{We demonstrate our calculation of the GWN field on non-watertight CAD models of varying size and complexity. 
        Each field is viewed through two 2D slices through the shape, the second of which is also mapped to a containment decision through a non-zero rule.}
\Description{Demonstration of 3D GWN field on non-watertight CAD shapes.}
        \label{fig:simple_results}
\end{figure*}

We now consider the full implementation of our algorithm, which solves Equation~\ref{eqn:adjusted_stokes} at arbitrary points up to arbitrary accuracy with a more sophisticated, error-controlled adaptive quadrature scheme.
In Figure~\ref{fig:simple_results}, we demonstrate the versatility of our algorithm by evaluating the GWN on several models adapted from the ABC dataset~\cite{koch-19-abcdataset}, a collection of CAD models of varying quality used to benchmark algorithms in computer graphics and machine learning.
The specific models used in this work can be found in~\cite{Axom-data}.
In order to exaggerate the effects of the GWN field on open 3D shapes, we have intentionally made each of these shapes non-watertight through the deletion of several surfaces.
We present qualitative results on additional watertight and non-watertight CAD models in the supplemental materials to this manuscript.

In general, a GWN-derived containment query is most useful in contexts where the objective is to faithfully represent input geometry, as opposed to when there is an unknown, ground-truth shape for which the provided boundaries are a subset.
The GWN field itself is uniquely determined by the set of input surfaces, and so it cannot correct the geometric errors in a surface or its trimming curves, which are themselves assumed to be inextricably defined within the input.
However, the subsequent containment query remains robust to such errors, as the containment result is not meaningfully changed outside a small region around the error.

This is demonstrated in Figure~\ref{fig:simple_results}, where, for each example, the GWN field neatly partitions the domain into exterior and interior regions far from open boundaries, while providing reasonable containment decisions elsewhere.
We can see this directly in the right-most column of Figure~\ref{fig:simple_results}, which shows how rounding the scalar GWN field accounts for small errors in the surface by defining an implicit boundary through the $0.5$-isosurface. 

To consider the accuracy of our method more closely, we introduce possible adaptations of existing GWN techniques on discretized surfaces to trimmed NURBS geometry.
First, we derive from the work of~\citet{Jacobson-13-winding} a strategy which triangulates the (possibly non-watertight) surface with a given level of linear deflection 
(as exposed through Open Cascade's \texttt{BRepMesh\_IncrementalMesh} triangulator~\cite{opencascade}) and evaluates the GWN field for the triangulated result.
Second, we derive from the work of~\citet{Barill-18-soupcloud} a strategy which takes a given number of samples from the surface of the shape model, along with the unit normal at these points, and evaluates the GWN field for the resulting point cloud.
For each method, we conservatively define a containment classification according to a 0.5-isosurface, where any point whose GWN value is greater than 0.5 is classified as interior, and any point whose GWN is less than 0.5 is classified as exterior.

\begin{figure*}[t]
    \centering
    \begin{tabular}{c}
    \includegraphics[width=0.7\linewidth]{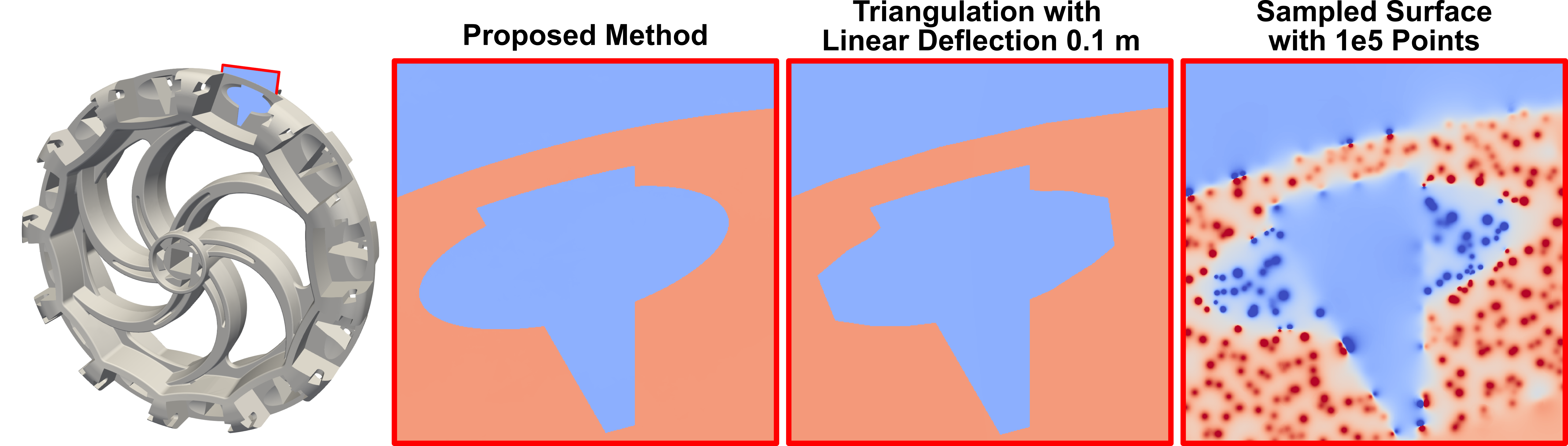}\\\\\cmidrule{1-1}\\
    \includegraphics[width=1.0\linewidth]{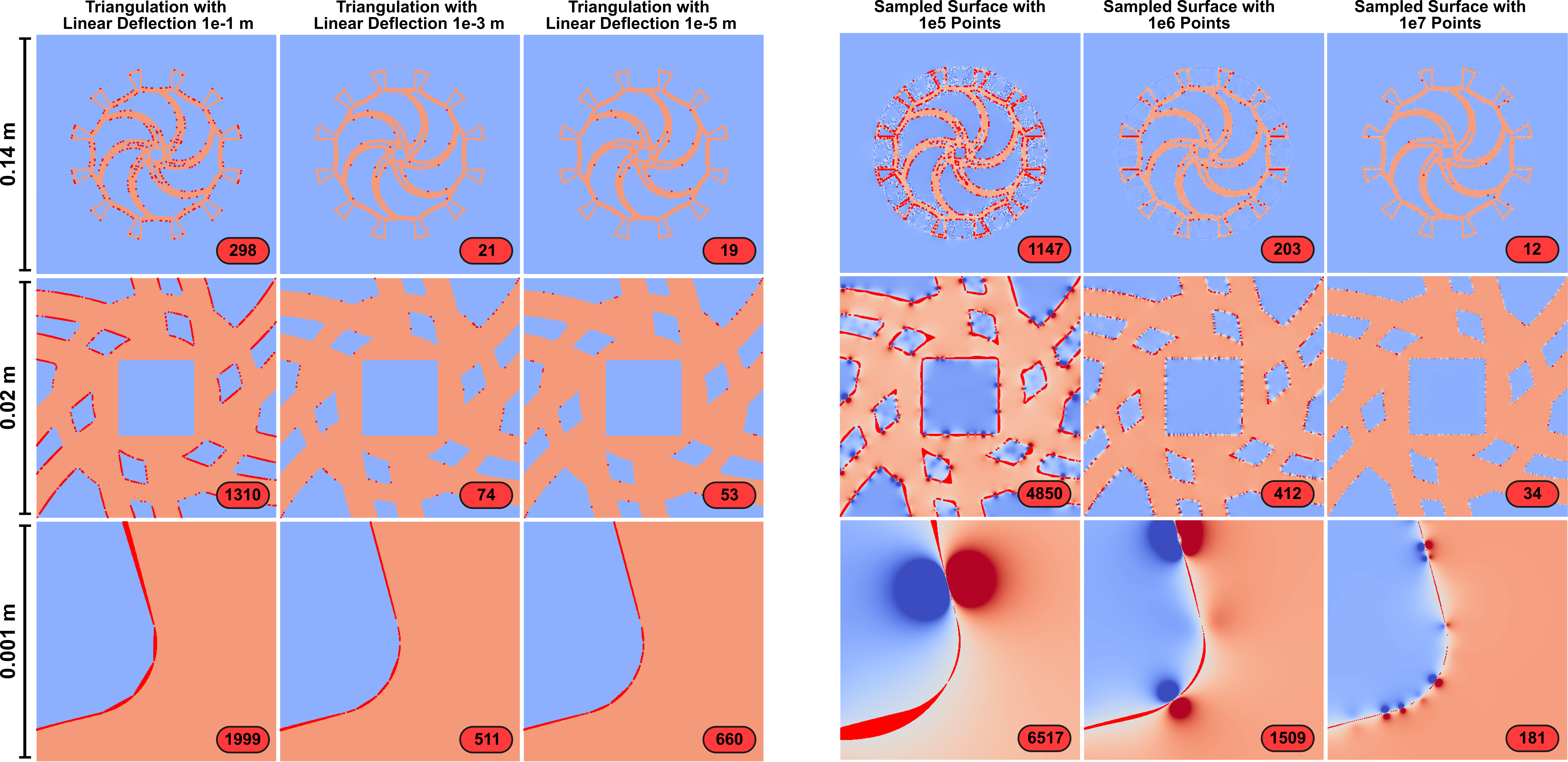}\\
    \includegraphics[width=0.25\linewidth]{figures/colorbars/gwn_colorbar_15.png}
    \end{tabular}
    \caption[Evaluation of GWN field on shape with complex geometry.]{
      Comparing our direct GWN method to discretized alternatives derived from triangulations~\cite{Jacobson-13-winding} and point clouds sampled from the surface~\cite{Barill-18-soupcloud} on a watertight ``Gear'' CAD model.
    To demonstrate the type of error that occurs with each method, we consider each with a relatively low level of discretization (top).
    At increasingly narrow 500$\times$500 slices of the shape, we directly mark in red the points which are misclassified according to a 0.5-surface (bottom).
    Increasing the resolution of the discretization (typically) decreases the number of misclassified points, but at increasing computational cost. 
    In contrast, our proposed method accurately classifies each of these points directly on the CAD geometry's trimmed surfaces.
    }
    \Description[GWN evaluation for proposed method vs. triangulation vs. point cloud sampling]{GWN evaluation for proposed method vs. triangulation vs. point cloud sampling, seen on three increasingly zoomed in views of a complex, watertight CAD shape.}
    \label{fig:zoomed}
\end{figure*}

In Figure~\ref{fig:zoomed} we compare our proposed algorithm to the above adaptations of conventional alternatives on subsets of a watertight ``Gear'' shape, with each view considered on a structured grid of 500$\times$500 sample points.
In Figure~\ref{fig:zoomed} (top), we consider the shape when it is coarsely, yet still meaningfully, discretized according to each method to see the primary characteristics of the resulting GWN field.
For the triangulation method, we see, as expected, that curved surfaces are made linear, resulting in unpredictable errors near the surface boundary while still producing integer-valued winding numbers for this watertight model.

In contrast, while the 0.5-isosurface of the scalar field provided by the point cloud method generally captures the curved features, the GWN field in its totality diverges significantly from the integer-valued ground truth.
Although the dipoles correctly place nearby query points on the correct side of the curved surface, their large magnitude results in scalar GWN values outside $[-0.5, 1.5]$.
This would lead to further misclassifications if containment were determined with the conventional strategy of rounding, followed by an even-odd or non-zero rule, 
which might be necessary on more complex shapes with nested surfaces.

To see this error more quantitatively, we record in Figure~\ref{fig:zoomed} (bottom) the total number of misclassifications on increasingly narrow views of the shape using the triangulation approach (left) and the point cloud approach (right), with each misclassified point in each 500$\times$500 grid marked in red.
While the number of misclassifications for each discretization strategy (generally) decreases as the resolution of the discretization increases, there is no level at which no misclassifications occur. 
Further, although the linear deflection parameter controls how far the triangulation can differ from the original shape, in effect ensuring that no points within a fixed absolute distance of the surface will be misclassified, it can be difficult to know \textit{a priori} what level of discretization is necessary for correct classification at user-specified query points.
This is well-illustrated in the narrowest view, in which the increased deflection parameter from $10^{-3}$ m to $10^{-5}$ m does not necessarily increase the accuracy of the method, as all points which are misclassified are by coincidence still closer than $10^{-5}$ m to the surface. 
This relationship is further obfuscated in the case of a point cloud representation, where it is similarly difficult to gauge the appropriate sampling density for a point cloud that will lead to accurate classifications. 
In either case, this can lead to unnecessary computational expense in the form of globally dense triangulations or sampling procedures. 
In contrast, operating directly on the trimmed surface allows our method to perform adaptively for each query to the requested level of accuracy and correctly classify the containment of all query points.

\begin{table*}[t]
    \centering
    \begin{tabular}{lrrrrr}
              & \includegraphics[trim={0 8cm 0 8cm},clip,width=0.08\linewidth]{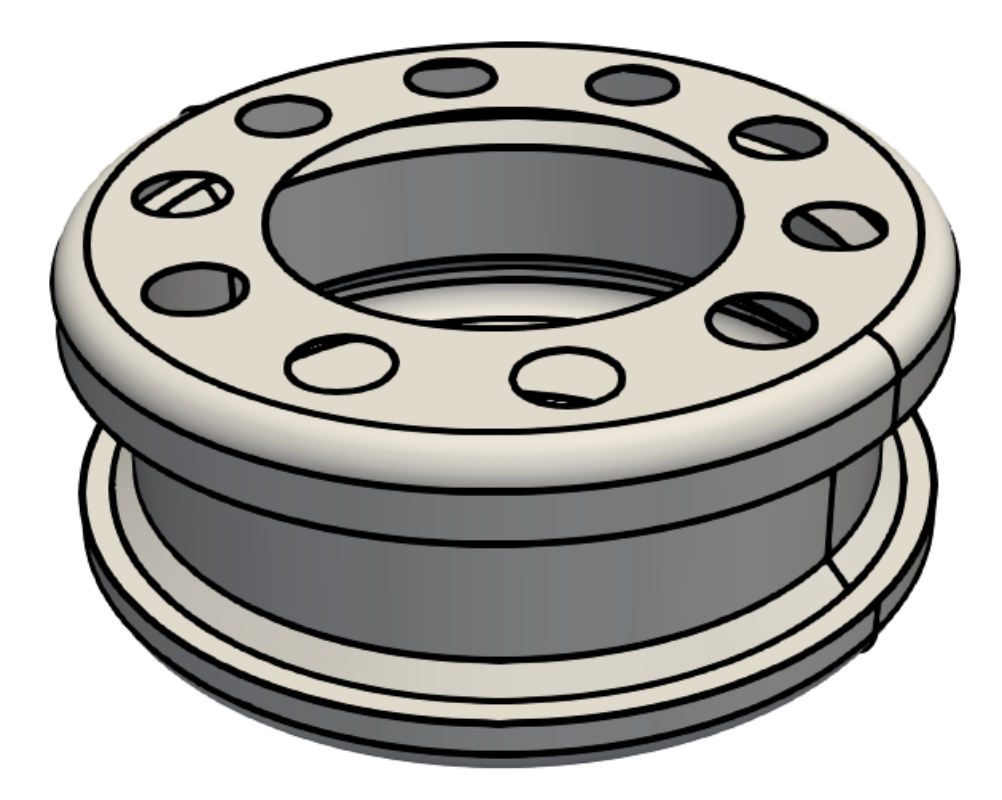} 
              & \includegraphics[trim={0 10cm 0 10cm},clip,width=0.12\linewidth]{figures/trailer/trailer.png} 
              & \includegraphics[width=0.07\linewidth]{figures/pipe/pipe.png}
              & \includegraphics[width=0.07\linewidth]{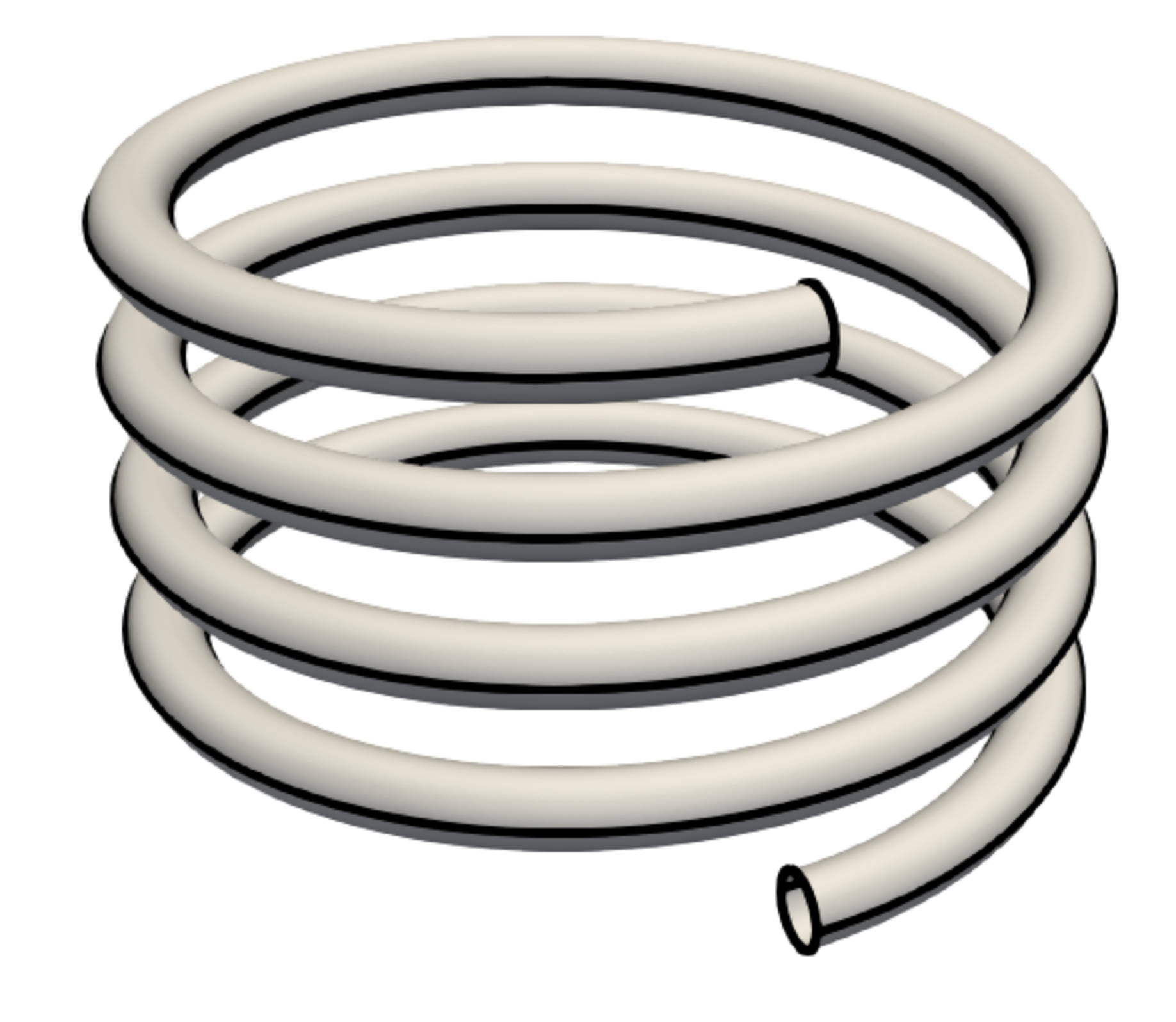}
              & \includegraphics[width=0.07\linewidth]{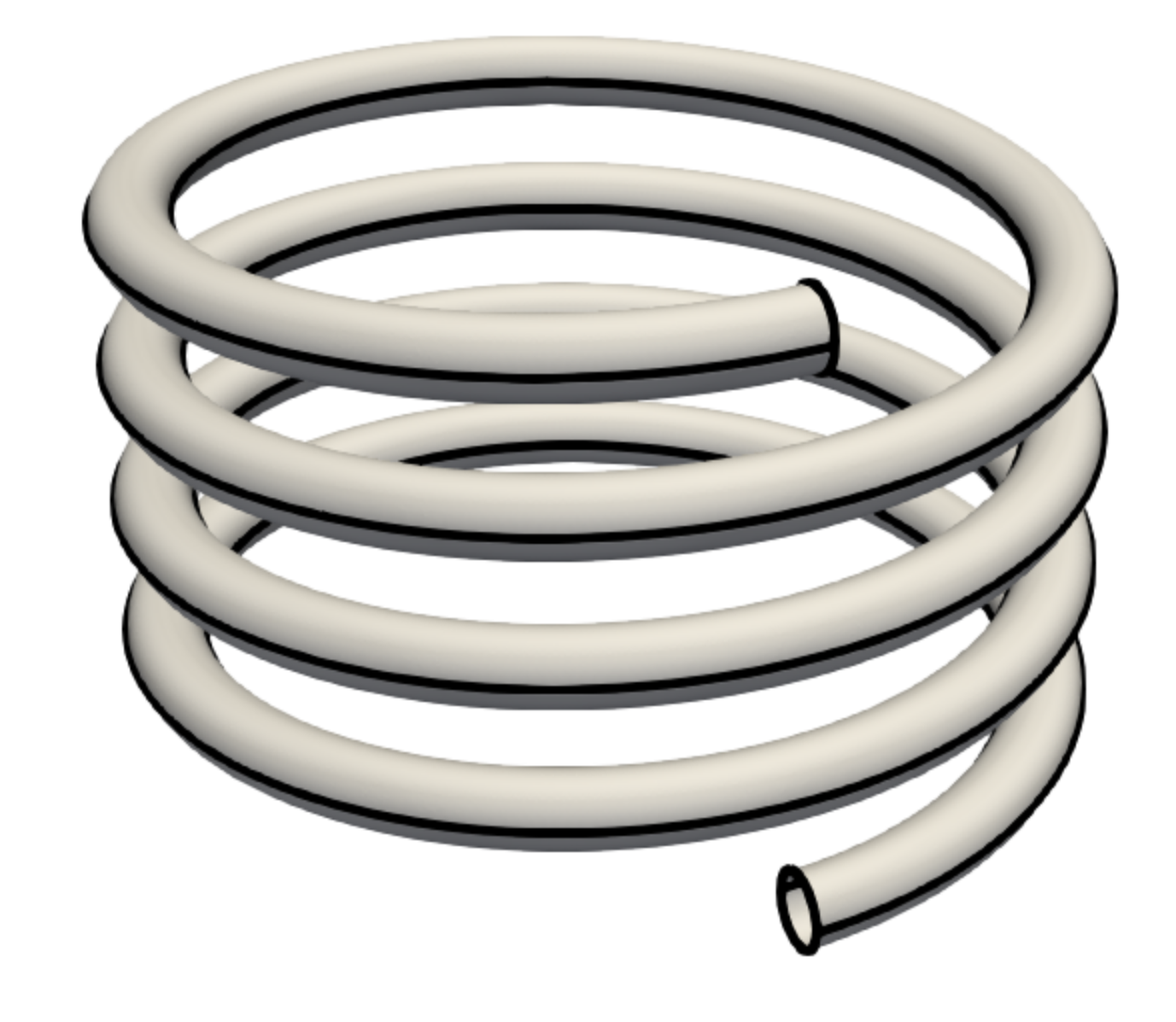}\\
        Shape & Bobbin & Trailer & Pipe & Spring & \makecell{Spring,\\Two-Patch} \\\cmidrule{1-6}
        Number of NURBS Patches         & 18        & 228      & 133     & 64       & 2        \\
        Number of Trimming Curves       & 76        & 1092     & 563     & 256      & 8        \\\cmidrule{1-6}
        \% Far-field Cases              & 92.56  \% & 99.8  \% & 99.02\% & 99.57 \% & 18.26 \% \\
        \% Near-field Cases             & 7.00 \%   & 0.116 \% & 0.81 \% & 0.42  \% & 76.15 \% \\
        \% Edge Cases                   & 0.42 \%   & 0.014 \% & 0.15 \% & 0.006 \% & 5.59  \% \\\cmidrule{1-6}
        Avg. Time per Query (ms)        & 1.11      & 1.18     & 21.1    & 0.332    & 11.07    \\\cmidrule{1-6}
        Avg. Far-field Case Time (ms)   & 0.0173    & 0.0046   & 0.0313  & 0.0043   & 0.0718   \\
        Avg. Near-field Case Time (ms)  & 0.0805    & 0.0585   & 0.916   & 0.1301   & 0.336    \\
        Avg. Edge Case Time (ms)        & 9.38      & 3.97     & 76.7    & 5.55     & 94.3     \\\bottomrule
    \end{tabular}
    \caption[Comparative performance of GWN algorithm on various open CAD models.]{
      We evaluate the GWN field at a uniform grid of query points for each shape, and report the average time spent to resolve each query. 
      We also show the number of query points that are treated by each of the three cases and the average time needed to resolve each of these cases.}
		\label{tab:bulk_timing_results}
\end{table*}

While the approximation accuracy with any of these discretizations may be tolerated within a particular application, the computational cost of generating and storing increasingly refined discretizations grows exponentially.
For example, the Gear shape has many flat regions which are trivial to triangulate, but still has 54k triangles at deflection $0.1$ m, increasing to 2 million at deflection $0.001$ m, and nearly 10 million at deflection $10^{-5}$ m.
The cost of constructing and accessing the spatial index through which these shapes are efficiently queried is similarly increasingly burdensome.
Indeed, applying the method of~\citet{Barill-18-soupcloud} in Figure~\ref{fig:zoomed} (bottom) to a cloud of $10^7$ points on our computing platform with 32 GB RAM using the \texttt{libigl} implementation~\cite{libigl-2018} required reducing the expansion order for clusters of points from $2$ to $1$, as it was otherwise not possible to store all 27 terms of the 2nd order expansion at each level in the spatial index.  
In contrast, our method is able to accurately classify the entire collection of points without the need for an expensive intermediate discretization, using only information provided directly from the 11MB STEP file.
\subsection{Performance Evaluation}

We now demonstrate the performance of Algorithm~\ref{alg:generalized_winding_number}. 
The following numerical experiments were conducted on an AMD Ryzen 7 5700 CPU with 32GB of RAM.
These results are computed serially, although we note that the evaluation of the GWN field at many points is a highly parallelizable task.

Although a triangulated B-Rep is composed of significantly more primitive components than the original collection of surfaces, 
methods that operate on CAD geometry are typically more expensive than equivalent operations on STL data types.
However, our use of cached per-curve quadrature nodes allows us to evaluate the GWN field at a practical rate, even without the use of more sophisticated hierarchical data structures, the implementation of which we largely consider to be future work.

In Table~\ref{tab:bulk_timing_results}, we show the average time needed to evaluate each query point on a uniform $50\times 50\times 50$ grid of points located in a bounding box of each shape, as well as a per-query average.
We also report the average time needed to evaluate the GWN for each surface in the shape (overall and by case), alongside the breakdown of all point-surface GWN calculations by case.
This table highlights several key features of the performance of our algorithm.
Most notably, although the total time needed to evaluate the GWN field in principle scales linearly with the number of trimming curves, there are many other factors which impact runtime.
For example, the overall distribution between point-surface case configurations significantly affects the overall runtime.
Naturally, far-field cases are the easiest to resolve, and so shapes for which most query points can be classified as far-field for most surfaces take the least amount of time to process. 
This is well-demonstrated by the ``Bobbin'' shape taking more time on average than the ``Spring'' case, despite the latter having many more surfaces and trimming curves.

Other differences come from more general characteristics of the component NURBS patches, which are best illustrated through comparison of the ``Trailer'' and ``Pipe'' examples.
Although the Trailer shape has more surfaces, the NURBS patches themselves are predominantly flat with simple trimming curves.
In contrast, the Pipe shape is more complicated, with many of its component NURBS patches being large spheres and cylinders which are trimmed and filleted.
Thus, while the times to process the individual far- and near-field cases are comparable, resolving the edge-cases of the Pipe example is more challenging, and therefore more expensive. 
We present similar timing tests on other watertight CAD shapes from the ABC dataset in the supplemental materials to this manuscript.

As a further point of comparison, we can consider a version of the Spring shape defined by only two NURBS surfaces.
Since the two patches share a bounding box, there are few query points that can be classified as far-field.
Of those that remain, the vast majority must be classified as edge cases which cannot benefit from our memoization strategy, an added cost which is further burdened by the resulting boundary curves being long and therefore poorly resolved by the quadrature at the initial refinement levels.
On the other hand, the original shape permits many more points to be classified and efficiently resolved as far-field.
It is worth noting that while reducing the size of each patch shifts the distribution of cases and processing times, our method recovers an identical GWN field for the two shapes
due to the exact subdivision of NURBS surfaces.

\section{Discussion}\label{sec:additional_algorithms}
\subsection{Additional Algorithms}
In this section, we highlight the computational components and tolerance parameters supporting our implementation and numerical experiments.
Although there are some specifications within these methods unique to our application, we expect that alternate approaches which meet these requirements would be suitable.

\subsubsection{Line-Surface Intersections}\label{sec:line_surface_intersections}

Accurately evaluating the near-field sum in Equation~\ref{eqn:adjusted_stokes} requires a robust intersection test between an arbitrary line and an input trimmed NURBS surface.

We use a geometric approach to identify the isoparametric coordinates of all signed intersections between a line and an untrimmed NURBS patch in the parameter space of the patch.
This involves recursively subdividing the patch geometry (either through \bezier\ extraction of NURBS patches or bisection of individual \bezier\ patches) and clipping away  portions of the surface where intersections provably cannot occur.
By maintaining a record of each subdivision patch's placement within the parameter space of the original surface, this method converges linearly to the set of parametric points $(u, v)$ at which the line intersects the untrimmed NURBS patch.

As an early stopping criterion for recursion, we treat patches which are ``approximately bilinear'' (i.e.\ have control points which are nearly coincident with the surface along a uniform grid) directly using the GARP algorithm (Geometric Approach to Ray/bilinear Patch)~\cite{reshetov-19-coolpatches}, which describes a closed-form solution for the intersection of a line with a bilinear patch.
This allows the intersection coordinates \textit{within} a sub-patch to be accurately mapped back to the parameter space of the original surface.

When the parametric coordinates of all intersections with the untrimmed patch are recorded, we supply them to a trim test based on the 2D GWN~\cite{spainhour_24_robustcontainment2d} to filter out intersection points which occur outside the parameter space defined by the trimming curves (see Algorithms~\ref{alg:line_patch_intersect} and~\ref{alg:line_patch_intersect_recursive}).

The use of this algorithm exposes the ``line-surface'' tolerance parameter $\epsilon_{ls}$, which determines when a sub-patch is considered to be approximately bilinear (see Algorithm~\ref{alg:is_approximately_bilinear}).
Naturally, a stricter $\epsilon_{ls}$ means more linear subdivisions are needed to capture the intersections, while a more relaxed $\epsilon_{ls}$ means the $(u, v)$ coordinates recorded by GARP may diverge more from the true points of intersection.
This may result in the query point being placed on the wrong side of the surface, and in turn, an incorrect correction term being applied in Equation~\ref{eqn:adjusted_stokes}.
However, we emphasize that this dynamic bilinear approximation is used \textit{only} to identify the parameter space coordinates of the intersection point, and is subsequently discarded.
All geometric quantities derived from these coordinates, such as the 3D intersection point and surface normal, are computed with respect to the original surface geometry.
In particular, we always compute the GWN with respect to the true surface boundary 
without further reference to this or any other approximation. 

\begin{figure}[tb]
    \centering
    \begin{tabular}{cc}
        \includegraphics[width=0.3\linewidth]{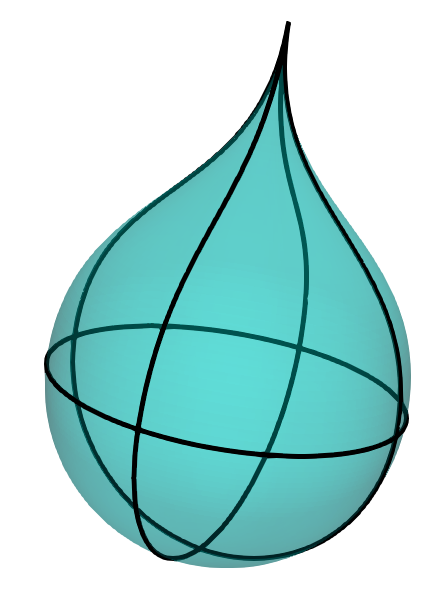} &
        \includegraphics[width=0.65\linewidth]{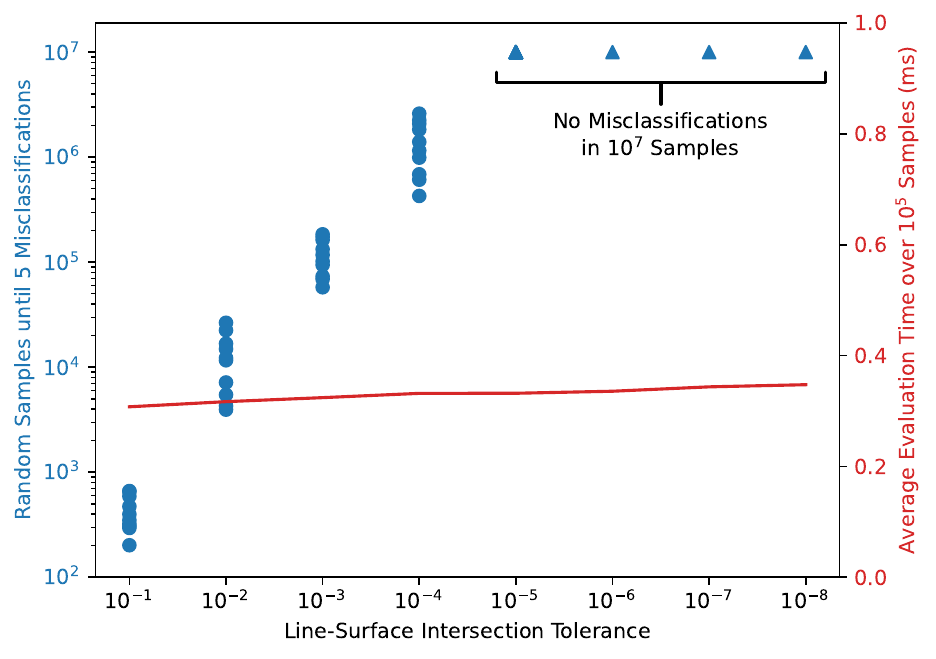}
    \end{tabular}
    \caption{Sensitivity study for the line-surface intersection threshold $\epsilon_{ls}$, evaluating the GWN on query points randomly sampled from an axis-aligned bounding box until 5 misclassifications are recorded.
    Beyond a certain threshold, the line-surface intersection routine is accurate enough to ensure there are no misclassifications in over $10^7$ sampled points, while the average run-time of the algorithm only slightly increases. 
    }
    \Description[Sensitivity study for line-surface intersection tolerance]{Example shape and line plot showing the sensitivity of accuracy to the tolerance for the line-surface intersection routine. As the tolerance decreases logarithmically, the number of randomly sampled points evaluated before any are misclassified increases to over ten million. The average evaluation cost increases slowly. }
    \label{fig:tolerances_ls}
\end{figure}

Within the context of the broader GWN algorithm, we further improve the robustness of the line-surface intersection routine by selecting a line for each configuration of surface and query point that is orthogonal or nearly-orthogonal to the surface at the point of intersection.
Doing so avoids some of the difficulties of identifying tangent or near-tangent points of intersection.
Directly identifying such a line would require additional processing, such as a closest point query, the cost of which may exceed that of the GWN calculation.
In this work, we heuristically select the line in the direction of an average surface normal~\cite{ueda-1996-meannormal}, which we have found to be very effective for the majority of B-Reps.

To explore the practical tradeoffs of the value for $\epsilon_{ls}$, we evaluate the GWN with respect to a challenging ``teardrop'' shape in Figure~\ref{fig:tolerances_ls}.
The bottom of this shape is spherical, but is composed of 4 degenerate bicubic patches rather than the biquartic patches used in the example of Figure~\ref{fig:quadrature_slices}. 
The top shape is generated by revolving a polynomial cubic \bezier\ curve around the $z$-axis, providing inflection points which pose additional challenge for the line-surface intersection test.
Around this shape, we randomly sample points using a uniform distribution within an axis-aligned bounding box and record if rounding the GWN at that point results in a containment misclassification according to an exact analytic containment formula.
In this example, we vary $\epsilon_{ls}$ and use a fixed adaptive quadrature tolerance ($\epsilon_q = 10^{-6}$) and extraction disk radius (1\% of the parameter space bounding box diagonal).

Because the probability that non-adversarially selected points are misclassified is quite low, particularly for strict values of $\epsilon_{ls}$, we continuously sample from this distribution until 5 points are misclassified.
We also evaluate the average runtime across a separate set of $10^5$ sample points, which is sufficiently many to effectively utilize cached quadrature nodes, allowing the computational cost to be amortized over the sample set.

As expected, the number of misclassified points rapidly decreases as the tolerance is made more strict, to a point where the likelihood of failure is less than $0.00001\%$.
On the other hand, the associated cost of the algorithm increases as more surface subdivisions are needed to evaluate intersection points, although this burden is marginal relative to the total cost of the algorithm. 
As a result of this experiment, and our other numerical examples, we consider a choice of $\epsilon_{ls} = 10^{-6}$ to offer a reasonable tradeoff between computational efficiency and accuracy.

\subsubsection{Adaptive Quadrature and Disk Radius}\label{sec:adaptive_quadrature_performance}

\begin{figure*}[tb]
    \centering
    \begin{tabular}{ccc}
        \includegraphics[width=0.3\linewidth]{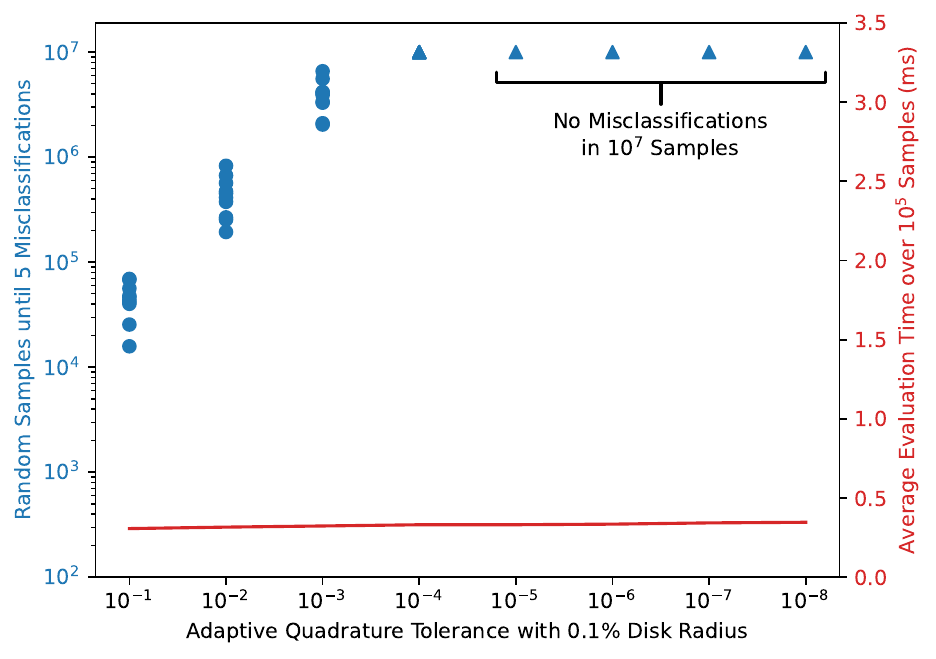} &
        \includegraphics[width=0.3\linewidth]{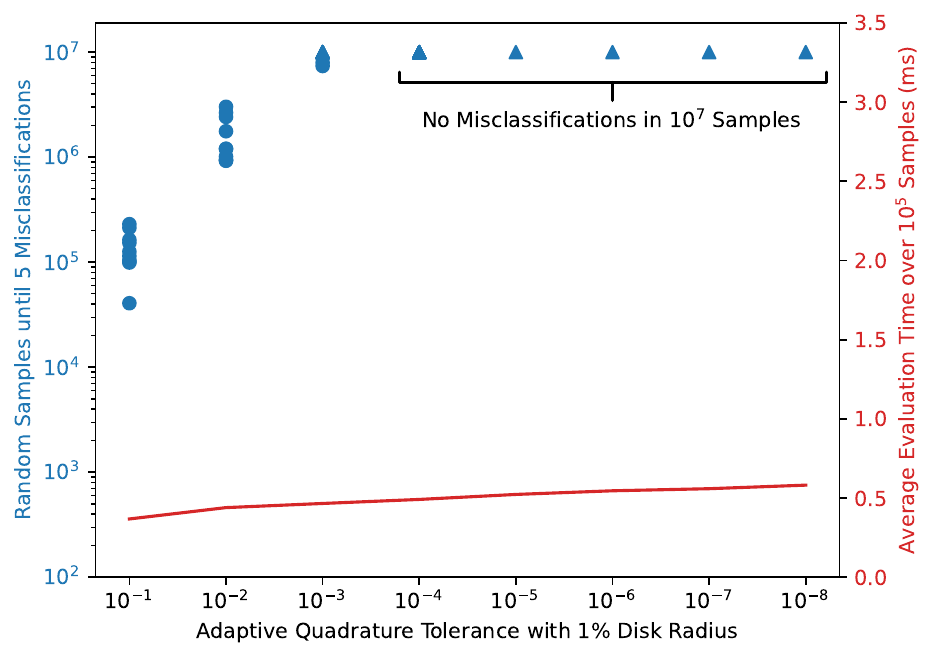} &
        \includegraphics[width=0.3\linewidth]{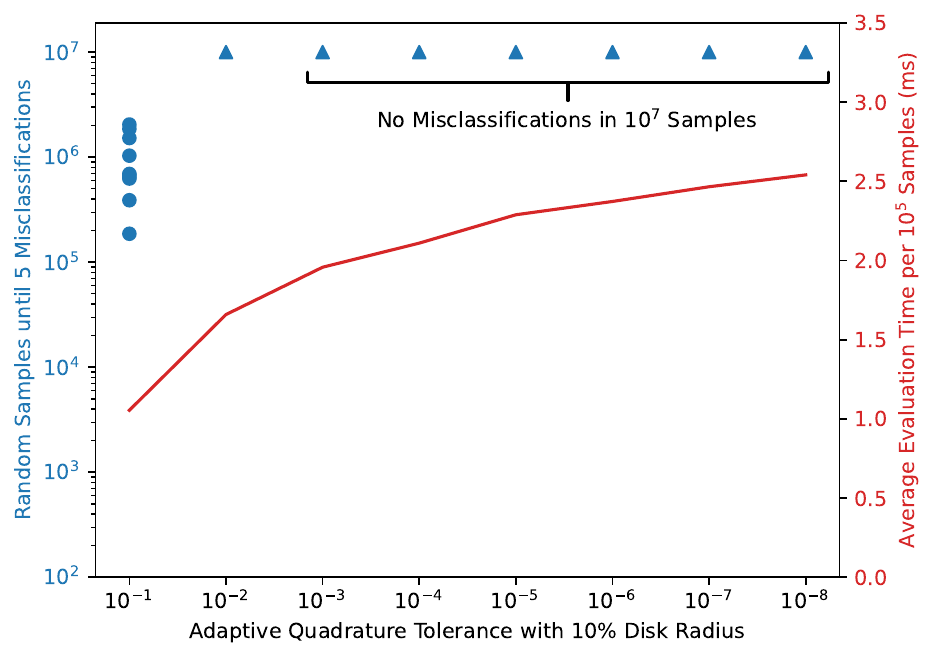}
    \end{tabular}
    \caption{Sensitivity study for extracted disk radius and quadrature tolerance parameter $\epsilon_{q}$, evaluating the GWN on query points randomly sampled from an axis-aligned bounding box until 5 misclassifications are recorded.
    Beyond a certain threshold, regardless of radius, the adaptive quadrature routine is accurate enough to ensure there are no misclassifications in over $10^7$ sampled points, 
    although there is a tradeoff between radius size and quadrature costs.}
    \label{fig:tolerances_quad}
    \Description[Sensitivity study for adaptive quadrature tolerance]{Three line plots showing the sensitivity of accuracy to the tolerance for the adaptive quadrature routine. As the tolerance decreases logarithmically, the number of randomly sampled points evaluated before any are misclassified increases to over ten million, while the cost of the procedure increases. As the disk radius increases from 0.1\% to 1\% to 10\%, the cost increases, but the tolerance at which no misclassifications are observed decreases.}
\end{figure*}

In our current implementation, memoized quadrature nodes cannot be reused if a patch's trimming curves change during evaluation of the GWN.
As a result, there is a strong correlation between overall computational performance and the choice of the disk radius used to identify and resolve near-field and edge-case configurations.
On the one hand, a smaller disk radius reduces the number of edge-cases (thereby improving reuse of cached quadrature nodes), since the disk is less likely to intersect existing trimming curves.
On the other hand, a smaller disk radius allows query points to be closer to the singularity in the boundary curve integrals, necessitating a stricter tolerance for $\epsilon_q$ to ensure consistent accuracy. 

To demonstrate how both the disk radius (measured as a percent of each patch's parameter space bounding box diagonal) and $\epsilon_q$ affects the accuracy and computational efficiency of the algorithm, we repeat the numerical experiment for $\epsilon_{ls}$ in Section~\ref{sec:line_surface_intersections} and report the results in Figure~\ref{fig:tolerances_quad}, varying $\epsilon_q$ during the evaluation of the GWN for disks of various sizes. 
In these examples, we set $\epsilon_{ls} = 10^{-6}$, and further note that $\epsilon_{ls}$ and the disk radius have an independent effect on the method's accuracy.
We see that as the radius decreases, a more strict tolerance for $\epsilon_q$ is needed to ensure there are exceptionally few misclassifications. 
Furthermore, we see that across disk radii, a stricter tolerance involves a greater computational cost, but this cost is marginal relative to the overall effect of disk radius.
It is for this reason that we use for our algorithm a disk radius that is 1\% of the parameter space bounding box diagonal, as it is small enough to allow full utilization of the memoization scheme, while still allowing for a somewhat lenient tolerance of $\epsilon_q = 10^{-6}$.

\subsubsection{Disk Extraction}\label{sec:disk_extraction}

\begin{figure}[tb]
    \centering
    \includegraphics[width=\linewidth]{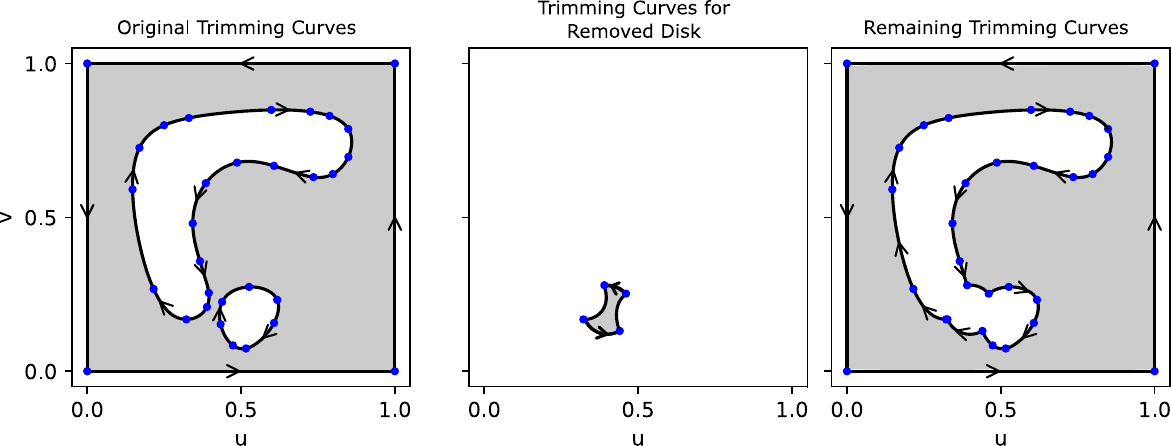}
    \caption[Disk extraction via trimming curves in parameter space.]{Disk extraction applied to the edge case in Figure~\ref{fig:edge_cases} (a).}
    \Description{Disk extraction via trimming curves in parameter space.}
    \label{fig:disk_subdivision}
\end{figure}

Handling edge cases efficiently requires a specific type of surface subdivision in which the surface is split according to a disk defined in parameter space around a given point (see Figure~\ref{fig:disk_subdivision}).
We accomplish this through the insertion of extra trimming curves around the given edge-case point. 
To the original surface, we add an extra trimming loop around such points in parameter space, and remove all existing trimming curves within the loop.
To a copy of the original surface, we remove all trimming curves \textit{exterior} to the loop in parameter space.
In each case, this processing requires computing the set of intersections between the initial set of trimming curves and a circle in the parameter space of the patch. 
However, this is greatly simplified by the use of 2D generalized winding numbers in our trim tests, as these trimming curves need not be explicitly connected to form closed loops.

\subsection{Robustness to Defective Trimming Curves}\label{sec:robustness}

\begin{figure*}[t]
    \centering
    \begin{tabular}{c ccc c ccc}
        \includegraphics[width=0.12\linewidth]{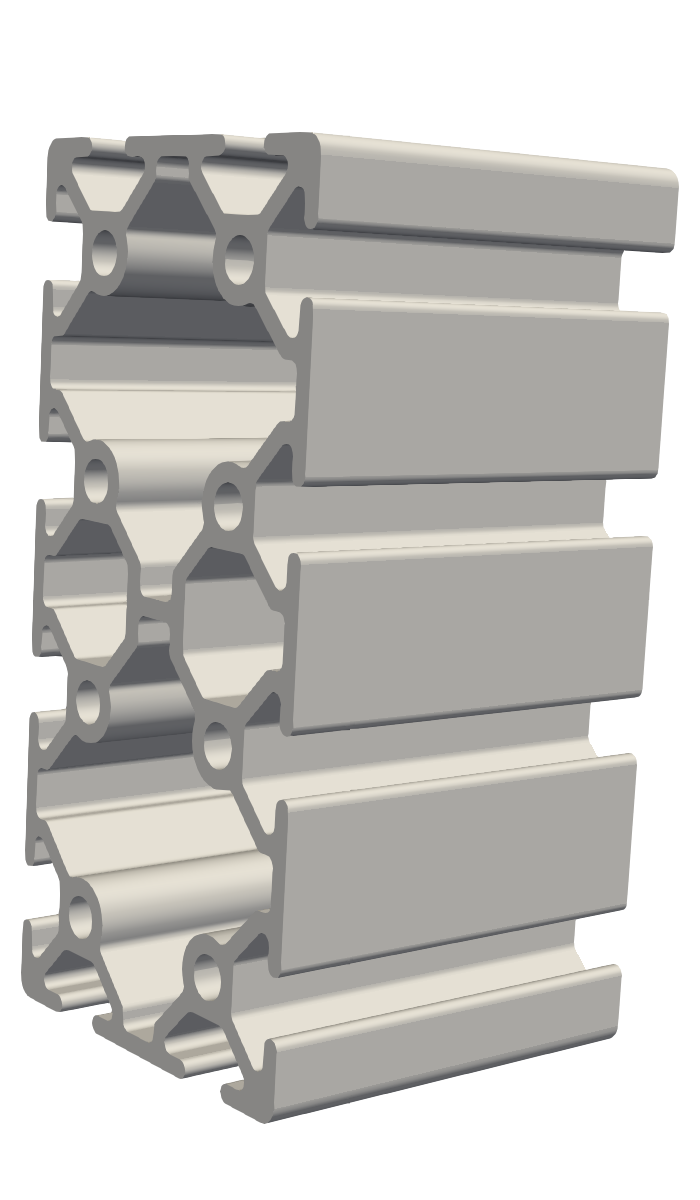} &
        \includegraphics[width=0.1235\linewidth]{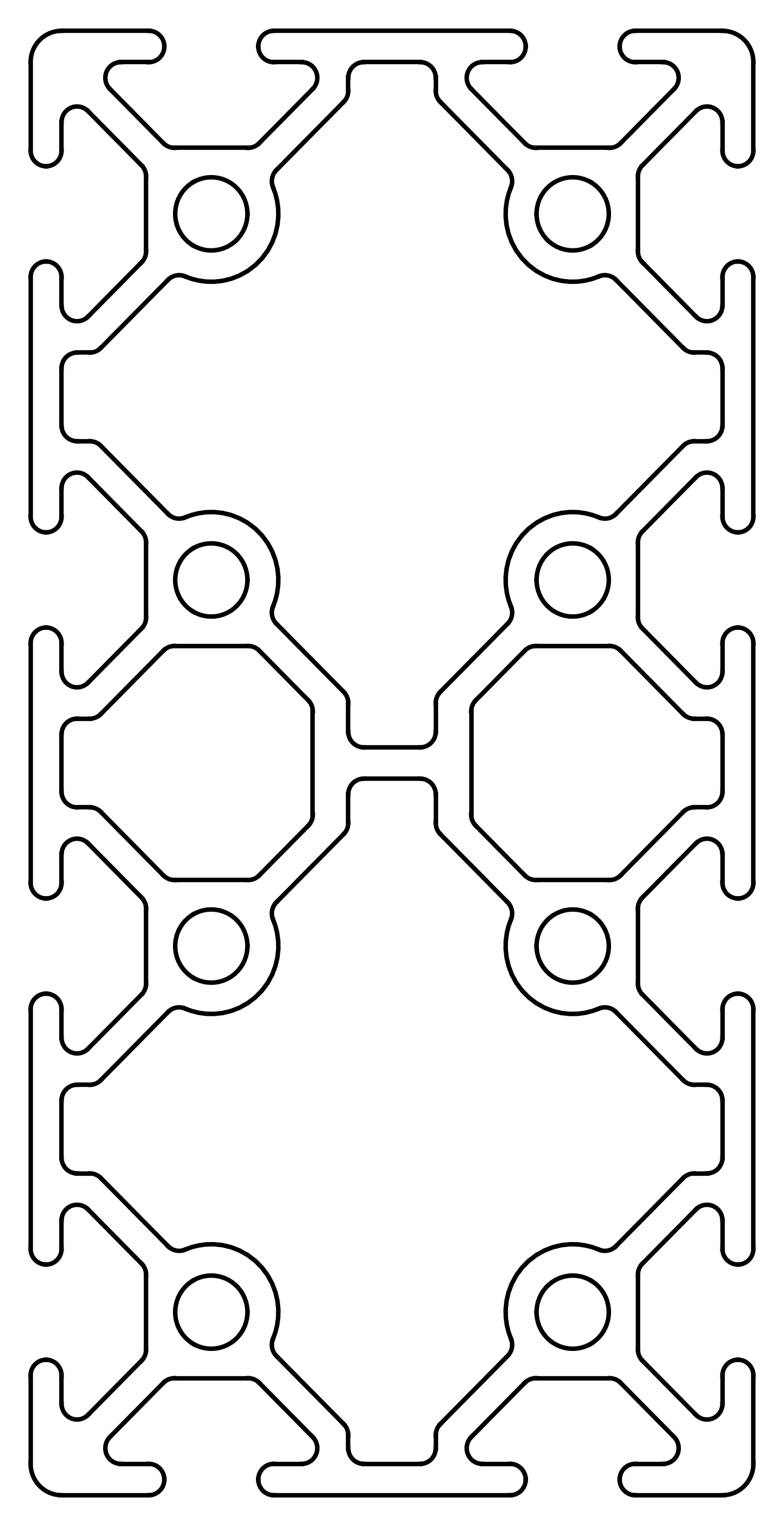} &
        \includegraphics[width=0.12\linewidth]{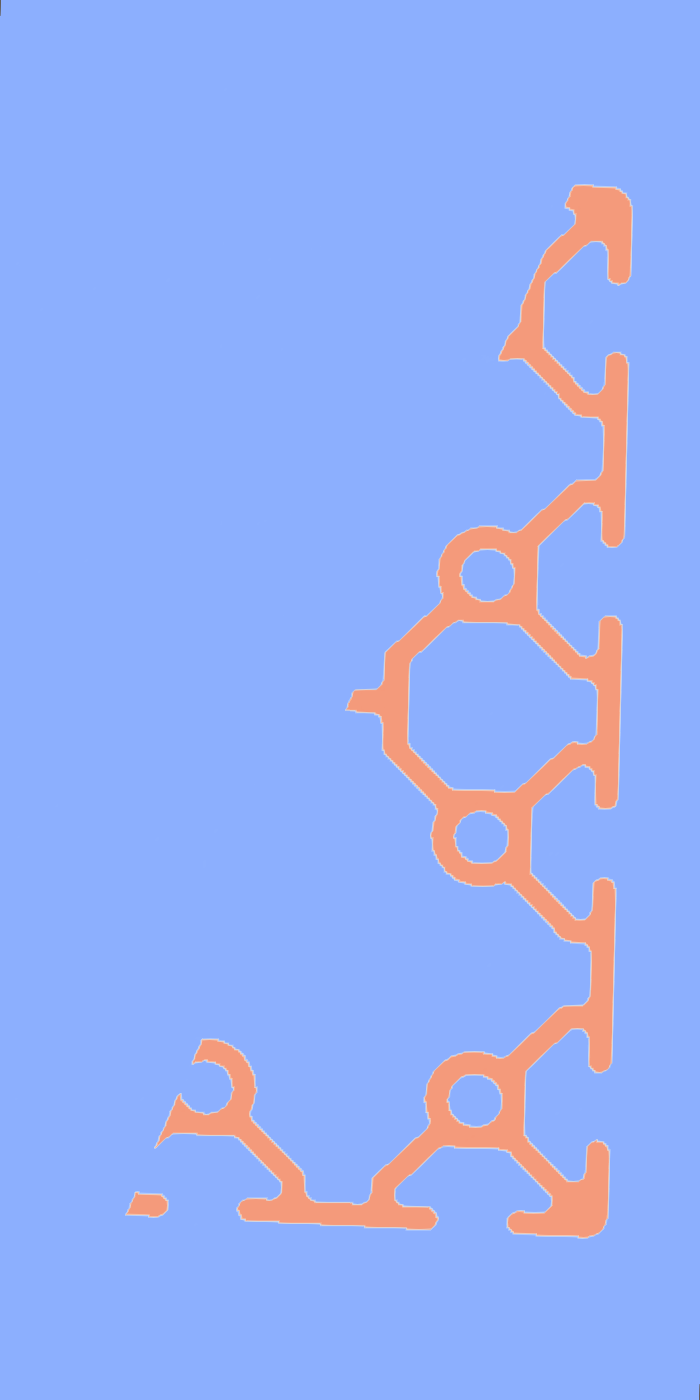} &
        \includegraphics[width=0.12\linewidth]{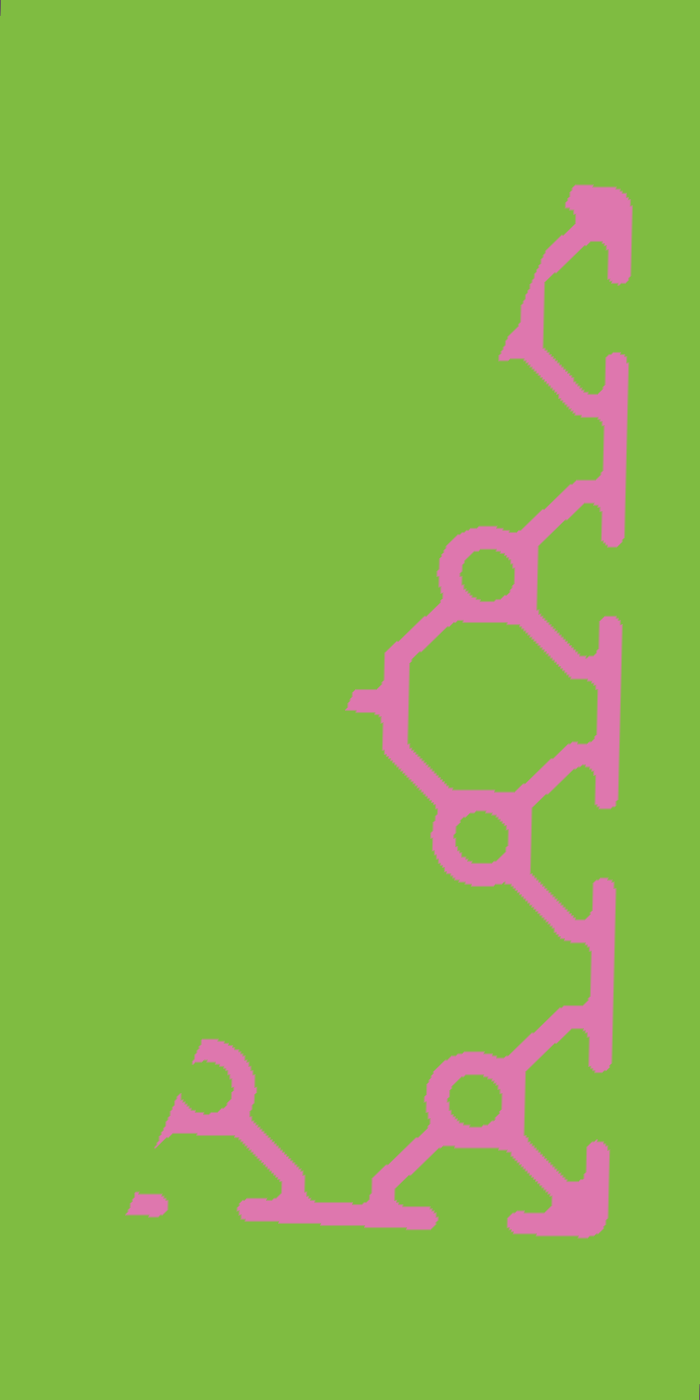} &
        &
        \includegraphics[width=0.1235\linewidth]{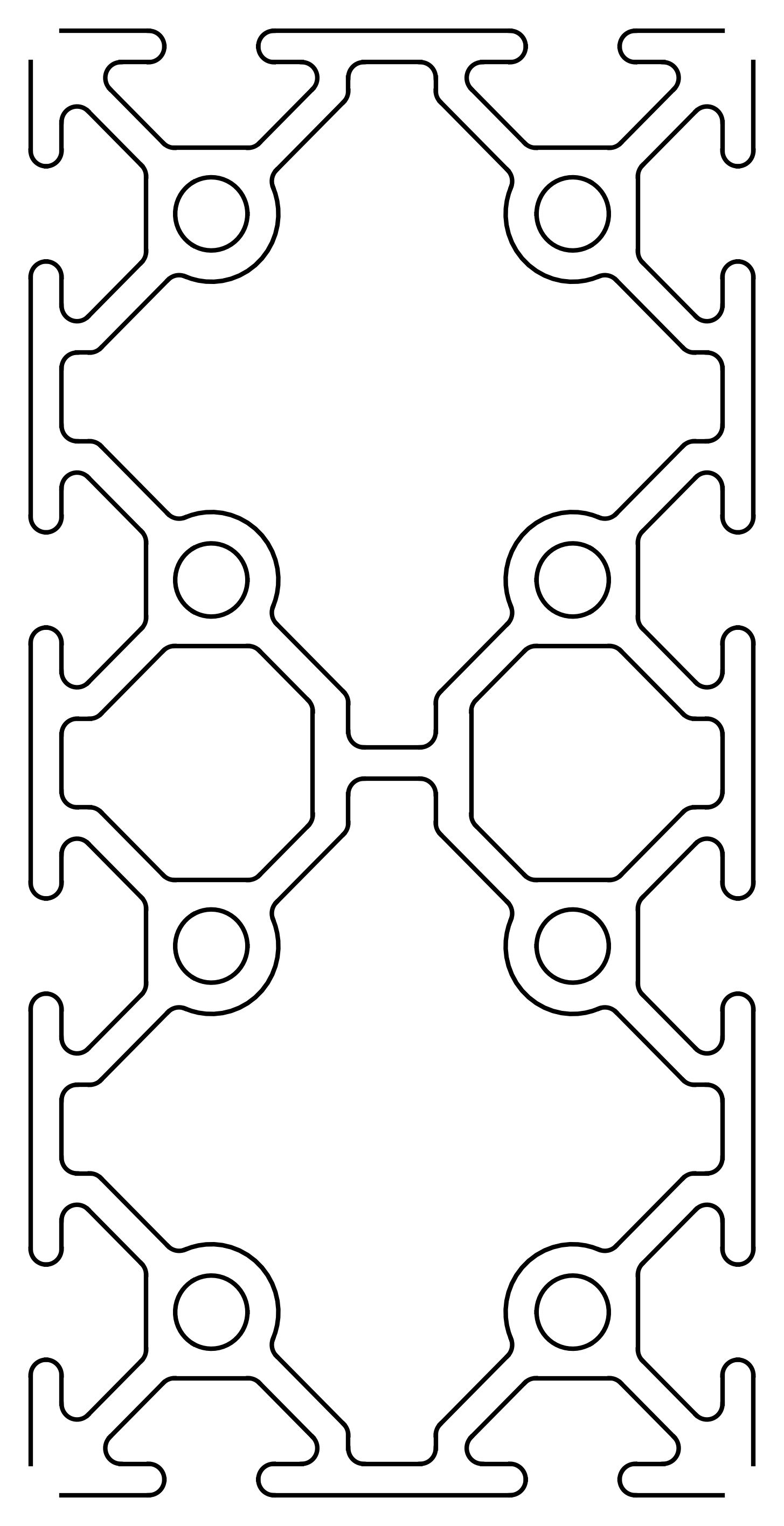} &
        \includegraphics[width=0.12\linewidth]{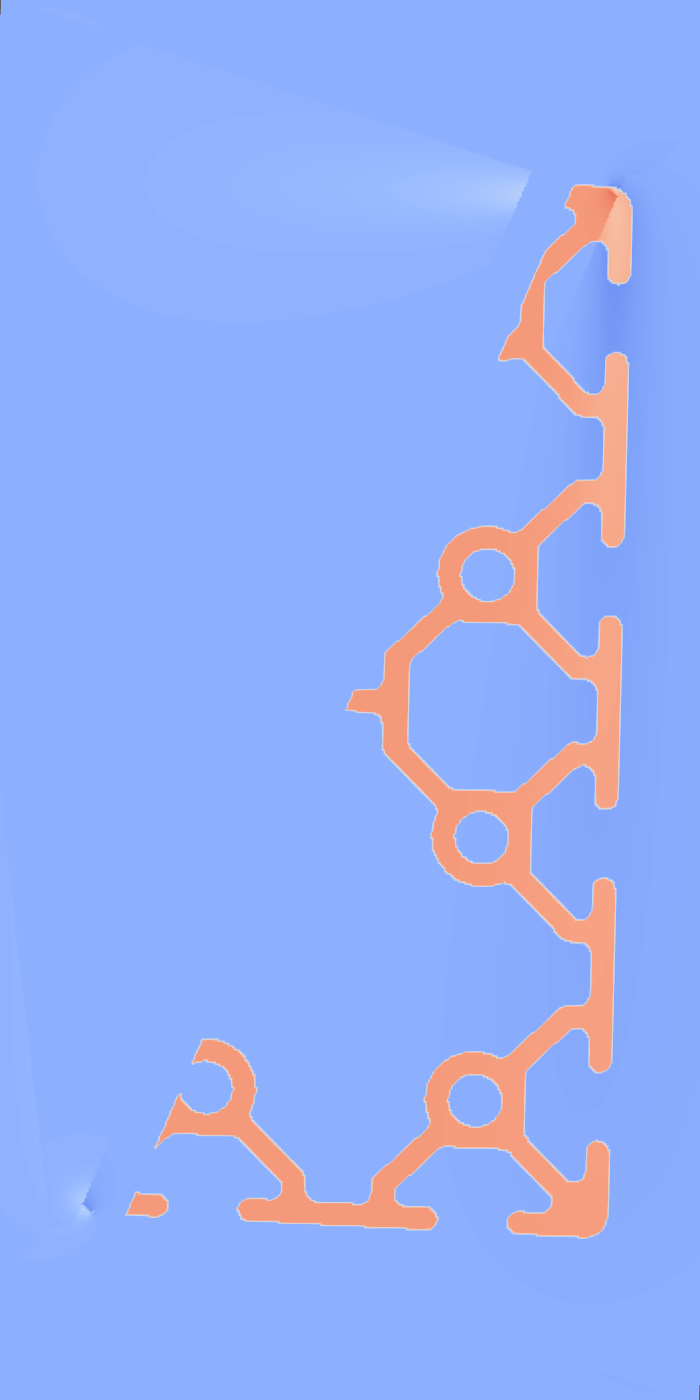} &
        \includegraphics[width=0.12\linewidth]{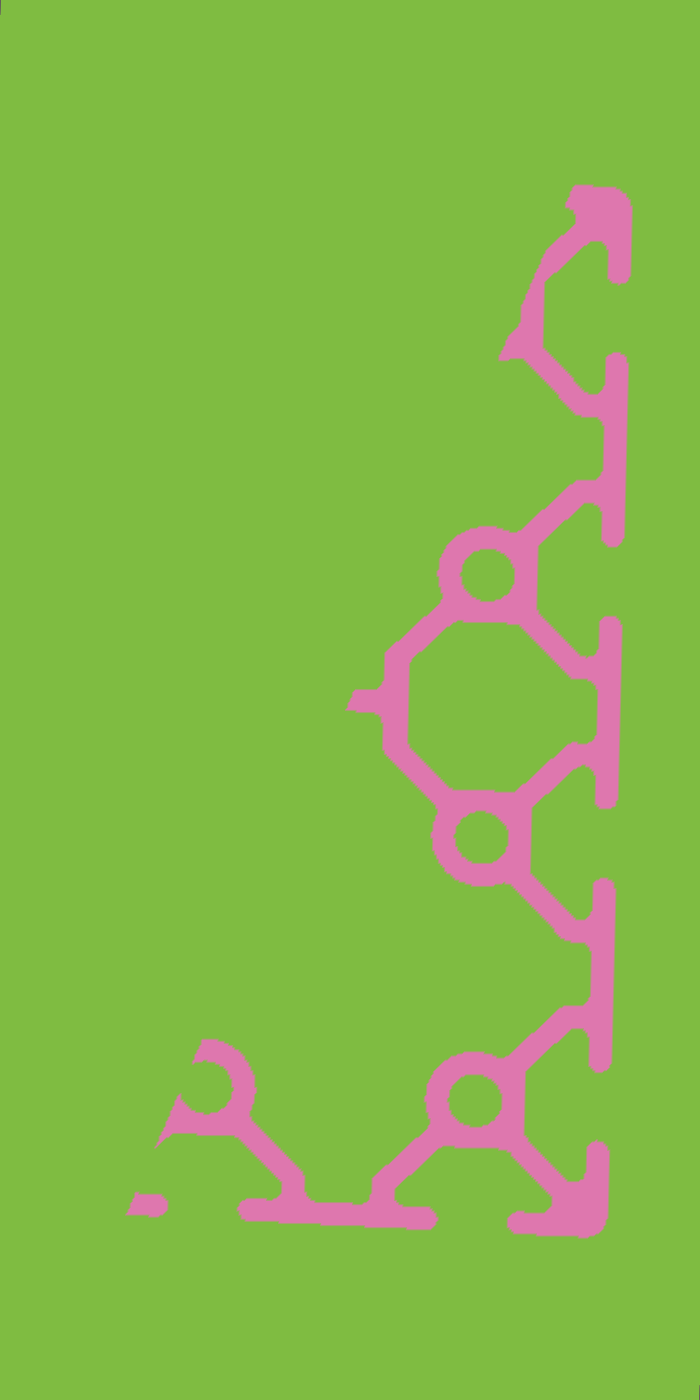} 
        \\
        \includegraphics[width=0.12\linewidth]{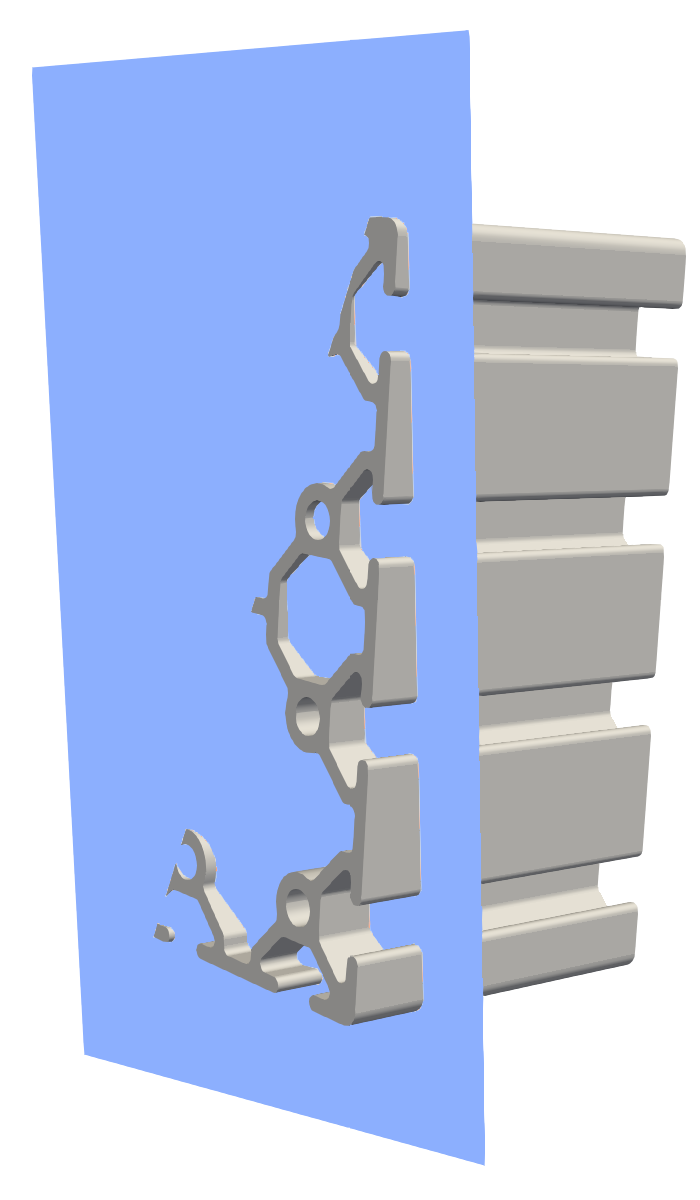} &
        \includegraphics[width=0.1235\linewidth]{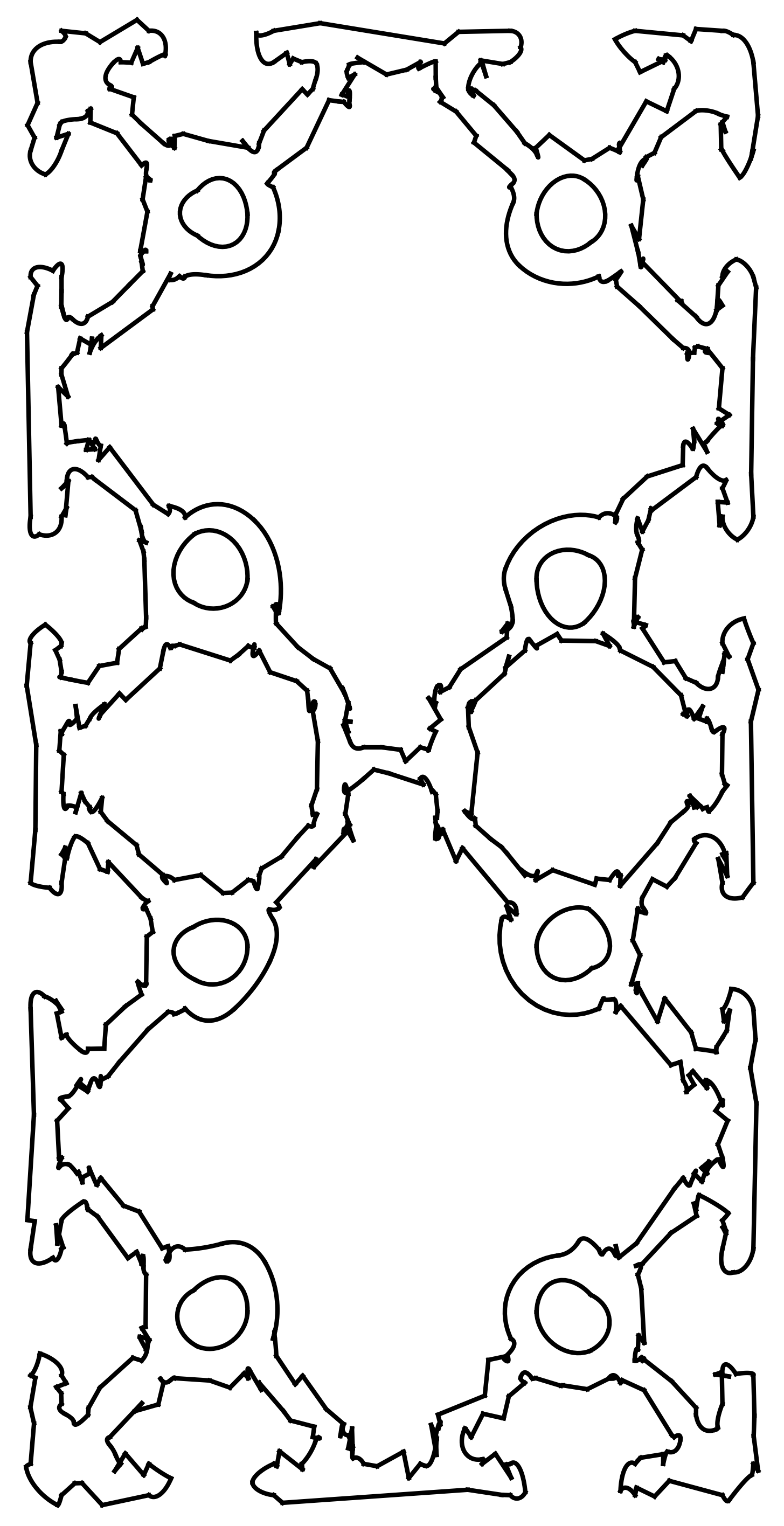} &
        \includegraphics[width=0.12\linewidth]{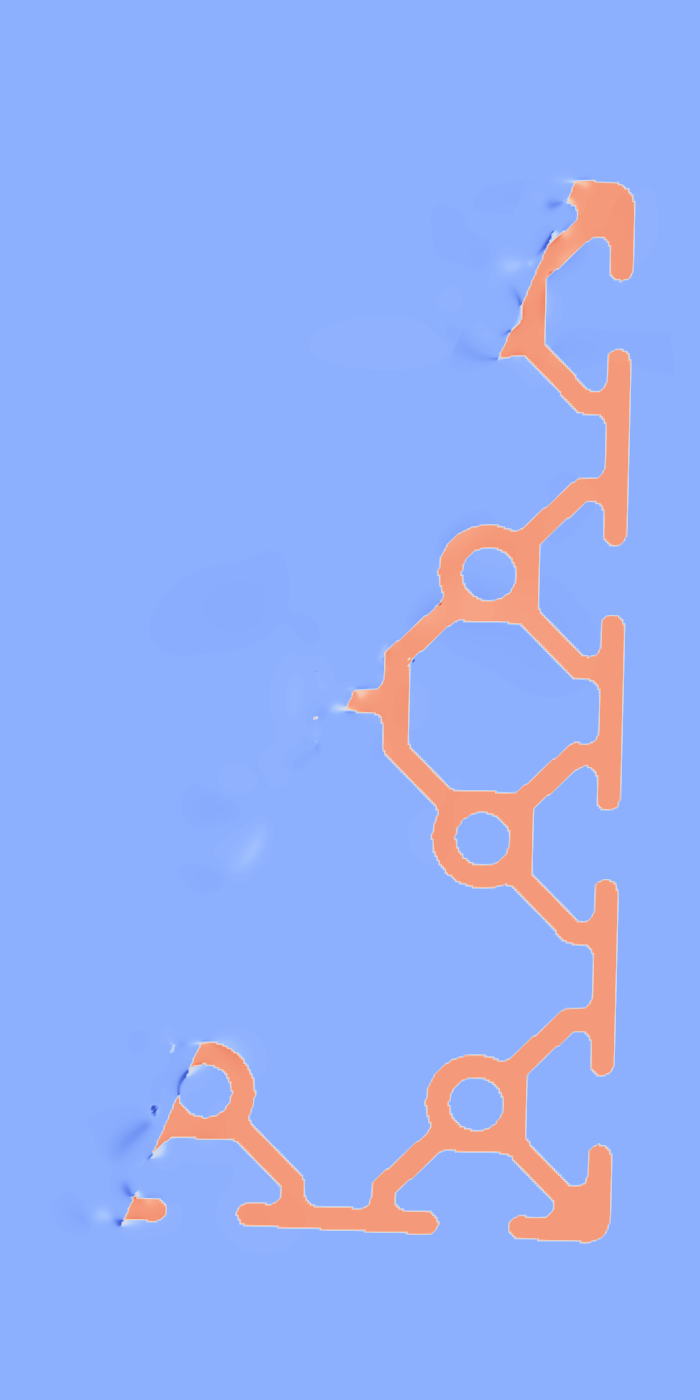} &
        \includegraphics[width=0.12\linewidth]{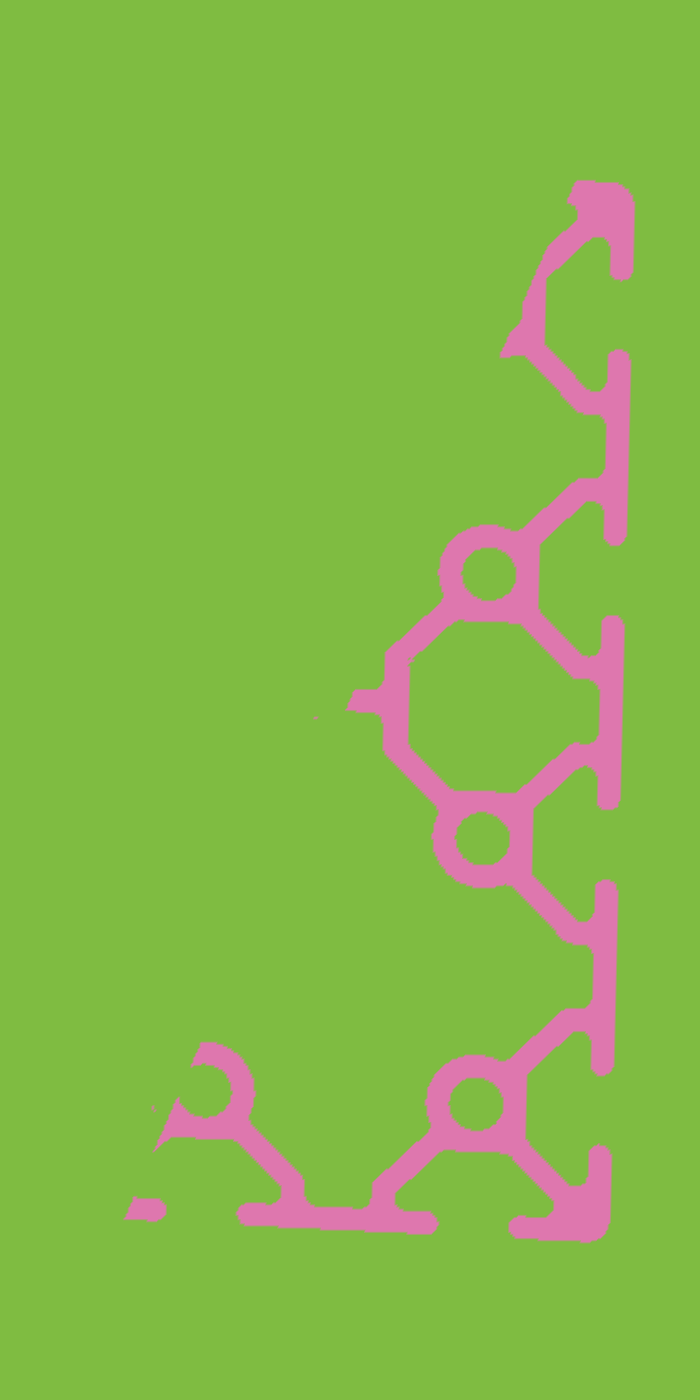} &
        &
        \includegraphics[width=0.1235\linewidth]{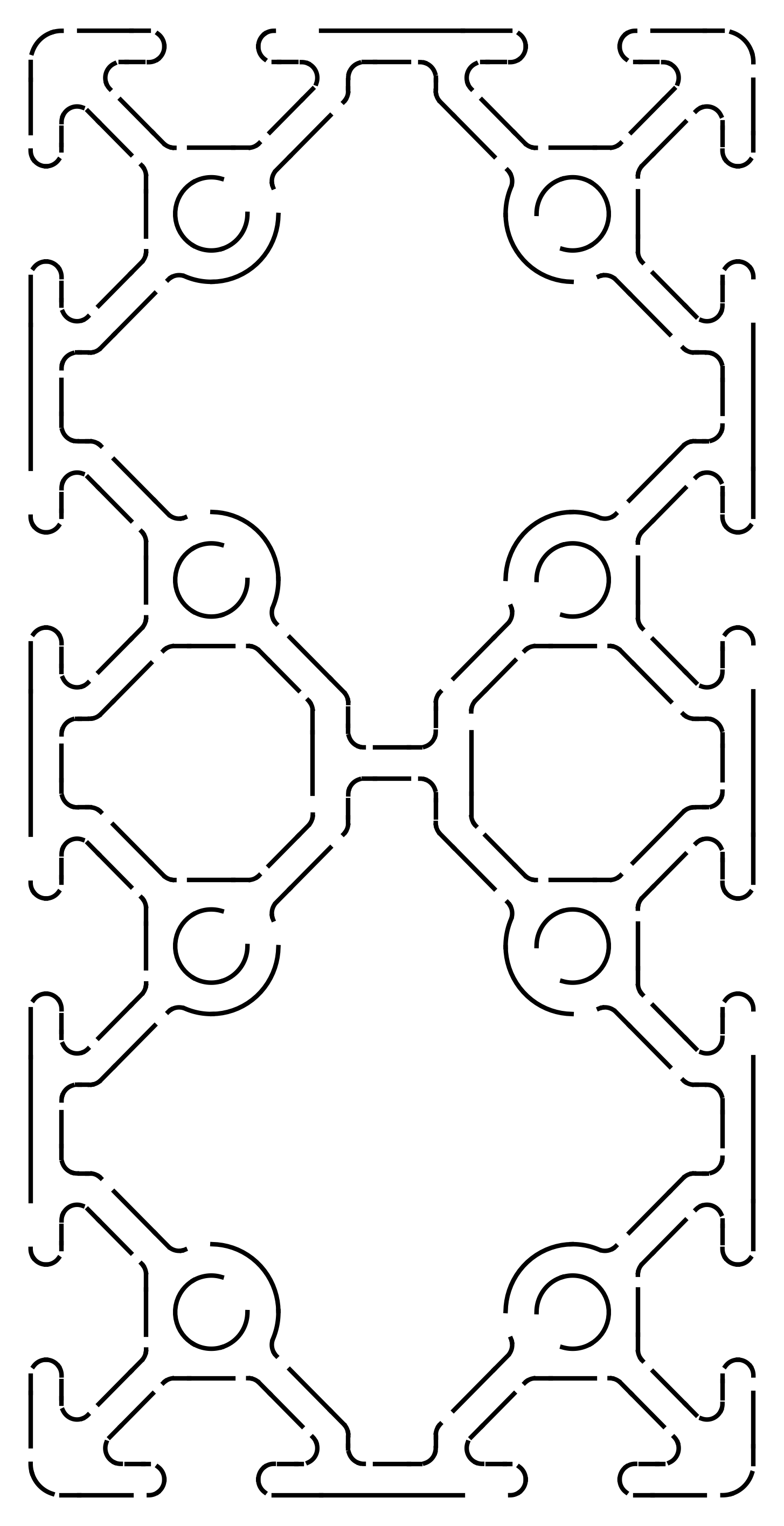} &
        \includegraphics[width=0.12\linewidth]{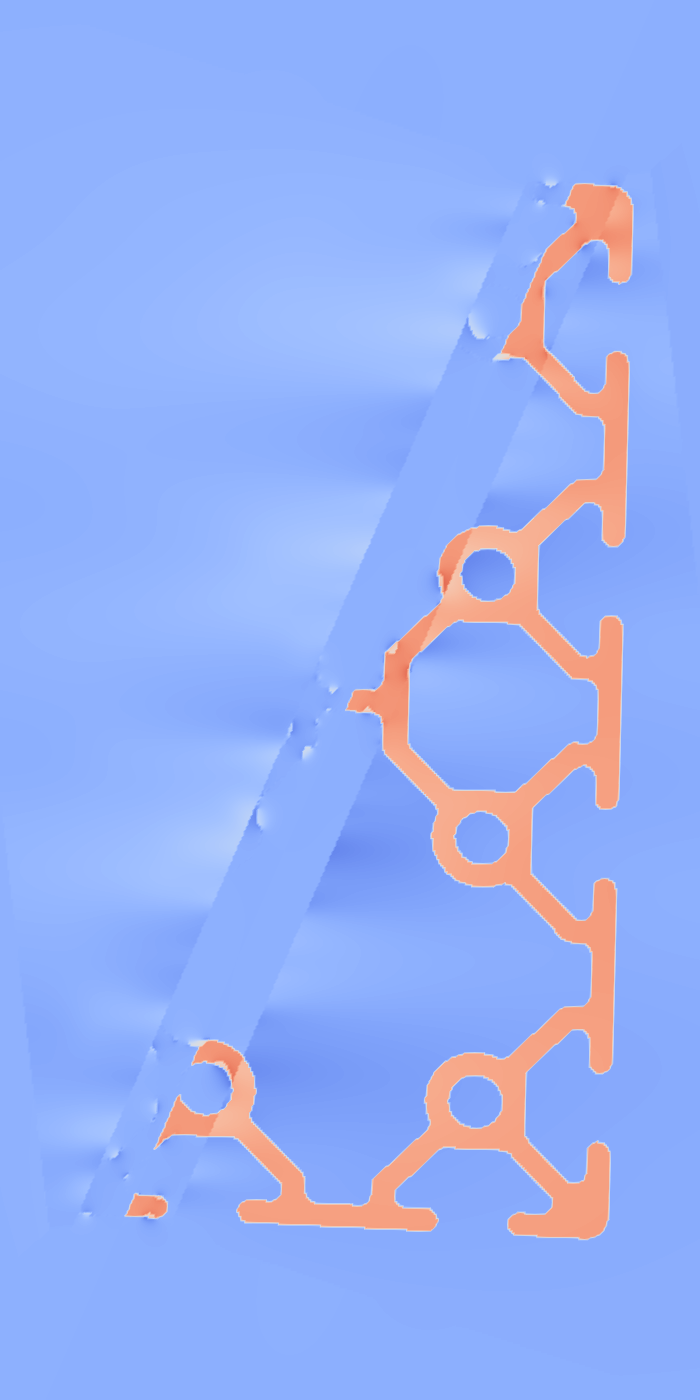} &
        \includegraphics[width=0.12\linewidth]{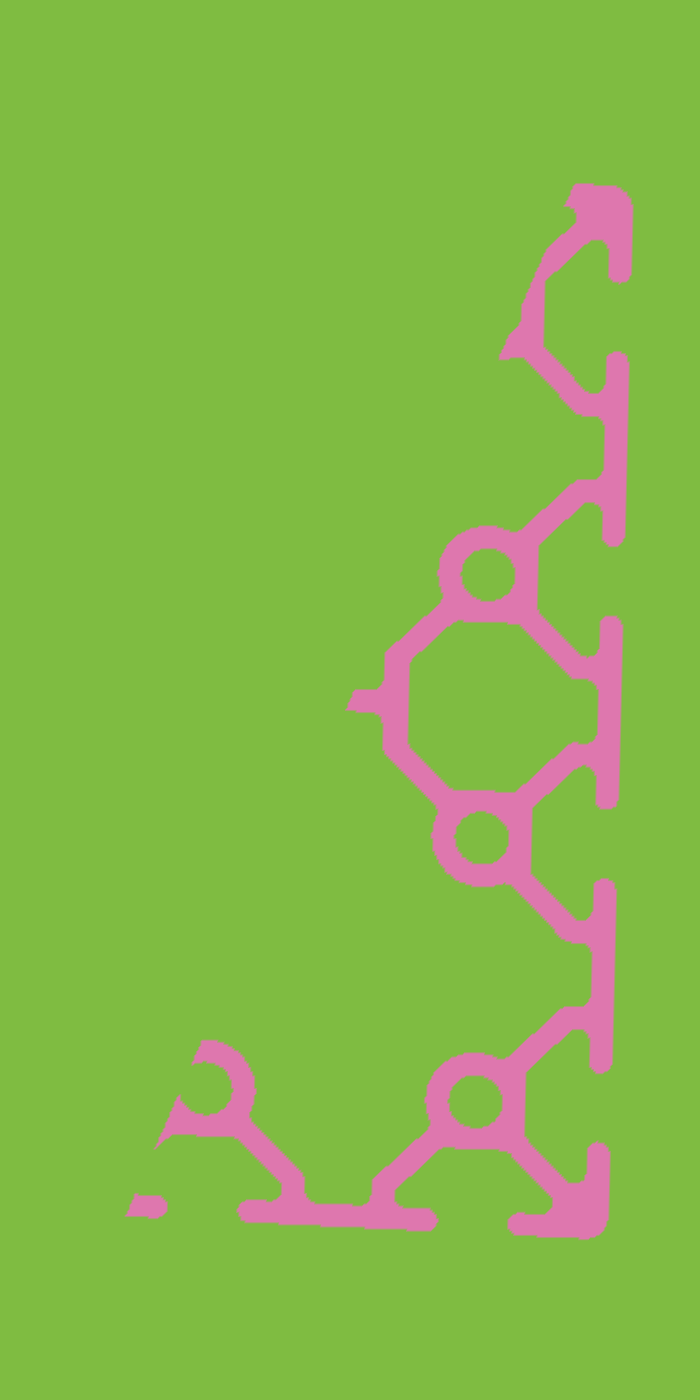} 
        \\
         & \multicolumn{3}{c}{\includegraphics[width=0.24\linewidth]{figures/colorbars/gwn_colorbar_15.png}} & & 
        \multicolumn{3}{c}{\includegraphics[width=0.24\linewidth]{figures/colorbars/containment_classification_colorbar.png}} 
    \end{tabular}
    \caption[Demonstration of algorithm robustness to invalid trimming curves]{We apply Algorithm~\ref{alg:generalized_winding_number} to a shape with a patch with complex, but defective trimming curves. The original collection of trimming curves (top left) is modified through perturbation of internal control points, while adjacent endpoints remain connected (bottom left), and by the removal of boundary components such that adjacent trimming curves no longer connect (less extreme on top right; more extreme on bottom right). We observe that while the GWN field for such shapes degrades non-locally in the presence of these errors, the containment decision remains consistent with expectations in all three cases.}
    \Description[Demonstration of algorithm robustness to invalid trimming curves]{The GWN field and containment classifications for four instances of trimming curve geometry manipulated to be invalid. Trimming curves are shown to be perturbed and non-watertight at varying levels of severity. With increasing severity the GWN field becomes more non-smooth, but the containment decisions are virtually unaffected.}
    \label{fig:robustness}
\end{figure*}

In the context of containment queries for messy CAD geometry, there are two relevant, yet distinct conceptions of robustness.

First is robustness to errors between the (valid) trimmed NURBS surfaces in the model, which is necessary when numerical gaps and overlaps between adjacent patches make conventional containment techniques like ray casting inaccurate.
By the linearity of the GWN, our algorithm is evaluated independently for each patch, ensuring our containment decision is reasonable for this kind of messy input.
We achieve this form of robustness through accurate and precise calculation of the GWN field at arbitrary 3D points, one surface at a time.
This property of our algorithm is exhibited on watertight and non-watertight examples throughout this paper (e.g.\ Figures~\ref{fig:graphical_abstract},~\ref{fig:vase_comparisons},~\ref{fig:simple_results} and~\ref{fig:zoomed}) as well as in the supplement.

Second is robustness to errors \textit{within} the representation of trimmed NURBS surfaces, which is necessary when numerical gaps and overlaps between adjacent trimming curves make the surface have an ill-defined boundary.
Such errors interfere with many primitive geometric operations, e.g.\ tessellation~\cite{xiong-23-eter} and calculation of surface areas~\cite{gunderman-21-trimmednurbsintegration} or surface normals~\cite{ueda-1996-meannormal}.
Other boundary-focused methods for the GWN specifically rely on surface boundaries forming closed loops, such the calculation of the GWN via the ``One-Shot'' approach~\cite{martens-2025-oneshot} or the adaptations of~\cite{Jacobson-13-winding} discussed in Section~\ref{sec:accuracy_results}.
More generally, trimming curve defects are challenging to resolve in this context because they cannot be directly addressed \textit{by} the GWN, as the scalar field itself becomes mathematically ill-defined.

In Algorithm~\ref{alg:generalized_winding_number} we make the standard assumption that the trimming curves for a given patch form simple, closed loops in parameter space, where each such loop corresponds to a single component of the surface boundary.
While we consider a rigorous definition and calculation of the GWN field for malformed surfaces to be a productive path for future research, we have nevertheless structured our implementation so that it can still meaningfully process surfaces which do not meet this assumption.
Perhaps most importantly, our trim tests, which determine if line-surface intersections lie in the visible region of a patch, are handled through the 2D GWN~\cite{spainhour_24_robustcontainment2d}, making our visibility queries indifferent to trimming curve watertightness.
Similarly, our strategy for resolving edge-cases through disk subdivision assumes no structure in the collection of trimming curves, further supported by our use of the 2D GWN.
Finally, our adaptive quadrature algorithm operates independently on each component of the boundary without reference to their overall arrangement.

Figure~\ref{fig:robustness} demonstrates how these decisions impact our calculation of the GWN field in the presence of a surface with poorly defined trimming curves, and in turn how this GWN field impacts the subsequent containment decision.
To the original shape in Figure~\ref{fig:robustness} (top left), we perturb the trimming curves of the front face while keeping adjacent curves connected.
The resulting set of trimming curves, shown in Figure~\ref{fig:robustness} (bottom left), is still closed and connected, but the perturbation has introduced self-intersections that form small clockwise loops.
The boundary formulation itself is indifferent to trimming curve orientation, and so for query points away from these self-intersections, Algorithm~\ref{alg:generalized_winding_number} continues to produce a GWN that degrades smoothly around the mismatch with the new boundary.
While this type of trimming curve geometry poses a challenge for our correction term (Equation~\ref{eqn:adjusted_stokes}) and edge-case strategy, resulting in a noisier GWN field, the containment decisions remain consistent with those of the original model.

Next, we consider collections of non-watertight trimming curves. 
For Figure~\ref{fig:robustness} (top right) we delete trimming curves at each corner in parameter space, and for Figure~\ref{fig:robustness} (bottom right) we delete a middle portion from \textit{each} curve.
In the presence of these defective trimming curves, we observe that the computed GWN field does not degrade smoothly in the same way that it does around open, yet well-defined edges. 
For example, applying our boundary formulation to the same malformed surface returns slightly different GWN values at the same query point depending on the chosen form of the line integral.
This is seen most prominently in the boundary between regions handled by different cases, such as between far-field and edge-case scenarios.
Nevertheless, we see that these discontinuities are concentrated near geometric errors, such that rounding and applying an even-odd rule to the calculated GWN field still achieves excellent containment decisions for these examples.

\section{Conclusions}
In this work, we have introduced the first algorithm to accurately and precisely compute generalized winding numbers over unstructured collections of trimmed NURBS patches.
We have used this to define a robust containment query that can be applied to general CAD models, even in the presence of non-watertightness, overlapping boundary components, and other geometric errors.

For points far from the surface, we evaluate the relevant surface integral using a novel application of Stokes' theorem, reducing it to a collection of line integrals defined by trimming curves.
We apply the same reformulation for points near the surface, alongside a correction term which respects the exact geometric fidelity of the original surface, resulting in accurate containment decisions at arbitrary points in space. 
In both cases, the line integrals themselves are solved in an error-controlled manner via adaptive quadrature, ensuring a precise calculation of the fractional value of the GWN.

This theoretical basis for our algorithm extends readily to other parametric surface representations (e.g.\ \bezier\ triangles, Coons patches, Gregory patches, curve networks), as our near-field correction term is general to entirely arbitrary surfaces. 
While manipulation of a NURBS surface and its boundaries is made simpler through access to control points and parametric trimming curves respectively, the basic geometric primitives on which our algorithm relies are readily available for other surfaces.
The only remaining barrier to practical usage is the treatment of edge-cases, which our implementation handles via manipulation of existing trimming curves. 
For other types of surfaces, we anticipate that clever usage of surface subdivision could be similarly applied to remove problematic portions of the surface.

Although the focus of this work is the theoretical underpinnings of our Stokes' theorem reformulation and its implications for the evaluation of the GWN for individual surface patches, we are interested in applying hierarchical approximation methods as in~\citet{Barill-18-soupcloud} to improve batch performance across collections of patches, particularly in the case of far-field evaluation.
Similarly, we are interested in potential performance improvements that come about from the use of more specialized integration techniques, such as singularity swapping~\cite{klinteberg-19-accurate} or more general fast multipole methods~\cite{bang-23-multipole}, which have been demonstrated as more stable than generic quadrature methods for evaluating harmonic, near-singular integrals such as the 2D GWN. 

We are also interested in rigorously extending the definition of the GWN to surfaces with ill-defined boundaries, such as NURBS surfaces with malformed trimming curves.
While our method is designed to produce reasonable containment decisions when standard assumptions are not met (e.g.\ trimming curve watertightness), the computed GWN field degrades irregularly near this type of geometric error.
This is in contrast to the true GWN field, which degrades smoothly and rapidly around the open edges of valid surfaces.

\begin{acks}
We would like to thank the anonymous reviewers for their thoughtful consideration and insightful feedback.

This work was performed under the auspices of the U.S. Department of Energy by Lawrence Livermore National Laboratory under Contract DE-AC52-07NA27344.
\end{acks}

\bibliographystyle{ACM-Reference-Format}
\bibliography{citations}

\appendix

\clearpage
\FloatBarrier

\section{Analytic discontinuity fix for Near-Field GWN}\label{sec:appendix}

In this section, we provide a formal proof of the correction term in Equation~\ref{eqn:adjusted_stokes}. 
Assuming the surface and query are shifted such that the query is at the origin, one can compute the GWN in the ``near-field'' using the Stokes' theorem boundary formulation, alongside a correction term which adds or subtracts $0.5$ for each signed intersection between the surface and the $z$-axis.

Without loss of generality, we assume there is only a single such point of intersection, as the following argument can be repeated for additional intersections.
We also assume, as in Section~\ref{sec:adjusted_stokes}, that this intersection is not on the surface boundary, has a well-defined normal vector on the surface at the point of intersection (i.e.\ does not occur on cusps of the surface), and that the $z$-axis is not tangent to the surface.
Furthermore, because the query point is not coincident with the surface, there necessarily exists a neighborhood $S_f \subset S$ around the point of intersection such that the query is in the ``far-field'' of $S_f$, i.e.\ there is some $\epsilon > 0$ such that $\norm{x - 0} > \epsilon$ for all $x \in S_f$.
This means that the GWN with respect to this neighborhood depends only on its boundary $\partial S_f$.

We use this property to deform the original surface within the boundary of the neighborhood $S_f$ to form a flat disk $D$ which is orthogonal to the $z$-axis.
This ``virtual'' procedure considerably simplifies the analytic evaluation of the GWN within subsequent steps of this proof, but never needs to be practically implemented to compute the GWN via Algorithm~\ref{alg:generalized_winding_number}.
Indeed, because the original surface matches the deformed surface at the boundary of the neighborhood $S_f$, this deformation does not change the value of the GWN evaluated over the entire surface $S$.

For notational simplicity, we continue to refer to this deformed surface as $S$, such that $D \subset S$.
To evaluate $w_S$, we now consider the relevant surface integral to be split between the disk $D$ and the remaining surface $S \setminus D$, such that
\begin{align}
    w_S &= \frac{1}{4\pi}\iint_S \frac{\vec{x} \cdot \hat{n}}{\norm{\vec{x}}^3}\,dS\\
    &= \frac{1}{4\pi}\left(\iint_D \frac{\vec{x} \cdot \hat{n}}{\norm{\vec{x}}^3}\,dS + \iint_{S \setminus D} \frac{\vec{x} \cdot \hat{n}}{\norm{\vec{x}}^3}\,dS\right).
\end{align}
This is essentially the procedure by which we resolve ``edge-cases'' in the main algorithm, but with the advantage of knowing the exact geometry of the disk which is extracted from the surface.

By construction, the punctured surface $S \setminus D$ does not intersect the $z$-axis, and so we can apply Stokes' theorem to this portion of the integral:
\begin{align}
    w_S &= \frac{1}{4\pi}\iint_D \frac{\vec{x} \cdot \hat{n}}{\norm{\vec{x}}^3}\,dS + \frac{1}{4\pi}\iint_{S \setminus D} \frac{\vec{x} \cdot \hat{n}}{\norm{\vec{x}}^3}\,dS\\
        &= \frac{1}{4\pi}\iint_D \frac{\vec{x} \cdot \hat{n}}{\norm{\vec{x}}^3}\,dS \\
        &+ \frac{1}{4\pi}\oint_{\partial(S \setminus D)}\left\langle \frac{yz}{(x^2+y^2)\,\norm{\vec{x}}}, \frac{-xz}{(x^2+y^2)\,\norm{\vec{x}}}, 0\right\rangle\cdot d\vec{\Gamma}\nonumber
\end{align}
We can split the integral over the boundary $\partial(S \setminus D)$ into two components: the boundary of the original surface $\partial S$ and the \textit{negatively} oriented boundary of the disk $\partial D^c$.
\begin{align}
        w_S &= \frac{1}{4\pi}\iint_D \frac{\vec{x} \cdot \hat{n}}{\norm{\vec{x}}^3}\,dS\nonumber\\
        &+ \frac{1}{4\pi}\oint_{\partial D^c}\left\langle \frac{yz}{(x^2+y^2)\,\norm{\vec{x}}}, \frac{-xz}{(x^2+y^2)\,\norm{\vec{x}}}, 0\right\rangle\cdot d\vec{\Gamma}\\
        &+ \frac{1}{4\pi}\oint_{\partial S}\left\langle \frac{yz}{(x^2+y^2)\,\norm{\vec{x}}}, \frac{-xz}{(x^2+y^2)\,\norm{\vec{x}}}, 0\right\rangle\cdot d\vec{\Gamma}.\nonumber
\end{align}
Of these three terms, we evaluate the integral over $\partial S$ through numerical integration, and indeed this is the \textit{only} integral evaluated through numerical integration, as described in Section~\ref{sec:methods_overview}.
For the remaining terms, we instead consider their behavior in the limit as the radius of the disk $D$ approaches zero. 

The more direct of the two is the surface integral over the disk $D$.
Because this value is exactly equal to $w_D$, i.e.\ the winding number of the surface $D$ with respect to the origin, it necessarily goes to 0 as the disk decreases in surface area.
Therefore, it remains to show that the line integral around $\partial D^c$ approaches $\pm 0.5$ as the radius of the disk approaches 0. 

By the construction of the deformed surface, we can parameterize the boundary of this disk on $t \in [0, 1]$ straightforwardly as
\begin{align}
    \Gamma(t) &= \left\langle r\cos(2\pi t), -r\sin(2\pi t), z_0\right\rangle,
\end{align}
where $z = z_0 \neq 0$ is the center of the disk and $r$ is its radius. 
We substitute this parameterization and the curve differential $d\vec{\Gamma}$ into the integral in question and after some simplification, we have:

\begin{align}
    &\frac{1}{4\pi}\oint_{\partial D^c}\left\langle \frac{yz}{(x^2+y^2)\,\norm{\vec{x}}}, \frac{-xz}{(x^2+y^2)\,\norm{\vec{x}}}, 0\right\rangle\cdot d\vec{\Gamma}\\
    &= \frac{1}{4\pi}\int_0^1\frac{\scriptstyle -r\sin(2\pi t)\cdot z_0 \cdot -2\pi r\sin(2\pi t) - r\cos(2\pi t) \cdot z_0 \cdot -2\pi r\cos(2\pi t) }{\scriptstyle (r^2\cos^2(2\pi t) + r^2\sin^2(2\pi t))\sqrt{r^2\cos^2(2\pi t) + r^2\sin^2(2\pi t) + z_0^2}}\,dt\nonumber\\
    &= \frac{1}{2}\int_0^1 \frac{z_0 r^2 (\sin^2(2\pi t) + \cos^2(2\pi t))}{r^2\sqrt{r^2 + z_0^2}}\,dt\\
    &= -\frac{z_0}{2}\int_0^1\frac{1}{\sqrt{r^2 + z_0^2}}\,dt\\
    &= -\frac{z_0}{2\sqrt{r^2 + z_0^2}}.
\end{align}

Clearly, as $r\to 0^+$, this value approaches $-0.5$ if $z_0 > 0$ and $0.5$ if $z_0 < 0$.
Altogether, in the case where there is an intersection between the $z$-axis and the \textit{original} surface $S$, we can write
\begin{align}
    w_S &= \frac{1}{4\pi}\oint_{\partial S}\left\langle \frac{yz}{(x^2+y^2)\,\norm{\vec{x}}}, \frac{-xz}{(x^2+y^2)\,\norm{\vec{x}}}, 0\right\rangle\cdot d\vec{\Gamma}\\
    &+\begin{cases}
     0.5  & \parbox[c]{.3\textwidth}{if the origin is \textit{above} the intersection point}, \\
     -0.5  & \parbox[c]{.3\textwidth}{if the origin is \textit{below} the intersection point}. \\
   \end{cases}\nonumber
\end{align}
Along with our handling of points coincident with the surface in Section~\ref{sec:coincident_winding_number}, this yields Equation~\ref{eqn:adjusted_stokes}, as desired.\hfill$\square$

\clearpage
\FloatBarrier

\section{Pseudocode for Additional Algorithms}\label{sec:additional_algorithms_pseudocode}

\begin{algorithm}
    \caption{\texttt{LinePatchIntersections} 
				Find all intersections between an oriented line and a trimmed \bezier\ or NURBS surface.
		}\label{alg:line_patch_intersect}
    \DontPrintSemicolon
    \KwIn{$L$: Line}
    \myinput{$S$: Trimmed \bezier\ or NURBS surface}
    \myinput{$\epsilon_{ls}$: Numerical tolerance for \texttt{isApproximatelyBilinear}}
    \KwOut{$t, u, v$: Lists of the parameters of intersections}
    \nonl\;
    \lIf{\textbf{not} $\texttt{intersects}(L, \texttt{BoundingBox}(S))$}{
        \Return $\{\}, \{\}, \{\}$\;
    }

    \tcc{Find all intersections of $L$ with untrimmed patch $S$}
    \uIf{$S$ is a \bezier\ surface}{
        $t, u, v \gets \texttt{LineBezierIntersectionsRecursive}(L, S)$\;
    }\ElseIf{$S$ is a NURBS surface}{
        $t, u, v \gets \{\}, \{\}, \{\}$\;
        \ForEach{\bezier\ surface $B_i$ in \texttt{ExtractBezier}($S$)}{
            $t_0, u_0, v_0 \gets \texttt{LineBezierIntersectionsRecursive}(L, B_i)$\;
            $u_0, v_0 \gets \texttt{RescaleToParameterSpace}(S, u_0, v_0)$\;
            $\texttt{concatenate}( \{t, t_0\}, \{u, u_0\}, \{v, v_0\})$\;
        }
    }
    $t, u, v \gets \texttt{RemoveDuplicates}(t, u, v)$\; 
    \nonl\;
    \tcc{Filter intersections using 2D GWN-based trim test}
    \ForEach{Intersection point $u_i, v_i$}{
        \If{$(u_i, v_i) \notin \texttt{TrimmingCurves}(S)$}{
            $\texttt{pop}( \{t, t_i\}, \{u, u_i\}, \{v, v_i\})$\;
        }
    }

    \Return $t, u, v$\;
\end{algorithm}

\FloatBarrier

\begin{algorithm}
    \caption{\texttt{LineBezierIntersectionsRecursive} Recursively find all intersections between an oriented line and a \bezier\ surface. For nearly bilinear surfaces, we use the GARP algorithm~\cite{reshetov-19-coolpatches} to directly find intersections.}
		\label{alg:line_patch_intersect_recursive}
    \DontPrintSemicolon
    \KwIn{$L$: Line}
    \myinput{$B$: Untrimmed \bezier\ surface}
    \myinput{$\epsilon_{ls}$: Numerical tolerance for $\texttt{isApproximatelyBilinear}$}
    \KwOut{$t, u, v$: Lists of the parameters of intersections}
    \nonl\;

    \lIf{\textbf{not} $\texttt{intersects}(L, \texttt{BoundingBox}(B))$}{
        \Return $\{\}, \{\}, \{\}$\;
    }

    \lIf{$\texttt{isApproximatelyBilinear}(B, \epsilon_{ls})$}{
        \Return $\texttt{GARP}(B, L)$\;
    }
        
    $t, u, v \gets \{\}, \{\}, \{\}$\;
    \ForEach{\bezier\ surface $B_i$ in $\texttt{split}(B)$}{
        $t_0, u_0, v_0 \gets \texttt{LineBezierIntersectionsRecursive}(L, B_i)$\;
        $u_0, v_0 \gets \texttt{RescaleToParameterSpace}(B, u_0, v_0)$\;
        $\texttt{concatenate}( \{t, t_0\}, \{u, u_0\}, \{v, v_0\})$\;
    }
    \nonl\;
    \Return $t, u, v$\;
\end{algorithm}

\FloatBarrier

\begin{algorithm}
    \caption{\texttt{isApproximatelyBilinear} Returns true if a \bezier\ surface $B$ is approximately bilinear according to the position of its control points $P_{i, j}$
		}\label{alg:is_approximately_bilinear}
    \DontPrintSemicolon
    \KwIn{$B$: Order $(p, q)$ Tensor product \bezier\ surface}
    \myinput{$\epsilon_{ls}$: Numerical tolerance}
    \nonl\;
    \uIf{$p \leq 1 \textbf{ and } q \leq 1$}{
        \Return True\;
    }
    $B_\ell(u, v) \gets \text{Bilinear surface defined by 4 vertex control points}$\;
    \For{$i = 0\dots p$}{
        \For{$j = 0\dots q$}{
            \tcc{Evaluate the bilinear surface at uniform points in parameter space}
            $\widetilde{P}_{i,j} = B\left(\tfrac{i}{p}, \tfrac{j}{q}\right)$\;
            $P_{i, j} = \texttt{ControlPoint}(B, i, j)$\;
            \If{$\texttt{squared\_distance}(P_{i,j}, \widetilde{P}_{i,j}) > \epsilon_{ls}$}{
                \Return False\;
            }
        }
    }
    
    \Return True\;
\end{algorithm}

\FloatBarrier

\begin{algorithm}
    \caption{\texttt{EvaluateLineIntegral} Numerically evaluate Equation~\ref{eqn:stokes_theorem_equation} with an adaptive Gaussian quadrature rule of fixed order. For a fixed curve and recursive depth, the quadrature nodes are cached and reused across queries.
		}
		\label{alg:adaptive_quadrature_recursive}
    \DontPrintSemicolon
    \KwIn{$C$: Trimming curve}
    \myinput{$q$: Query point}
    \myinput{$\epsilon_q$: Numerical tolerance for absolute error}
    \nonl\;

    $w = \texttt{GaussianQuadrature}(C, q)$\;
    \If{$q \notin \texttt{BoundingBox(q)}$}{
        \Return $w$\;
    }
    \nonl\;
        
    $C_1, C_2 \gets \texttt{bisect}(C)$\;
    $w_1 = \texttt{GaussianQuadrature}(C_1, q)$\;
    $w_2 = \texttt{GaussianQuadrature}(C_2, q)$\;
    
    \eIf{$|w - (w_1 + w_2)| < \epsilon_q$}{
        \Return $w_1 + w_2$\;
    }{
        \tcc{Repeat the quadrature on each half}
        \Return $\texttt{EvaluateLineIntegral}(C_1, q, \epsilon_q) + \texttt{EvaluateLineIntegral}(C_2, q, \epsilon_q)$\;
    }
\end{algorithm}

\FloatBarrier

\onecolumn
\clearpage
\FloatBarrier

\section{Supplemental Material}

In this supplemental material, we evaluate our 3D GWN algorithm qualitatively and quantitatively on several additional watertight and non-watertight CAD models.
While most of these additional shapes are taken directly from the ABC dataset~\cite{koch-19-abcdataset}, others were either constructed by hand in modeling software or sourced from other online repositories. 
The specific models used in this work can be found in~\cite{Axom-data}.

For each shape, we provide summary statistics in the format of Table 1 from the primary manuscript: We compute the GWN over a uniform grid of $50\times50\times50$ points within the shown bounding box and report the average time, in milliseconds, needed to evaluate the GWN for each surface in the shape, alongside the breakdown of all point-surface GWN calculations by case.
We also visualize the 3D GWN scalar field along different 2D slices, and show the nearest integer value to each scalar. 
These integers can be straightforwardly mapped to a containment decision using either a non-zero rule, or an even-odd rule.

\begin{figure}
\begin{minipage}{0.5\textwidth}
\centering
\begin{tabular}{lr}
    \includegraphics[width=0.45\linewidth]{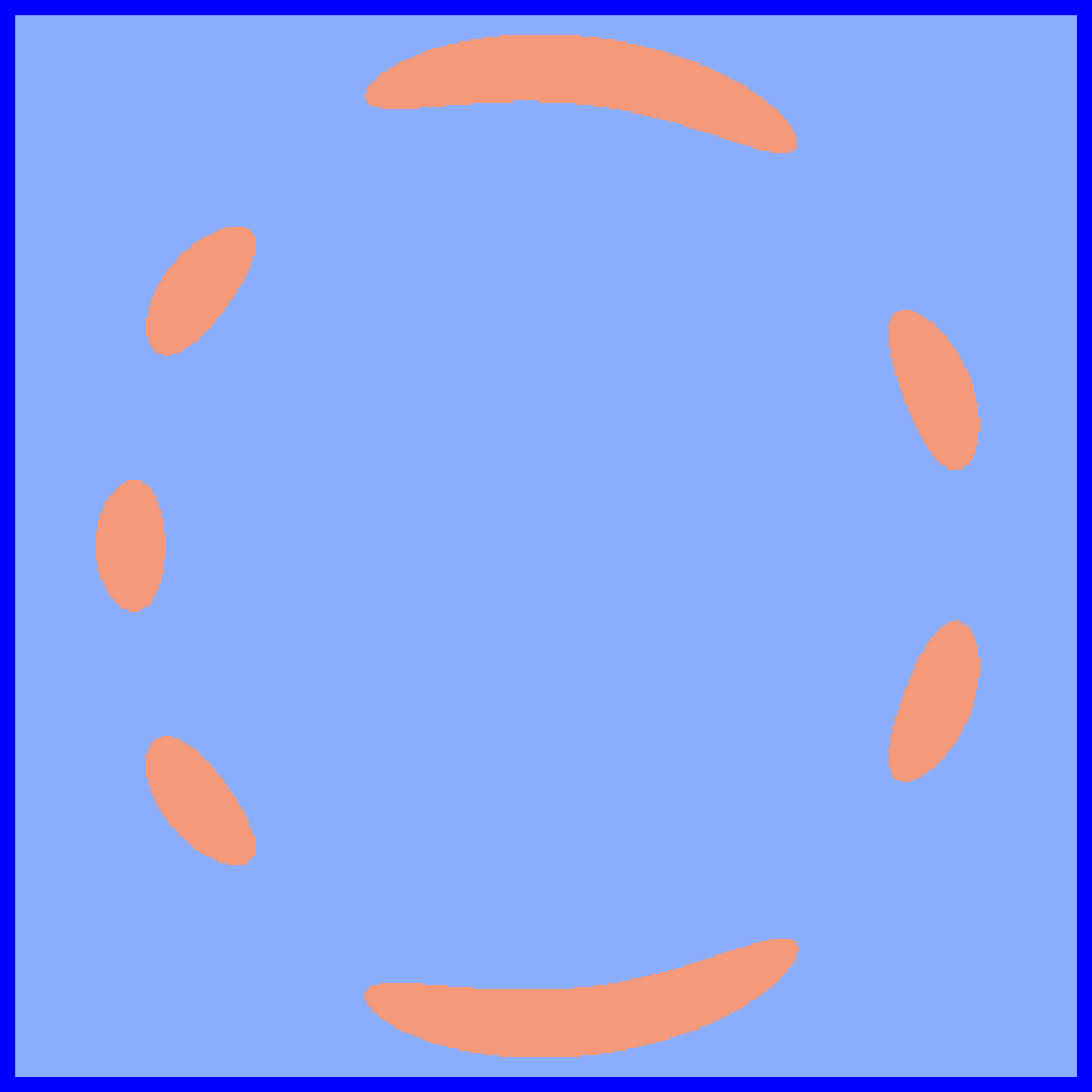} &
    \includegraphics[width=0.45\linewidth]{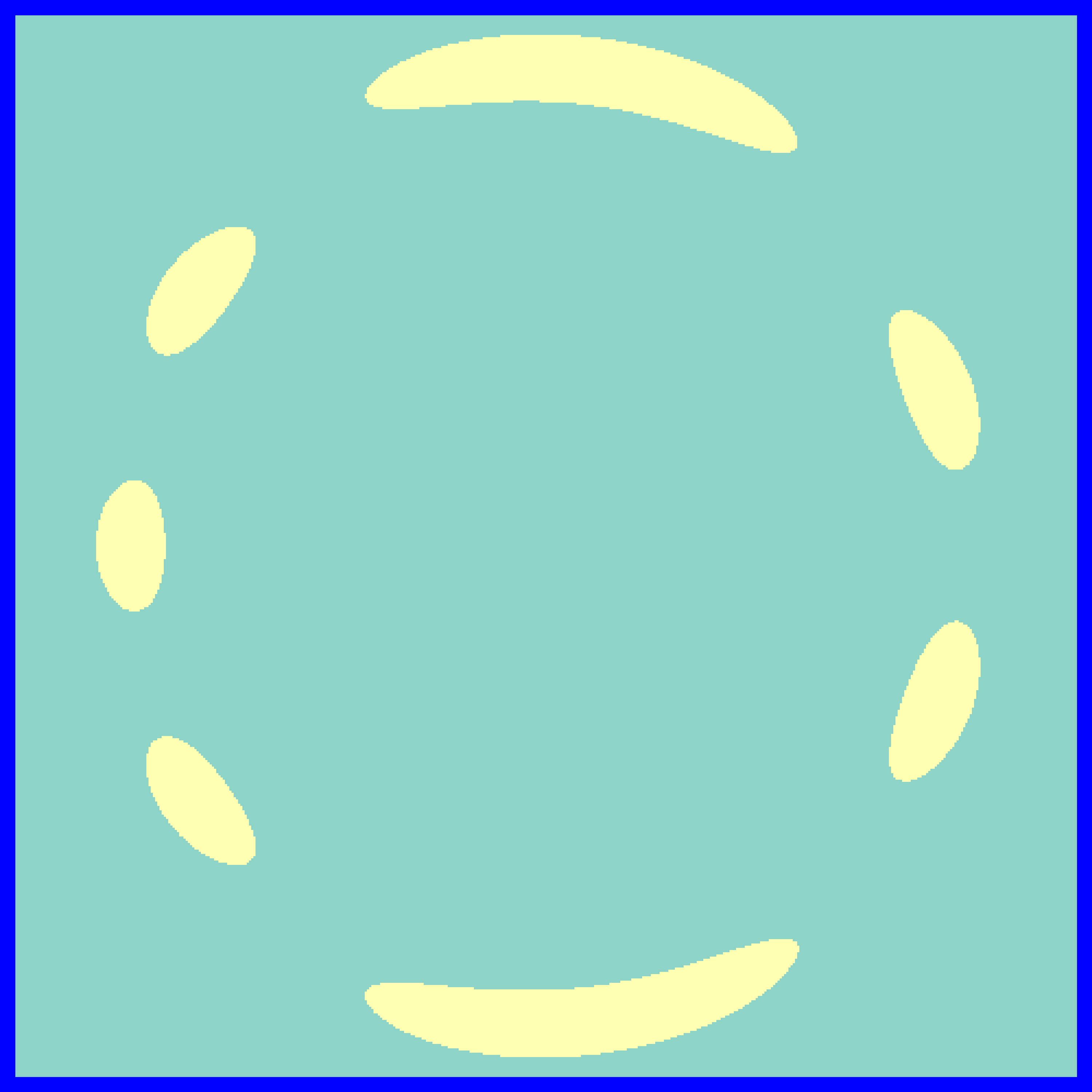} \\
    \includegraphics[width=0.45\linewidth]{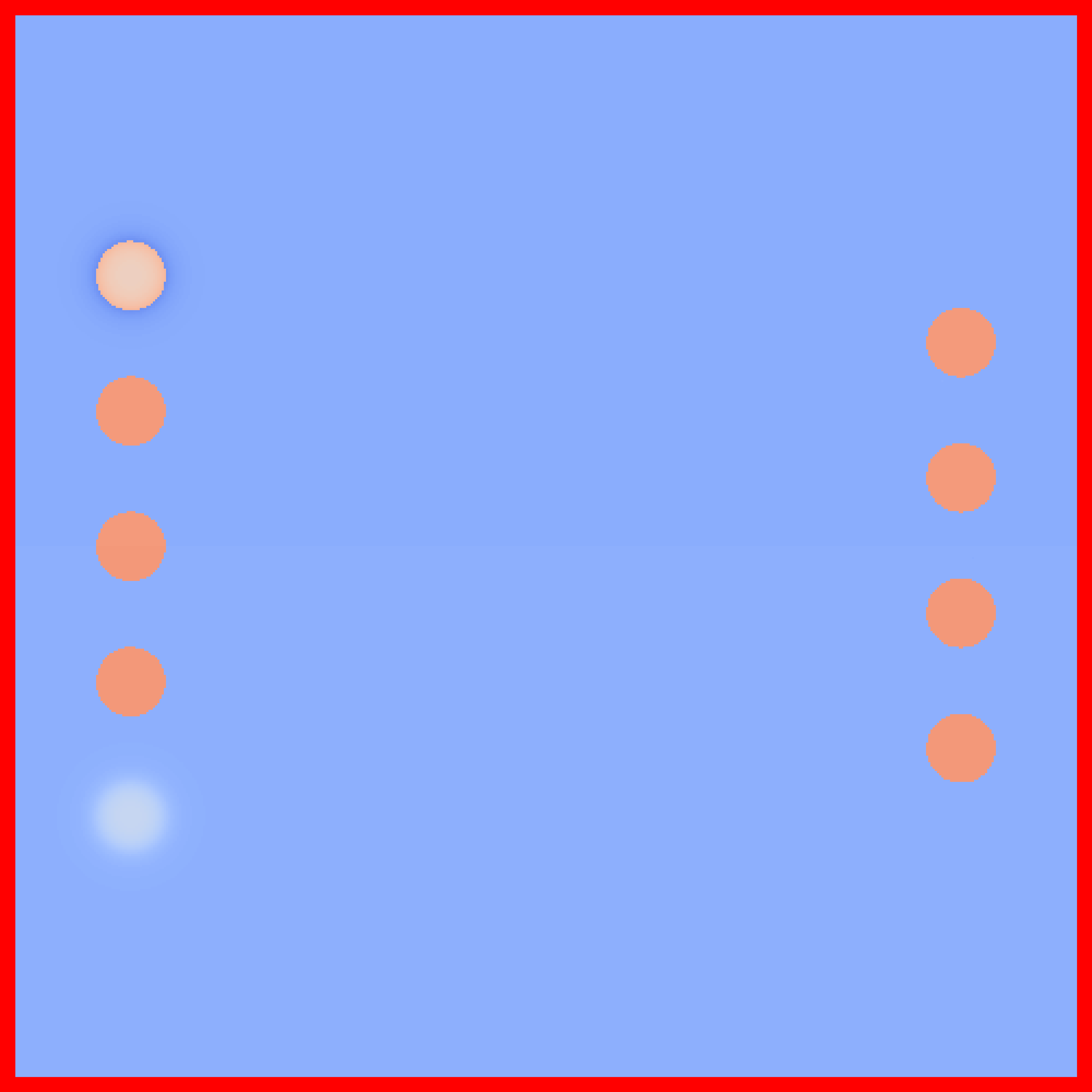} &
    \includegraphics[width=0.45\linewidth]{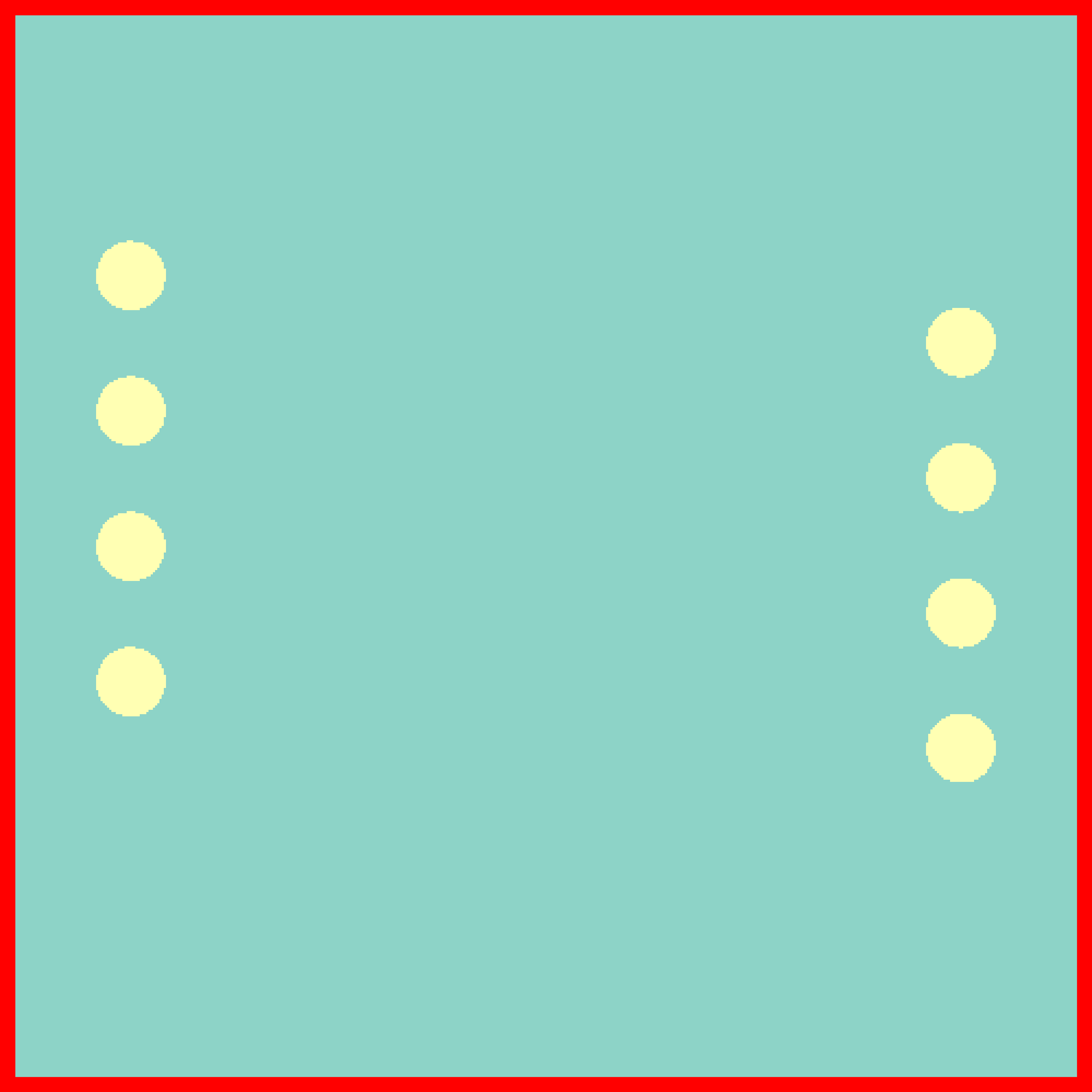} \\
    \includegraphics[width=0.45\linewidth]{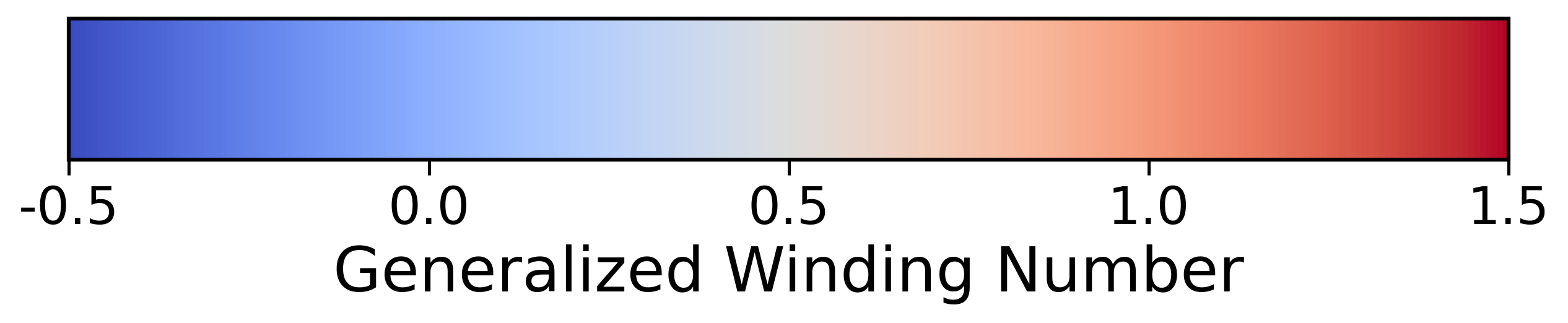} &
    \includegraphics[width=0.45\linewidth]{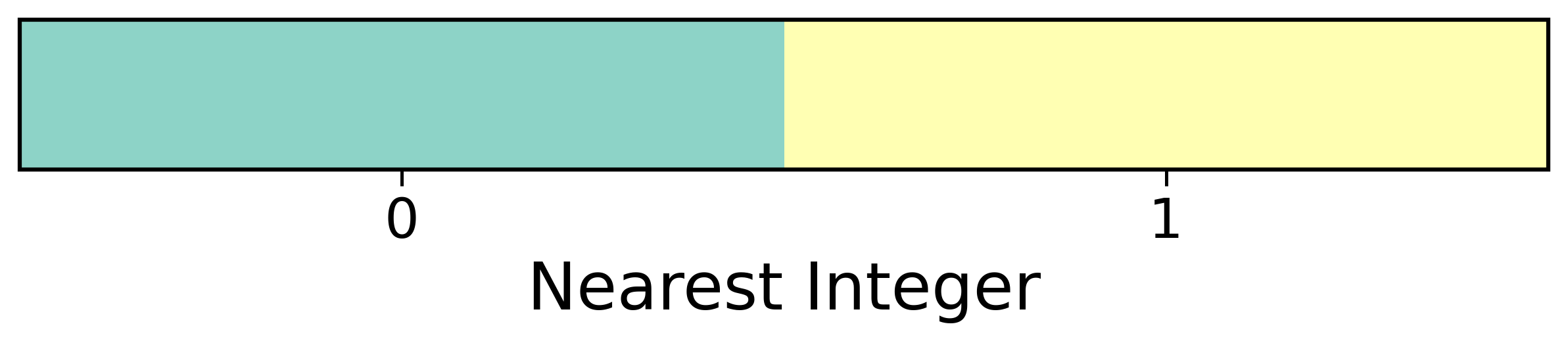} \\
\end{tabular}
\end{minipage}%
\begin{minipage}{0.5\textwidth}
\centering
\begin{tabular}{lr}
    \includegraphics[width=0.4\linewidth]{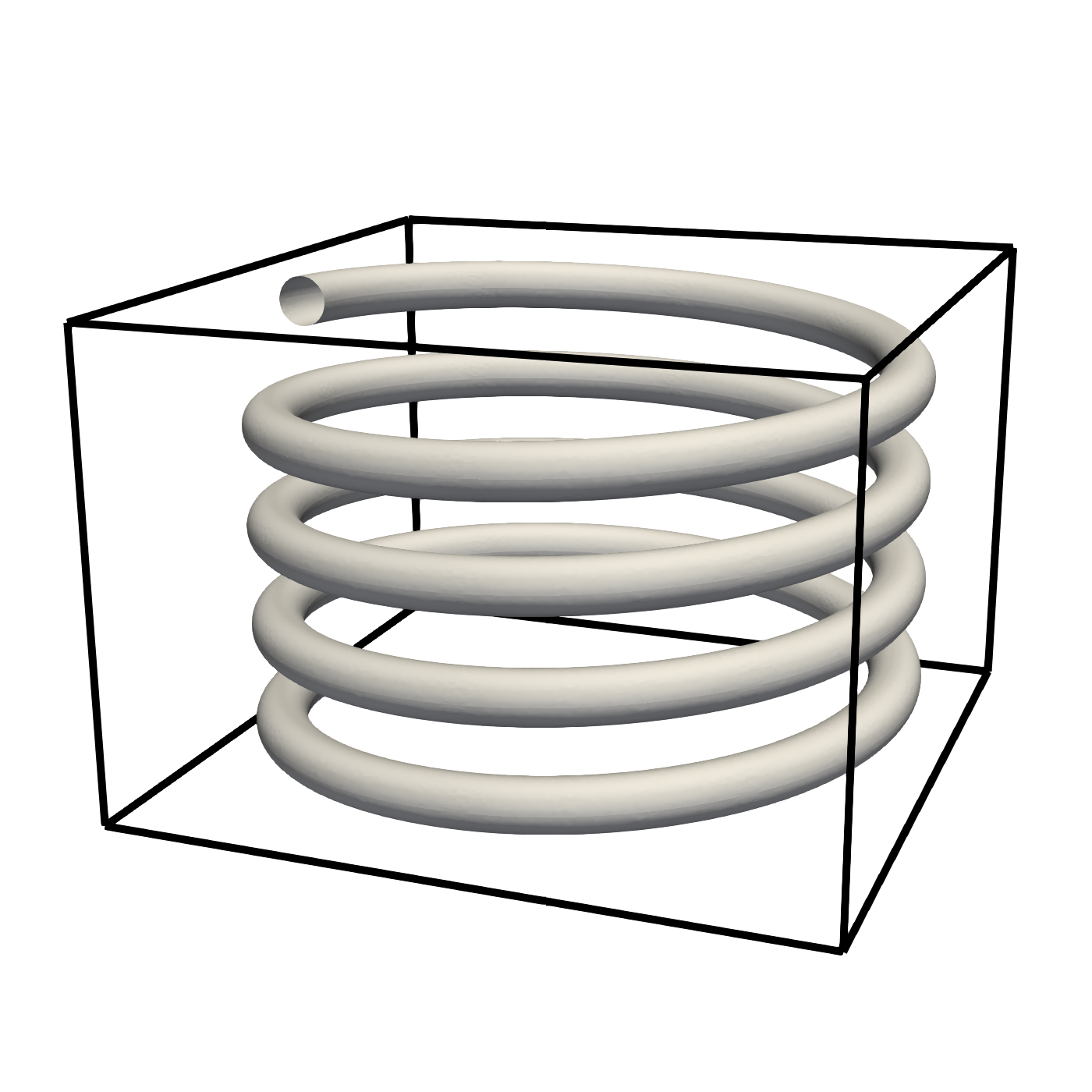}
    &
    \includegraphics[width=0.4\linewidth]{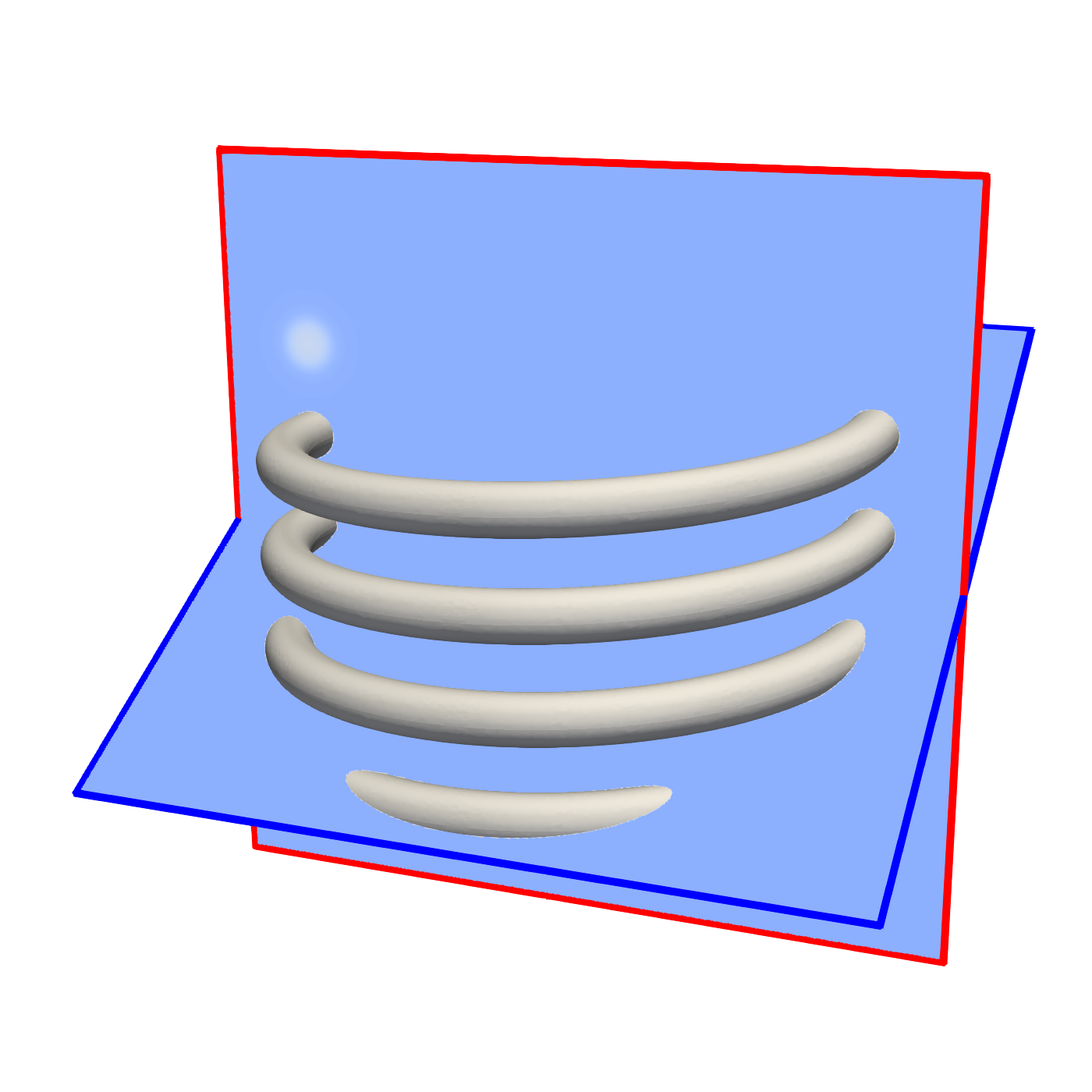}
\end{tabular}
\begin{tabular}{lrr}
                              &   & Two-Patch  \\
    Shape                          & Spring  & Spring \\\cmidrule{1-3}
    Number of NURBS Patches        & 64      & 2           \\
    Number of Trimming Curves      & 256     & 8           \\\cmidrule{1-3}
    \% Far-field Cases             & 99.57 \% & 18.26 \% \\
    \% Near-field Cases            & 0.42  \% & 76.15 \% \\
    \% Edge Cases                  & 0.006 \% & 5.58 \% \\\cmidrule{1-3}
    Avg. Time per Query (ms)       & 0.332    & 11.07      \\\cmidrule{1-3}
    Avg. Far-field Case Time (ms)  & 0.0043   & 0.0718        \\
    Avg. Near-field Case Time (ms) & 0.1301   & 0.336       \\
    Avg. Edge Case Time (ms)       & 5.55     & 94.3      \\
    \end{tabular}
\end{minipage}
\caption{This ``Spring'' shape is derived from the ABC dataset model with index 86. 
		We modified this originally watertight shape by removing the caps at each end. 
		In addition to the 64-patch model defined by untrimmed \bezier\ patches, we also consider a version composed of only 2 NURBS patches with multiple knot spans each.
		Interestingly, the 64-patch model has better overall performance due to the differing distribution of cases.
}
\end{figure}

\begin{figure}
\begin{minipage}{0.5\textwidth}
\centering
\begin{tabular}{lr}
    \includegraphics[width=0.45\linewidth]{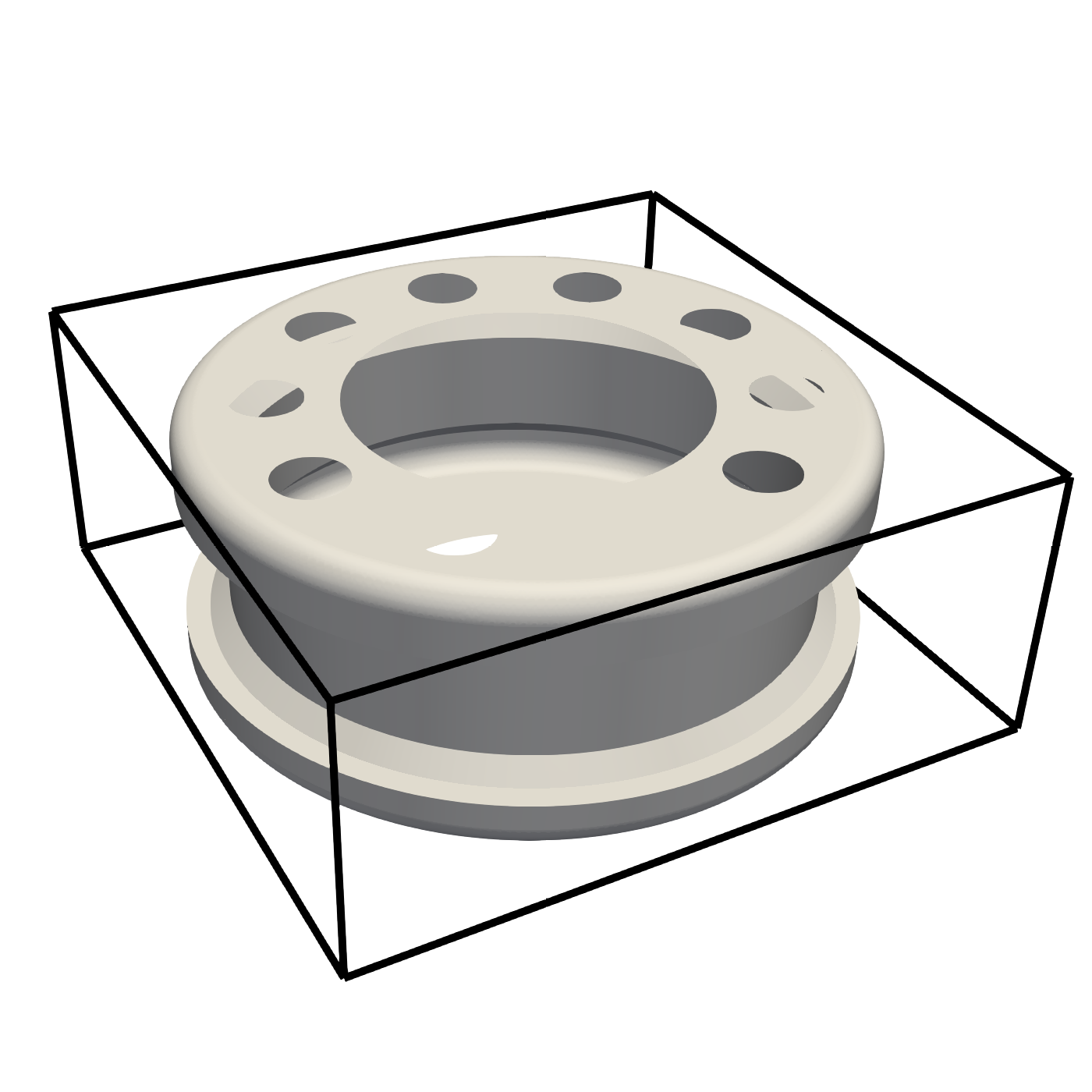}
    &
    \includegraphics[width=0.45\linewidth]{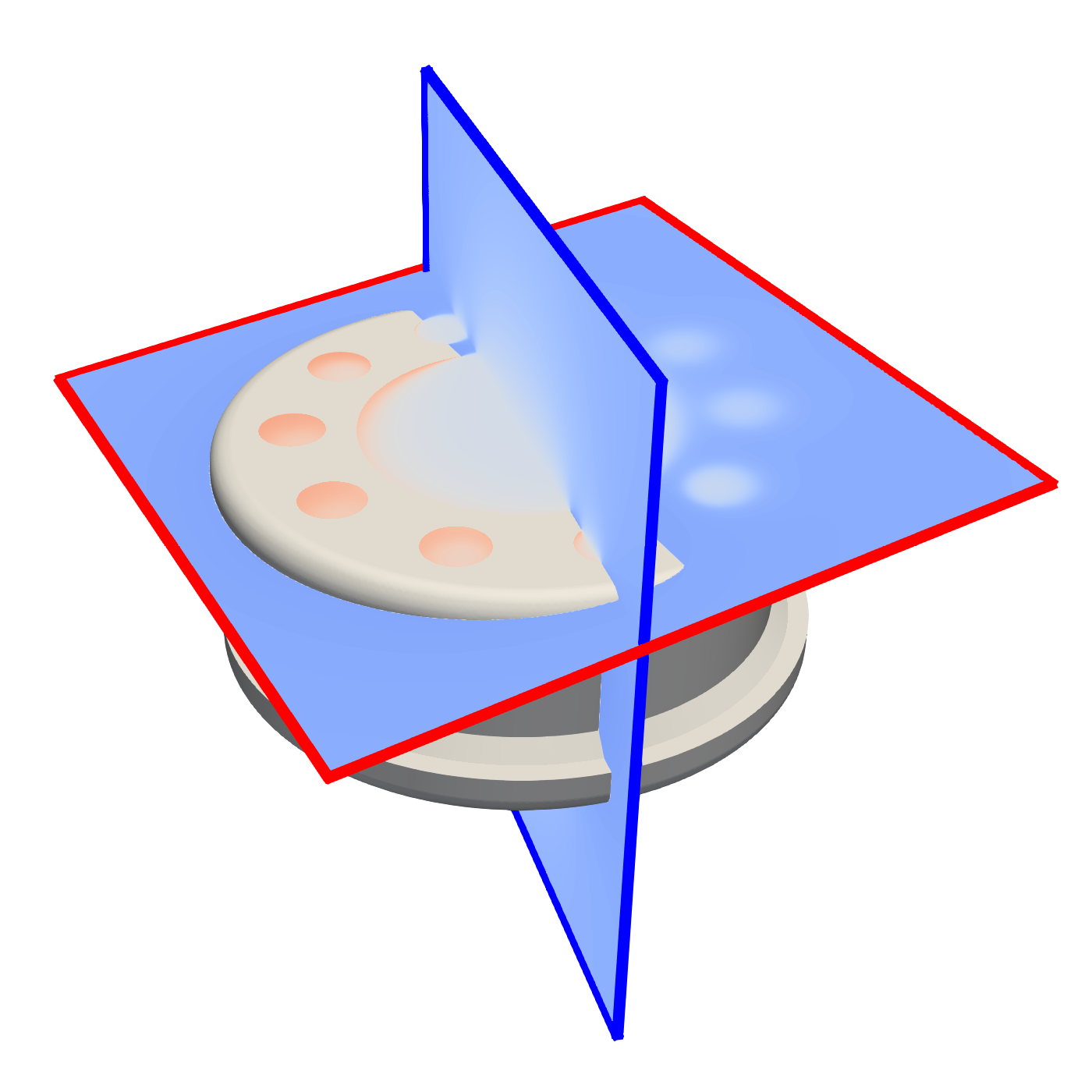}
\end{tabular}
\begin{tabular}{lr}
    Shape                          & Bobbin                        \\\cmidrule{1-2}
    Number of NURBS Patches        & 18                 \\
    Number of Trimming Curves      & 76                \\\cmidrule{1-2}
    \% Far-field Cases             & 92.56  \%  \\
    \% Near-field Cases            & 7.00 \%  \\
    \% Edge Cases                  & 0.42 \%  \\\cmidrule{1-2}
    Avg. Time per Query (ms)       & 1.12          \\\cmidrule{1-2}
    Avg. Far-field Case Time (ms)  & 0.0173           \\
    Avg. Near-field Case Time (ms) & 0.0805          \\
    Avg. Edge Case Time (ms)       & 9.38          \\
    \end{tabular}
\end{minipage}%
\begin{minipage}{0.5\textwidth}
\centering
\begin{tabular}{lr}
    \includegraphics[width=0.45\linewidth]{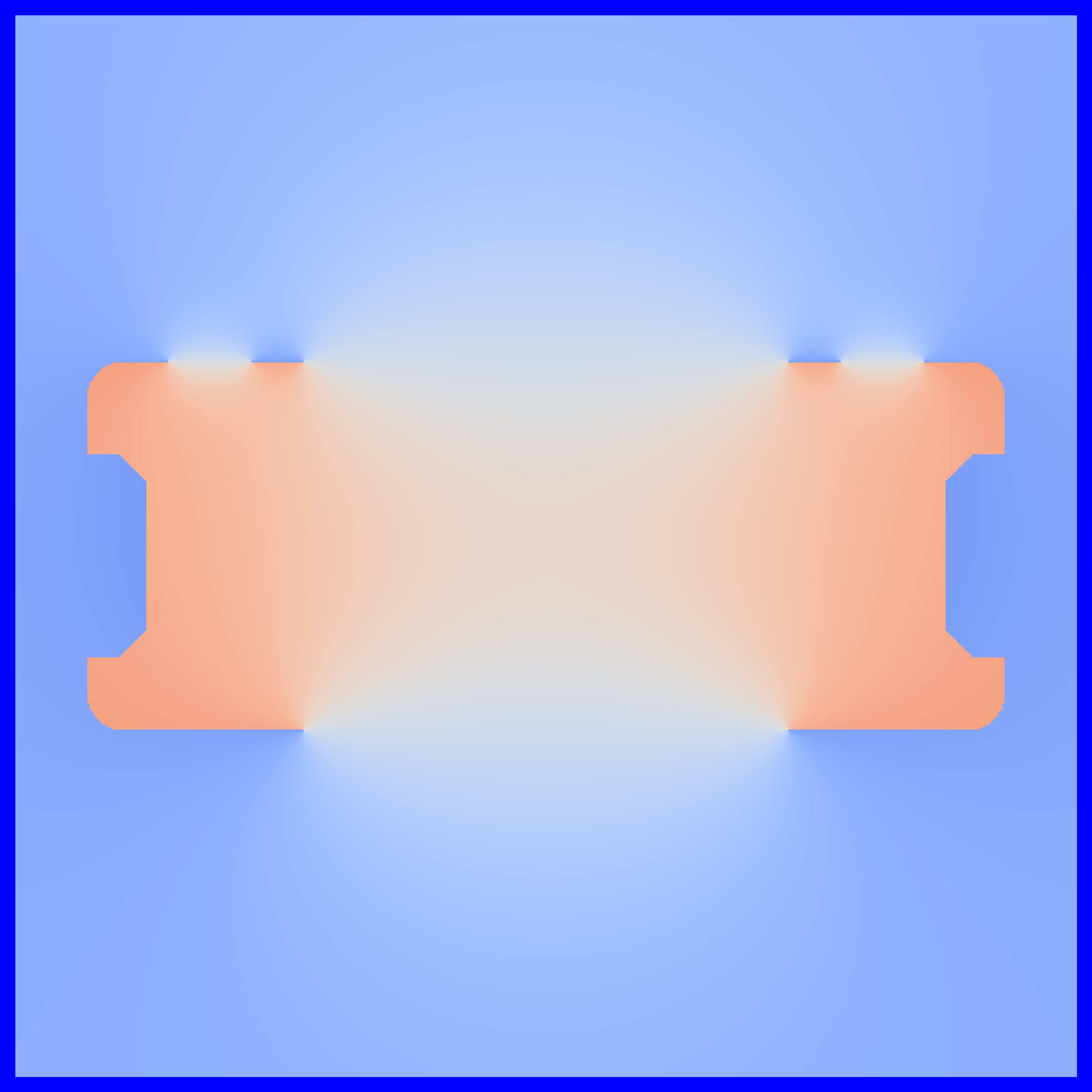} &
    \includegraphics[width=0.45\linewidth]{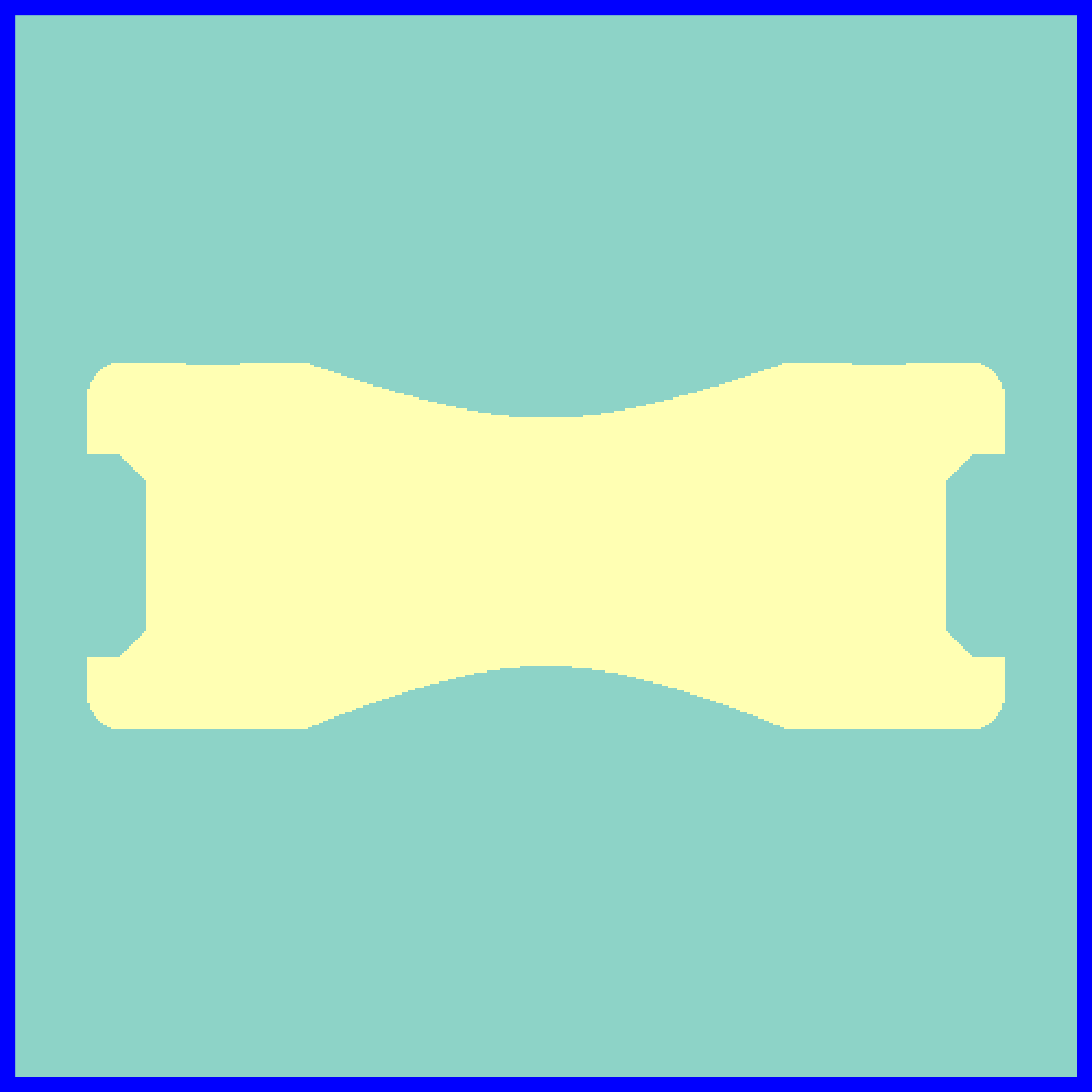} \\
    \includegraphics[width=0.45\linewidth]{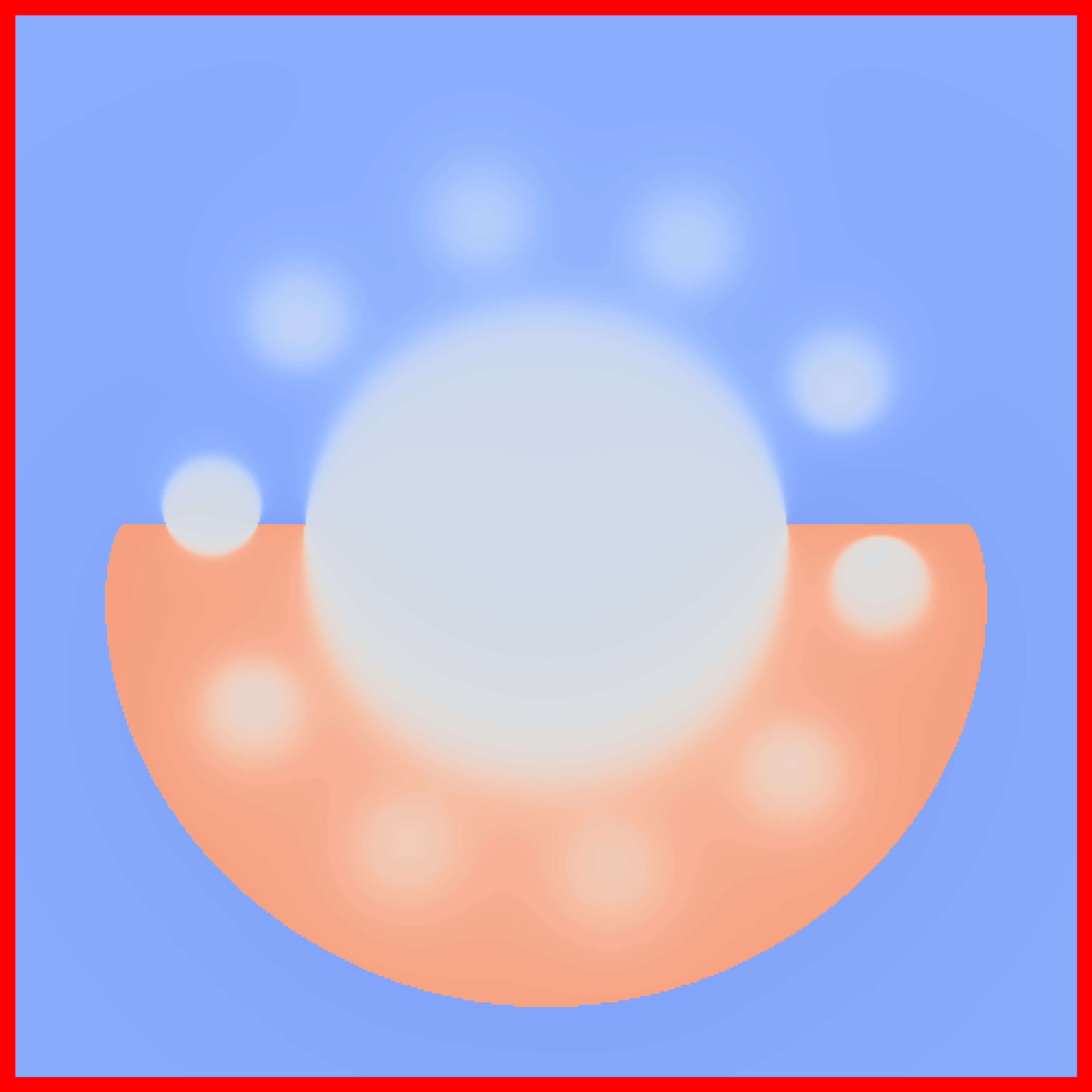} &
    \includegraphics[width=0.45\linewidth]{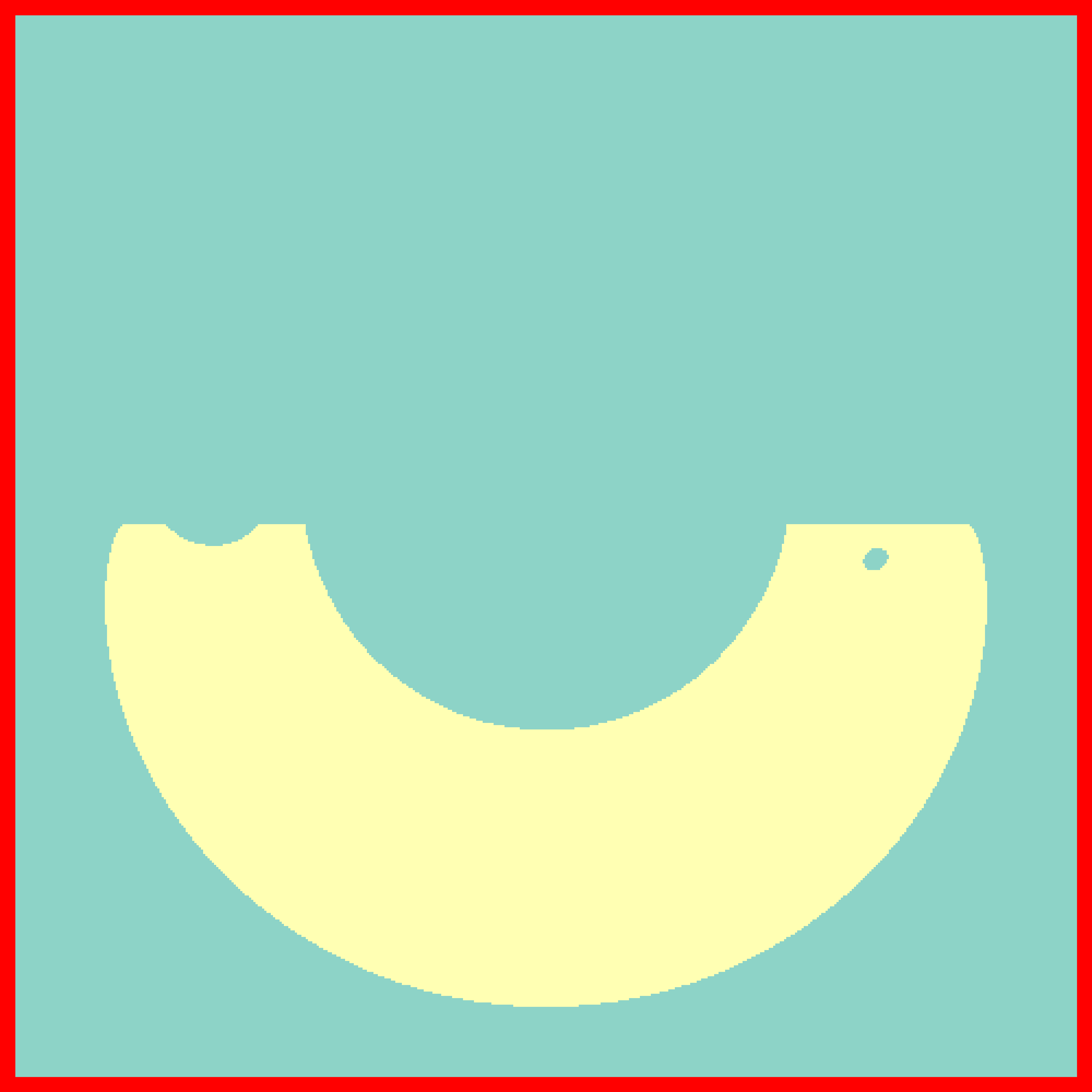} \\
    \includegraphics[width=0.45\linewidth]{summary_figures/gwn_colorbar_15.png} &
    \includegraphics[width=0.45\linewidth]{summary_figures/categories_colorbar_2.png} \\
\end{tabular}
\end{minipage}
\caption{This ``Bobbin'' shape is derived from the ABC dataset model with index 933. We modified this originally watertight shape by removing the interior portion and some additional features to reveal holes on the top face.}
\end{figure}

\vspace{1cm}

\begin{figure}

\begin{minipage}{0.5\textwidth}
\centering
\begin{tabular}{lr}
    \includegraphics[width=0.45\linewidth]{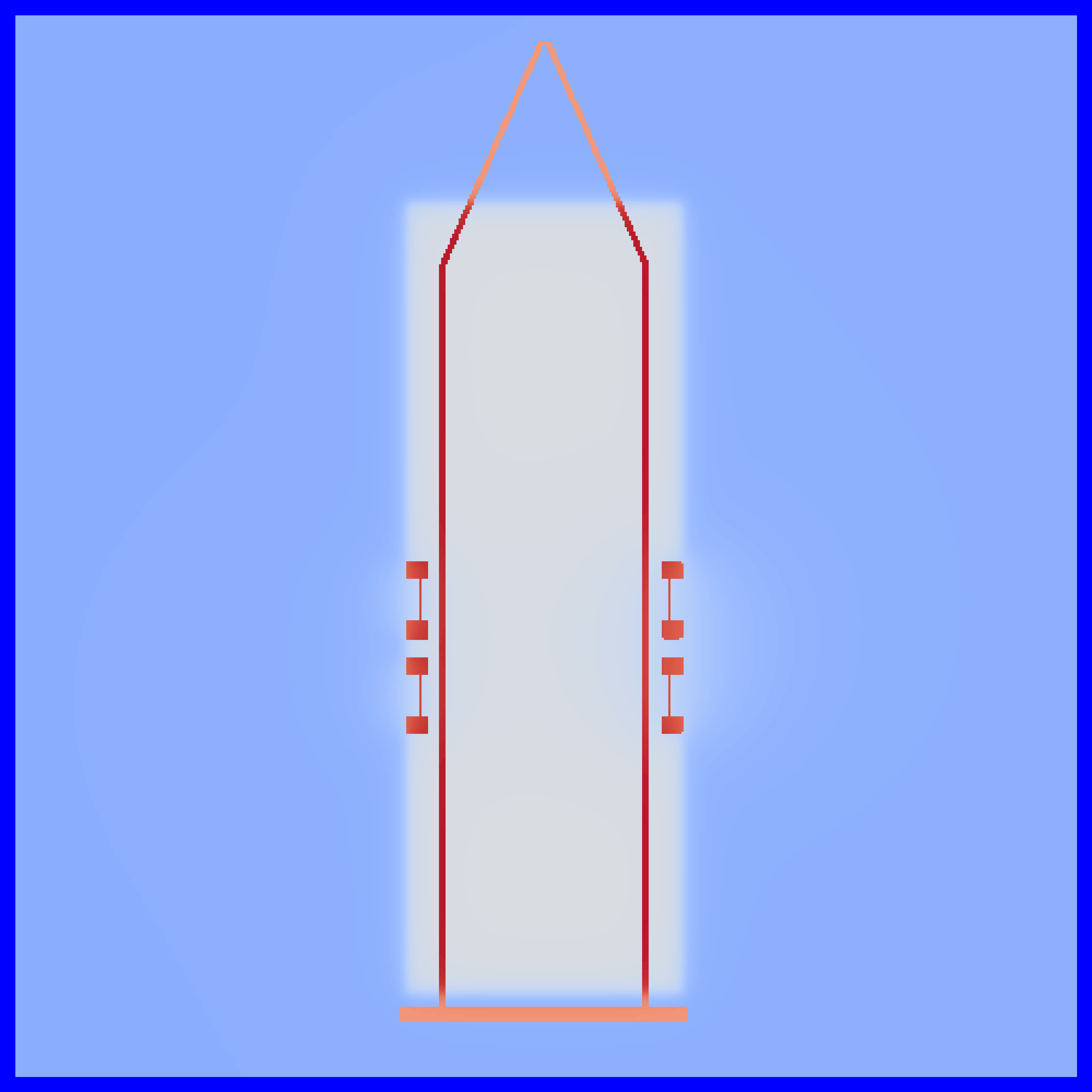} &
    \includegraphics[width=0.45\linewidth]{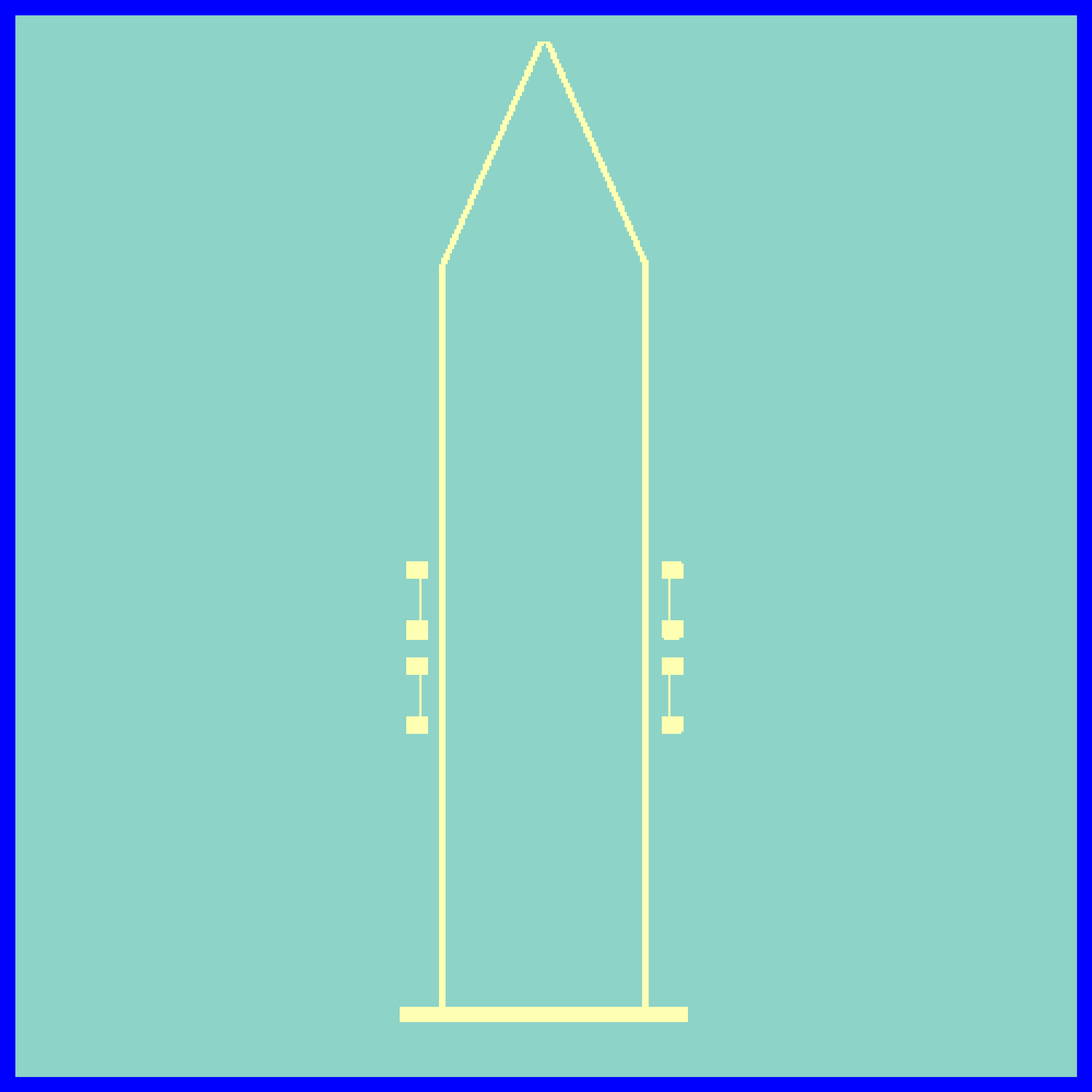} \\
    \includegraphics[width=0.45\linewidth]{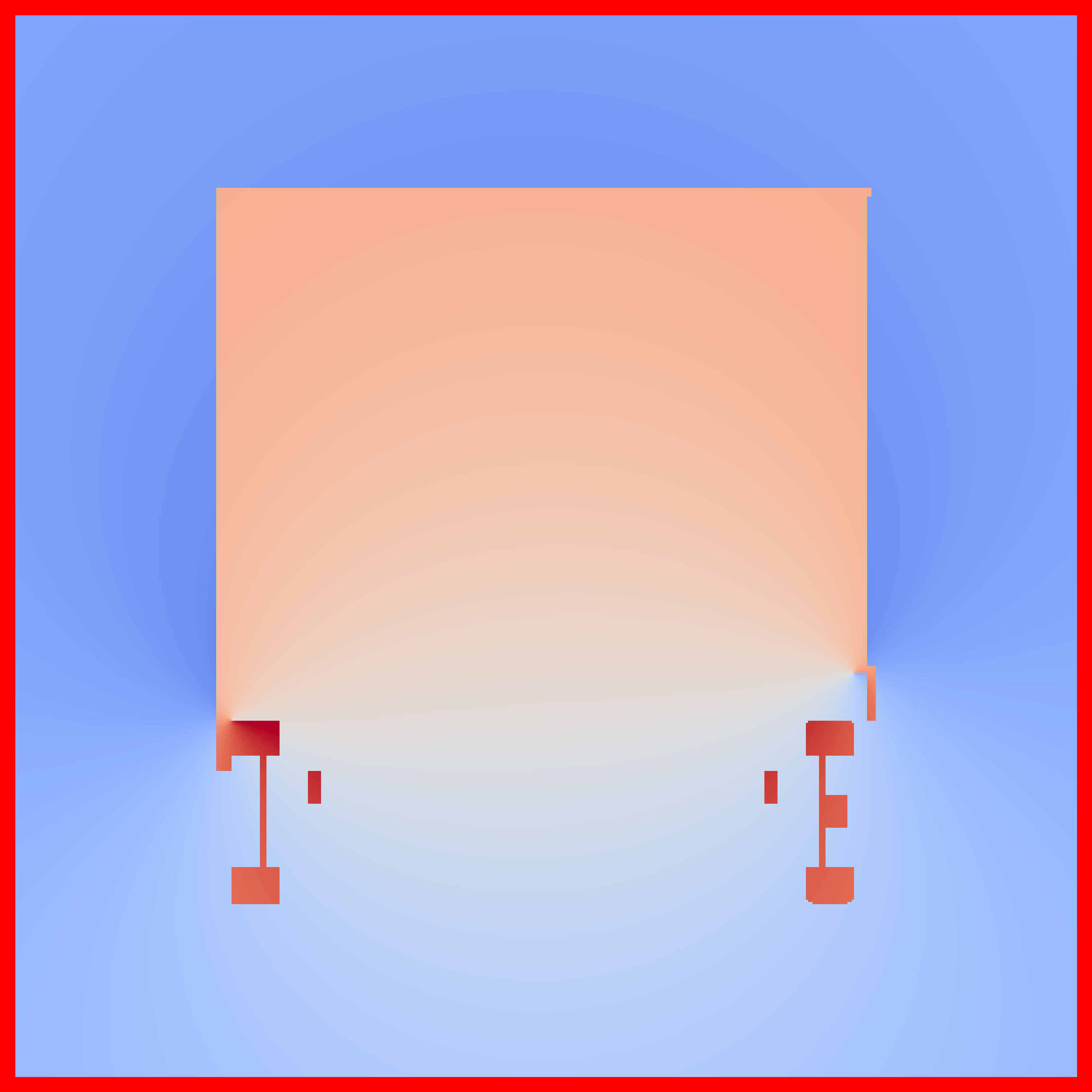} &
    \includegraphics[width=0.45\linewidth]{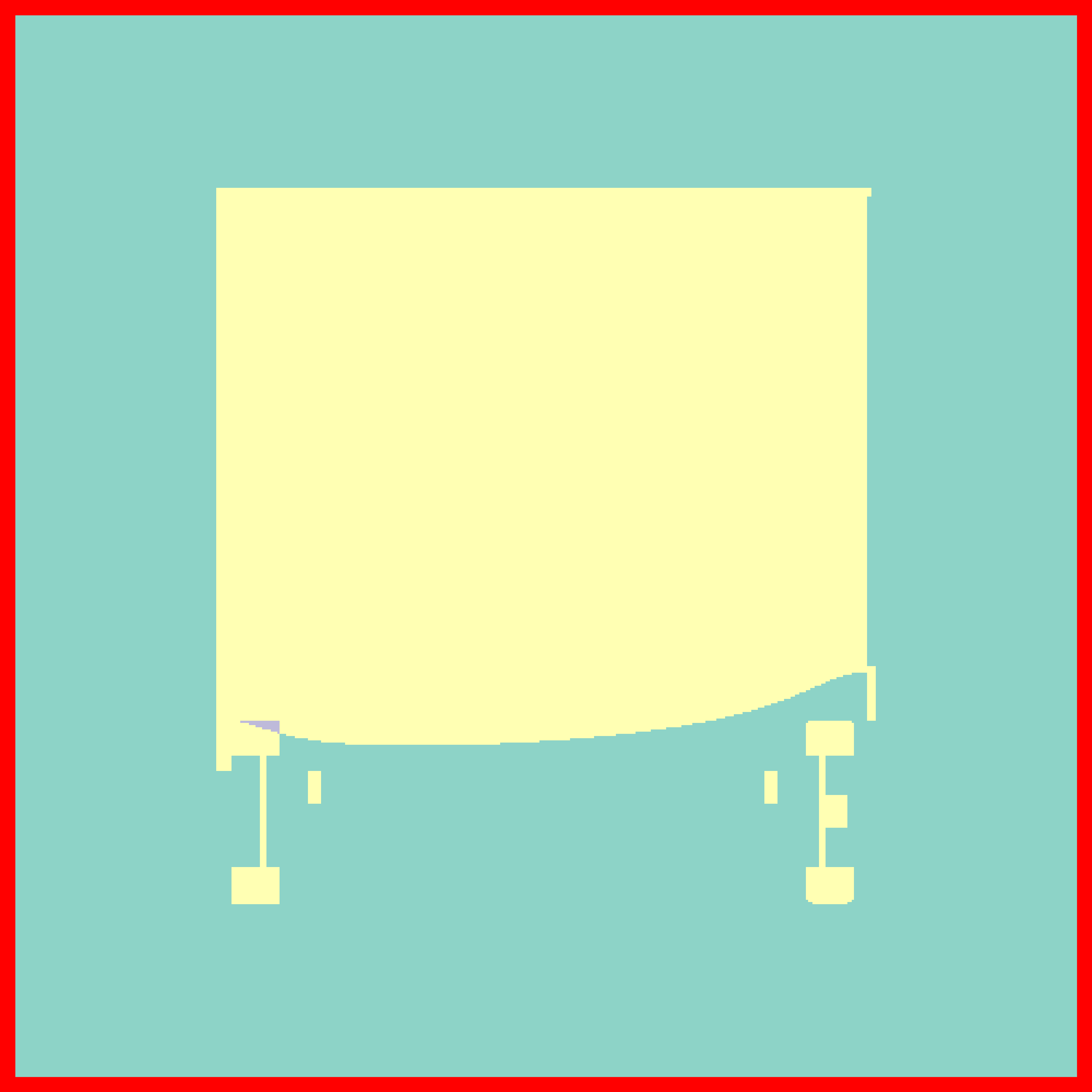} \\
    \includegraphics[width=0.45\linewidth]{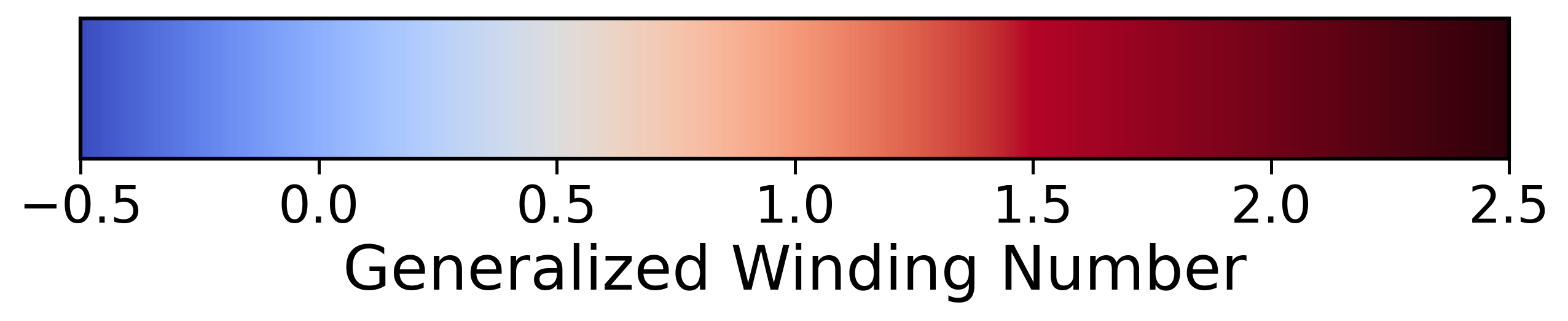} &
    \includegraphics[width=0.45\linewidth]{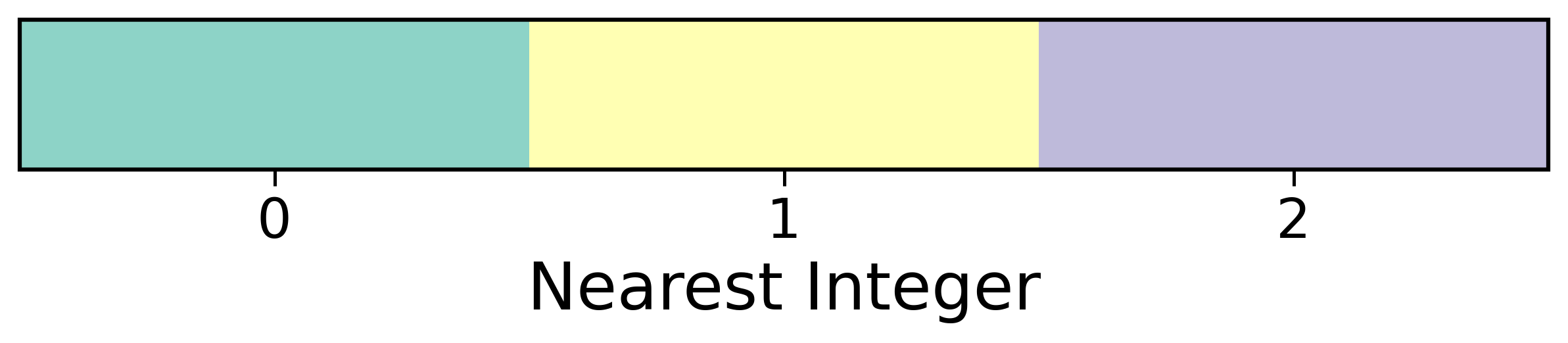} \\
\end{tabular}
\end{minipage}%
\begin{minipage}{0.5\textwidth}
\centering
\begin{tabular}{lr}
    \includegraphics[width=0.45\linewidth]{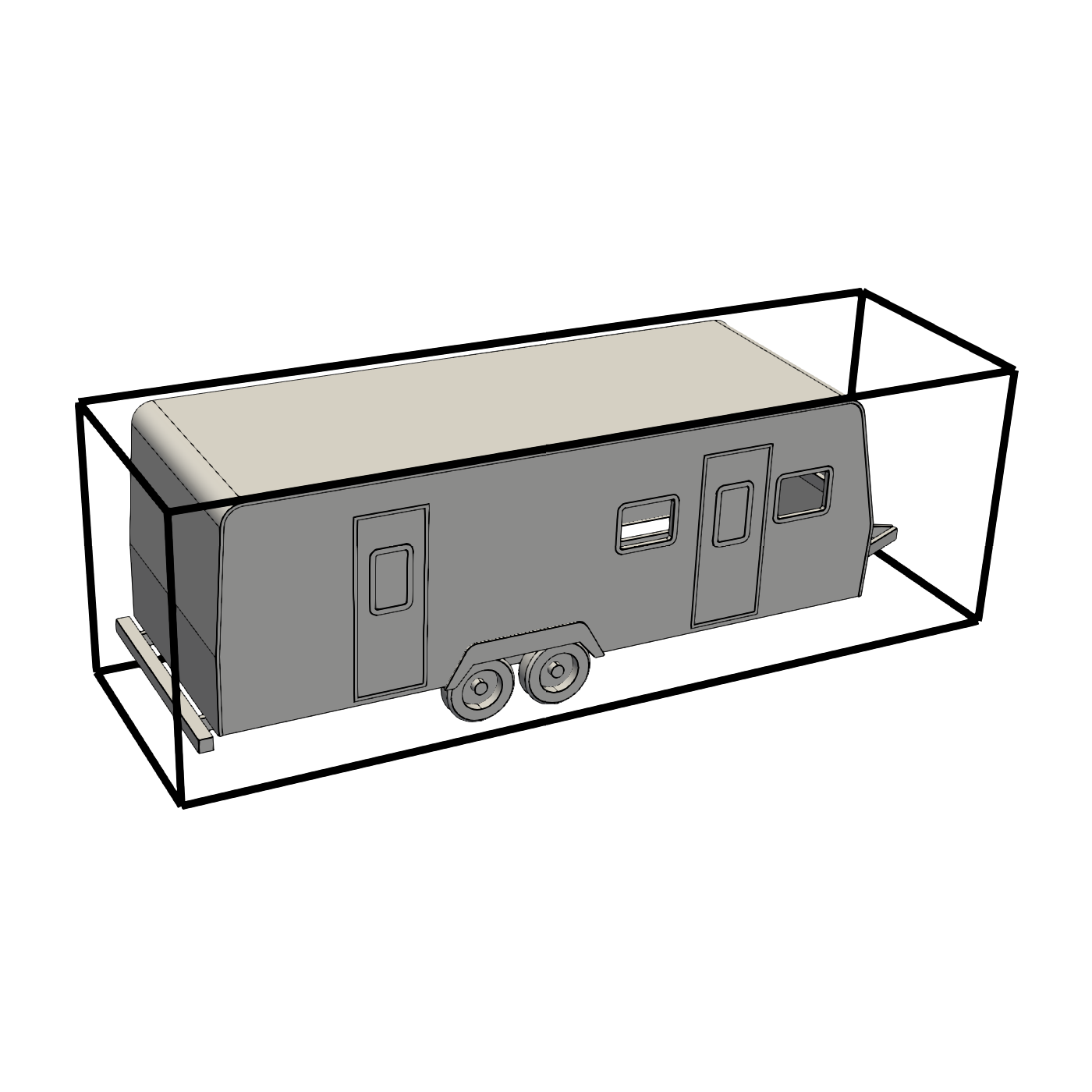}
    &
    \includegraphics[width=0.45\linewidth]{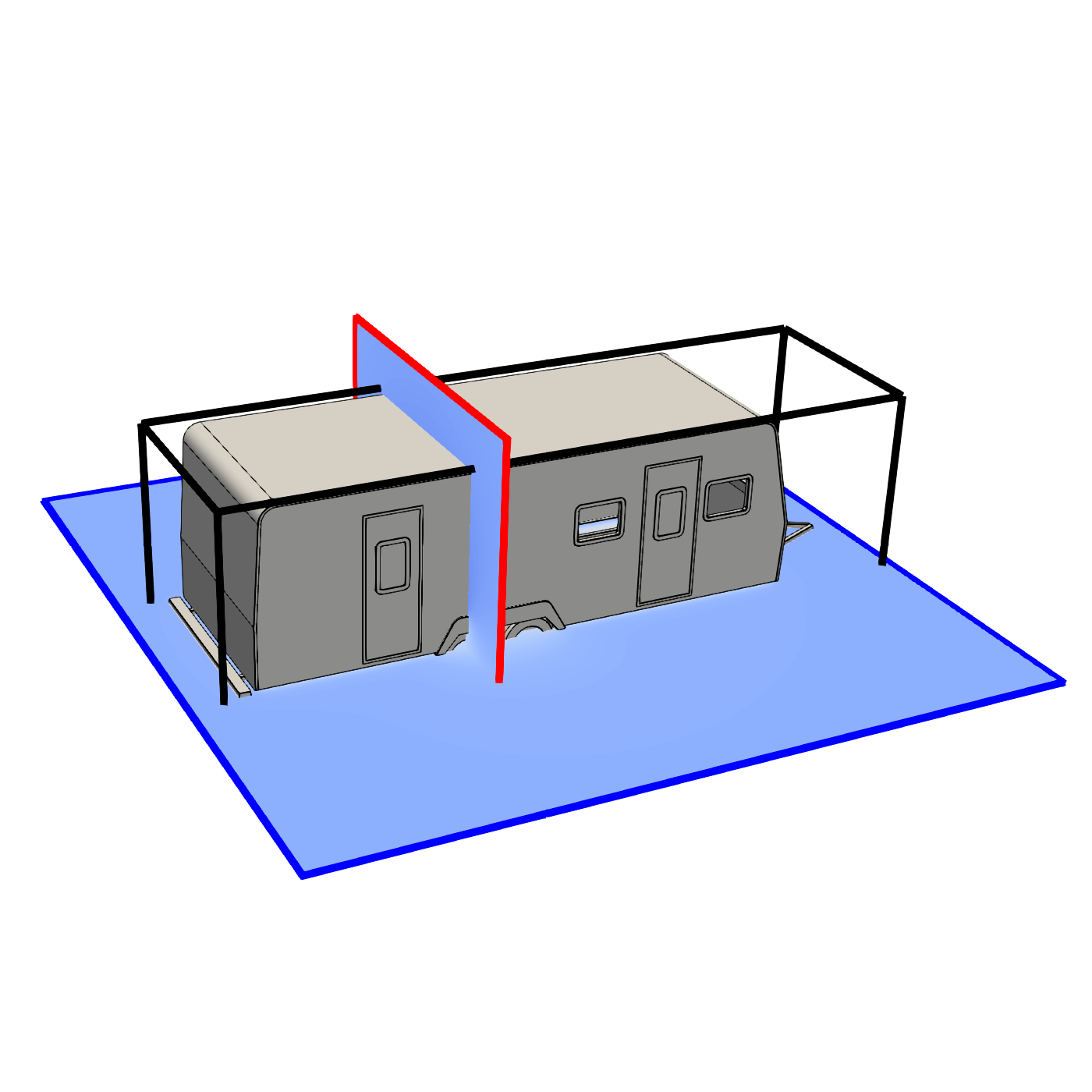}
\end{tabular}
\begin{tabular}{lr}
    Shape                          & Trailer                        \\\cmidrule{1-2}
    Number of NURBS Patches        & 228                 \\
    Number of Trimming Curves      & 1092                \\\cmidrule{1-2}
    \% Far-field Cases             & 99.8  \%  \\
    \% Near-field Cases            & 0.116 \%  \\
    \% Edge Cases                  & 0.014 \%  \\\cmidrule{1-2}
    Avg. Time per Query (ms)       & 1.18          \\\cmidrule{1-2}
    Avg. Far-field Case Time (ms)  & 0.0046           \\
    Avg. Near-field Case Time (ms) & 0.0585          \\
    Avg. Edge Case Time (ms)       & 3.97          \\
    \end{tabular}
\end{minipage}
\caption{This ``Trailer'' shape is derived from the ABC dataset model with index 4192. We modified this originally watertight shape by removing the bottom face, as well as some interior details and the surface making up the windows. 
				 Note that this shape, even in its original watertight version, contains features which overlap with one another, and therefore result in winding numbers greater than 1.}
\end{figure}

\vspace{1cm}

\begin{figure}

\begin{minipage}{0.5\textwidth}
\centering
\begin{tabular}{lr}
    \includegraphics[width=0.45\linewidth]{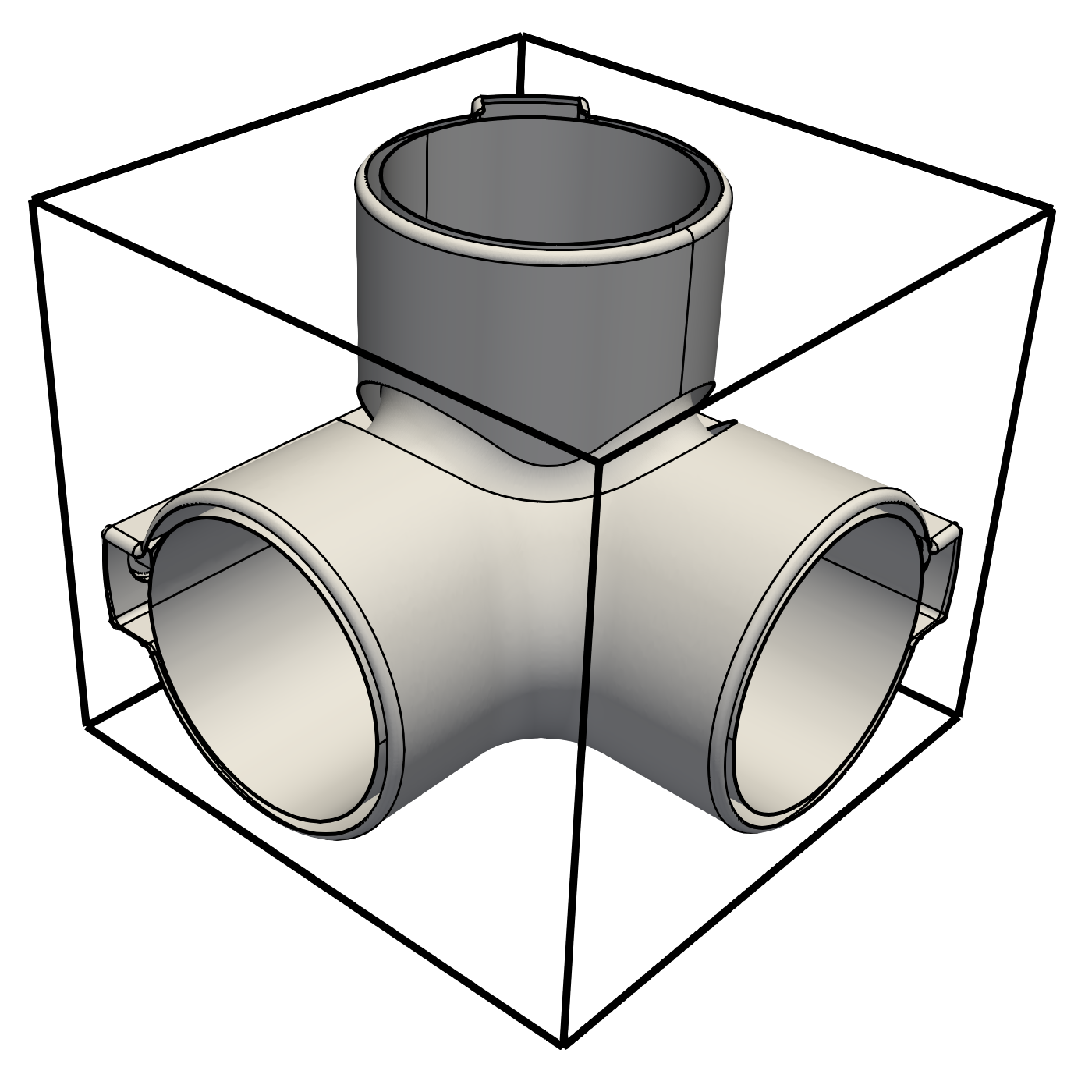}
    &
    \includegraphics[width=0.45\linewidth]{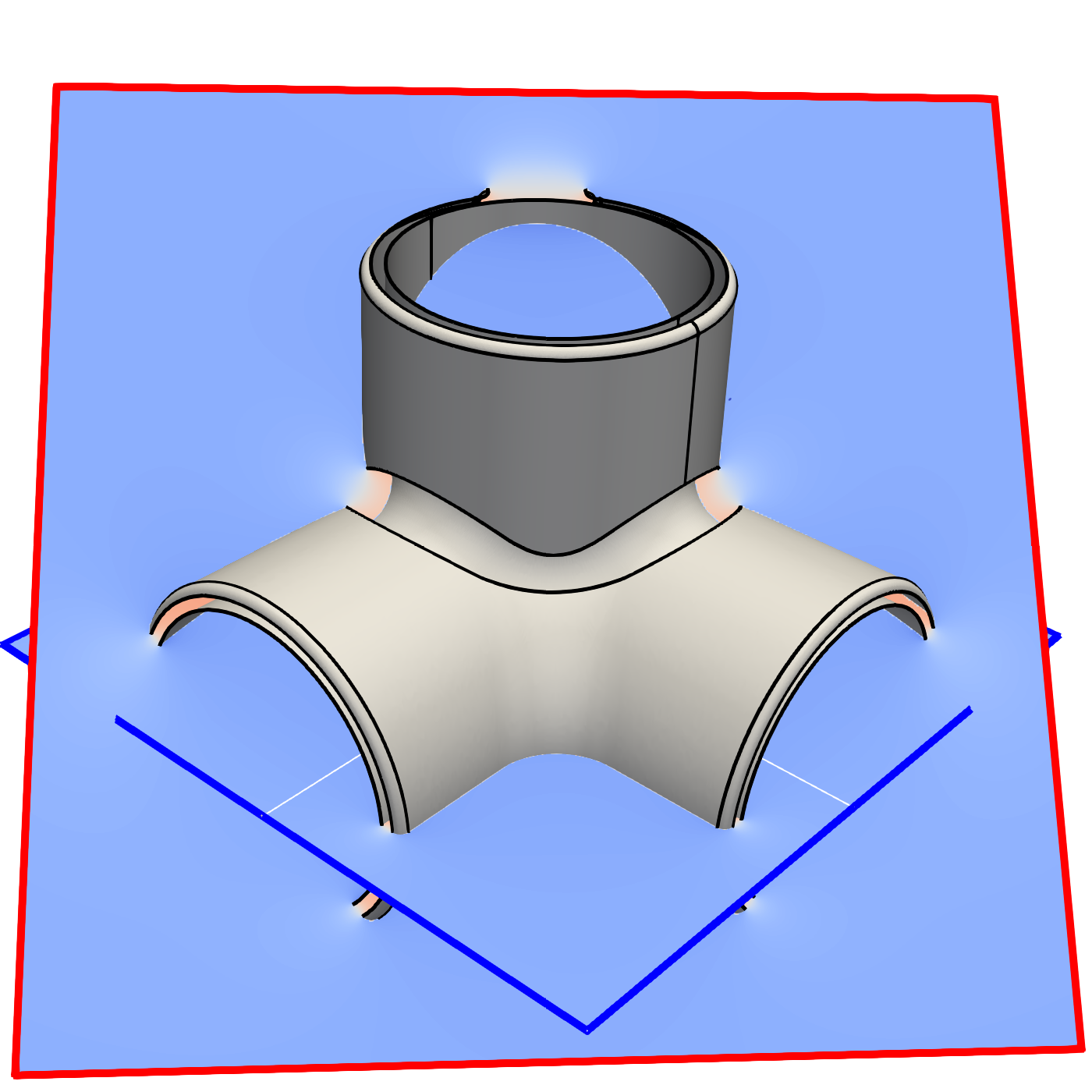}
\end{tabular}
\begin{tabular}{lr}
    Shape                          & Pipe                        \\\cmidrule{1-2}
    Number of NURBS Patches        & 133                 \\
    Number of Trimming Curves      & 563                \\\cmidrule{1-2}
    \% Far-field Cases             & 99.02\%  \\
    \% Near-field Cases            & 0.81 \%  \\
    \% Edge Cases                  & 0.15 \%  \\\cmidrule{1-2}
    Avg. Time per Query (ms)       & 21.1          \\\cmidrule{1-2}
    Avg. Far-field Case Time (ms)  & 0.0313           \\
    Avg. Near-field Case Time (ms) & 0.916          \\
    Avg. Edge Case Time (ms)       & 76.7          \\
    \end{tabular}
\end{minipage}%
\begin{minipage}{0.5\textwidth}
\centering
\begin{tabular}{lr}
    \includegraphics[width=0.45\linewidth]{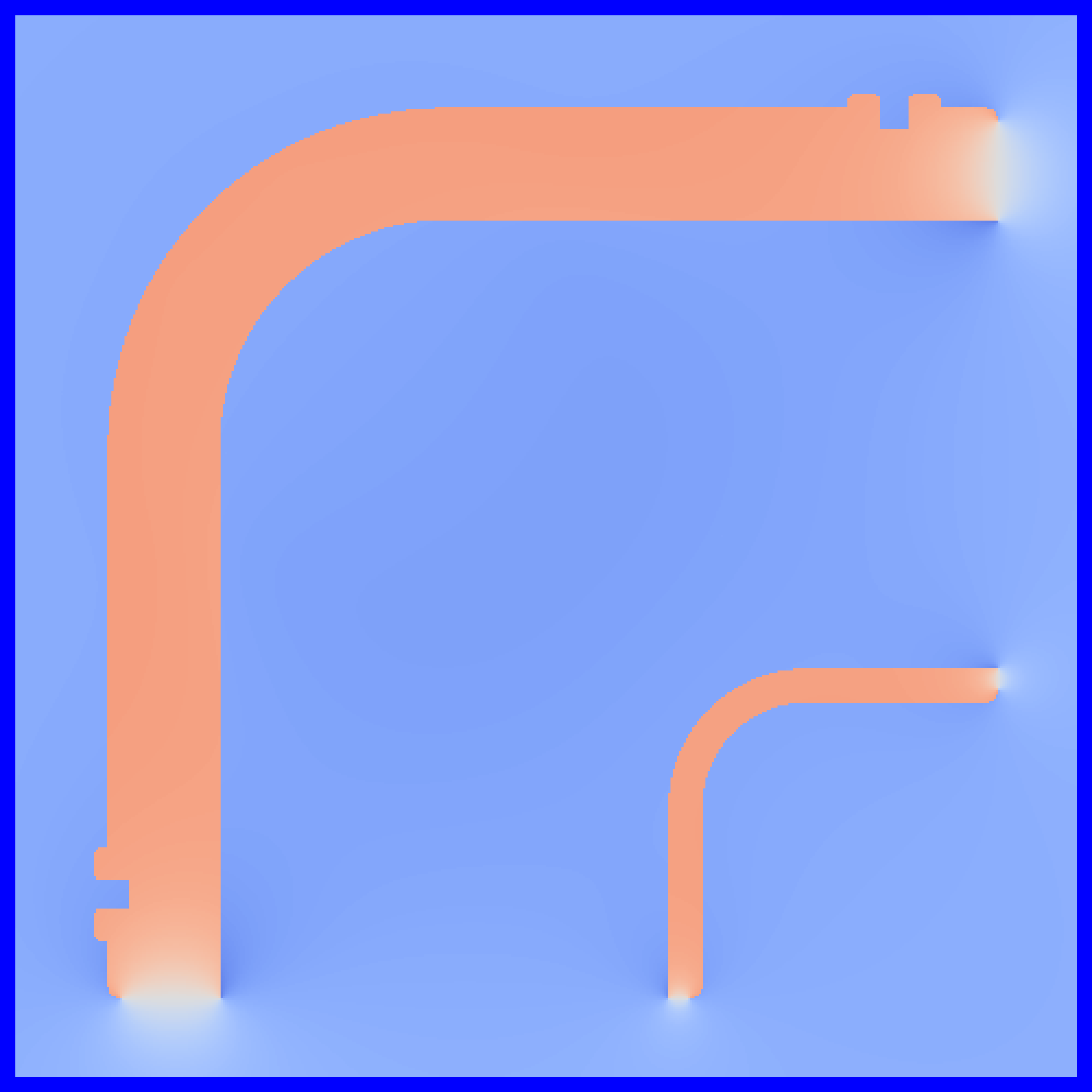} &
    \includegraphics[width=0.45\linewidth]{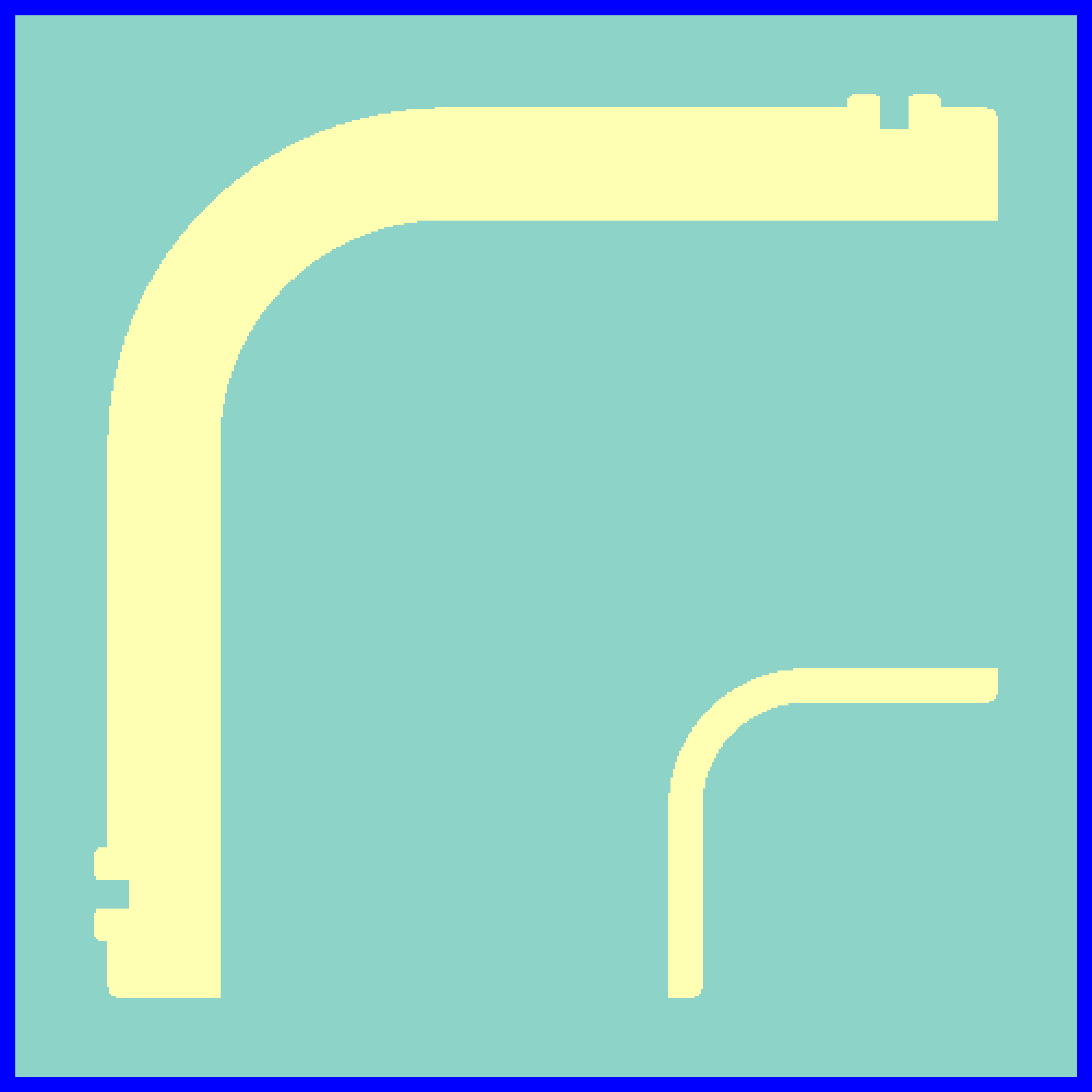} \\
    \includegraphics[width=0.45\linewidth]{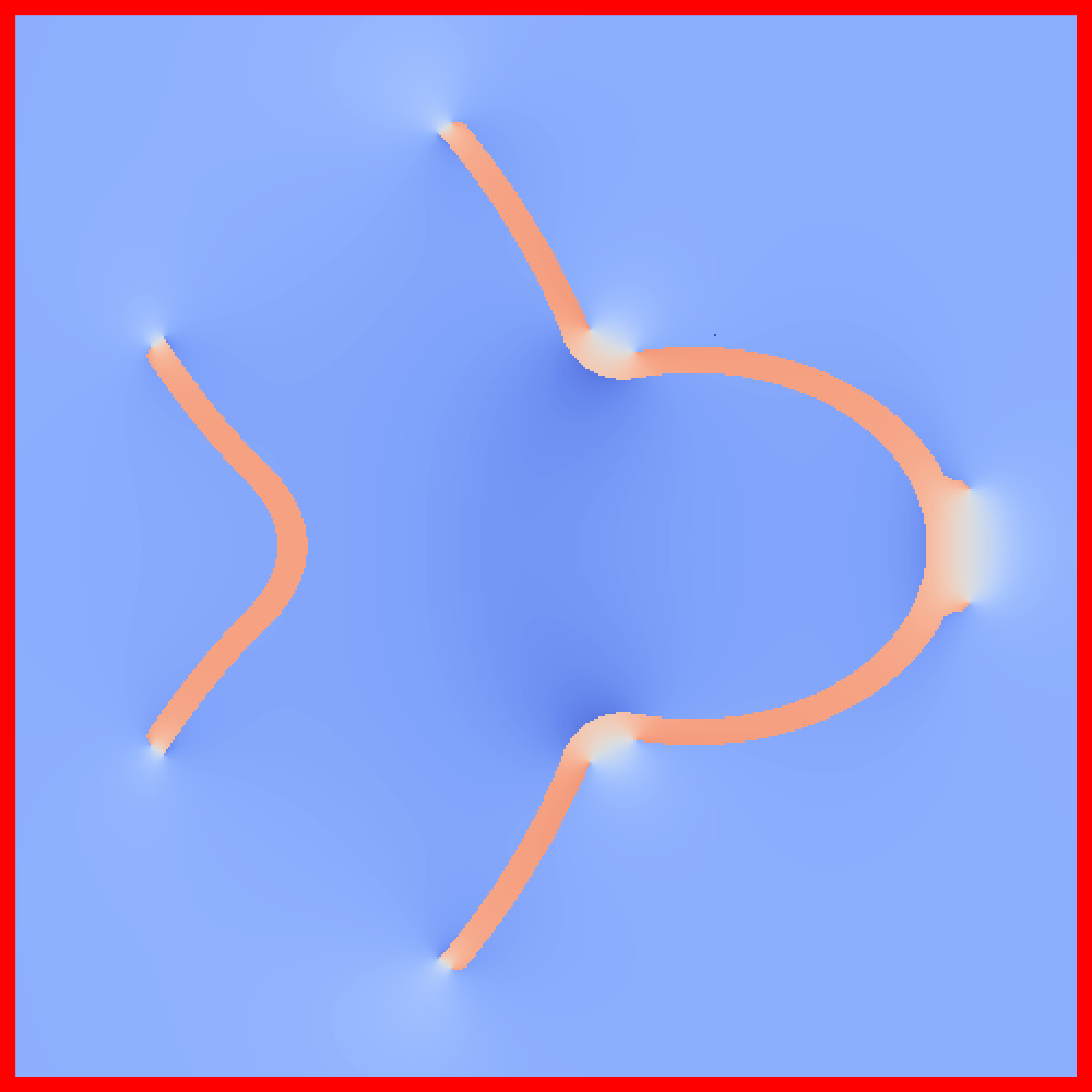} &
    \includegraphics[width=0.45\linewidth]{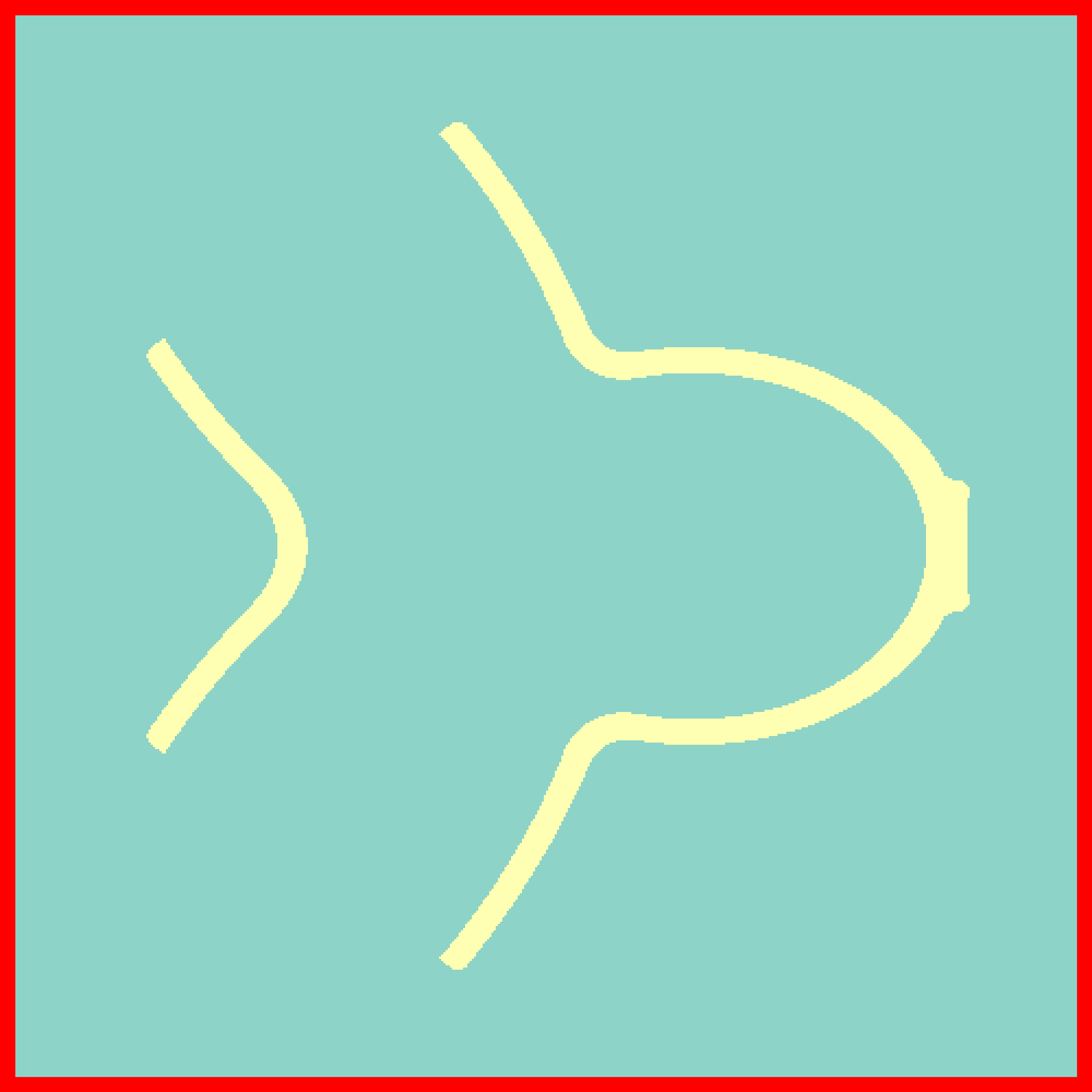} \\
    \includegraphics[width=0.45\linewidth]{summary_figures/gwn_colorbar_15.png} &
    \includegraphics[width=0.45\linewidth]{summary_figures/categories_colorbar_2.png} \\
\end{tabular}
\end{minipage}
\caption{This ``Pipe'' shape is derived from the ABC dataset model with index 9992. We modified this originally watertight shape by removing the front face of each opening and a patch in the interior.}
\end{figure}

\vspace{1cm}

\begin{figure}

\begin{minipage}{0.5\textwidth}
\centering
\begin{tabular}{lr}
    \includegraphics[width=0.45\linewidth]{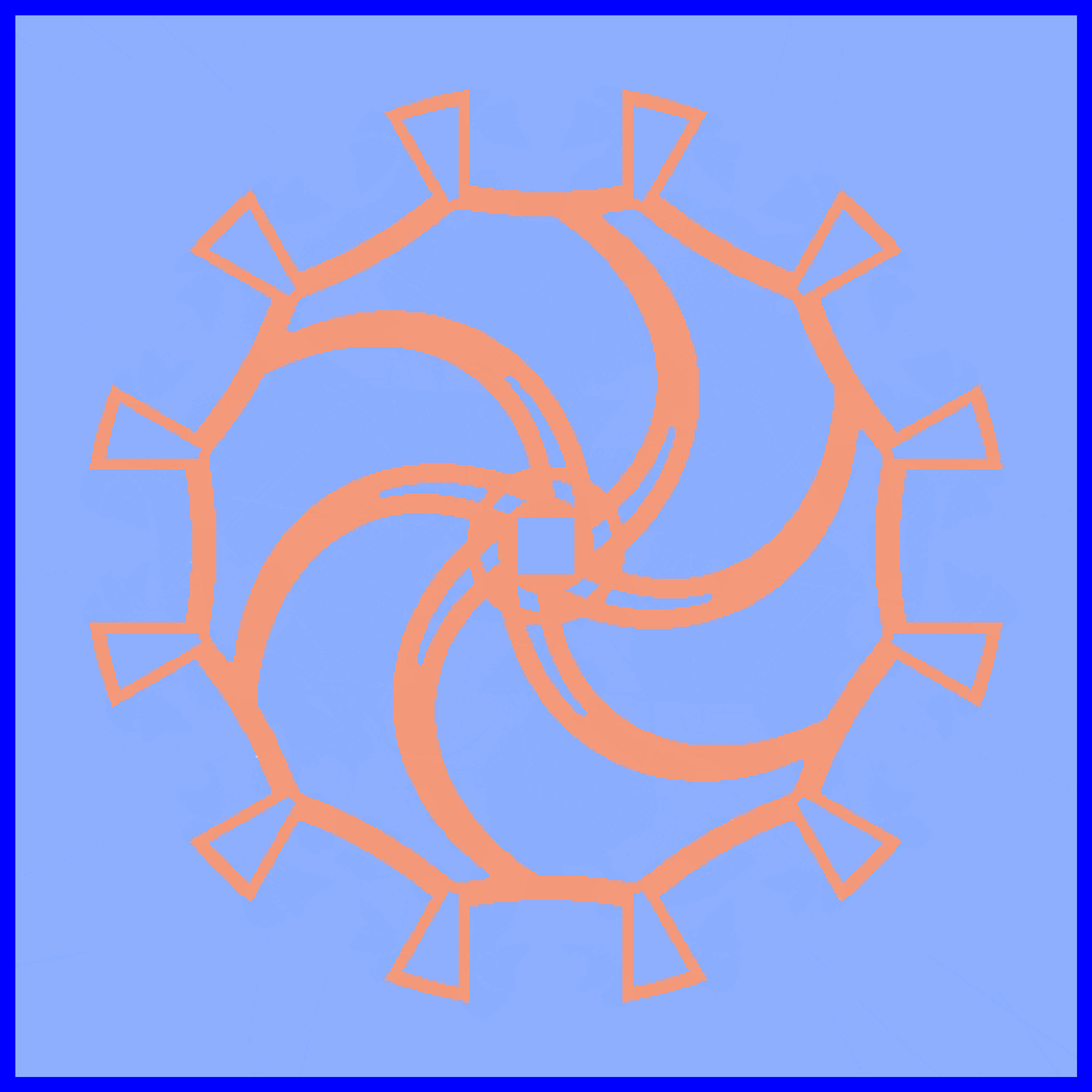} &
    \includegraphics[width=0.45\linewidth]{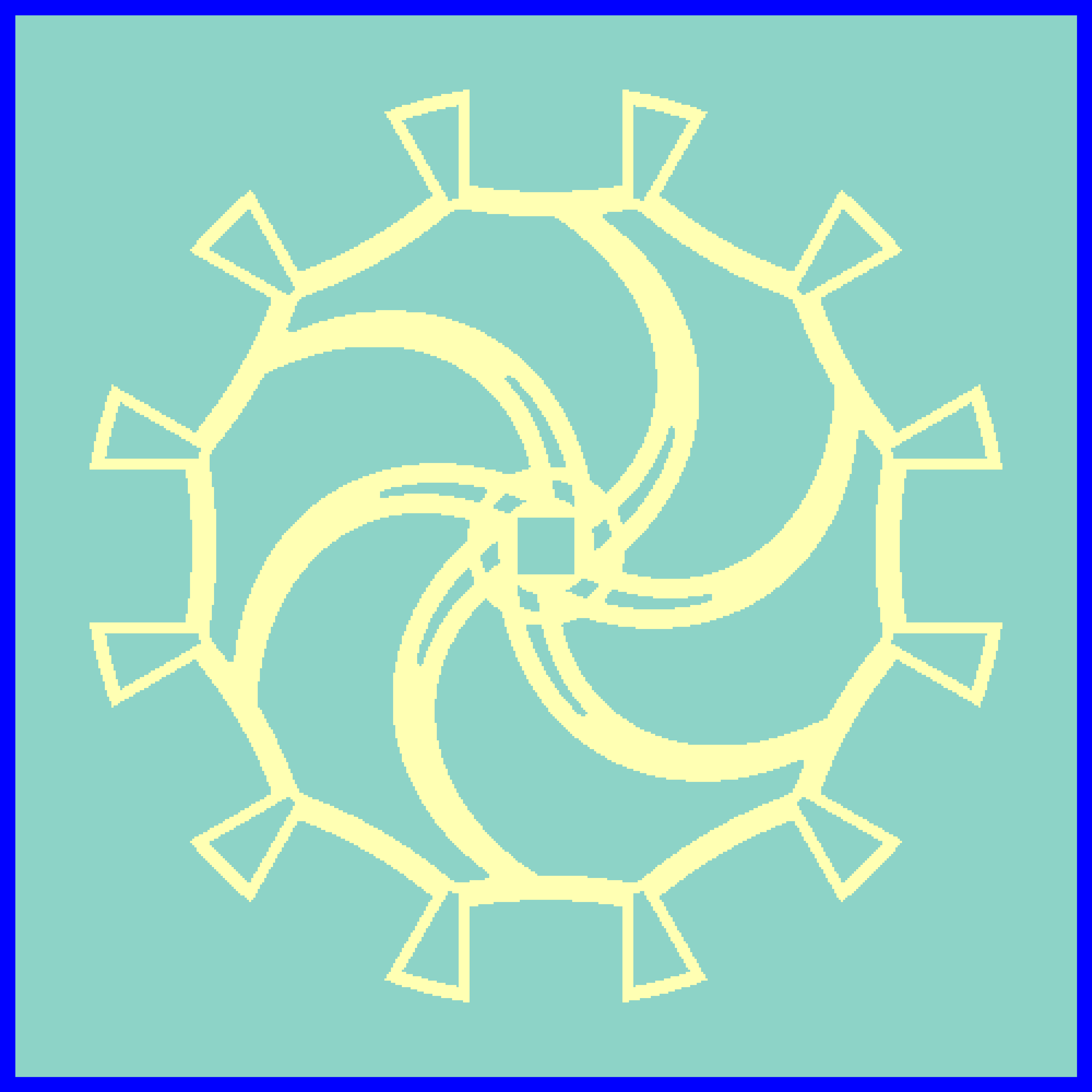} \\
    \includegraphics[width=0.45\linewidth]{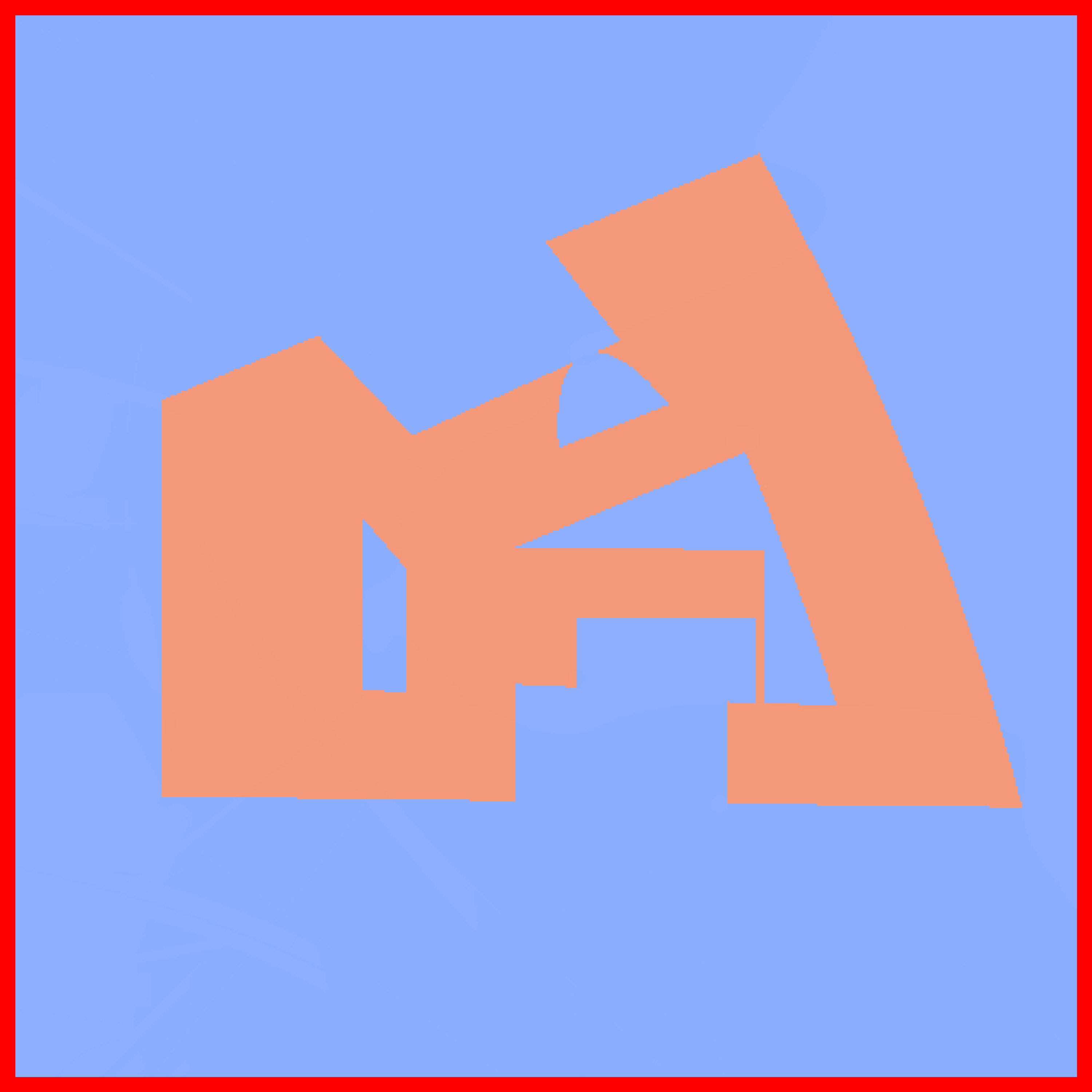} &
    \includegraphics[width=0.45\linewidth]{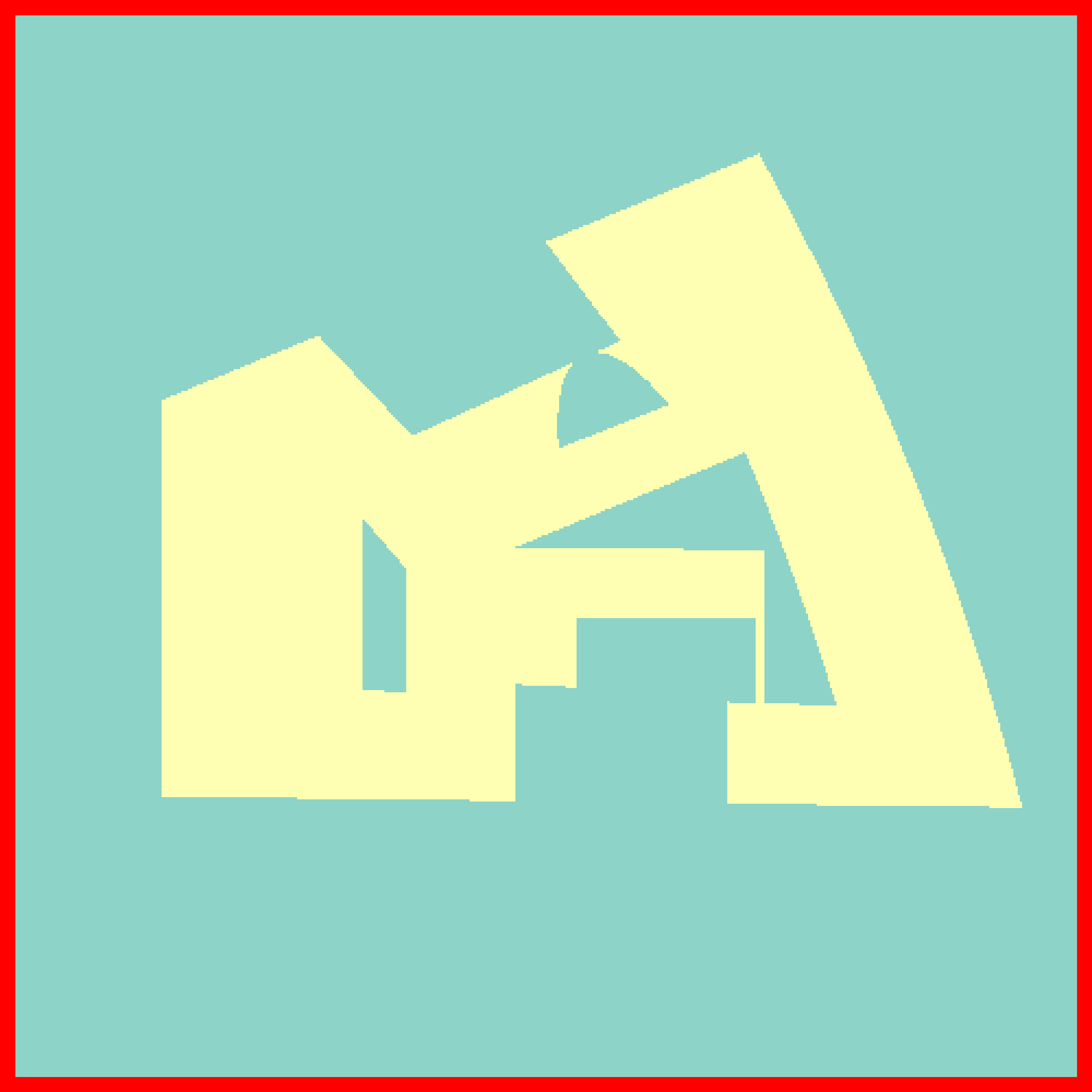} \\
    \includegraphics[width=0.45\linewidth]{summary_figures/gwn_colorbar_15.png} &
    \includegraphics[width=0.45\linewidth]{summary_figures/categories_colorbar_2.png} \\
\end{tabular}
\end{minipage}%
\begin{minipage}{0.5\textwidth}
\centering
\begin{tabular}{lr}
    \includegraphics[width=0.45\linewidth]{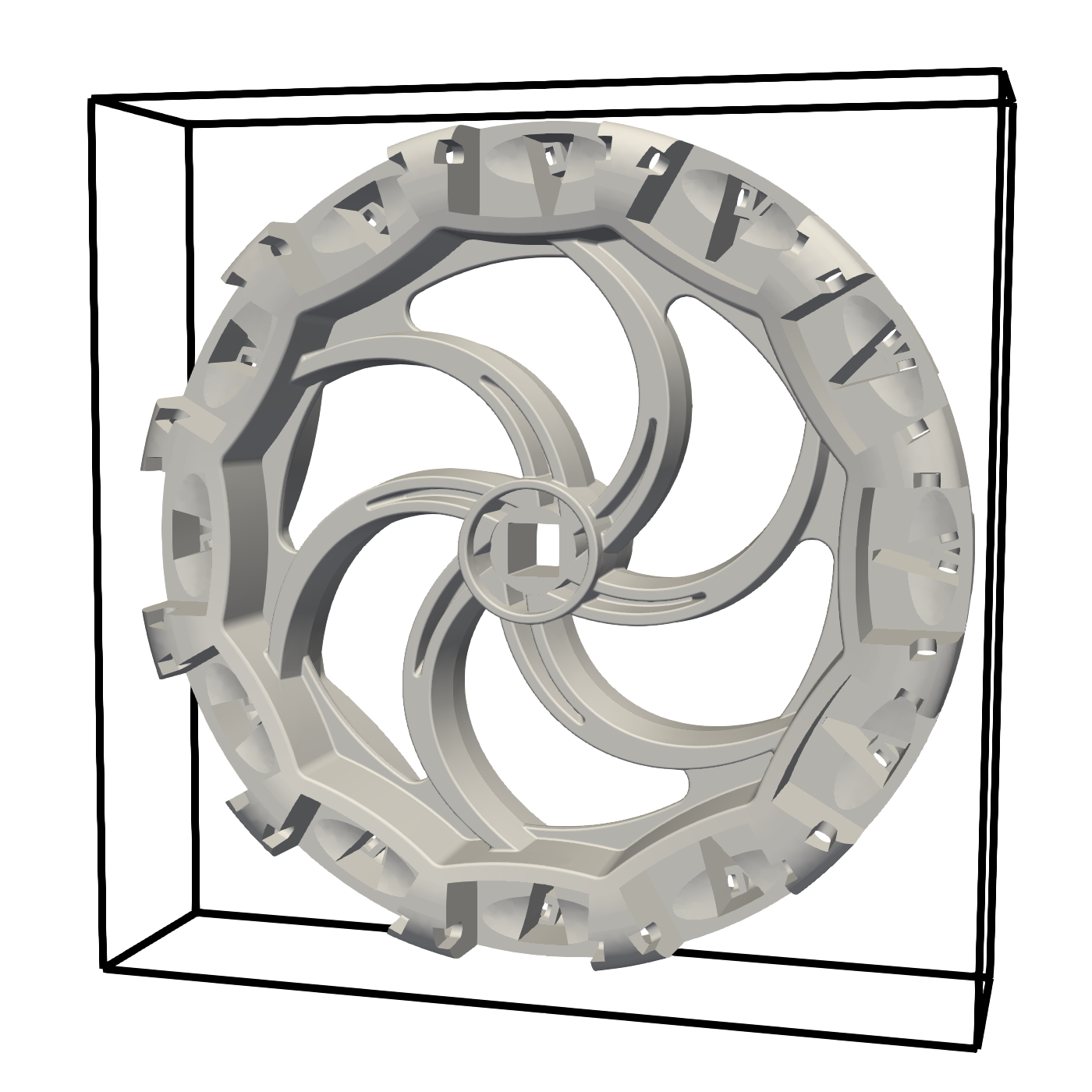}
    &
    \includegraphics[width=0.45\linewidth]{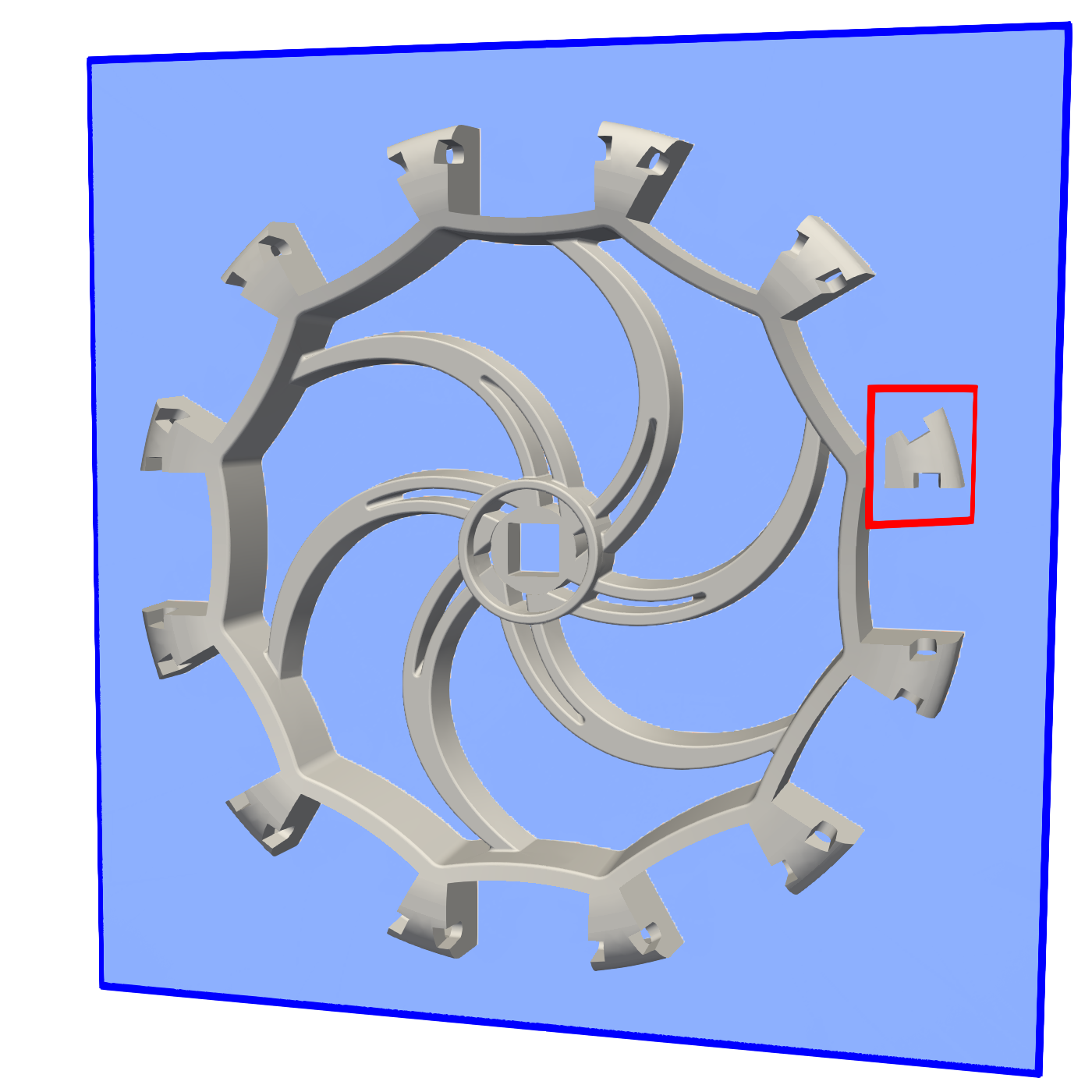}
\end{tabular}
\begin{tabular}{lr}
    Shape                          & Gear                        \\\cmidrule{1-2}
    Number of NURBS Patches        & 1784                 \\
    Number of Trimming Curves      & 10256                \\\cmidrule{1-2}
    \% Far-field Cases            & 99.92  \%  \\
    \% Near-field Cases           & 0.066 \%  \\
    \% Edge Cases                 & 0.0094 \%  \\\cmidrule{1-2}
    Avg. Time per Query (ms)       & 55.02          \\\cmidrule{1-2}
    Avg. Far-field Case Time (ms)  & 0.0289           \\
    Avg. Near-field Case Time (ms) & 0.253          \\
    Avg. Edge Case Time (ms)       & 18.4          \\
    \end{tabular}
\end{minipage}
\caption{This watertight ``Gear'' shape is derived from the ABC dataset model with index 9979.}
\end{figure}

\vspace{1cm}

\begin{figure}

\begin{minipage}{0.5\textwidth}
\centering
\begin{tabular}{lr}
    \includegraphics[width=0.45\linewidth]{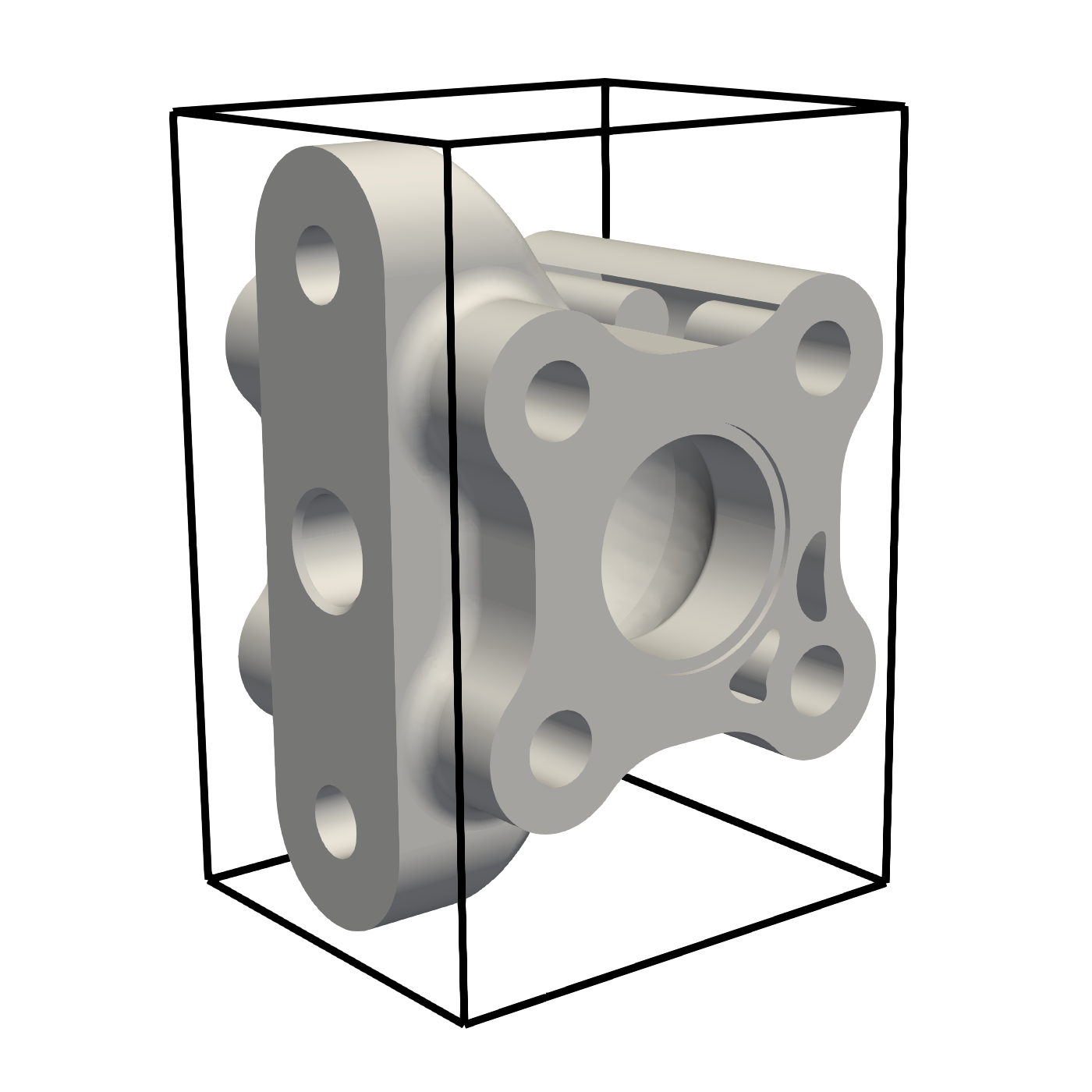}
    &
    \includegraphics[width=0.45\linewidth]{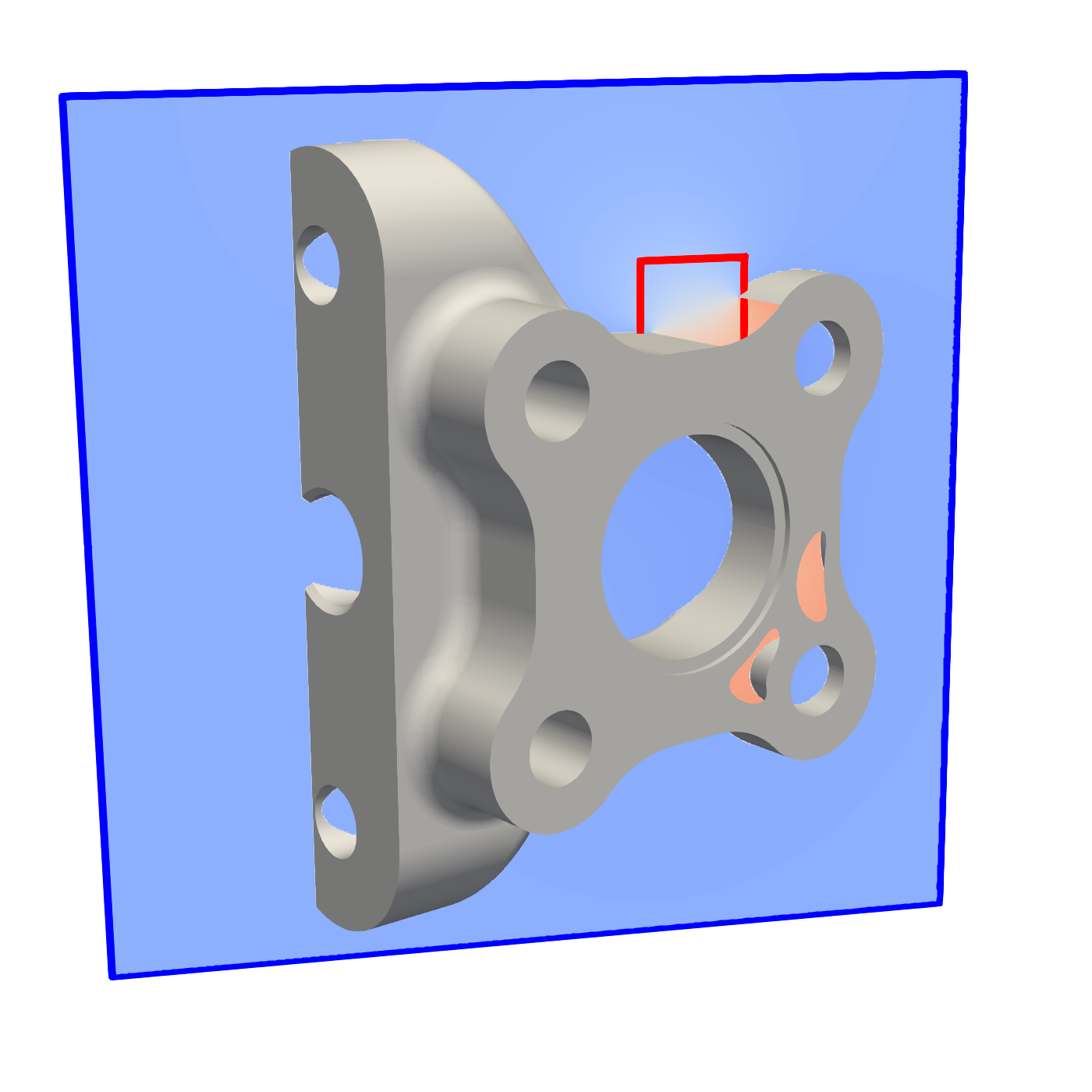}
\end{tabular}
\begin{tabular}{lr}
    Shape                          & Joint                        \\\cmidrule{1-2}
    Number of NURBS Patches        & 117                 \\
    Number of Trimming Curves      & 564                \\\cmidrule{1-2}
    \% Far-field Cases             & 99.5  \%  \\
    \% Near-field Cases            & 0.435 \%  \\
    \% Edge Cases                  & 0.046 \%  \\\cmidrule{1-2}
    Avg. Time per Query (ms)       & 1.86          \\\cmidrule{1-2}
    Avg. Far-field Case Time (ms)  & 0.0134           \\
    Avg. Near-field Case Time (ms) & 0.118          \\
    Avg. Edge Case Time (ms)       & 6.69          \\
    \end{tabular}
\end{minipage}%
\begin{minipage}{0.5\textwidth}
\centering
\begin{tabular}{lr}
    \includegraphics[width=0.45\linewidth]{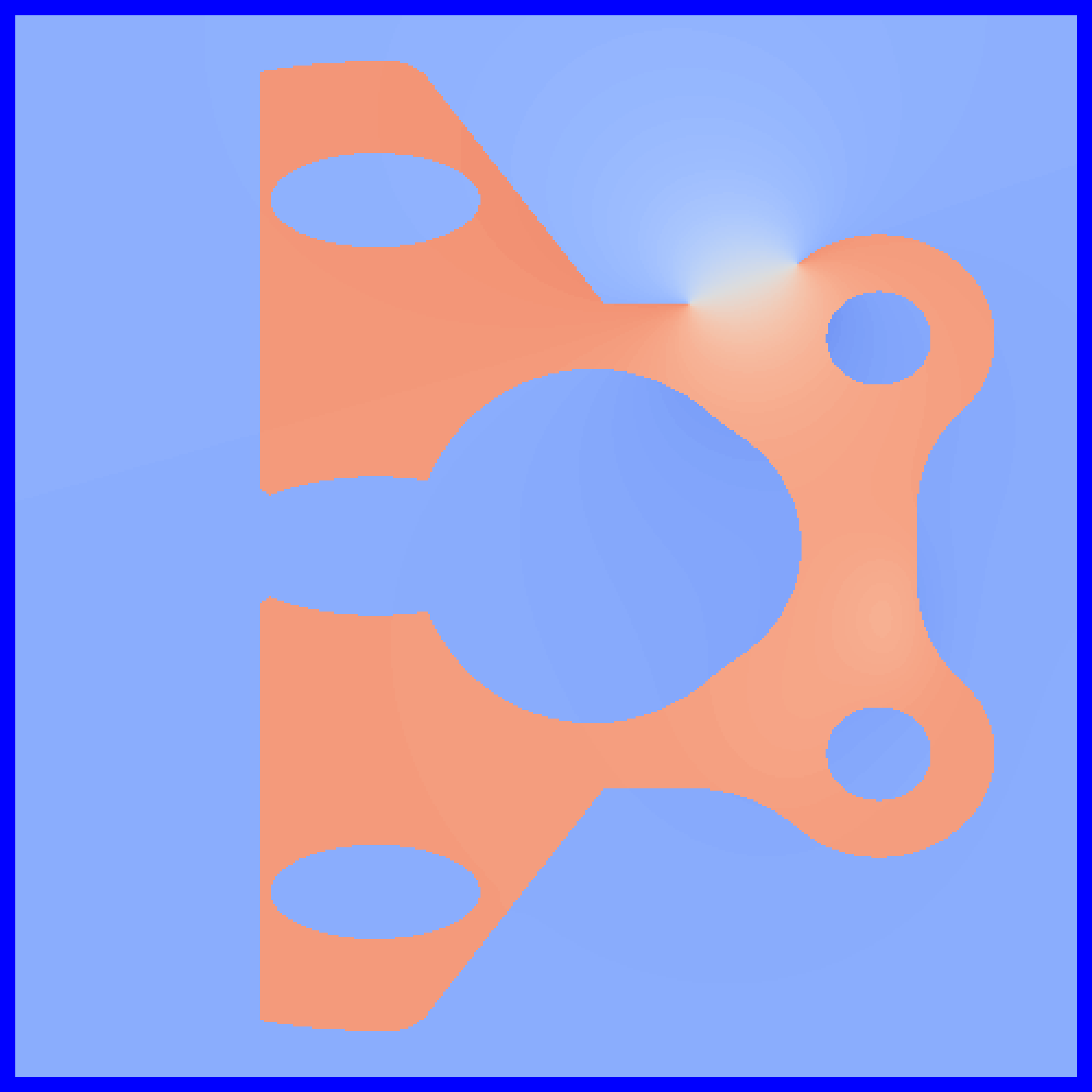} &
    \includegraphics[width=0.45\linewidth]{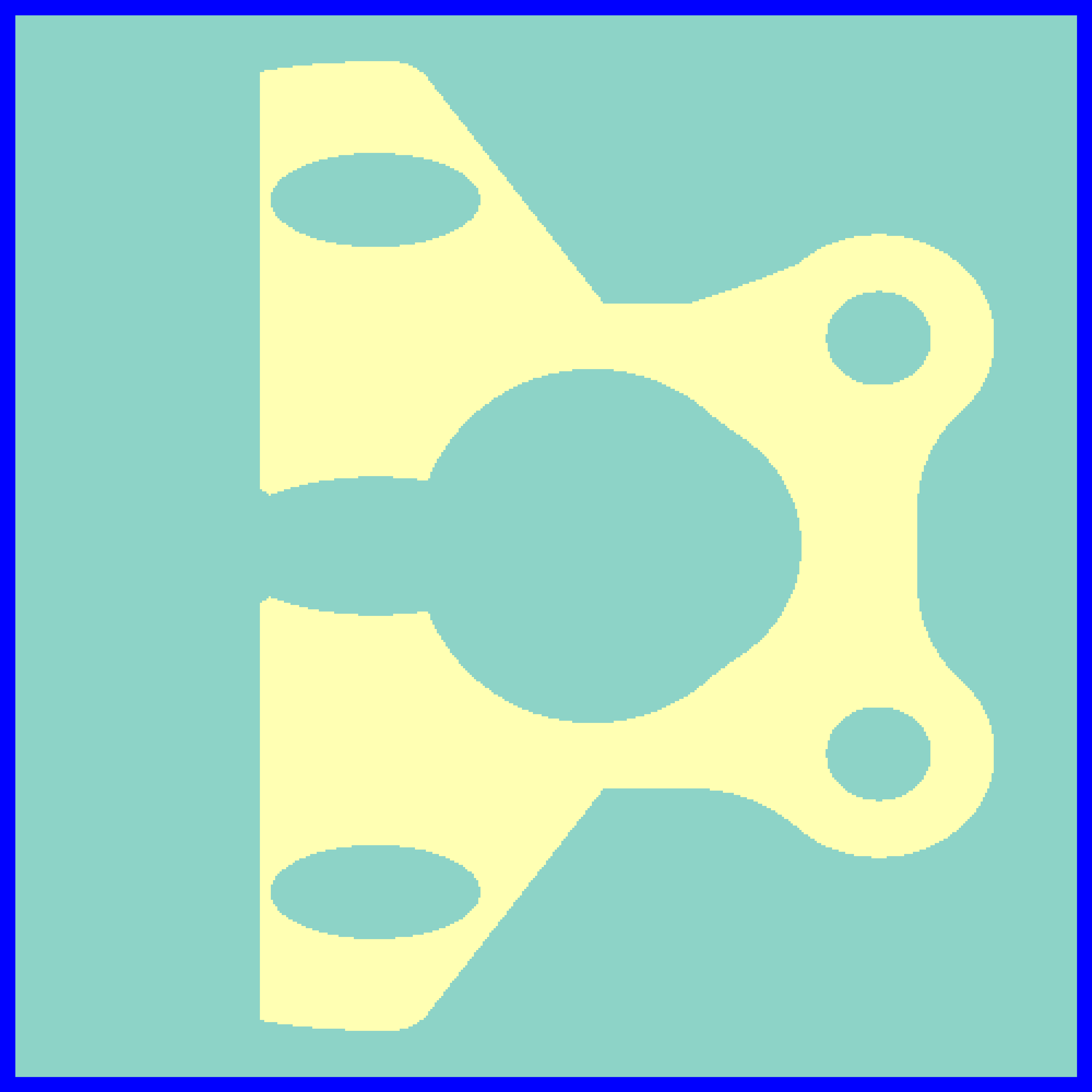} \\
    \includegraphics[width=0.45\linewidth]{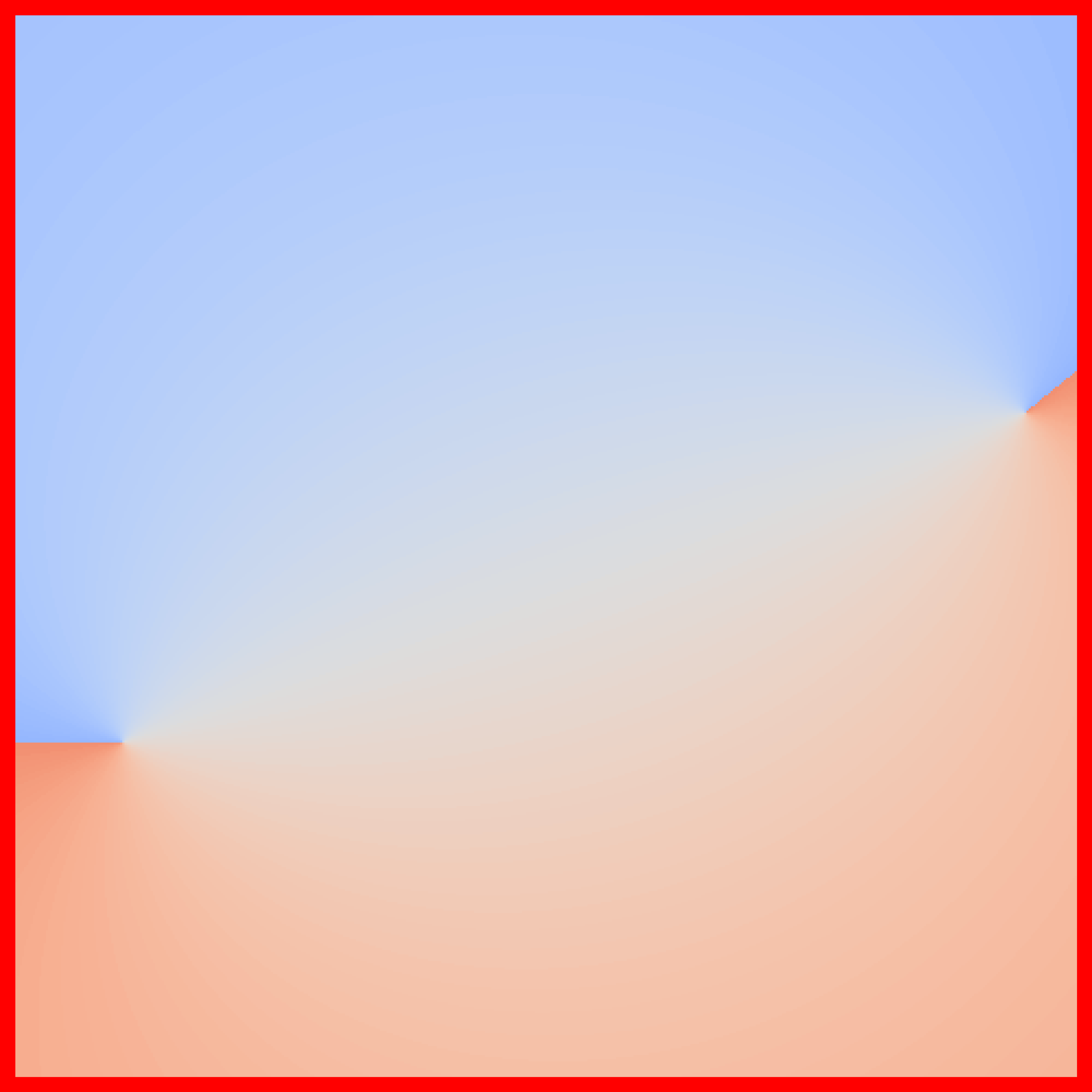} &
    \includegraphics[width=0.45\linewidth]{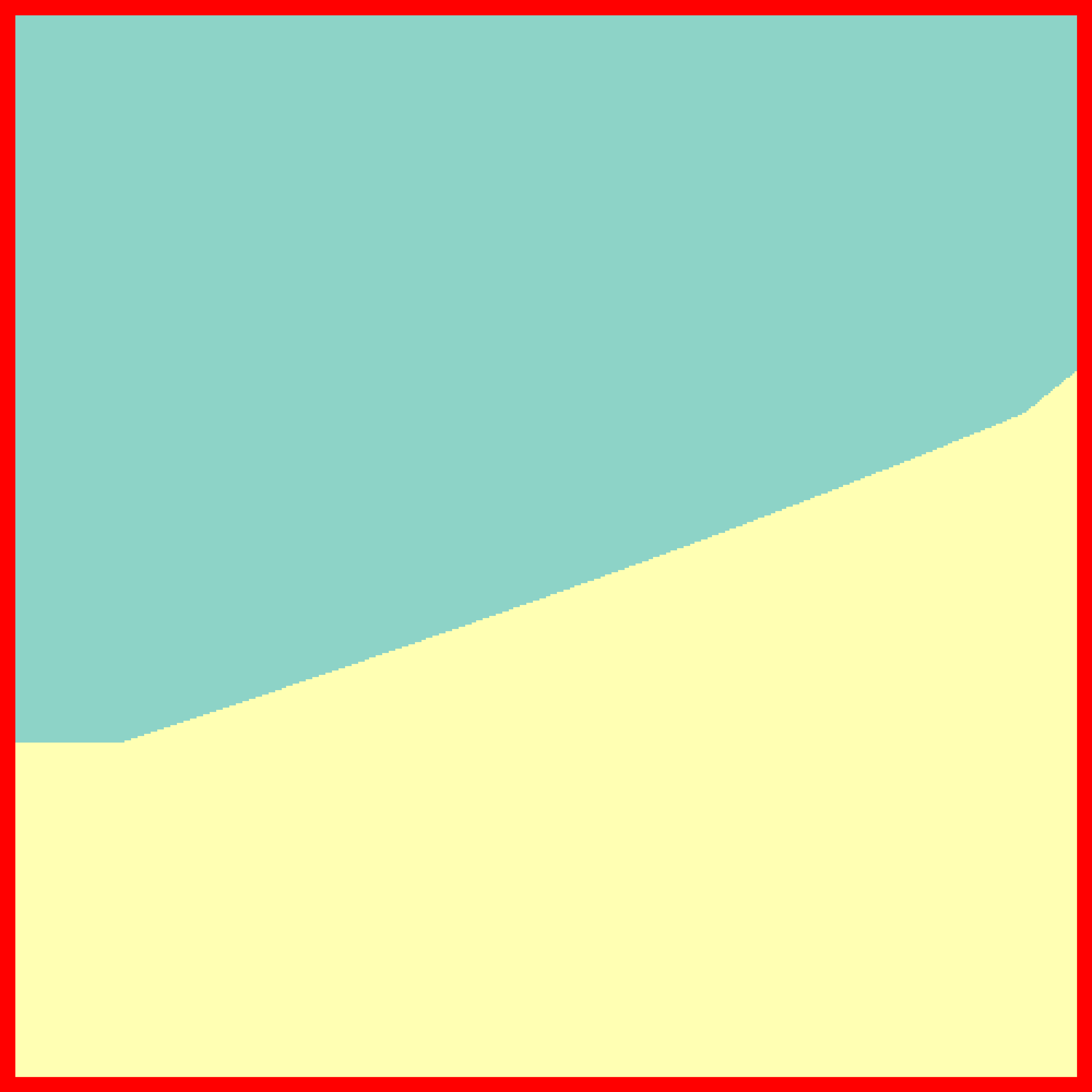} \\
    \includegraphics[width=0.45\linewidth]{summary_figures/gwn_colorbar_15.png} &
    \includegraphics[width=0.45\linewidth]{summary_figures/categories_colorbar_2.png} \\
\end{tabular}
\end{minipage}
\caption{This ``Joint'' shape is derived from the ABC dataset model with index 13. We modified this originally watertight shape by adding trimming curves to the front face, and removing a patch on the top of the shape.}
\end{figure}

\vspace{1cm}

\begin{figure}

\begin{minipage}{0.5\textwidth}
\centering
\begin{tabular}{lr}
    \includegraphics[width=0.45\linewidth]{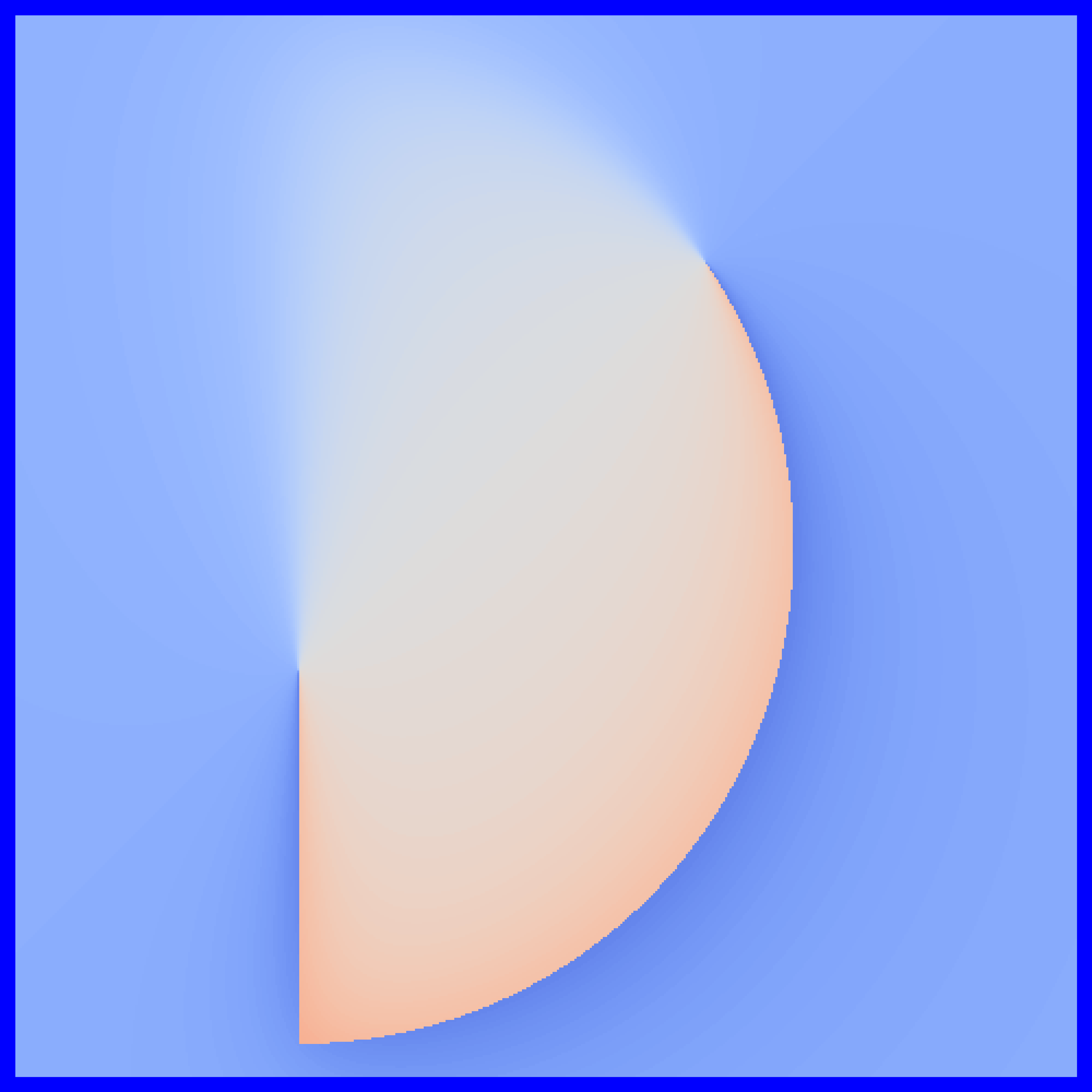} &
    \includegraphics[width=0.45\linewidth]{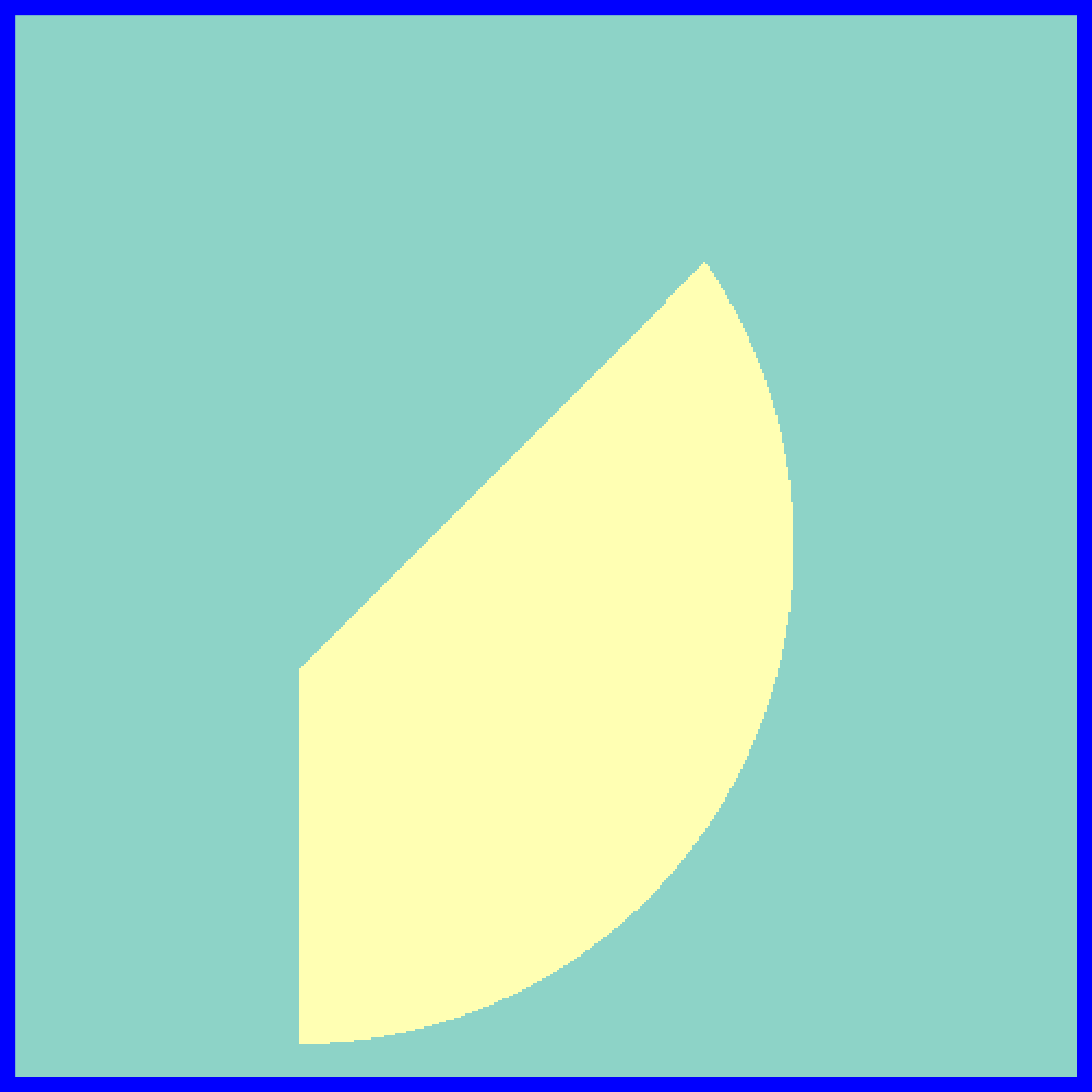} \\
    \includegraphics[width=0.45\linewidth]{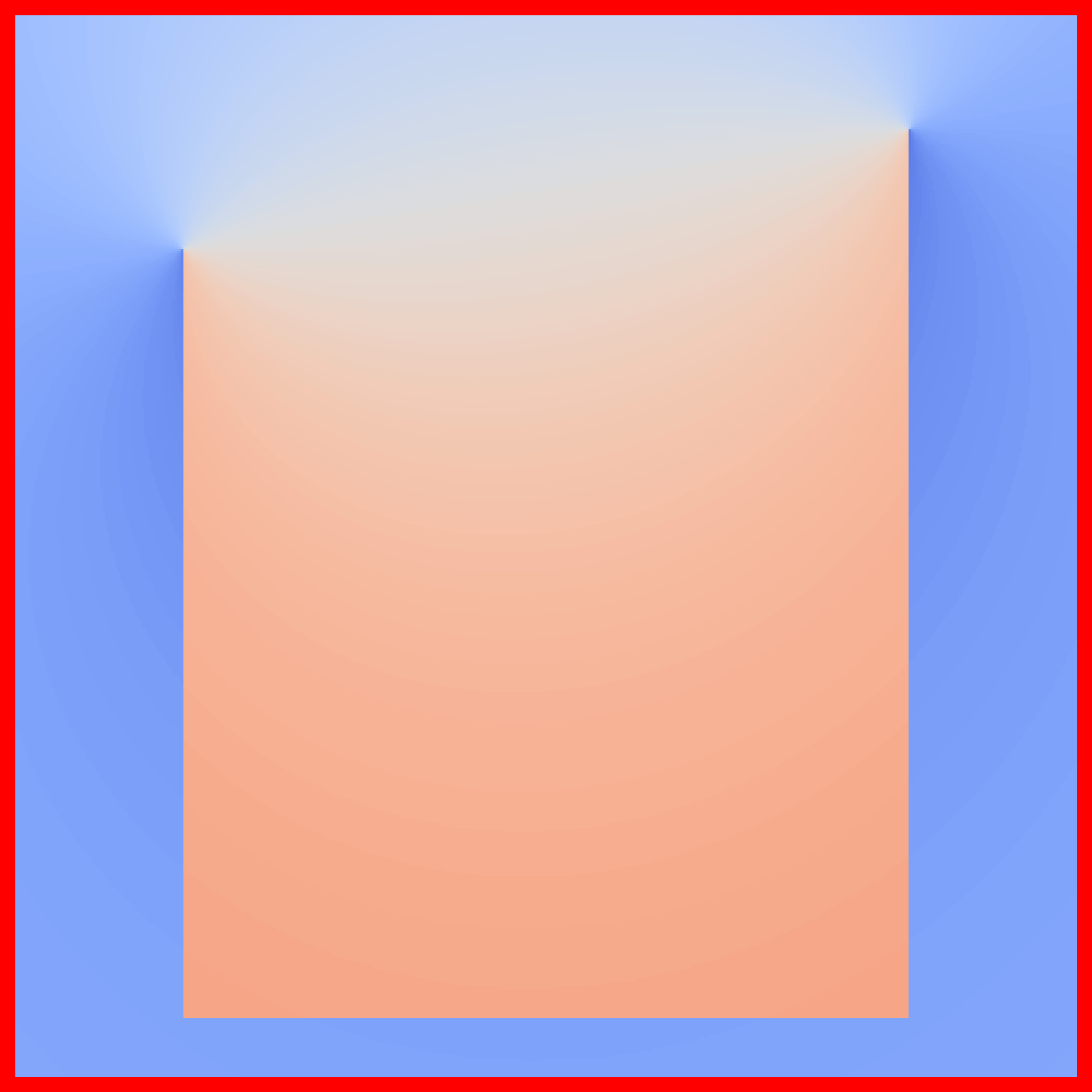} &
    \includegraphics[width=0.45\linewidth]{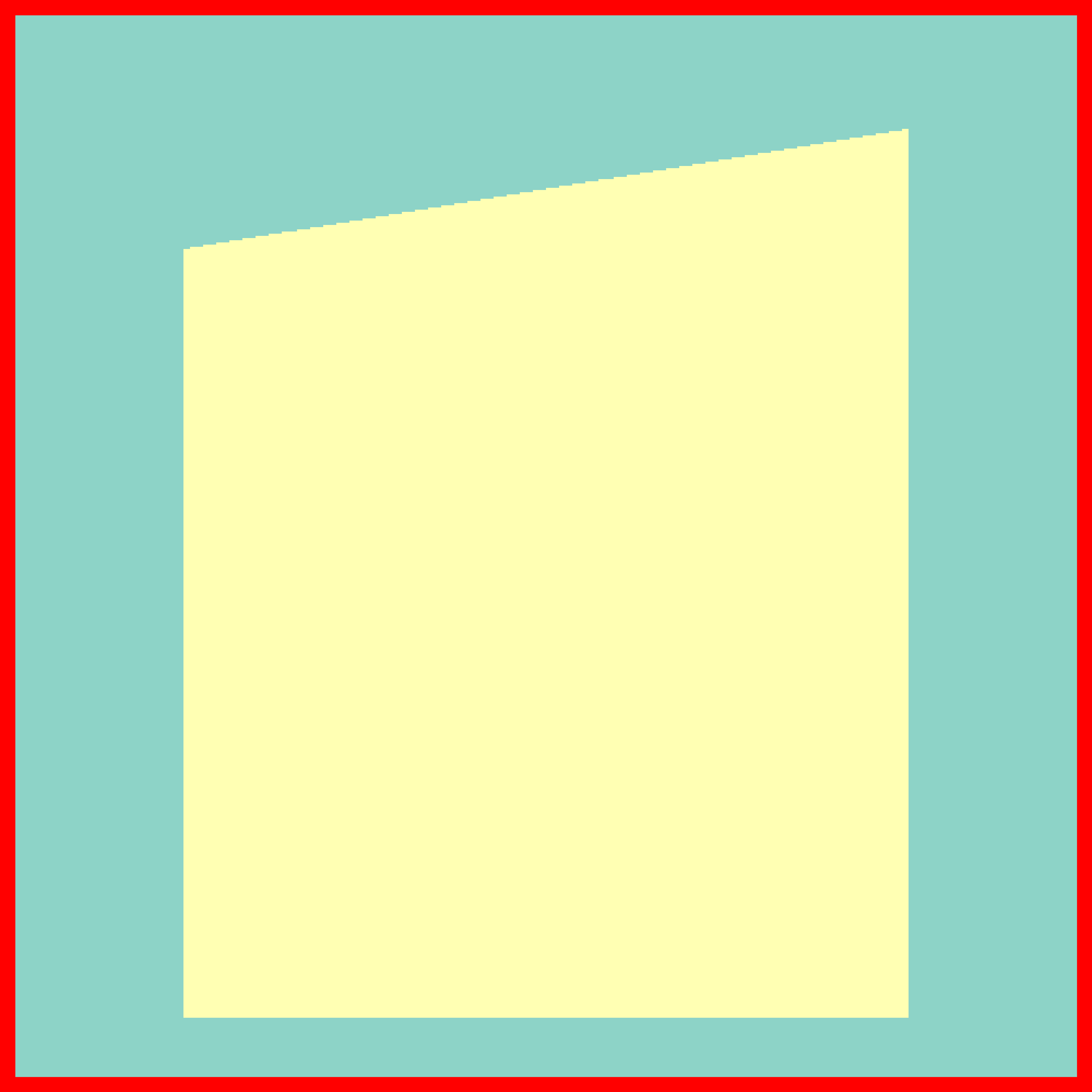} \\
    \includegraphics[width=0.45\linewidth]{summary_figures/gwn_colorbar_15.png} &
    \includegraphics[width=0.45\linewidth]{summary_figures/categories_colorbar_2.png} \\
\end{tabular}
\end{minipage}%
\begin{minipage}{0.5\textwidth}
\centering
\begin{tabular}{lr}
    \includegraphics[width=0.45\linewidth]{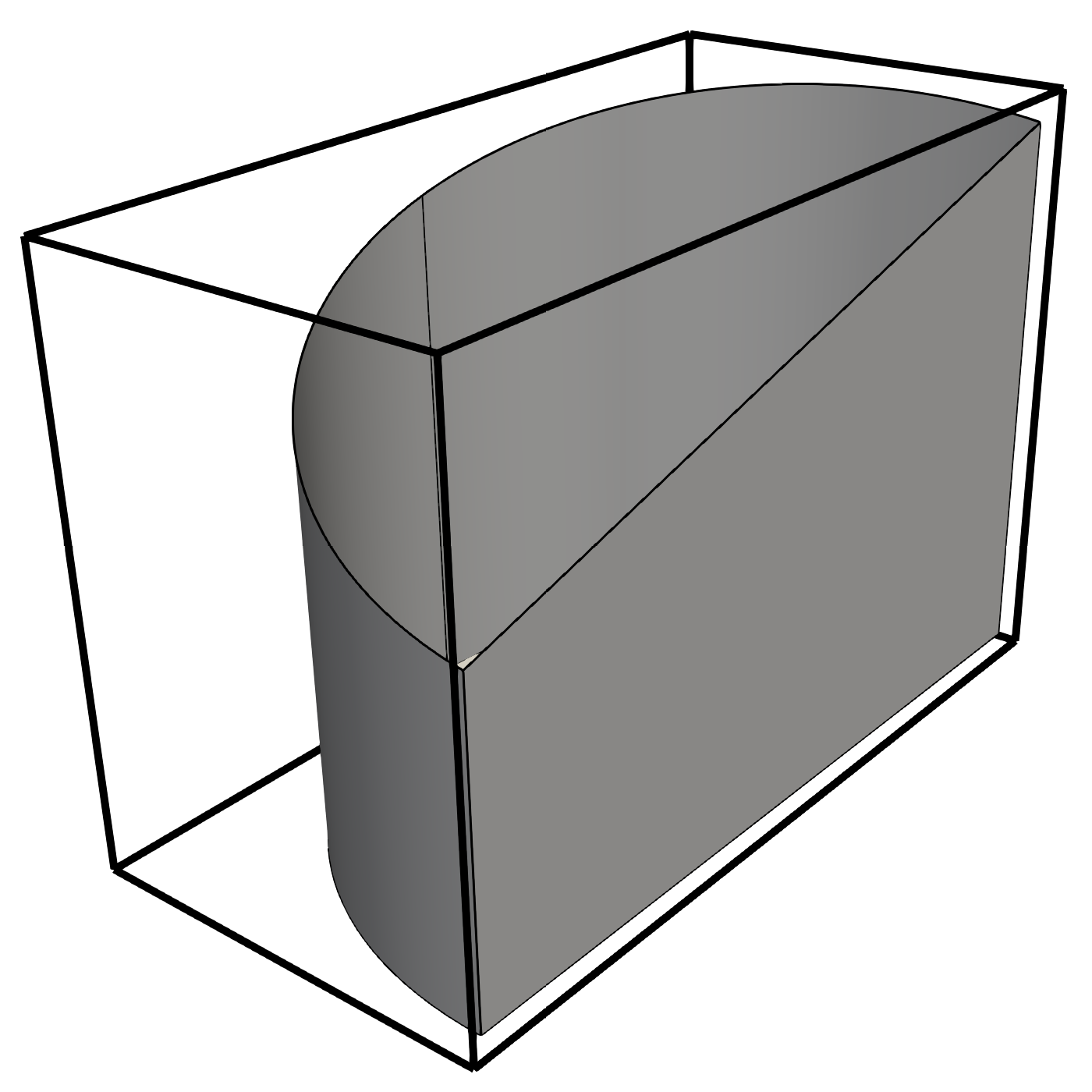}
    &
    \includegraphics[width=0.45\linewidth]{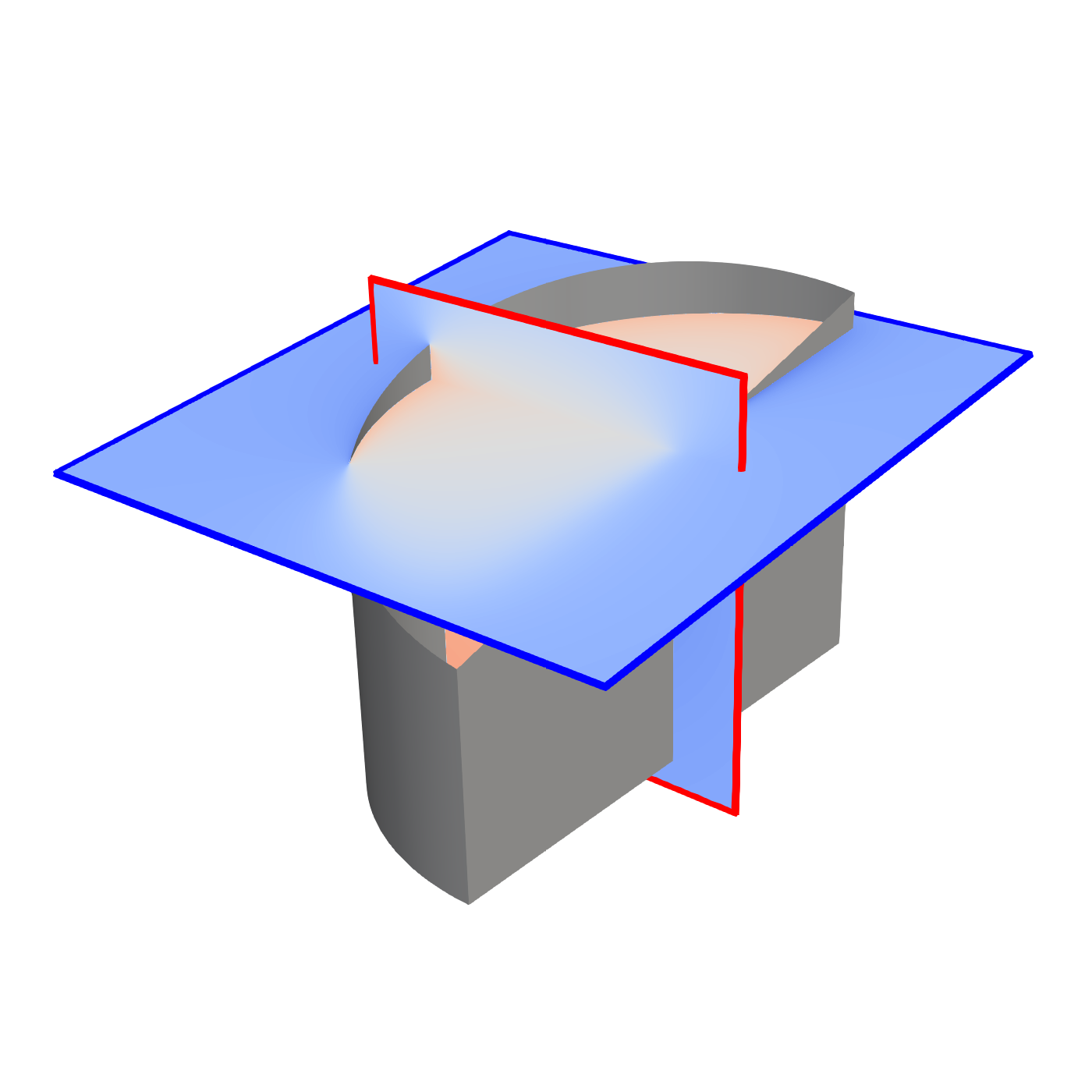}
\end{tabular}
\begin{tabular}{lr}
    Shape                          & Cylinder                        \\\cmidrule{1-2}
    Number of NURBS Patches        & 4                 \\
    Number of Trimming Curves      & 15                \\\cmidrule{1-2}
    \% Far-field Cases            & 81.9  \%  \\
    \% Near-field Cases           & 17.6 \%  \\
    \% Edge Cases                 & 0.504 \%  \\\cmidrule{1-2}
    Avg. Time per Query (ms)       & 0.183          \\\cmidrule{1-2}
    Avg. Far-field Case Time (ms)  & 0.00973           \\
    Avg. Near-field Case Time (ms) & 0.0842          \\
    Avg. Edge Case Time (ms)       & 4.59          \\
    \end{tabular}
\end{minipage}
\caption{We modeled this open ``Sliced-Cylinder'' shape after an example from~\cite{marussig-17-isogeometricreview} using Rhino3D.}
\end{figure}

\vspace{1cm}

\begin{figure}

\begin{minipage}{0.5\textwidth}
\centering
\begin{tabular}{lr}
    \includegraphics[width=0.45\linewidth]{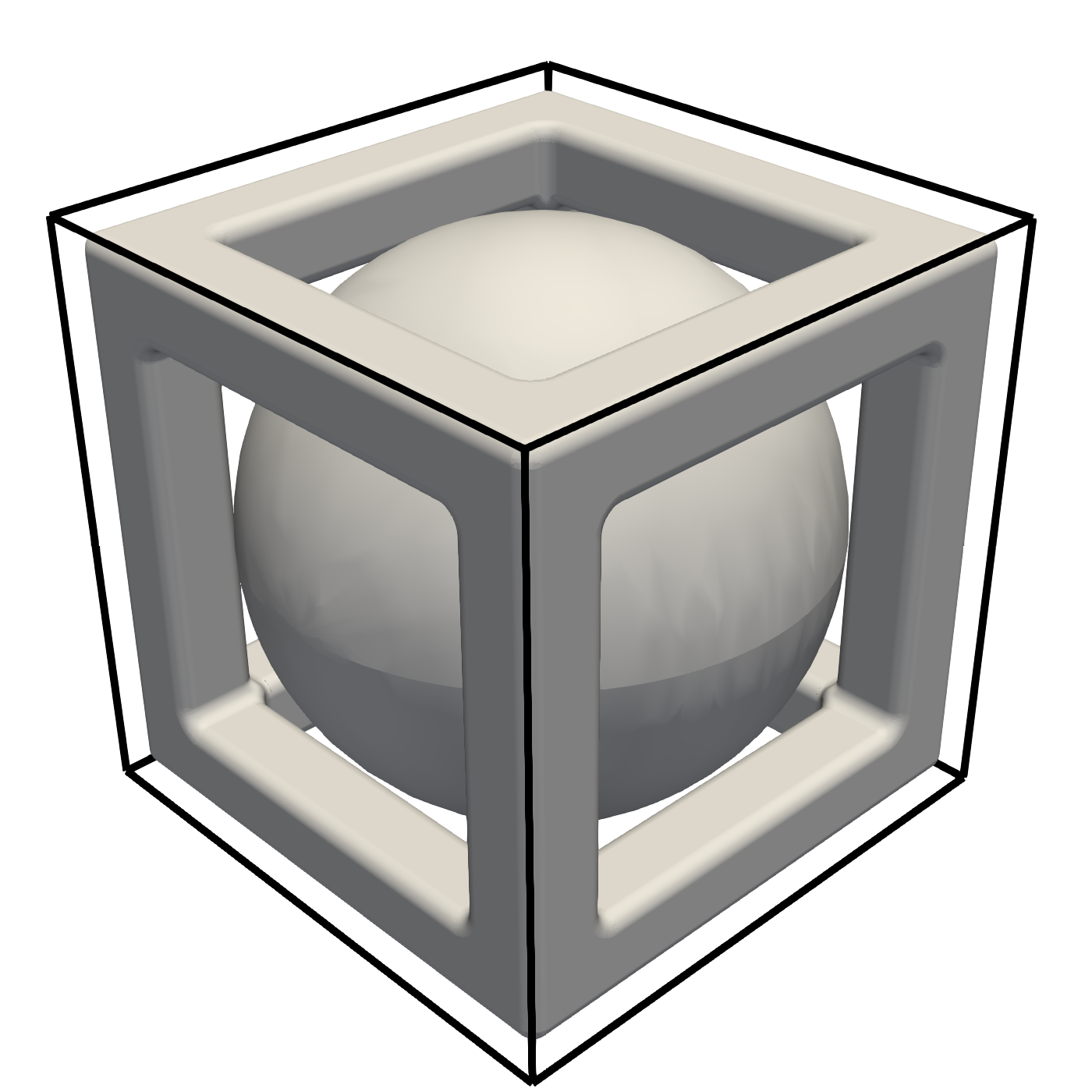}
    &
    \includegraphics[width=0.45\linewidth]{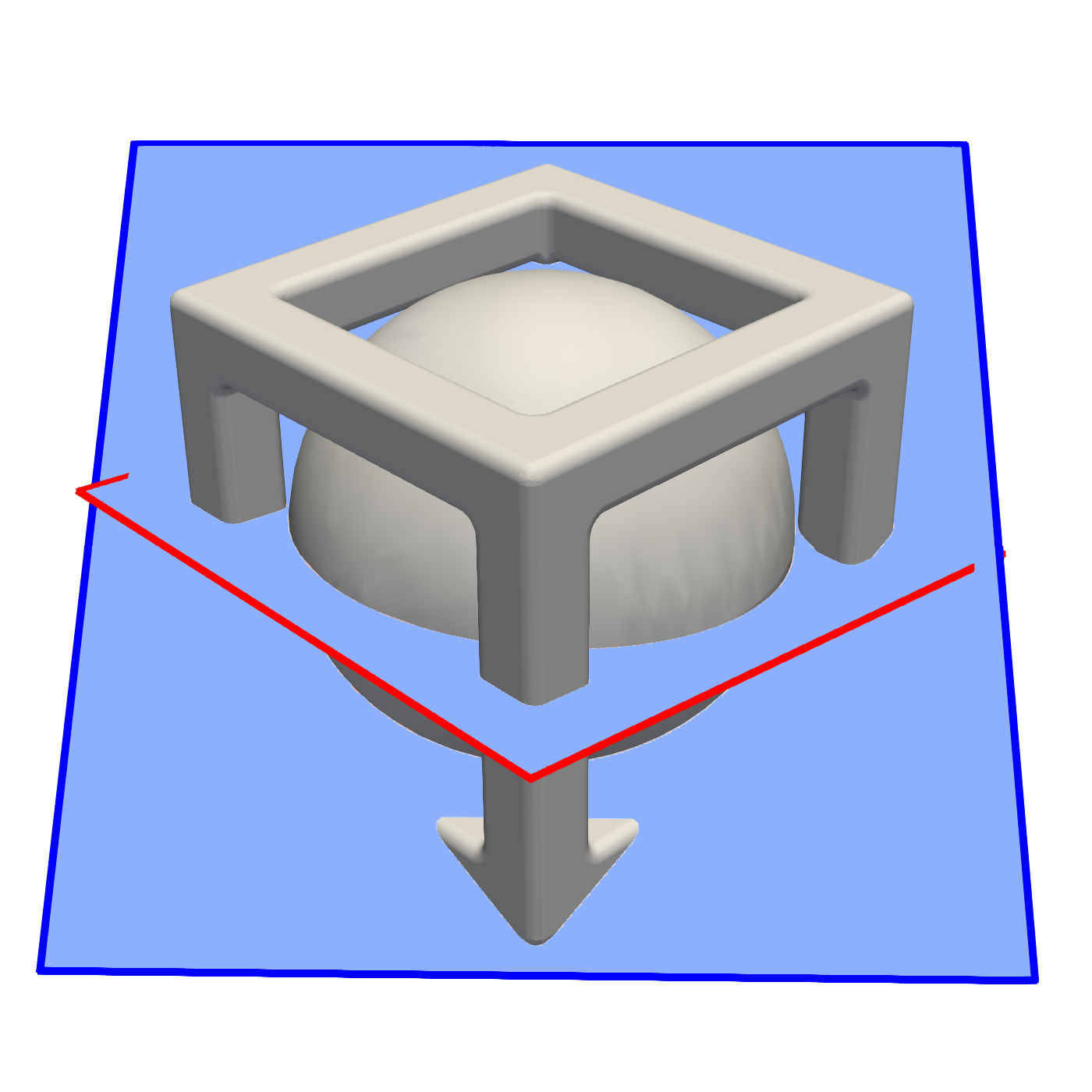}
\end{tabular}
\begin{tabular}{lr}
    Shape                          & Box-Sphere \\  \cmidrule{1-2}
    Number of NURBS Patches        & 192        \\  
    Number of Trimming Curves      & 968        \\  \cmidrule{1-2}
    \% Far-field Cases             & 99.43 \%   \\  
    \% Near-field Cases            & 0.554 \%   \\  
    \% Edge Cases                  & 0.012 \%   \\  \cmidrule{1-2}
    Avg. Time per Query (ms)       & 1.46       \\  \cmidrule{1-2}
    Avg. Far-field Case Time (ms)  & 0.0064     \\  
    Avg. Near-field Case Time (ms) & 0.0844     \\  
    Avg. Edge Case Time (ms)       & 6.40       \\  
\end{tabular}
\end{minipage}%
\begin{minipage}{0.5\textwidth}
\centering
\begin{tabular}{lr}
    \includegraphics[width=0.45\linewidth]{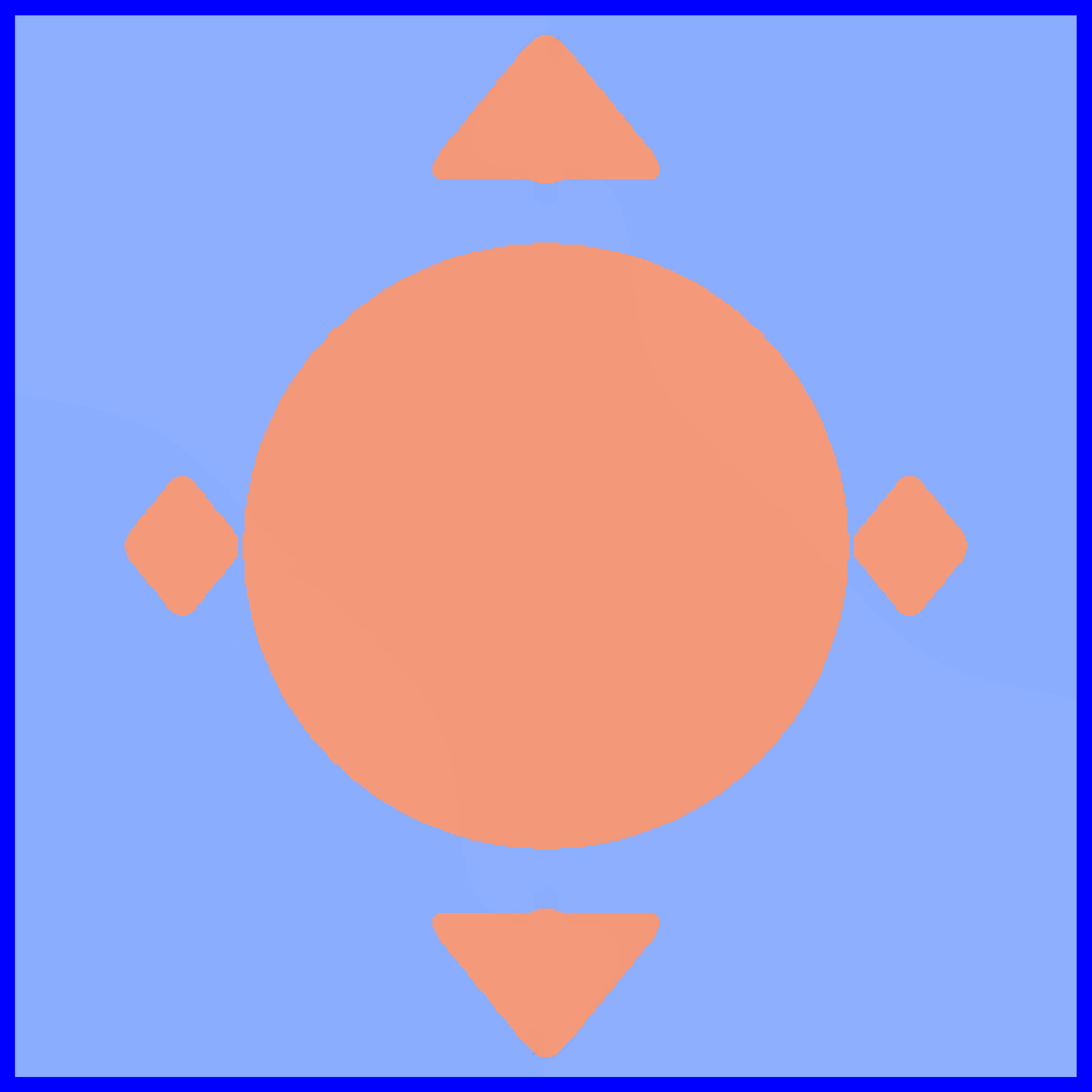} &
    \includegraphics[width=0.45\linewidth]{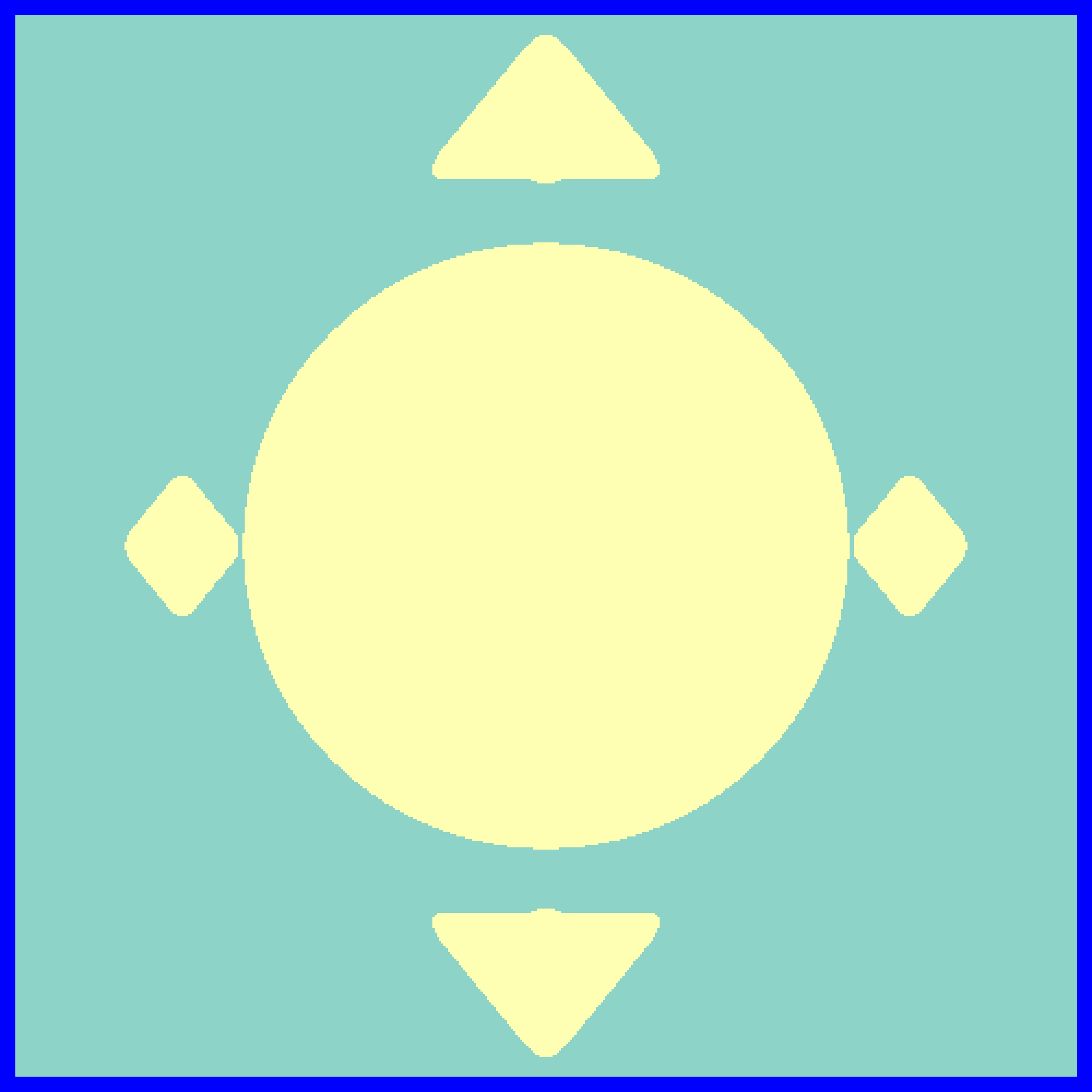} \\
    \includegraphics[width=0.45\linewidth]{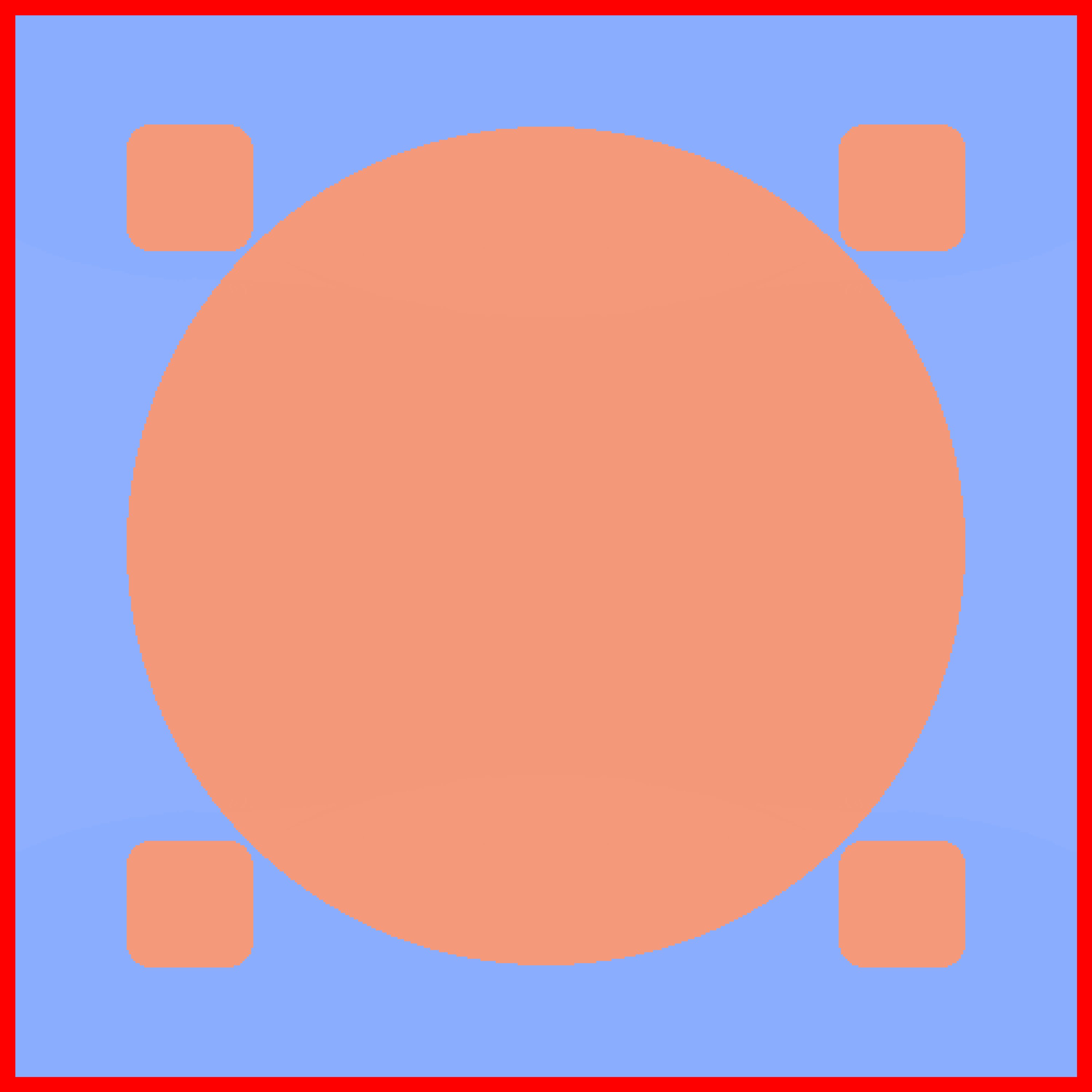} &
    \includegraphics[width=0.45\linewidth]{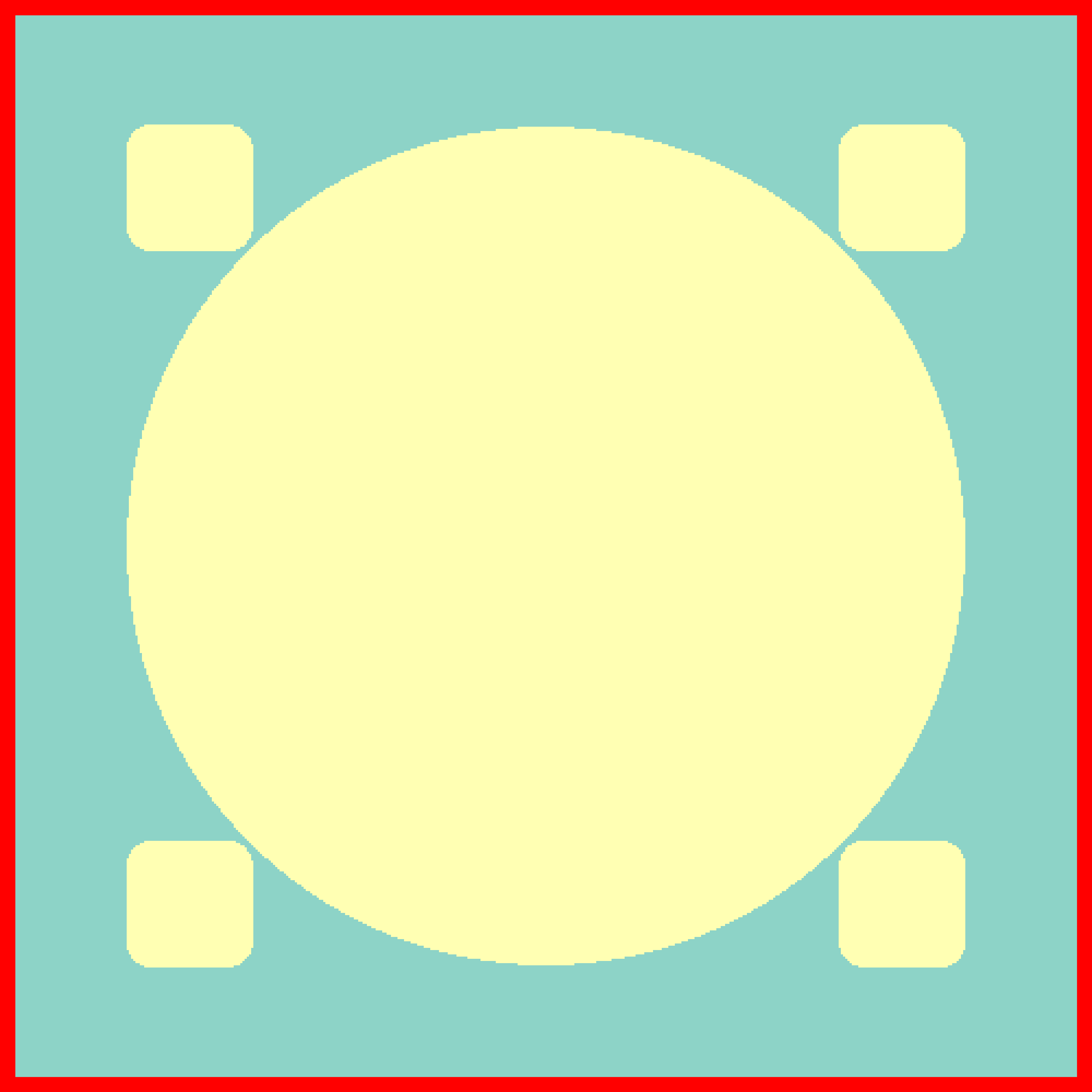} \\
    \includegraphics[width=0.45\linewidth]{summary_figures/gwn_colorbar_15.png} &
    \includegraphics[width=0.45\linewidth]{summary_figures/categories_colorbar_2.png} \\
\end{tabular}
\end{minipage}
\caption{We modeled this watertight ``Box-Sphere'' shape after an existing STL mesh using Rhino3D. Although the example looks relatively simple, the fileted edges and corners add considerable complexity to the shape. }
\end{figure}

\vspace{1cm}

\begin{figure}


\begin{minipage}{0.5\textwidth}
\centering
\begin{tabular}{lr}
    \includegraphics[width=0.45\linewidth]{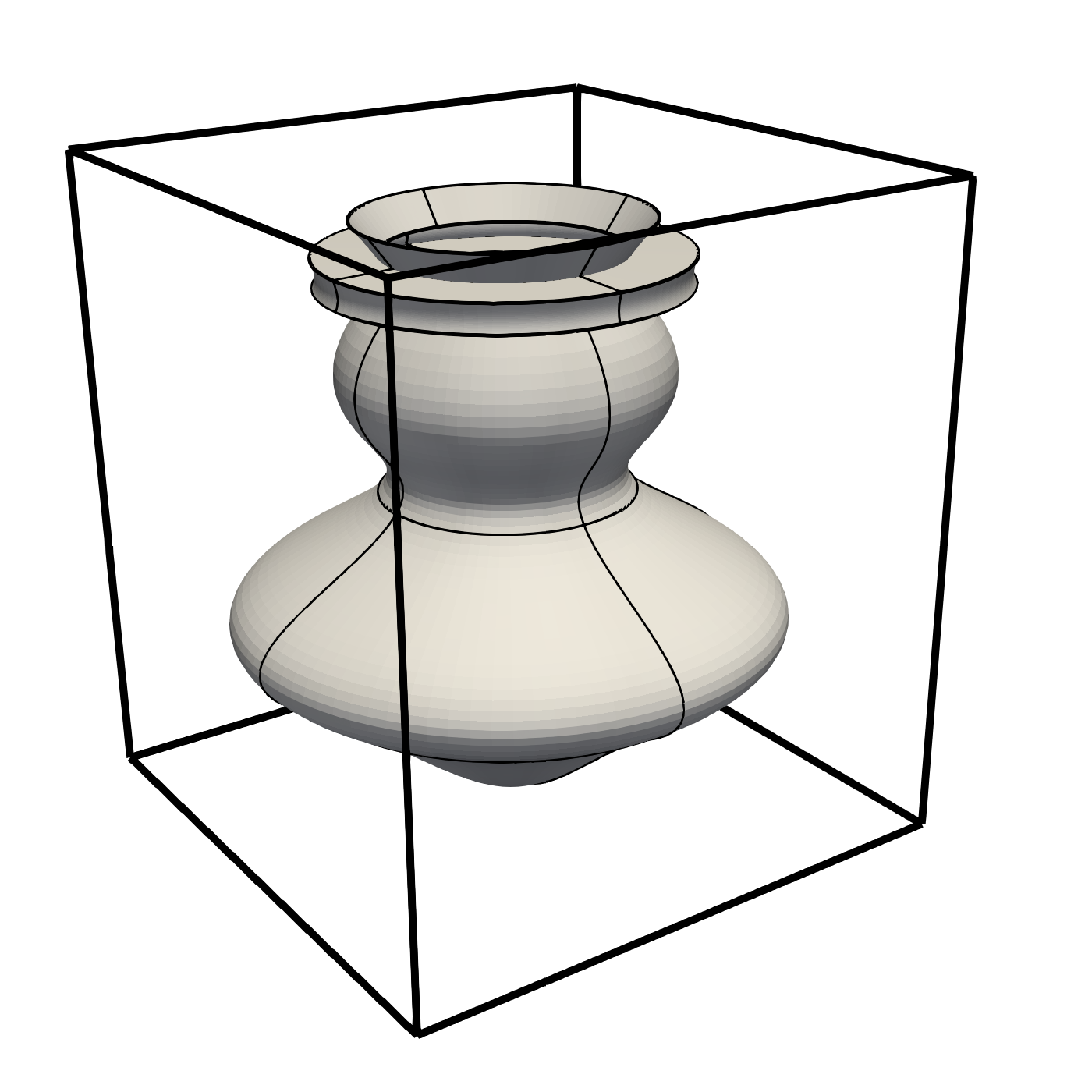}
    &
    \includegraphics[width=0.45\linewidth]{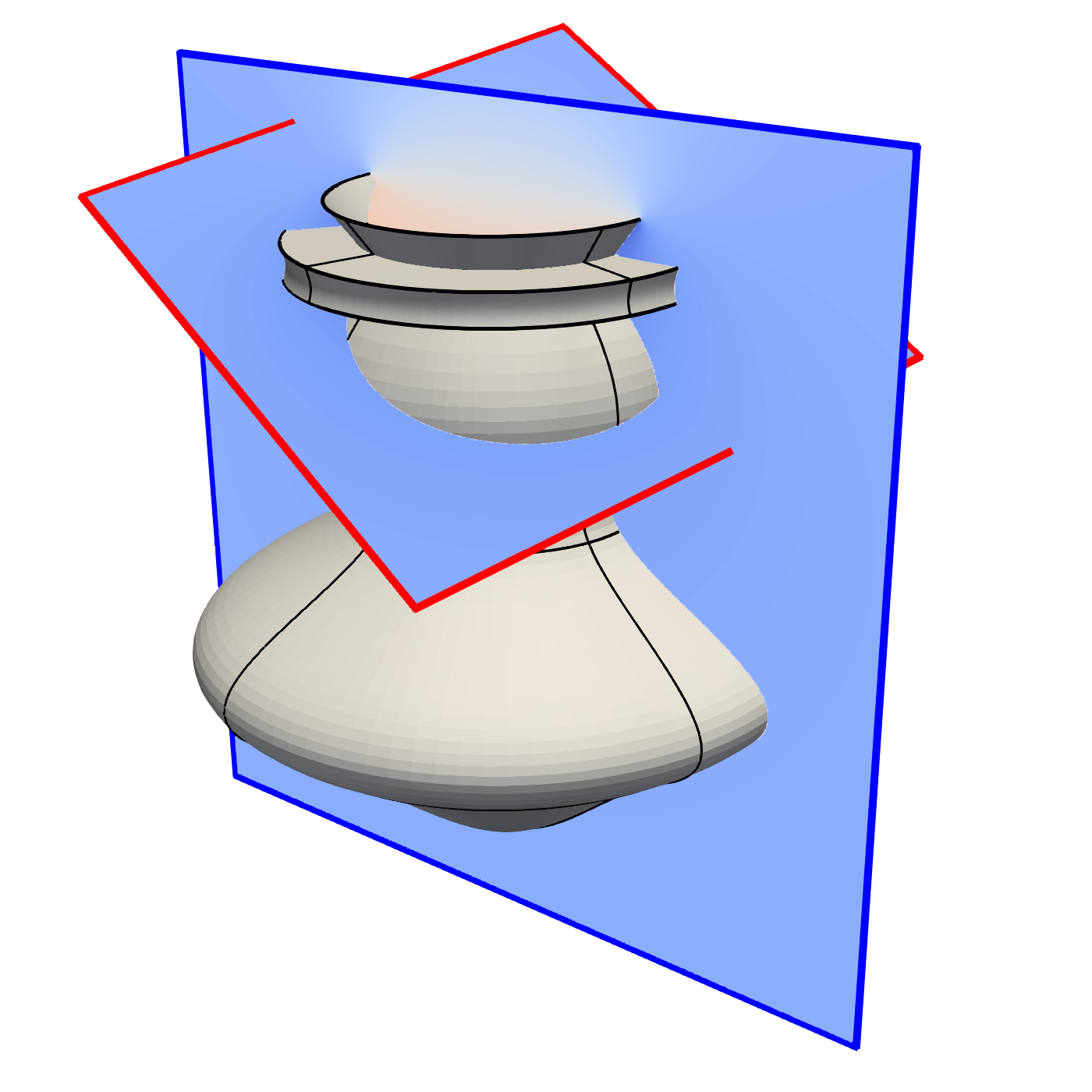}
\end{tabular}
\begin{tabular}{lr}
    Shape                          & Vase                        \\\cmidrule{1-2}
    Number of NURBS Patches        & 28                 \\
    Number of Trimming Curves      & 112                \\\cmidrule{1-2}
    \% Far-field Cases             & 98.00  \%  \\
    \% Near-field Cases            & 1.96 \%  \\
    \% Edge Cases                  & 0.032 \%  \\\cmidrule{1-2}
    Avg. Time per Query (ms)       & 0.192          \\\cmidrule{1-2}
    Avg. Far-field Case Time (ms)  & 0.00313           \\
    Avg. Near-field Case Time (ms) & 0.130          \\
    Avg. Edge Case Time (ms)       & 3.80          \\
    \end{tabular}
\end{minipage}%
\begin{minipage}{0.5\textwidth}
\centering
\begin{tabular}{lr}
    \includegraphics[width=0.45\linewidth]{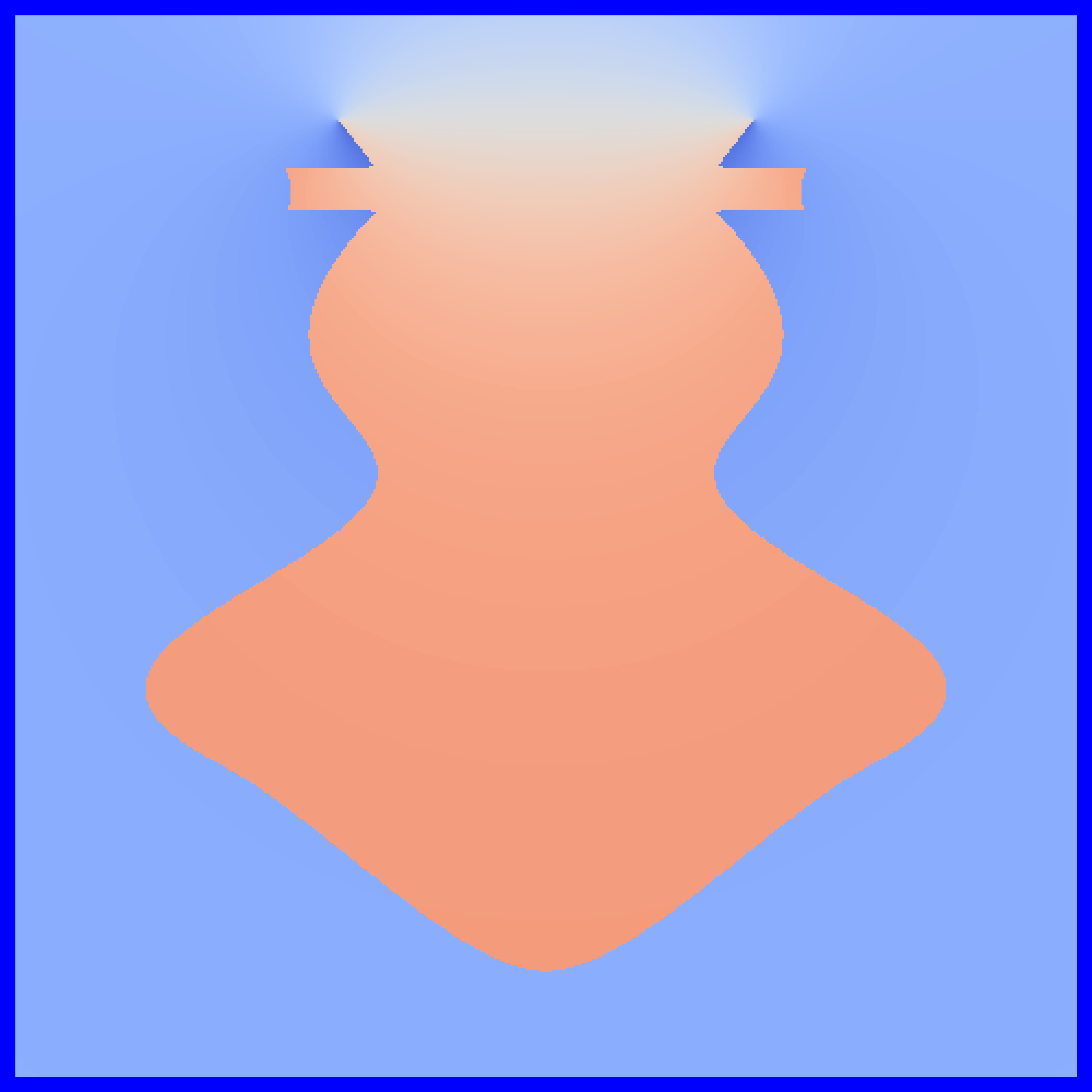} &
    \includegraphics[width=0.45\linewidth]{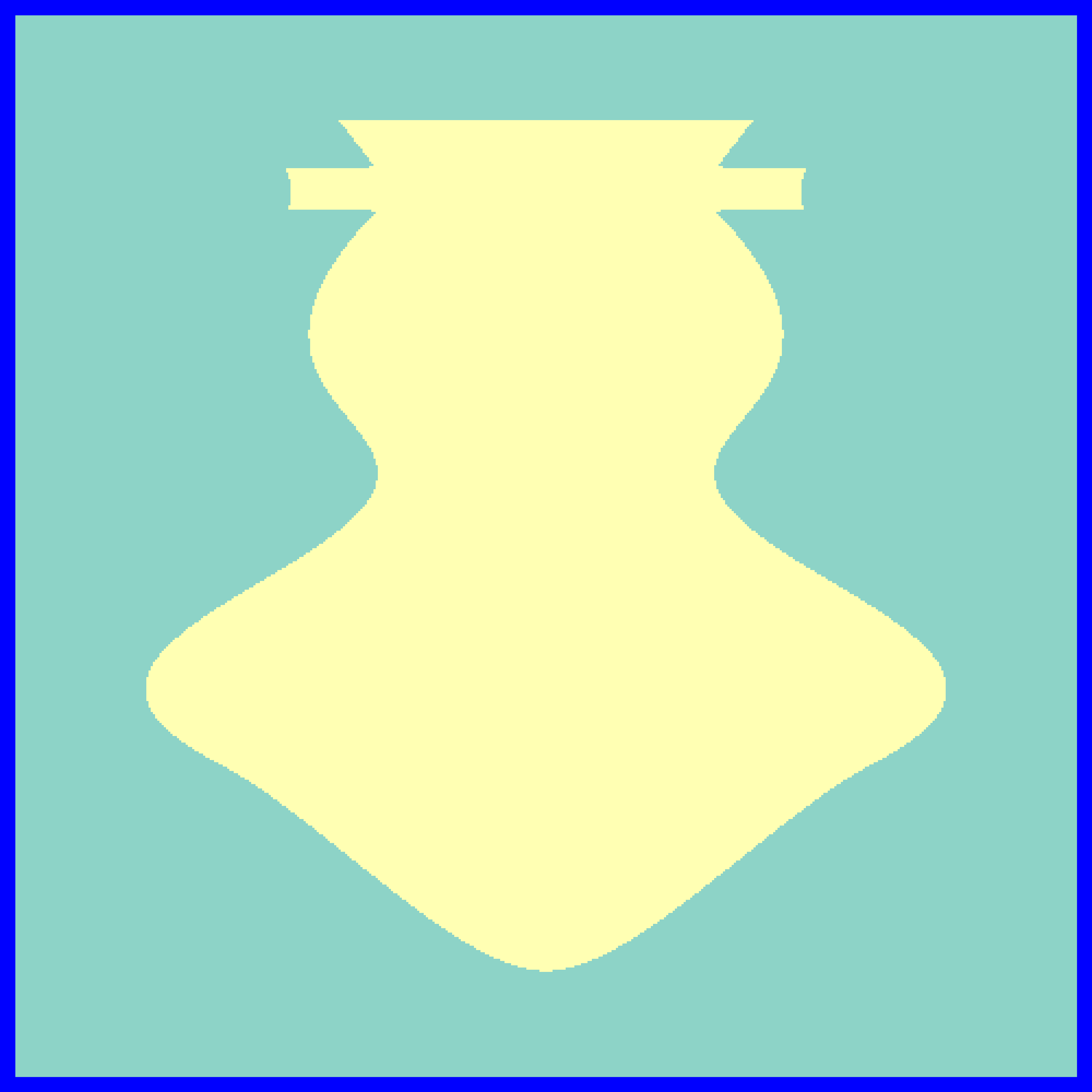} \\
    \includegraphics[width=0.45\linewidth]{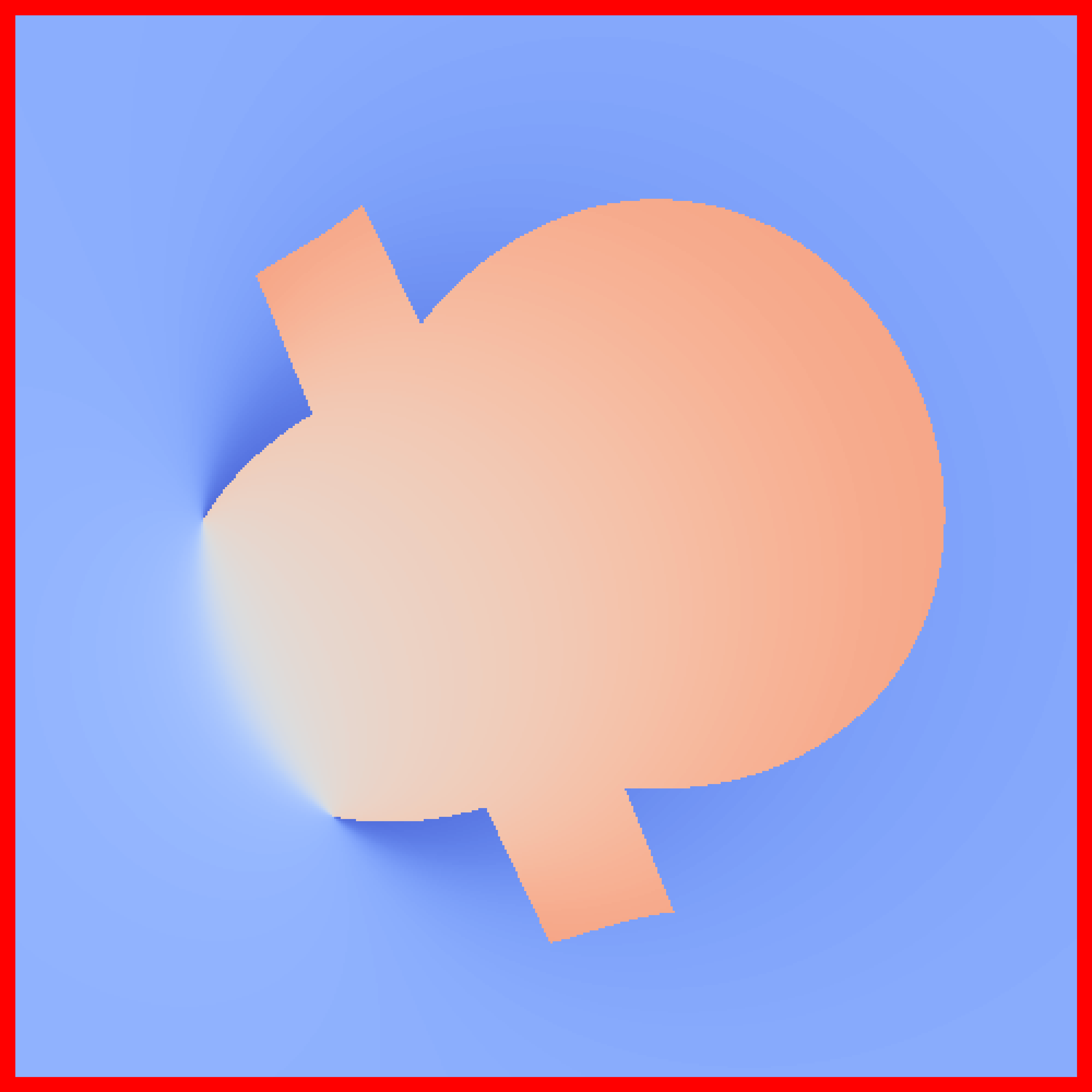} &
    \includegraphics[width=0.45\linewidth]{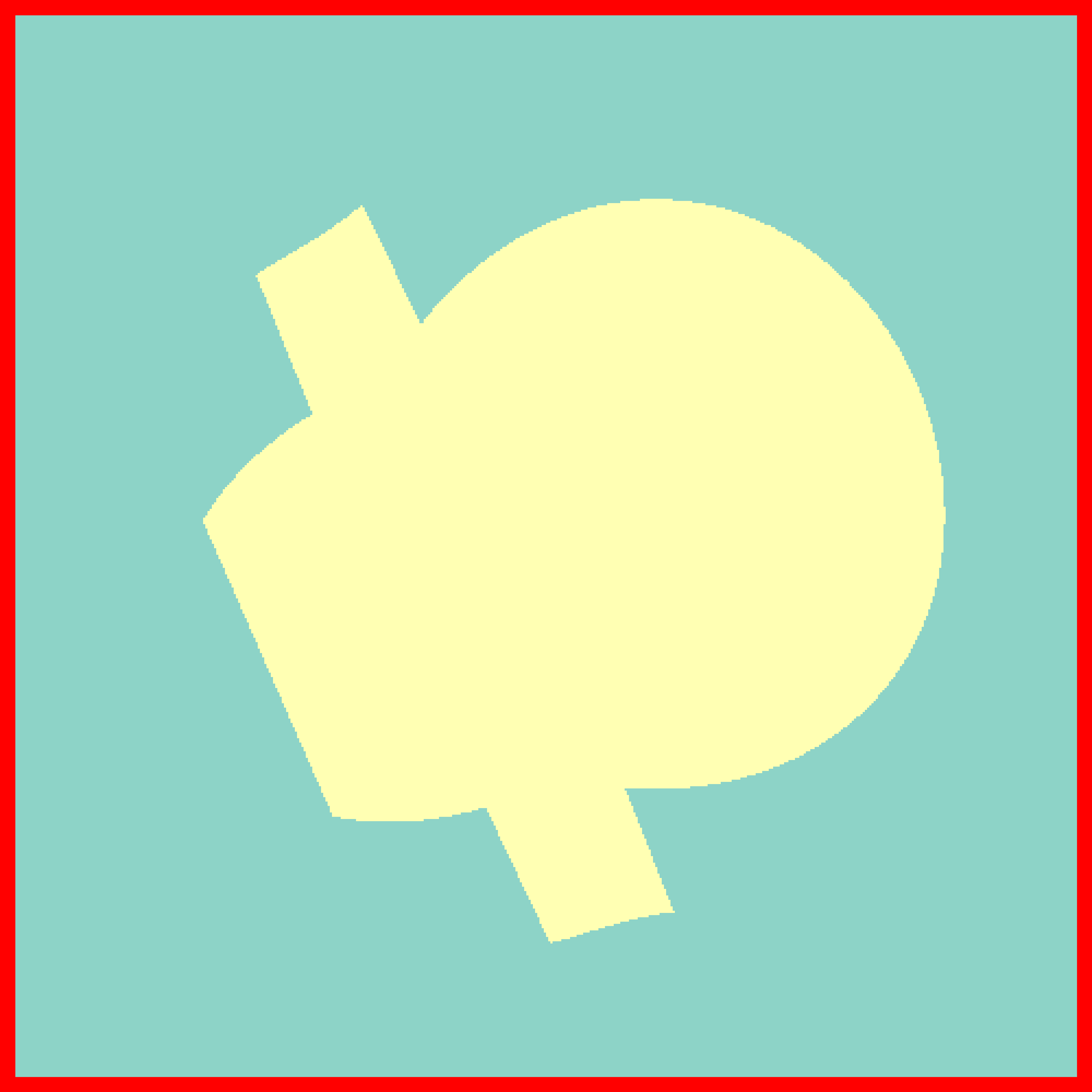} \\
    \includegraphics[width=0.45\linewidth]{summary_figures/gwn_colorbar_15.png} &
    \includegraphics[width=0.45\linewidth]{summary_figures/categories_colorbar_2.png} \\
\end{tabular}
\end{minipage}
\caption{This ``Vase'' shape is adapted from an example in~\cite{martens-2025-oneshot}, and was created as a surface of revolution from 7 cubic \bezier\ curves around the z-axis with circular cross-sections composed of 4 rational \bezier\ curves.}
\end{figure}

\vspace{1cm}

\begin{figure}

\begin{minipage}{0.5\textwidth}
\centering
\begin{tabular}{lr}
    \includegraphics[width=0.45\linewidth]{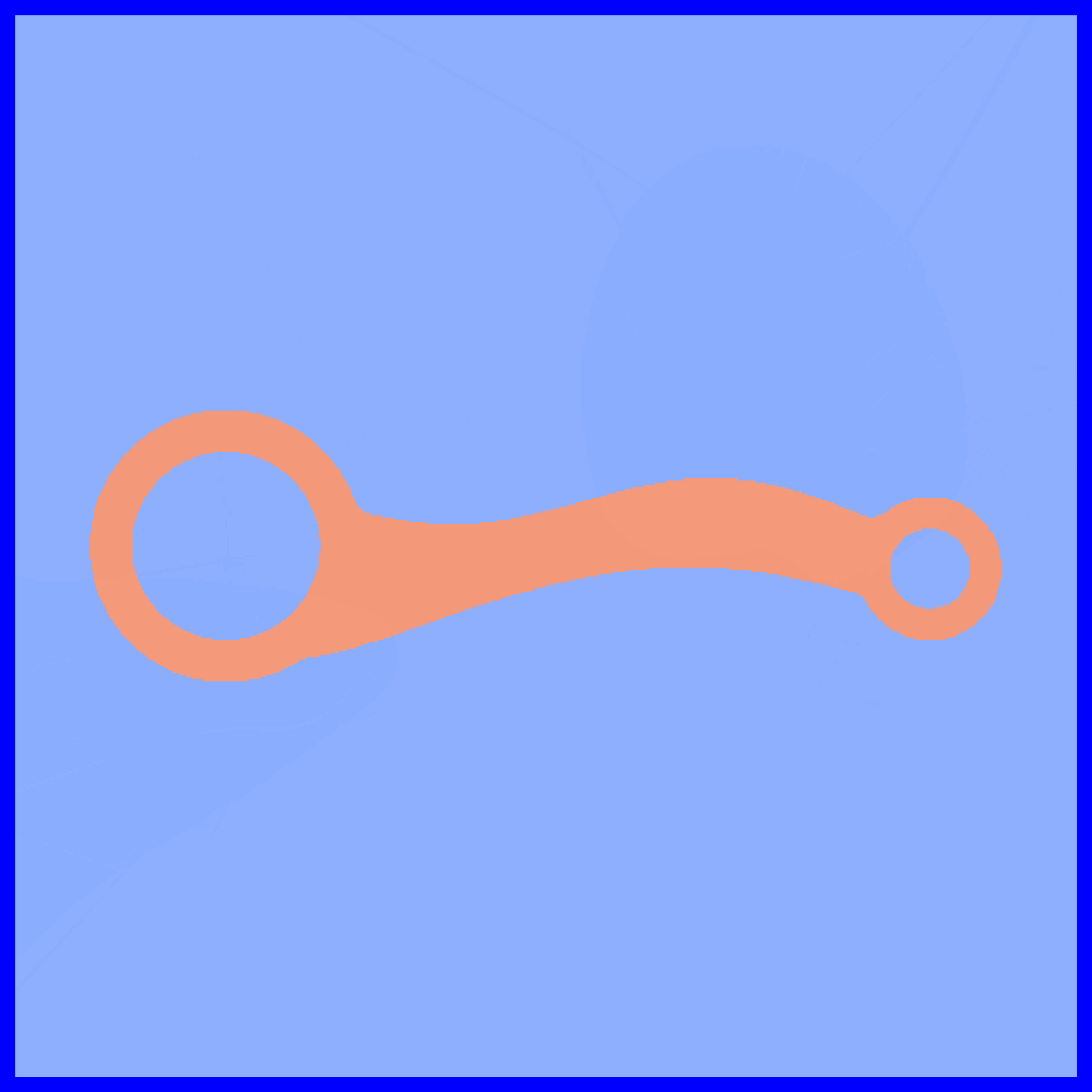} &
    \includegraphics[width=0.45\linewidth]{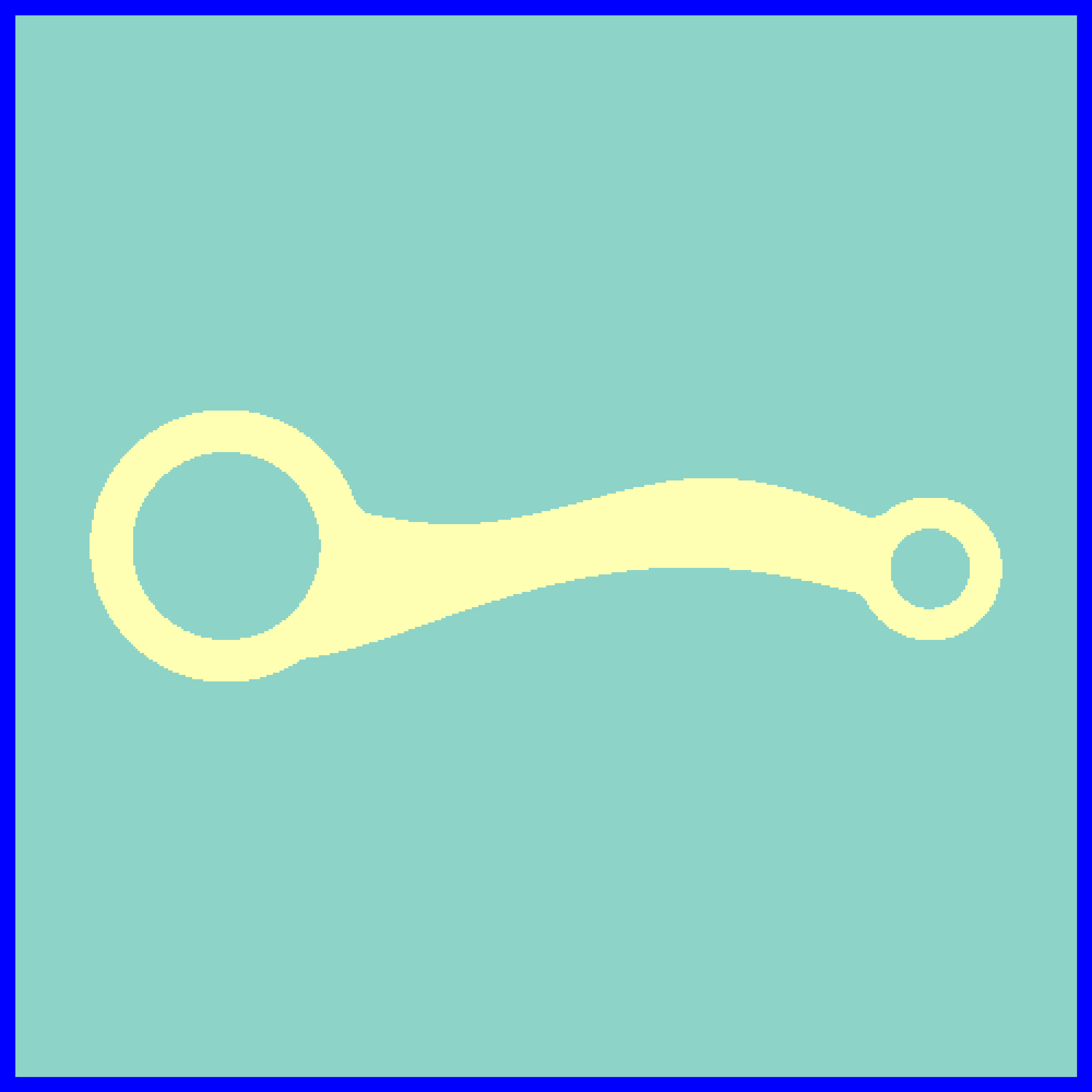} \\
    \includegraphics[width=0.45\linewidth]{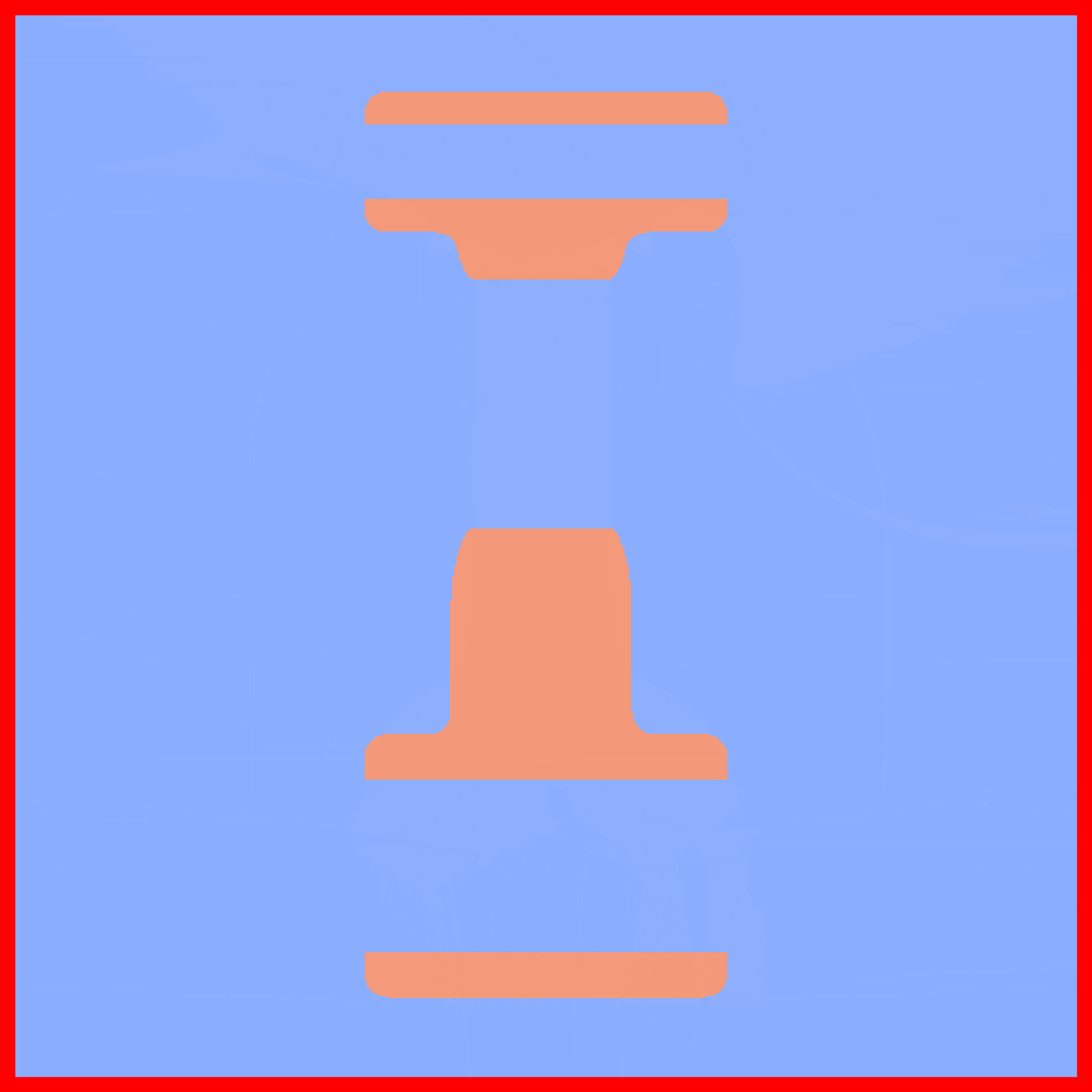} &
    \includegraphics[width=0.45\linewidth]{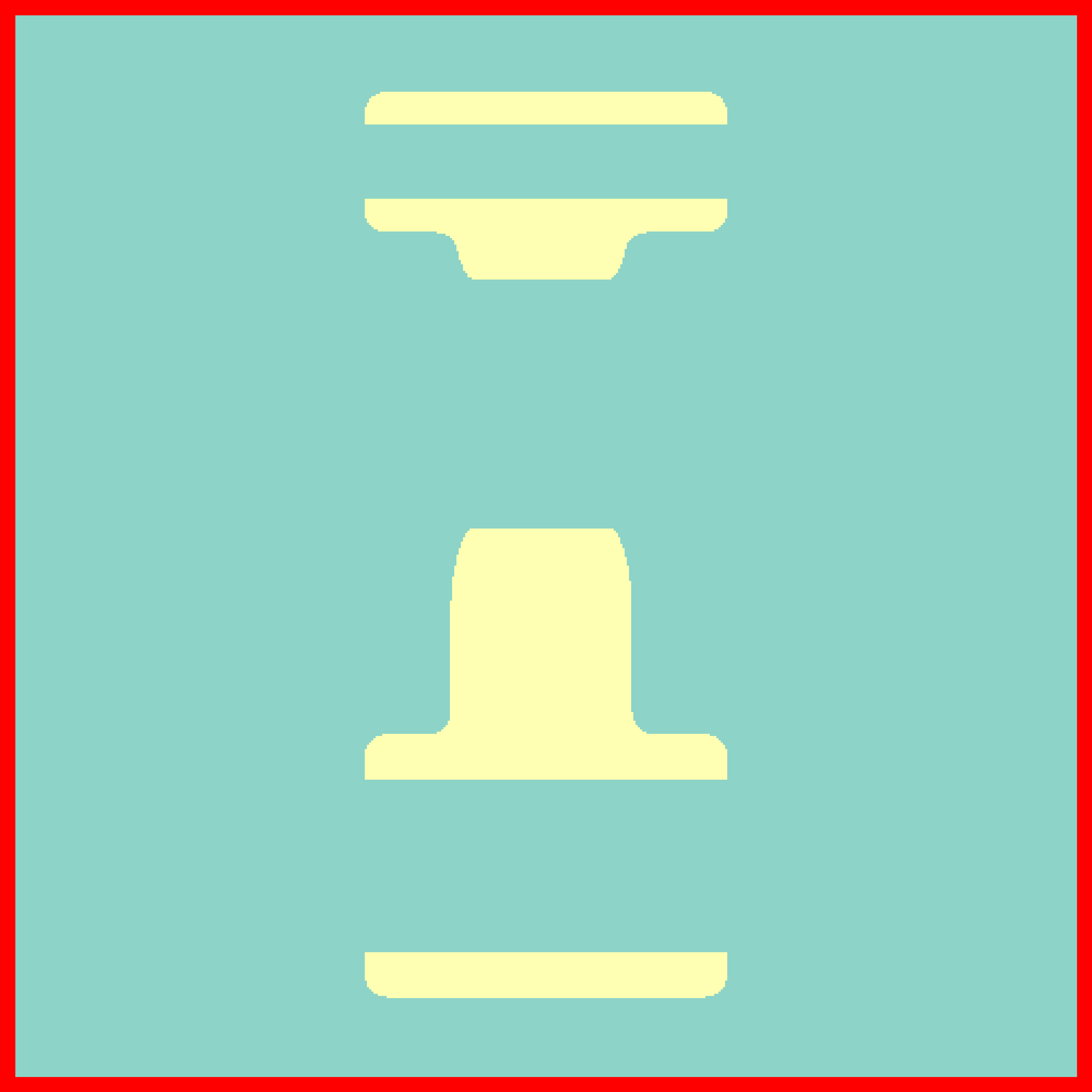} \\
    \includegraphics[width=0.45\linewidth]{summary_figures/gwn_colorbar_15.png} &
    \includegraphics[width=0.45\linewidth]{summary_figures/categories_colorbar_2.png} \\
\end{tabular}
\end{minipage}%
\begin{minipage}{0.5\textwidth}
\centering
\begin{tabular}{lr}
    \includegraphics[width=0.45\linewidth]{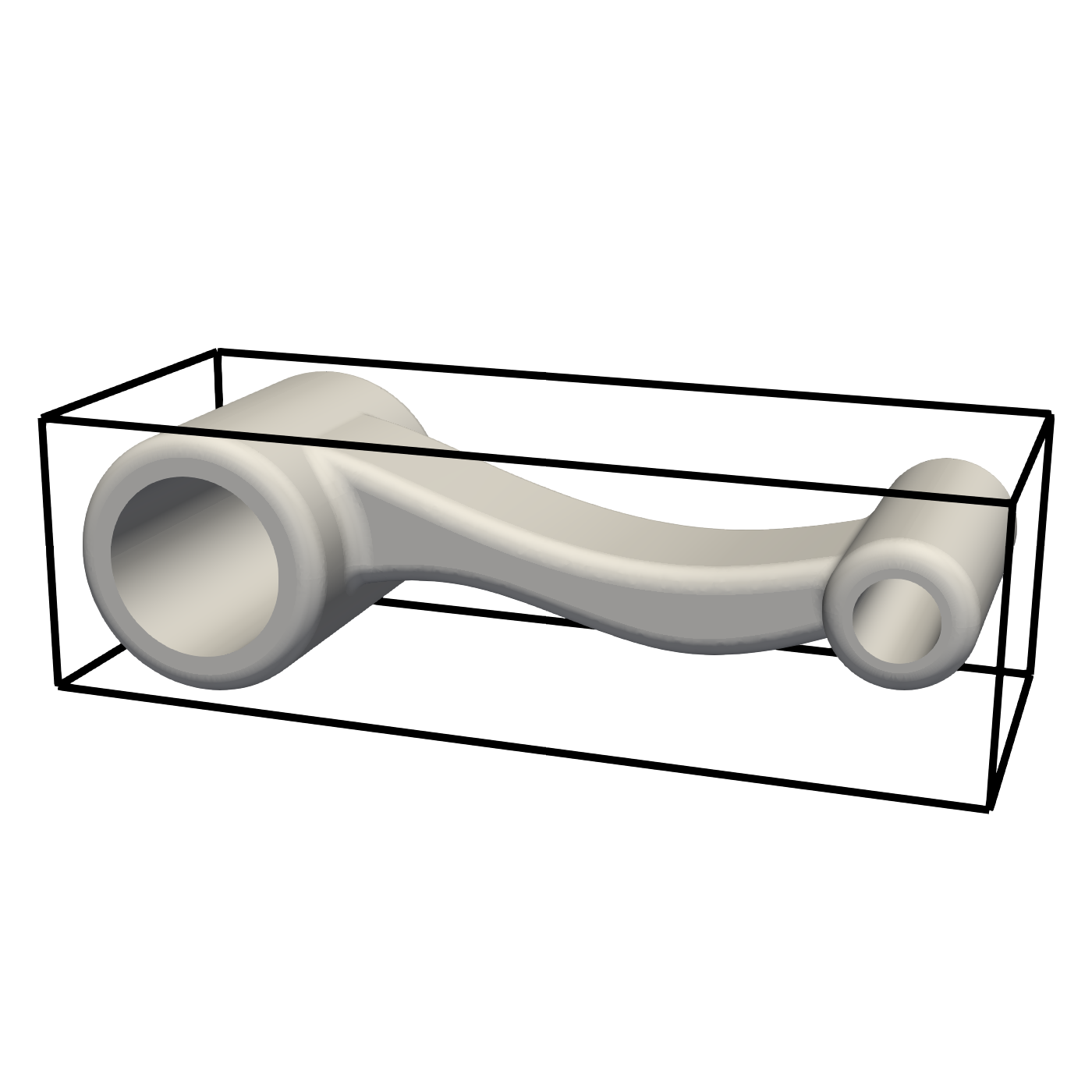}
    &
    \includegraphics[width=0.45\linewidth]{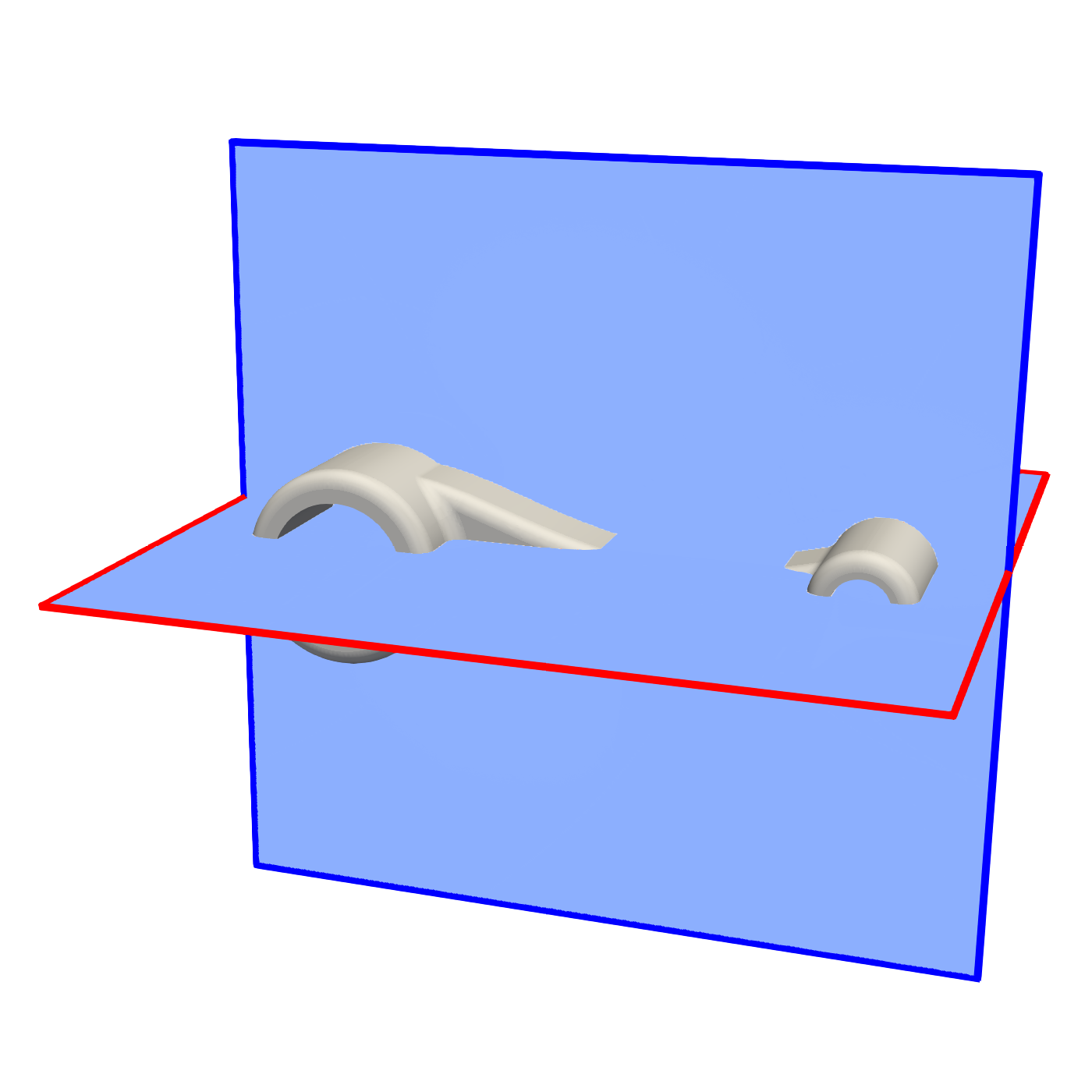}
\end{tabular}
\begin{tabular}{lr}
    Shape                          & Connector                        \\\cmidrule{1-2}
    Number of NURBS Patches        & 60                 \\
    Number of Trimming Curves      & 334                \\\cmidrule{1-2}
    \% Far-field Cases             & 98.8  \%  \\
    \% Near-field Cases            & 1.17 \%  \\
    \% Edge Cases                  & 0.0778 \%  \\\cmidrule{1-2}
    Avg. Time per Query (ms)       & 5.54          \\\cmidrule{1-2}
    Avg. Far-field Case Time (ms)  & 0.0584           \\
    Avg. Near-field Case Time (ms) & 0.376          \\
    Avg. Edge Case Time (ms)       & 38.9          \\
    \end{tabular}
\end{minipage}
\caption{This ``Connector'' model is bundled with releases of OpenCascade~\cite{opencascade}.}
\end{figure}

\vspace{1cm}

\begin{figure}

\begin{minipage}{0.5\textwidth}
\centering
\begin{tabular}{lr}
    \includegraphics[width=0.45\linewidth]{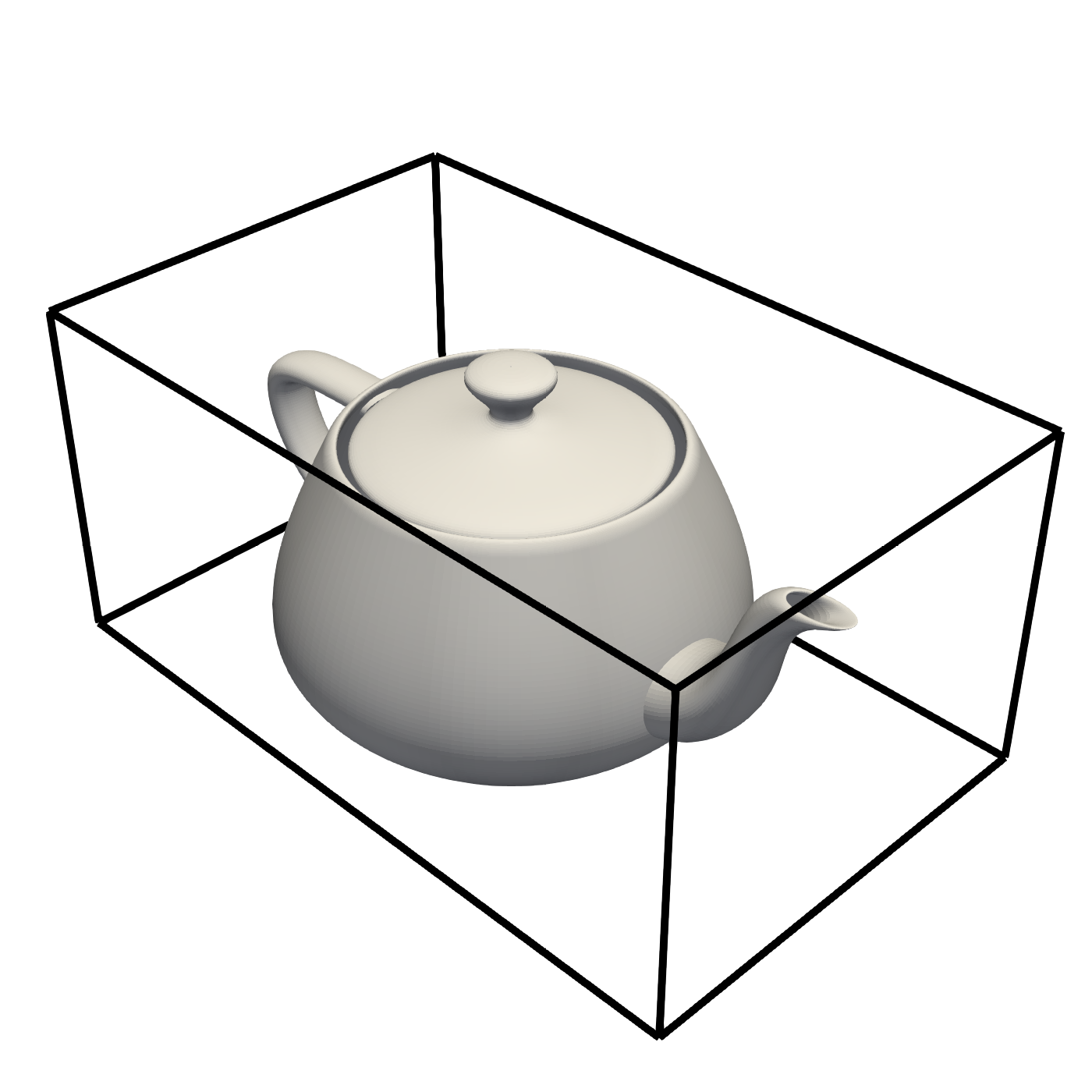}
    &
    \includegraphics[width=0.45\linewidth]{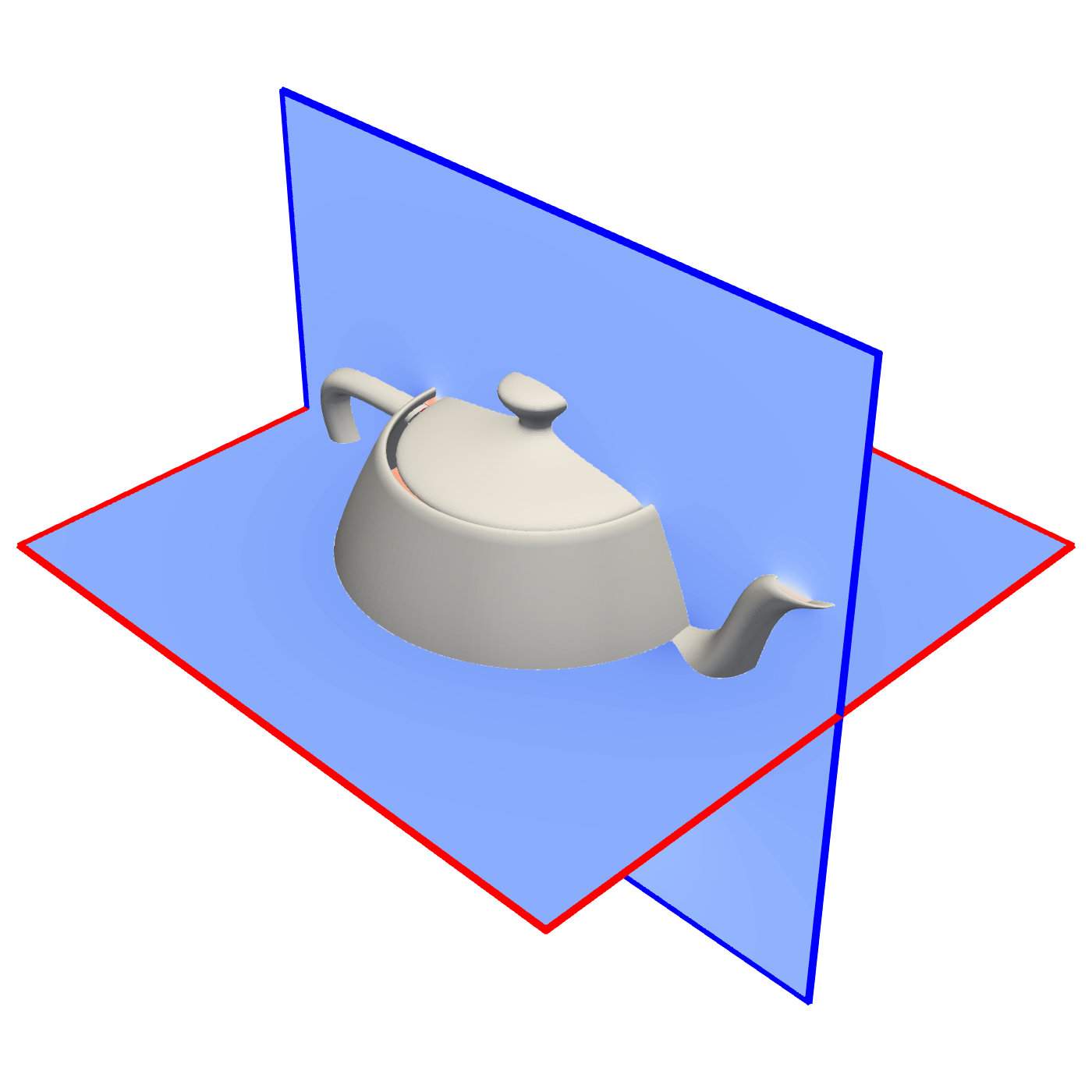}
\end{tabular}
\begin{tabular}{lr}
    Shape                          & Utah Teapot                        \\\cmidrule{1-2}
    Number of NURBS Patches        & 28                 \\
    Number of Trimming Curves      & 112                \\\cmidrule{1-2}
    \% Far-field Cases             & 98.7  \%  \\
    \% Near-field Cases            & 1.23 \%  \\
    \% Edge Cases                  & 0.0271 \%  \\\cmidrule{1-2}
    Avg. Time per Query (ms)       & 0.152          \\\cmidrule{1-2}
    Avg. Far-field Case Time (ms)  & 0.00310           \\
    Avg. Near-field Case Time (ms) & 0.103          \\
    Avg. Edge Case Time (ms)       & 4.27          \\
    \end{tabular}
\end{minipage}%
\begin{minipage}{0.5\textwidth}
\centering
\begin{tabular}{lr}
    \includegraphics[width=0.45\linewidth]{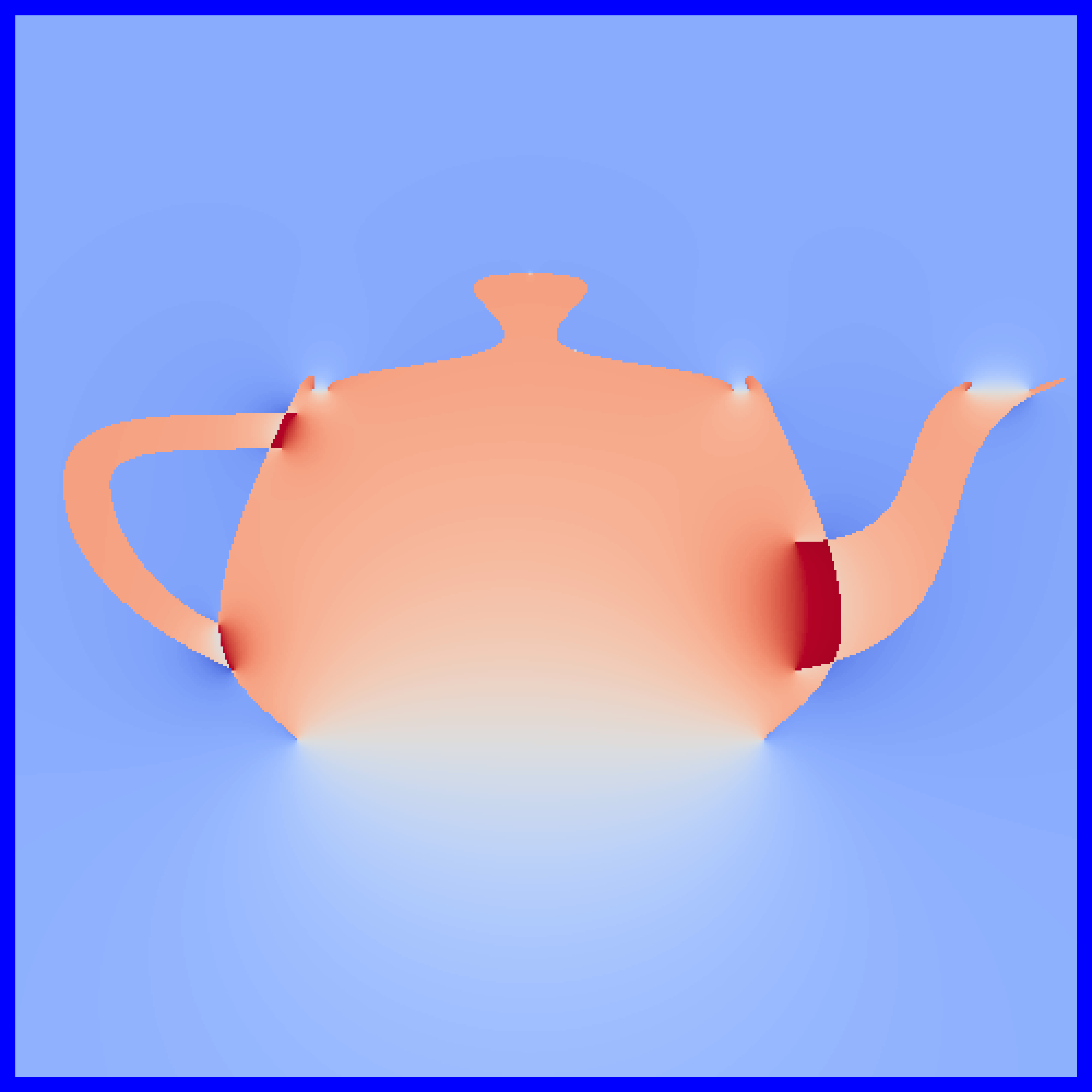} &
    \includegraphics[width=0.45\linewidth]{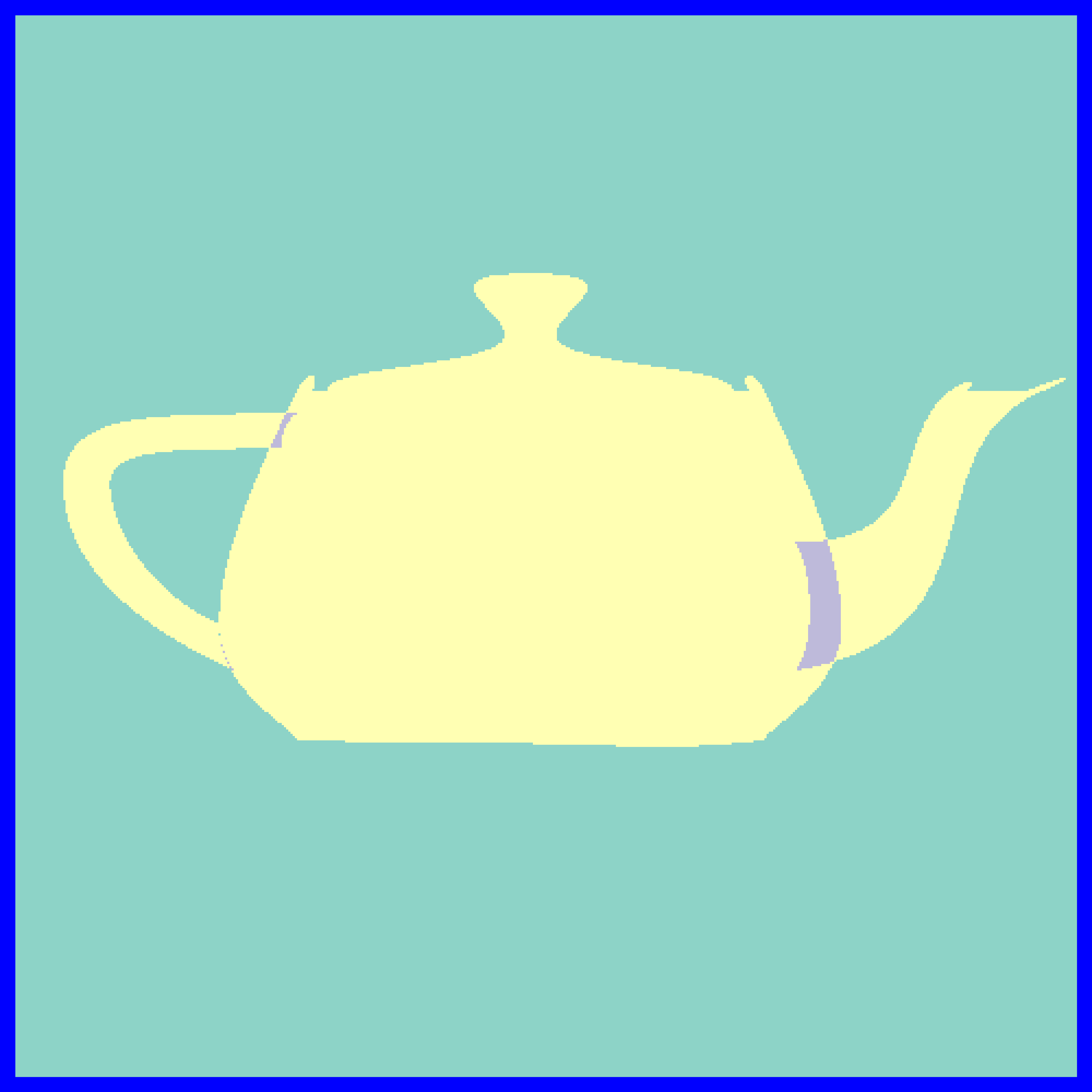} \\
    \includegraphics[width=0.45\linewidth]{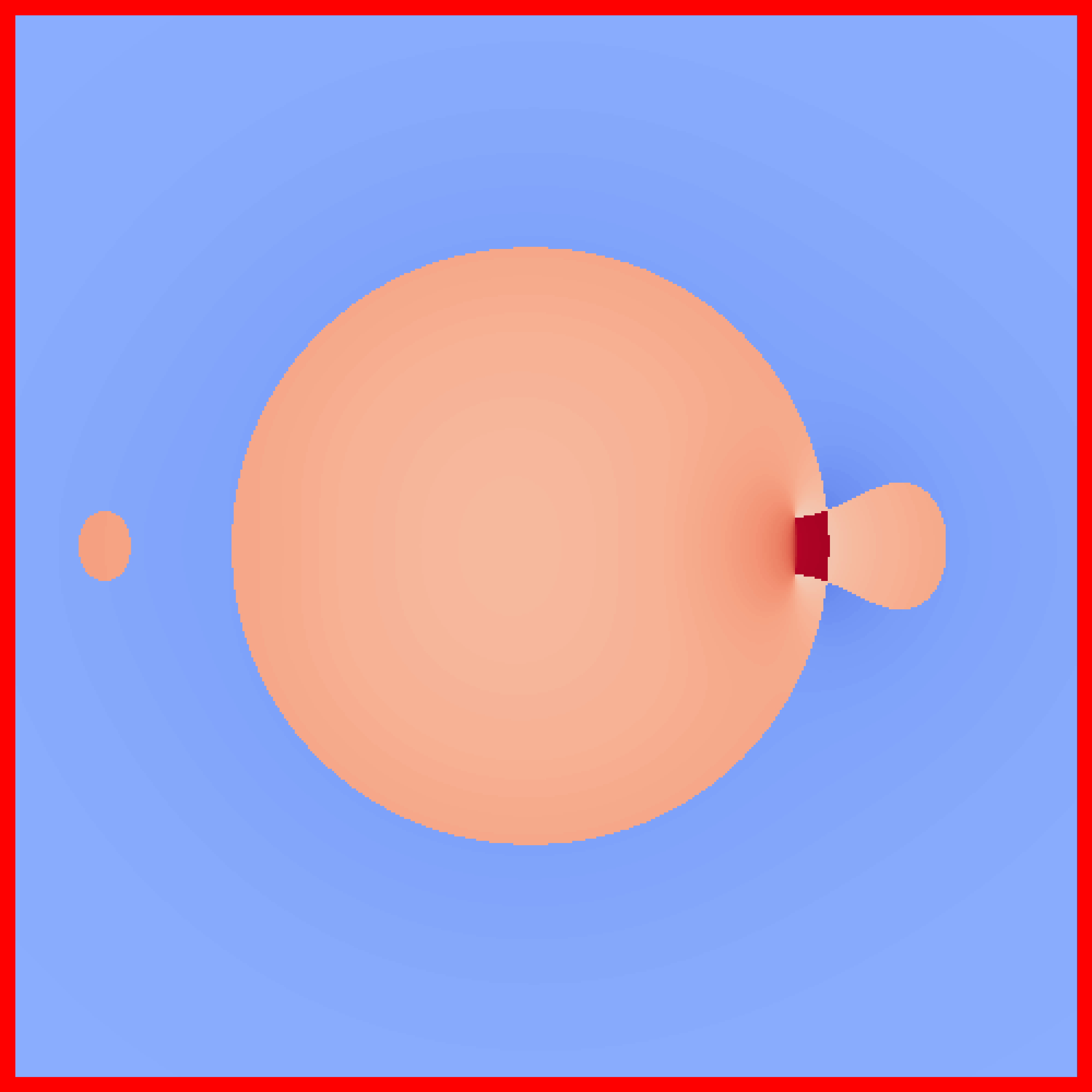} &
    \includegraphics[width=0.45\linewidth]{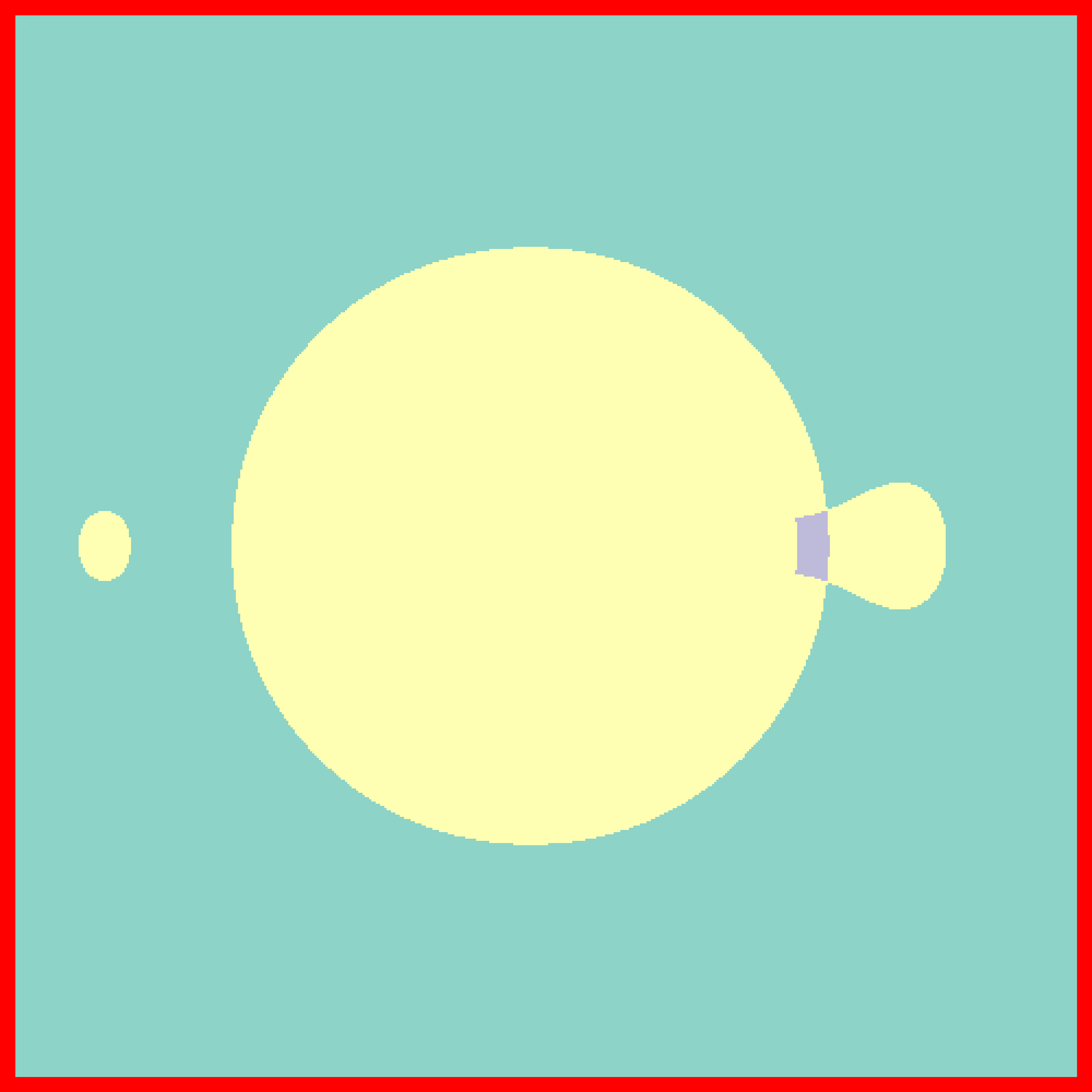} \\
    \includegraphics[width=0.45\linewidth]{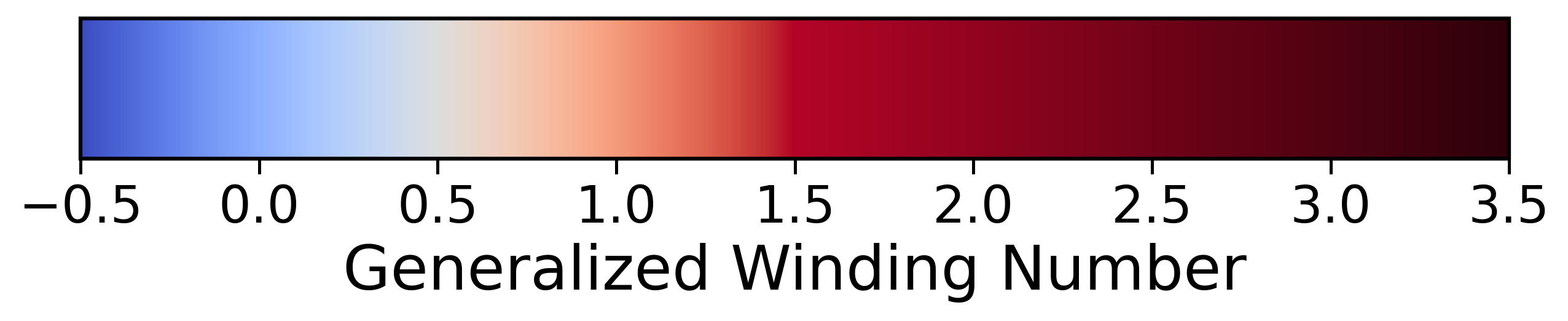} &
    \includegraphics[width=0.45\linewidth]{summary_figures/categories_colorbar_3.png} \\
\end{tabular}
\end{minipage}
\caption{This original ``Utah Teapot''~\cite{johnson-2000-utahteapot} is composed of \bezier\ patches, many of which overlap. Note that the lid to the teapot is disconnected from the body, and the shape has no bottom.}
\end{figure}

\vspace{1cm}

\begin{figure}

\begin{minipage}{0.5\textwidth}
\centering
\begin{tabular}{lr}
    \includegraphics[width=0.45\linewidth]{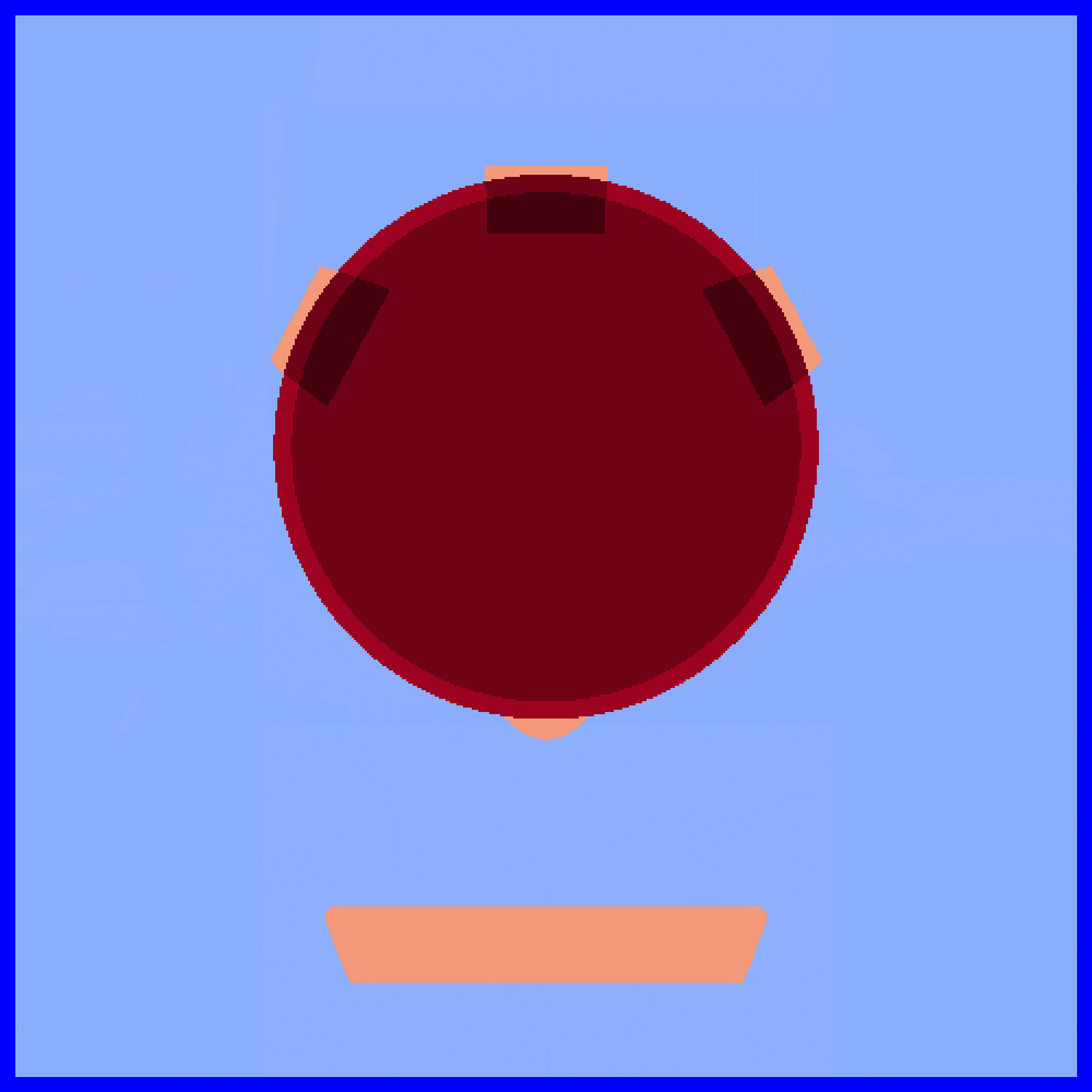} &
    \includegraphics[width=0.45\linewidth]{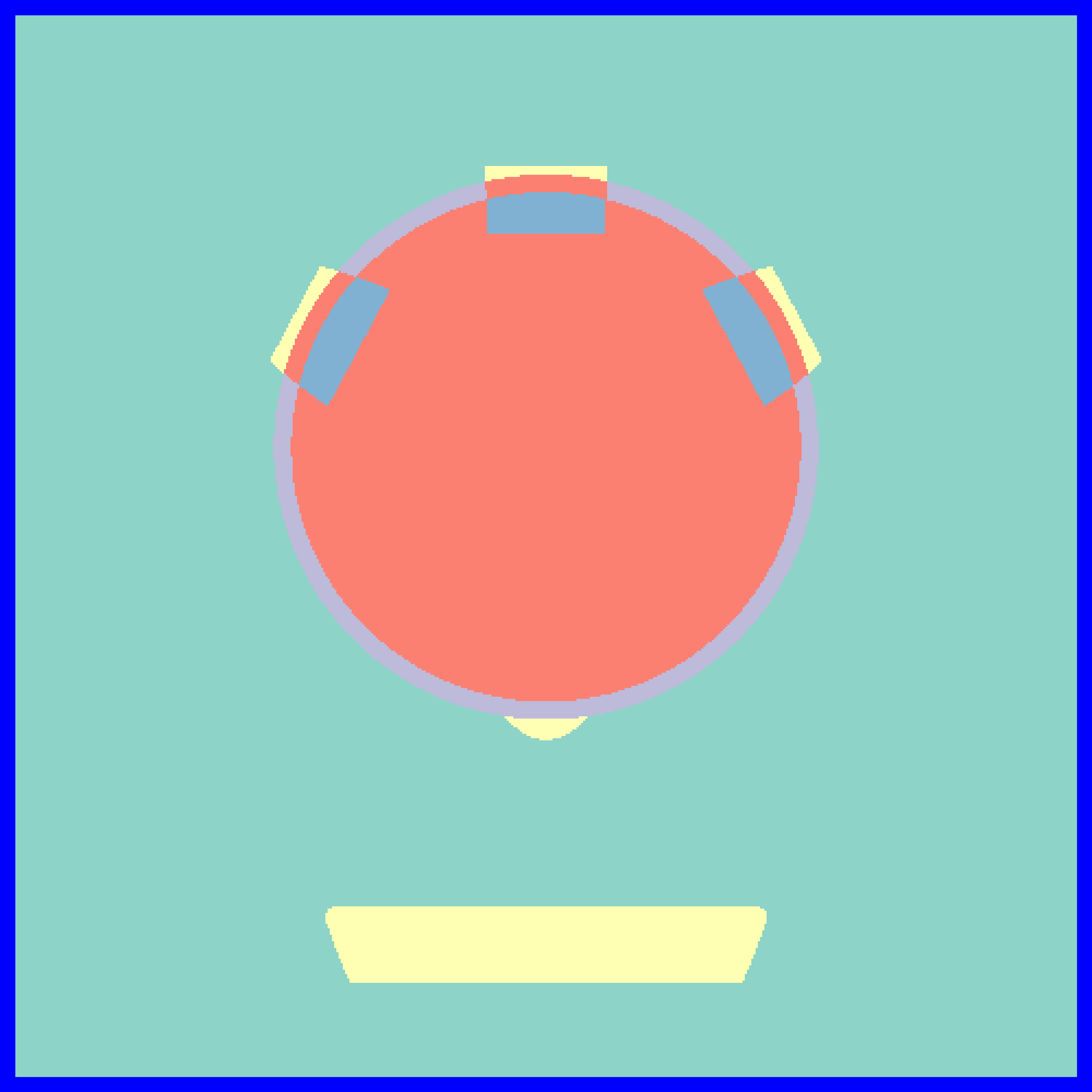} \\
    \includegraphics[width=0.45\linewidth]{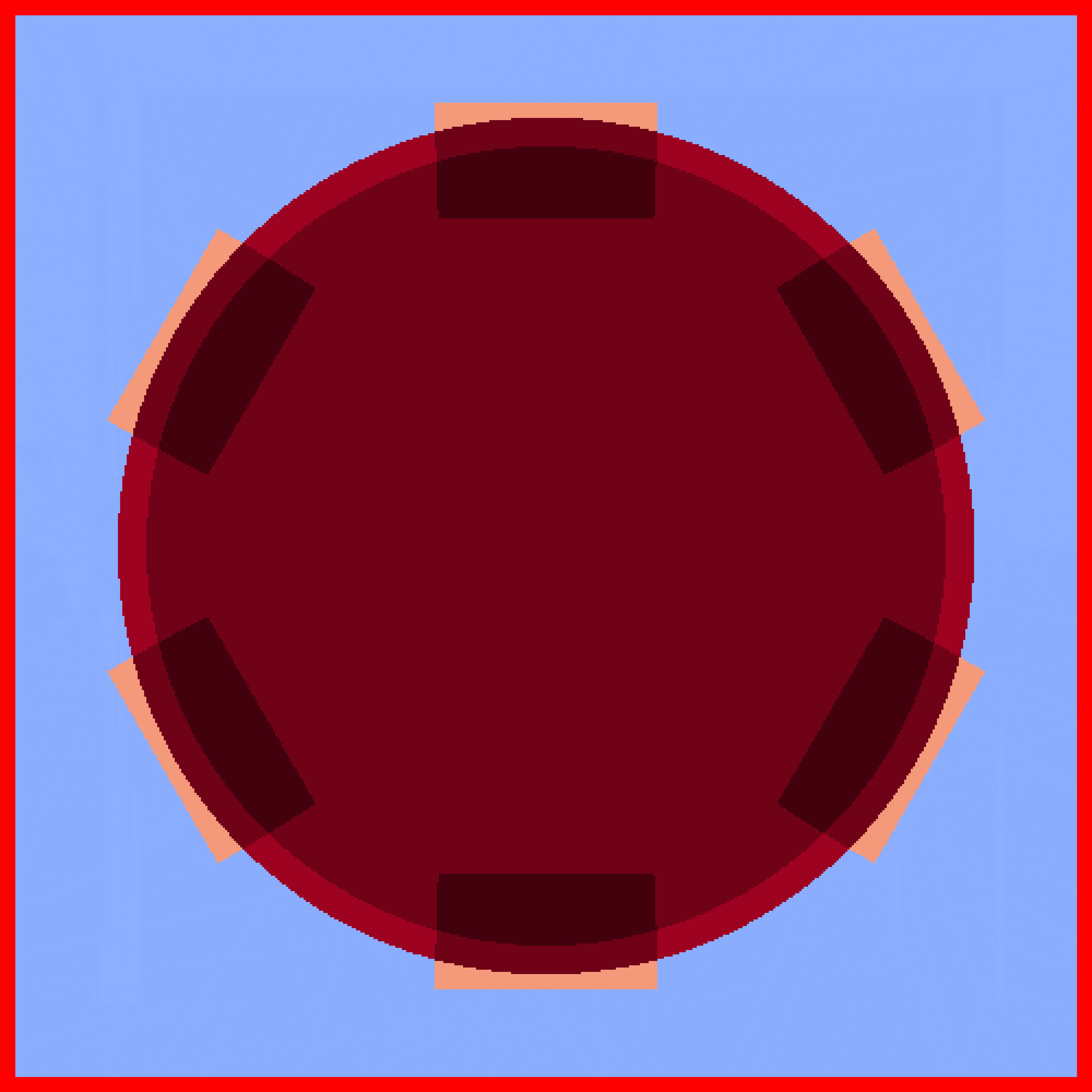} &
    \includegraphics[width=0.45\linewidth]{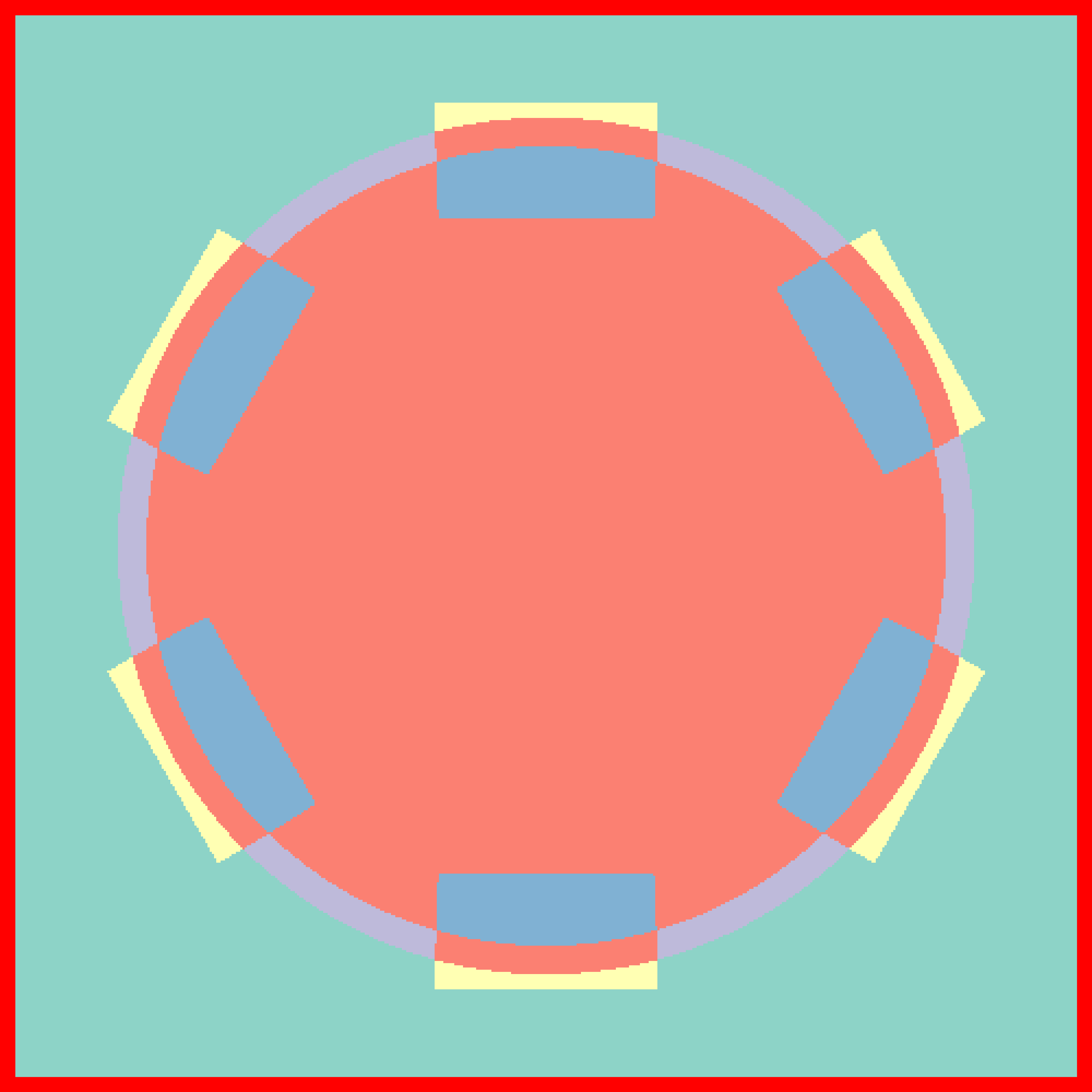} \\
    \includegraphics[width=0.45\linewidth]{summary_figures/gwn_colorbar_35.png} &
    \includegraphics[width=0.45\linewidth]{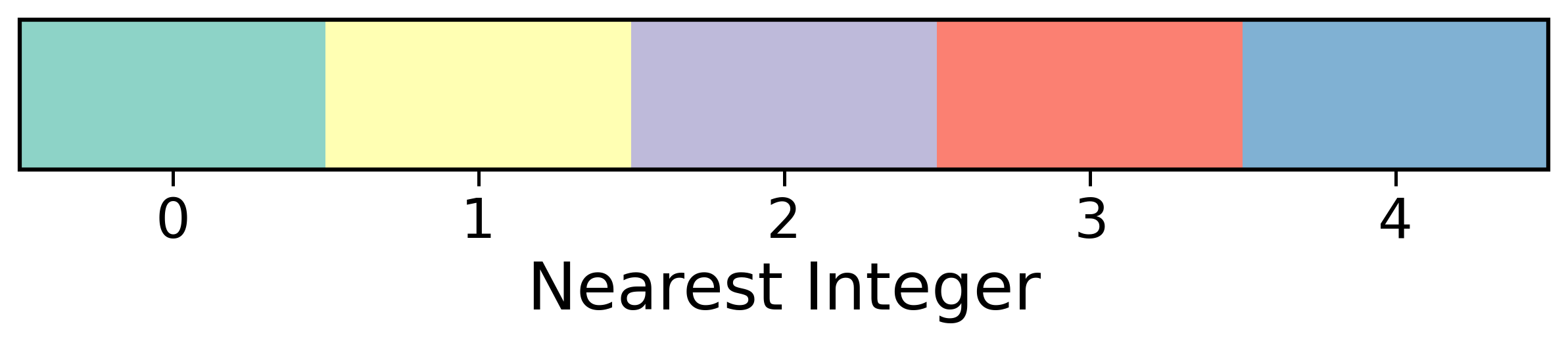} \\
\end{tabular}
\end{minipage}%
\begin{minipage}{0.5\textwidth}
\centering
\begin{tabular}{lr}
    \includegraphics[width=0.45\linewidth]{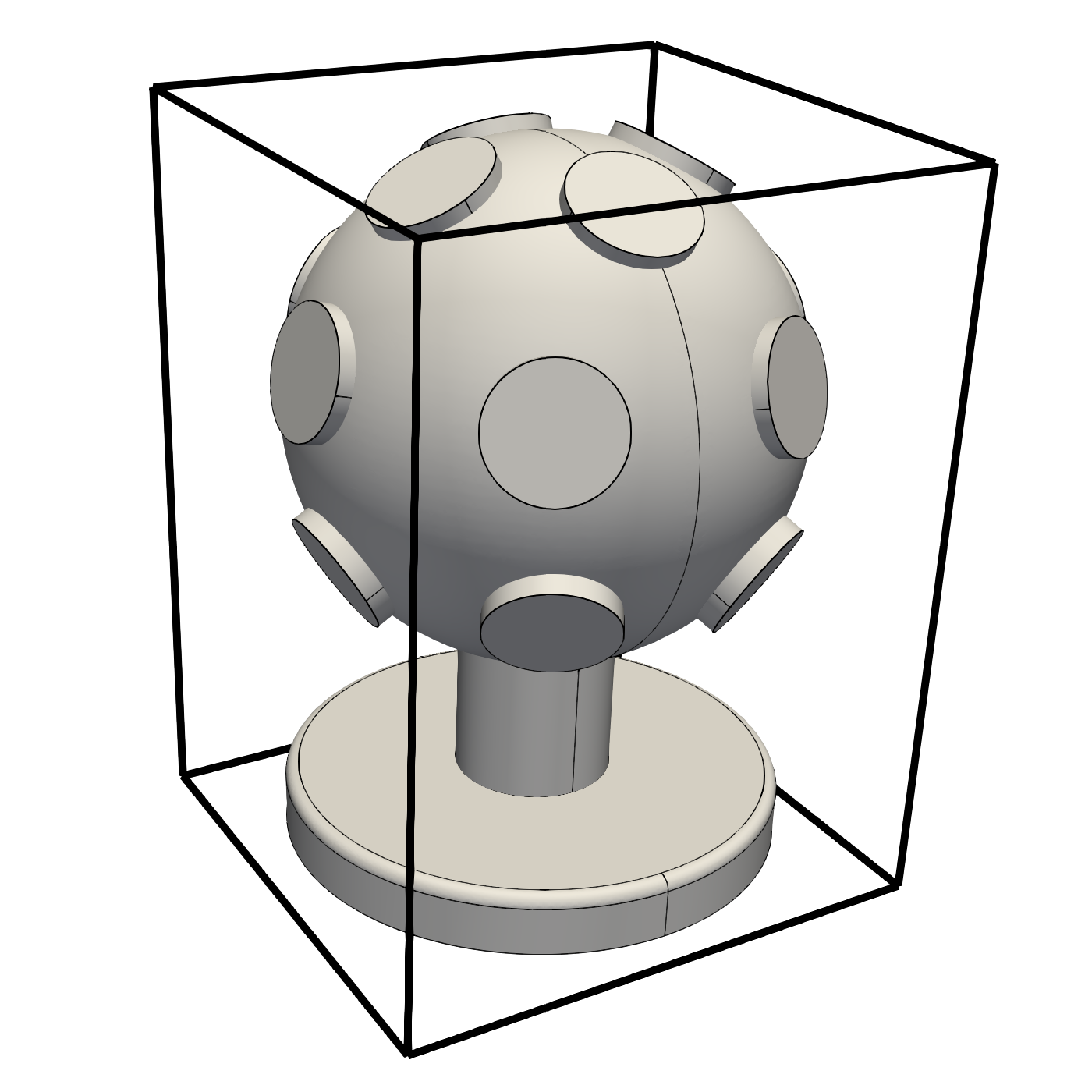}
    &
    \includegraphics[width=0.45\linewidth]{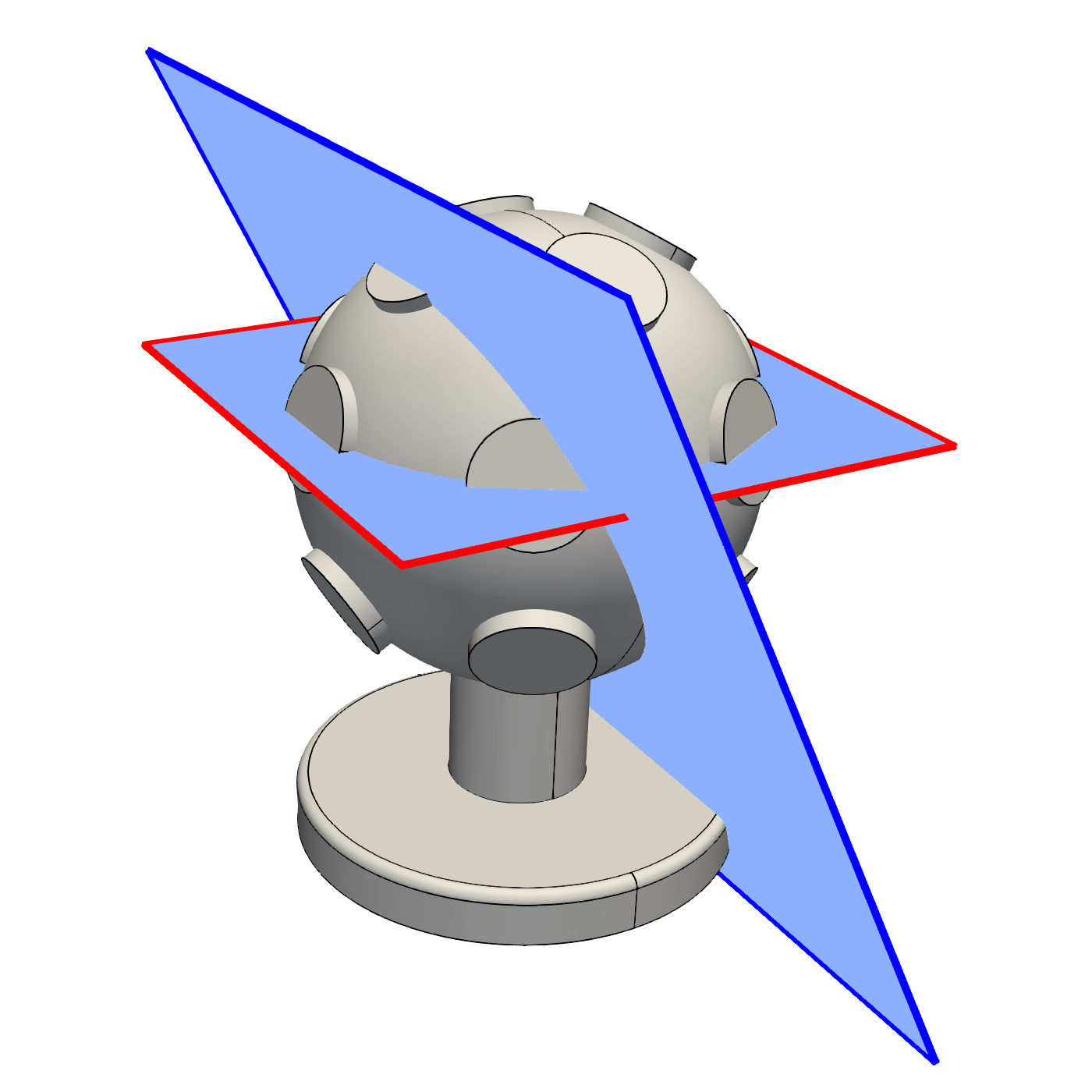}
\end{tabular}
\begin{tabular}{lr}
    Shape                          & Lamp                        \\\cmidrule{1-2}
    Number of NURBS Patches        & 78                 \\
    Number of Trimming Curves      & 214                \\\cmidrule{1-2}
    \% Far-field Cases             & 97.4  \%  \\
    \% Near-field Cases            & 2.56 \%  \\
    \% Edge Cases                  & 0.0093 \%  \\\cmidrule{1-2}
    Avg. Time per Query (ms)       & 0.836         \\\cmidrule{1-2}
    Avg. Far-field Case Time (ms)  & 0.00862           \\
    Avg. Near-field Case Time (ms) & 0.0753          \\
    Avg. Edge Case Time (ms)       & 4.09          \\
    \end{tabular}
\end{minipage}
\caption{This ``Lamp'' shape is taken from the ABC dataset model with index 3800. Note that the central sphere of the shape has several overlapping watertight components, which increase the winding number by integer values.}
\end{figure}

\vspace{1cm}

\begin{figure}

\begin{minipage}{0.5\textwidth}
\centering
\begin{tabular}{lr}
    \includegraphics[width=0.45\linewidth]{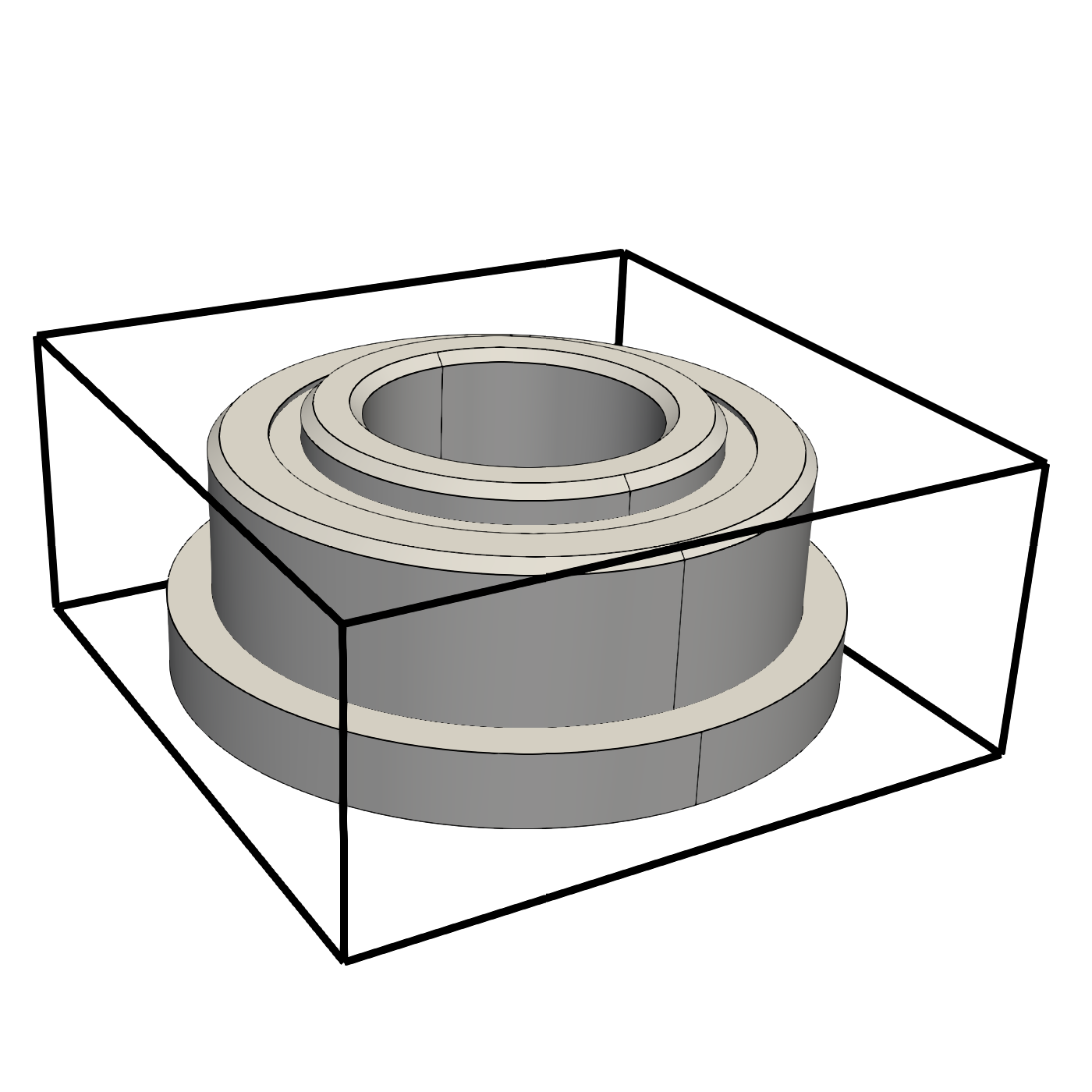}
    &
    \includegraphics[width=0.45\linewidth]{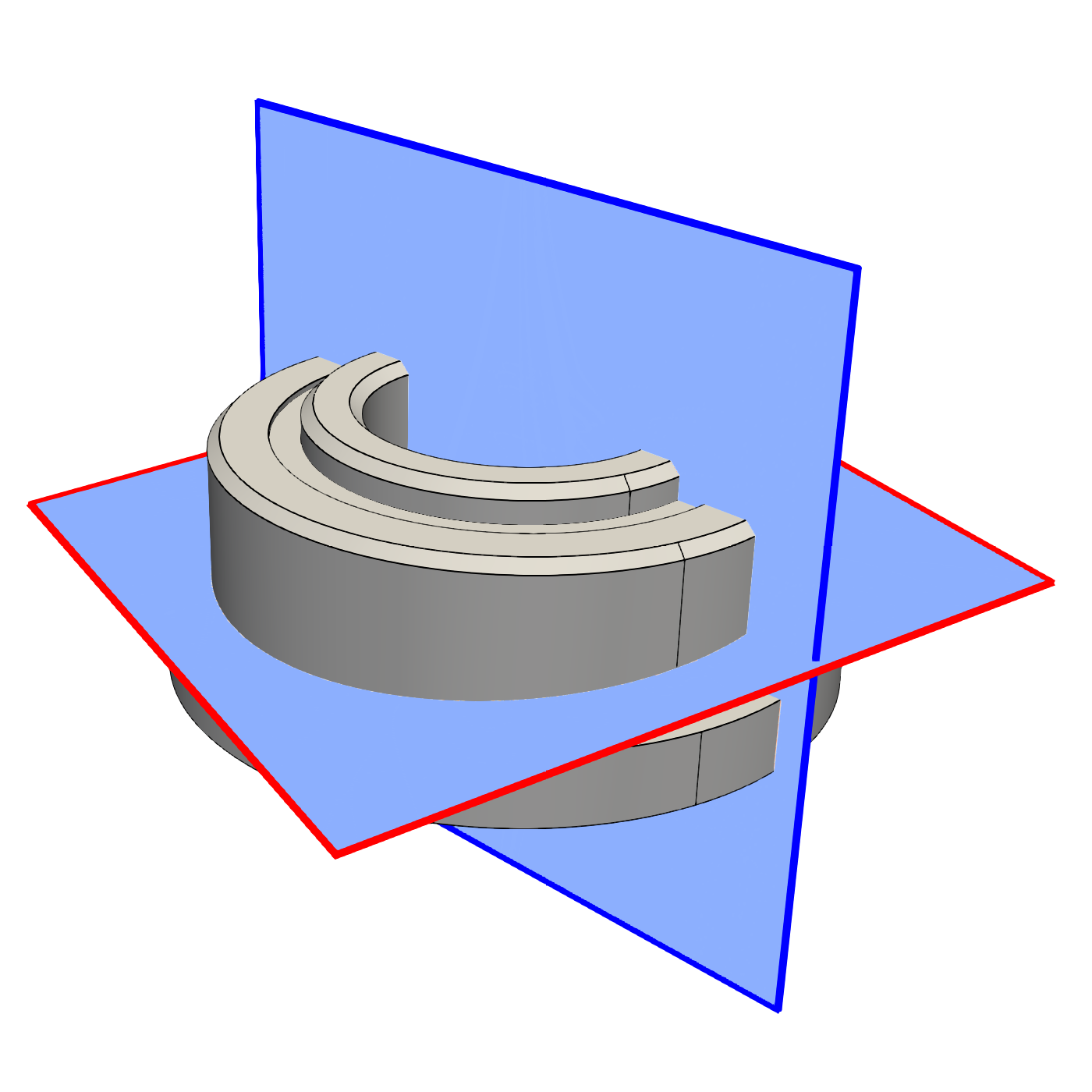}
\end{tabular}
\begin{tabular}{lr}
    Shape                          & Bearings                        \\\cmidrule{1-2}
    Number of NURBS Patches        & 66                 \\
    Number of Trimming Curves      & 248                \\\cmidrule{1-2}
    \% Far-field Cases             & 96.4  \%  \\
    \% Near-field Cases            & 3.38  \%  \\
    \% Edge Cases                  & 0.145 \%  \\\cmidrule{1-2}
    Avg. Time per Query (ms)       & 1.23          \\\cmidrule{1-2}
    Avg. Far-field Case Time (ms)  & 0.00882          \\
    Avg. Near-field Case Time (ms) & 0.0628         \\
    Avg. Edge Case Time (ms)       & 5.53          \\
    \end{tabular}
\end{minipage}%
\begin{minipage}{0.5\textwidth}
\centering
\begin{tabular}{lr}
    \includegraphics[width=0.45\linewidth]{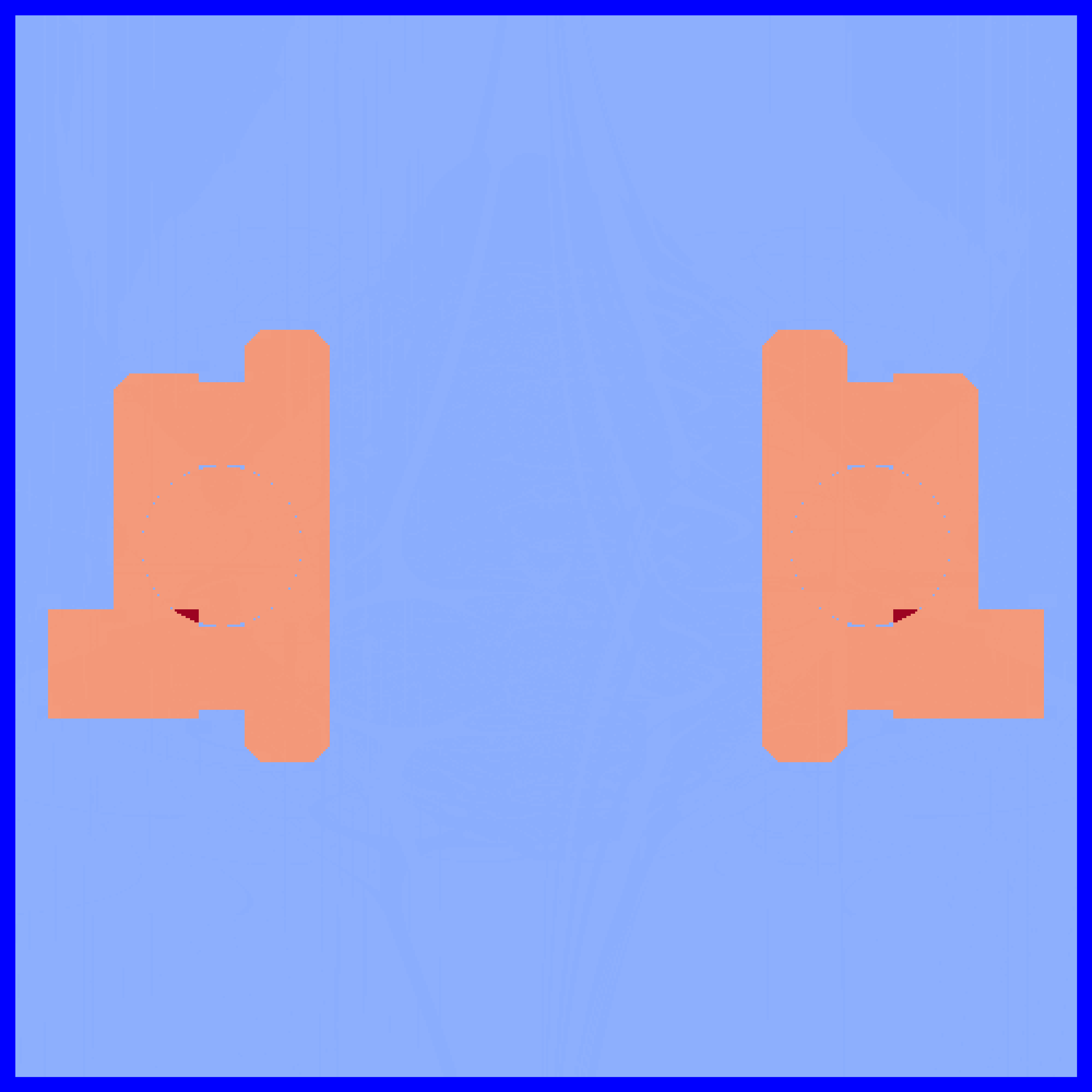} &
    \includegraphics[width=0.45\linewidth]{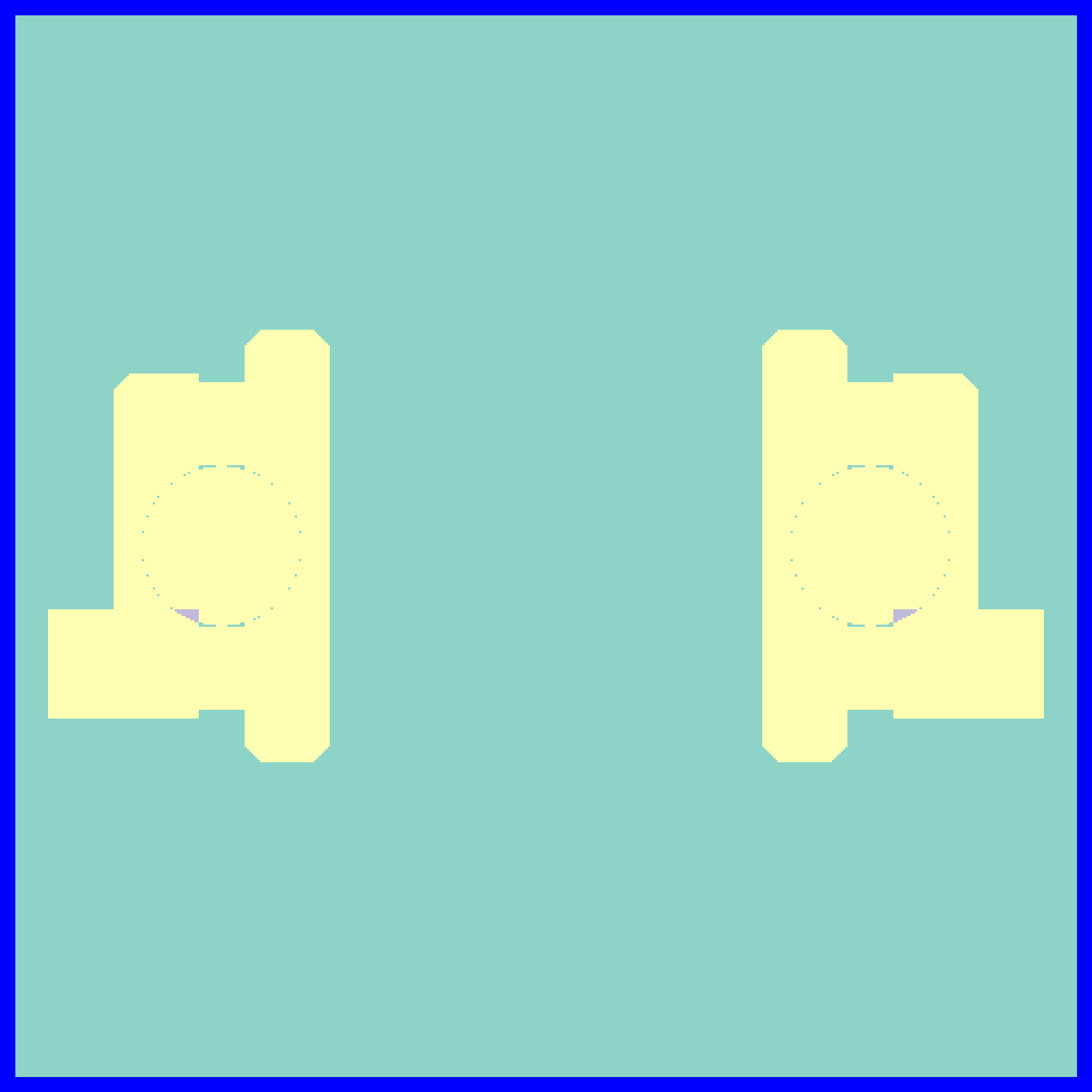} \\
    \includegraphics[width=0.45\linewidth]{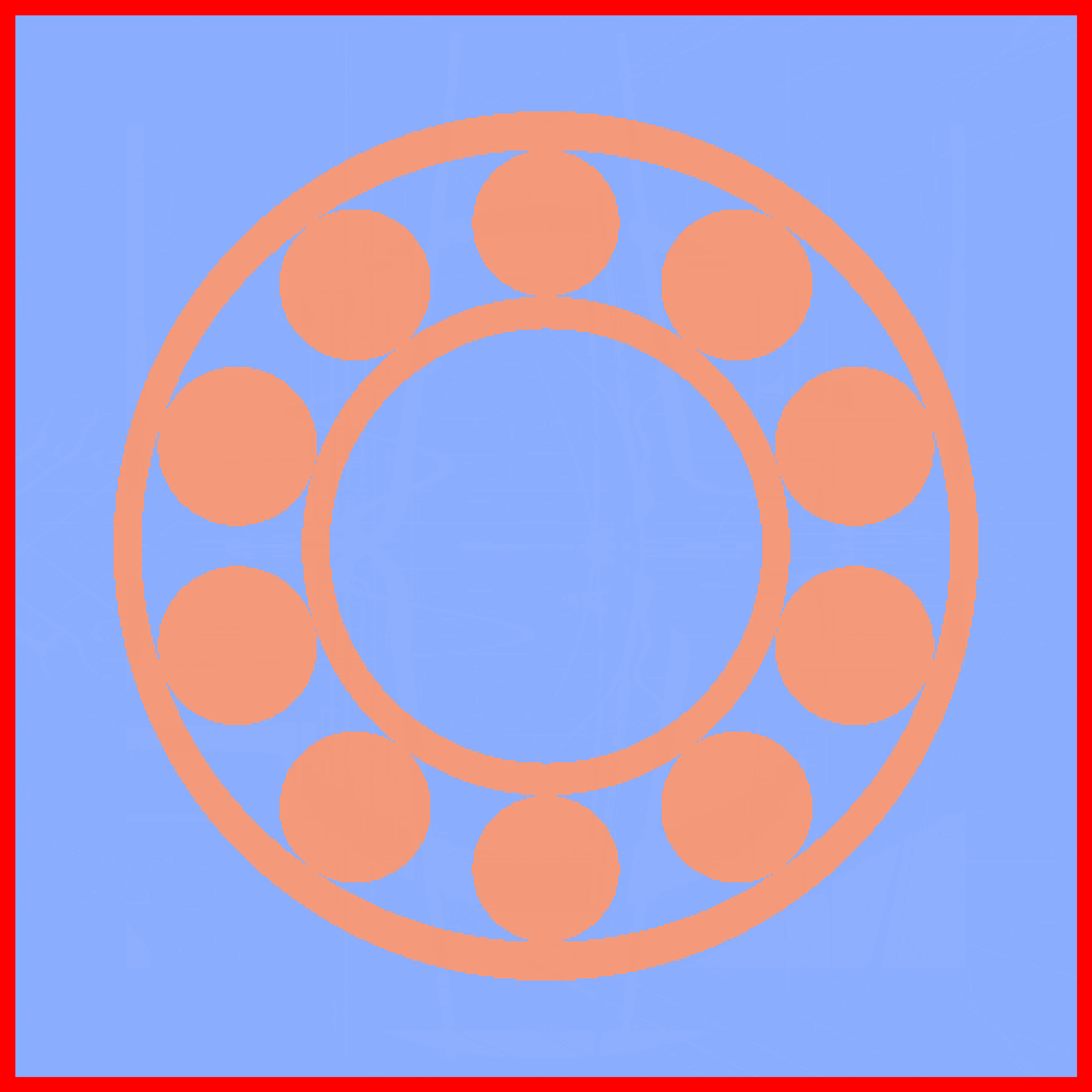} &
    \includegraphics[width=0.45\linewidth]{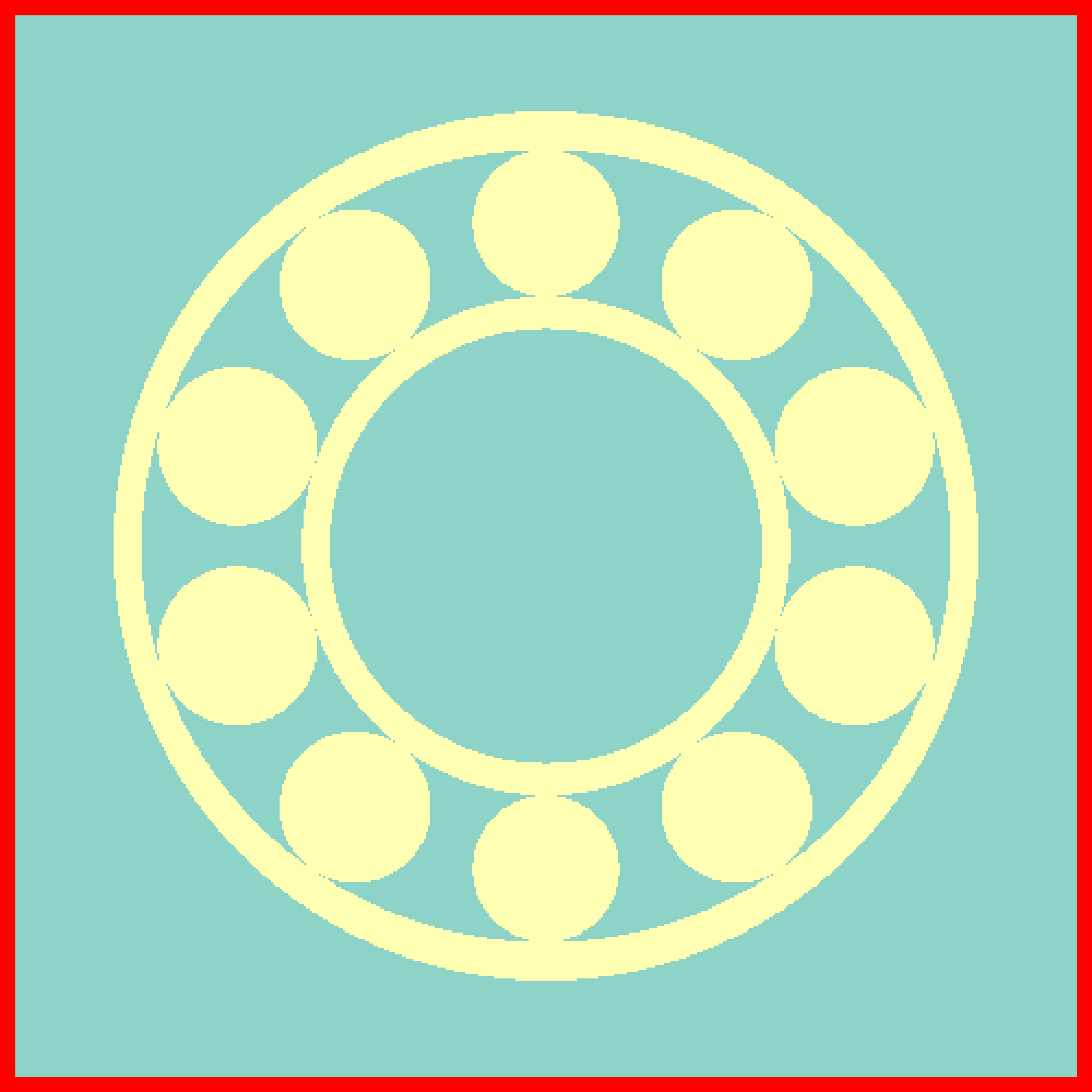} \\
    \includegraphics[width=0.45\linewidth]{summary_figures/gwn_colorbar_25.png} &
    \includegraphics[width=0.45\linewidth]{summary_figures/categories_colorbar_3.png} \\
\end{tabular}
\end{minipage}
\caption{This ``Bearings'' shape is derived from the ABC dataset model with index 7963. Note that there is a small section of the outer ring that intersects with the inner ball bearings.}
\end{figure}

\vspace{1cm}

\begin{figure}

\begin{minipage}{0.5\textwidth}
\centering
\begin{tabular}{lr}
    \includegraphics[width=0.45\linewidth]{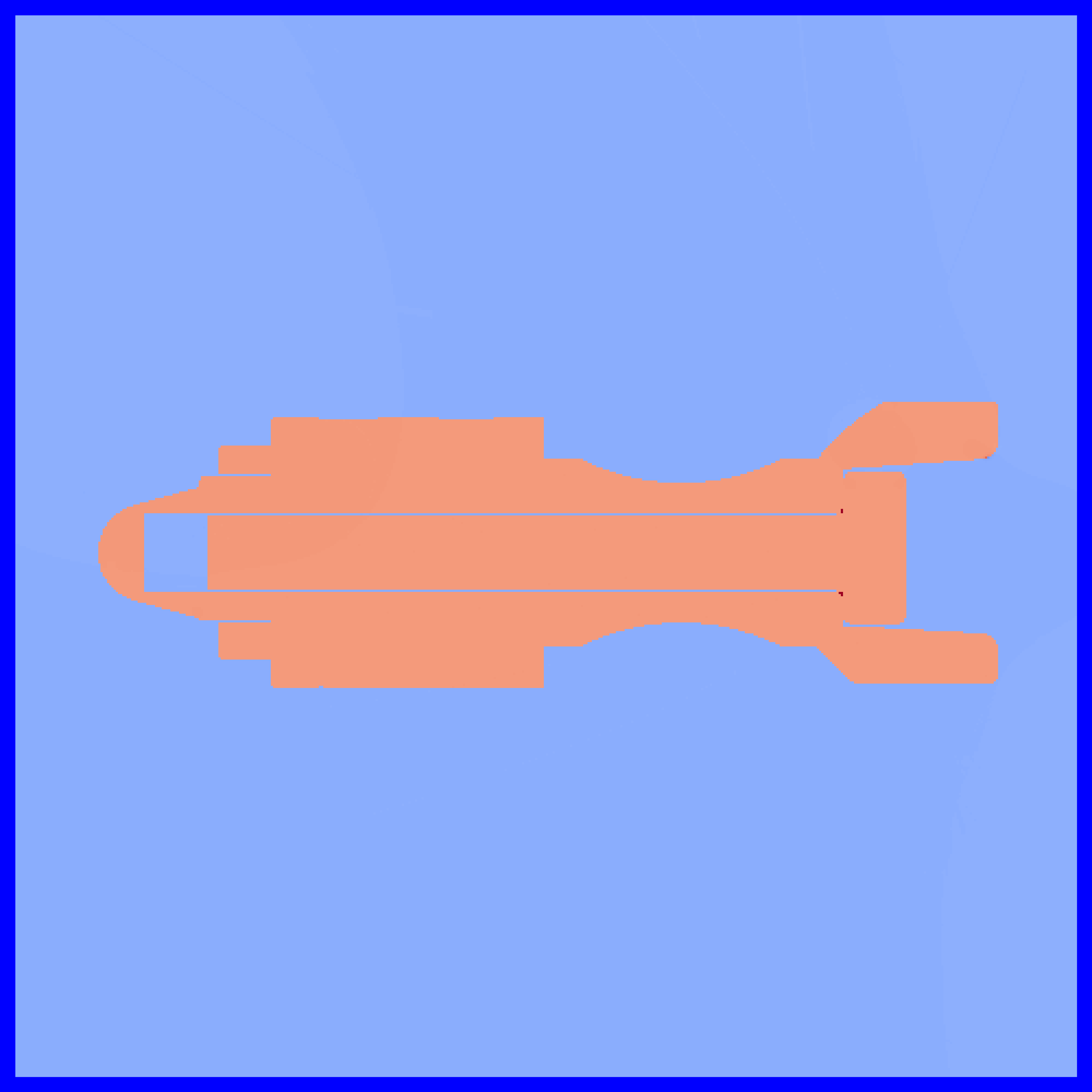} &
    \includegraphics[width=0.45\linewidth]{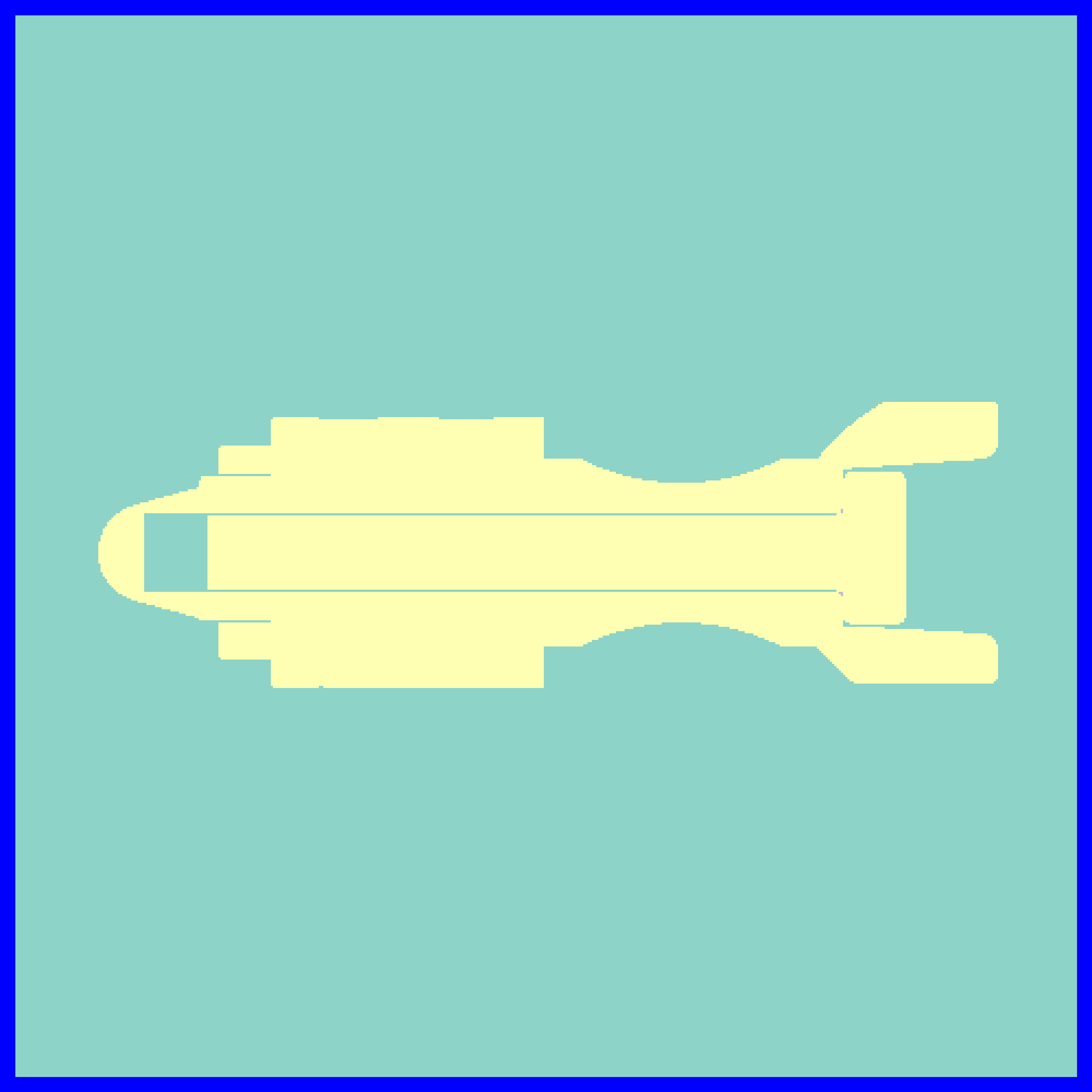} \\
    \includegraphics[width=0.45\linewidth]{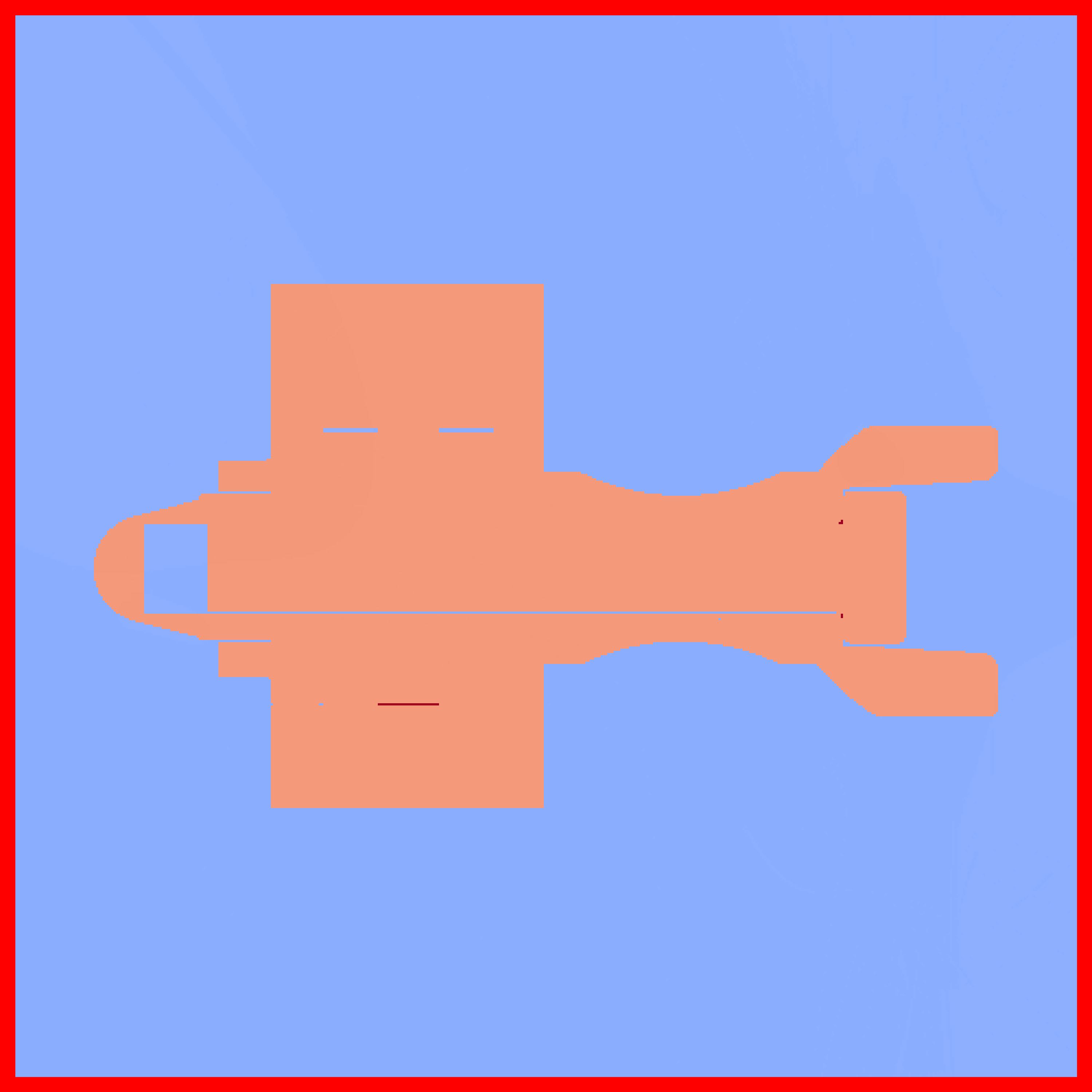} &
    \includegraphics[width=0.45\linewidth]{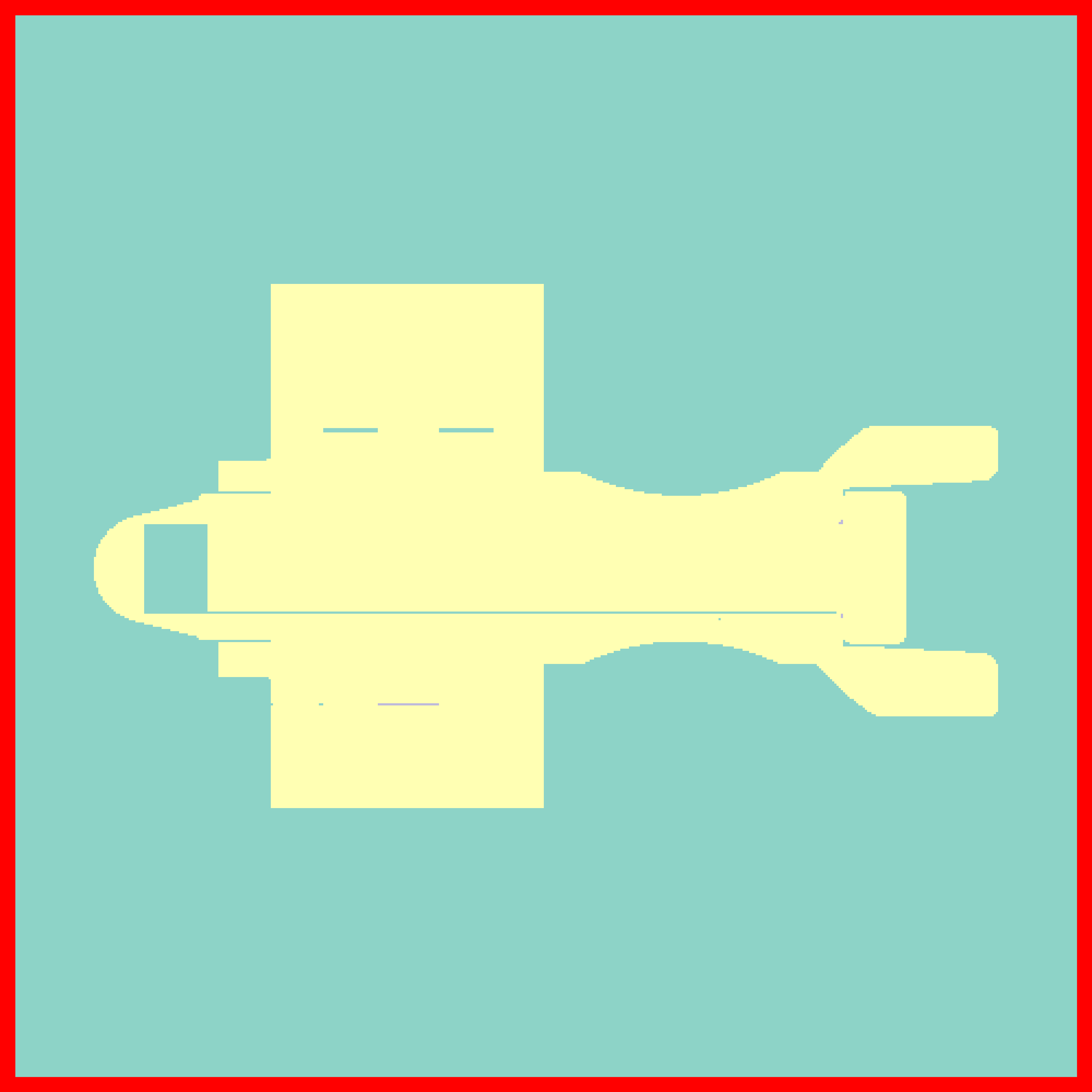} \\
    \includegraphics[width=0.45\linewidth]{summary_figures/gwn_colorbar_25.png} &
    \includegraphics[width=0.45\linewidth]{summary_figures/categories_colorbar_3.png} \\
\end{tabular}
\end{minipage}%
\begin{minipage}{0.5\textwidth}
\centering
\begin{tabular}{lr}
    \includegraphics[width=0.45\linewidth]{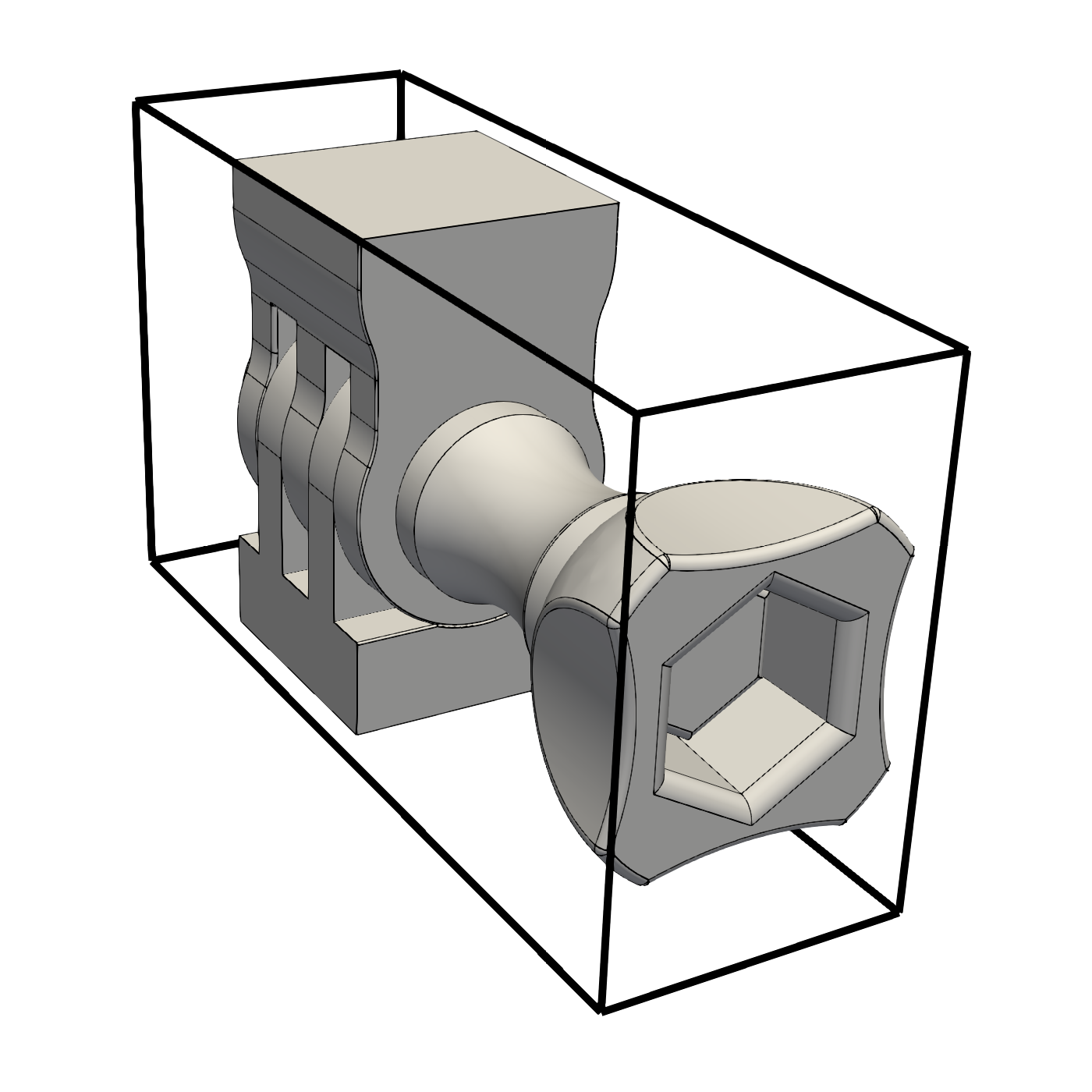}
    &
    \includegraphics[width=0.45\linewidth]{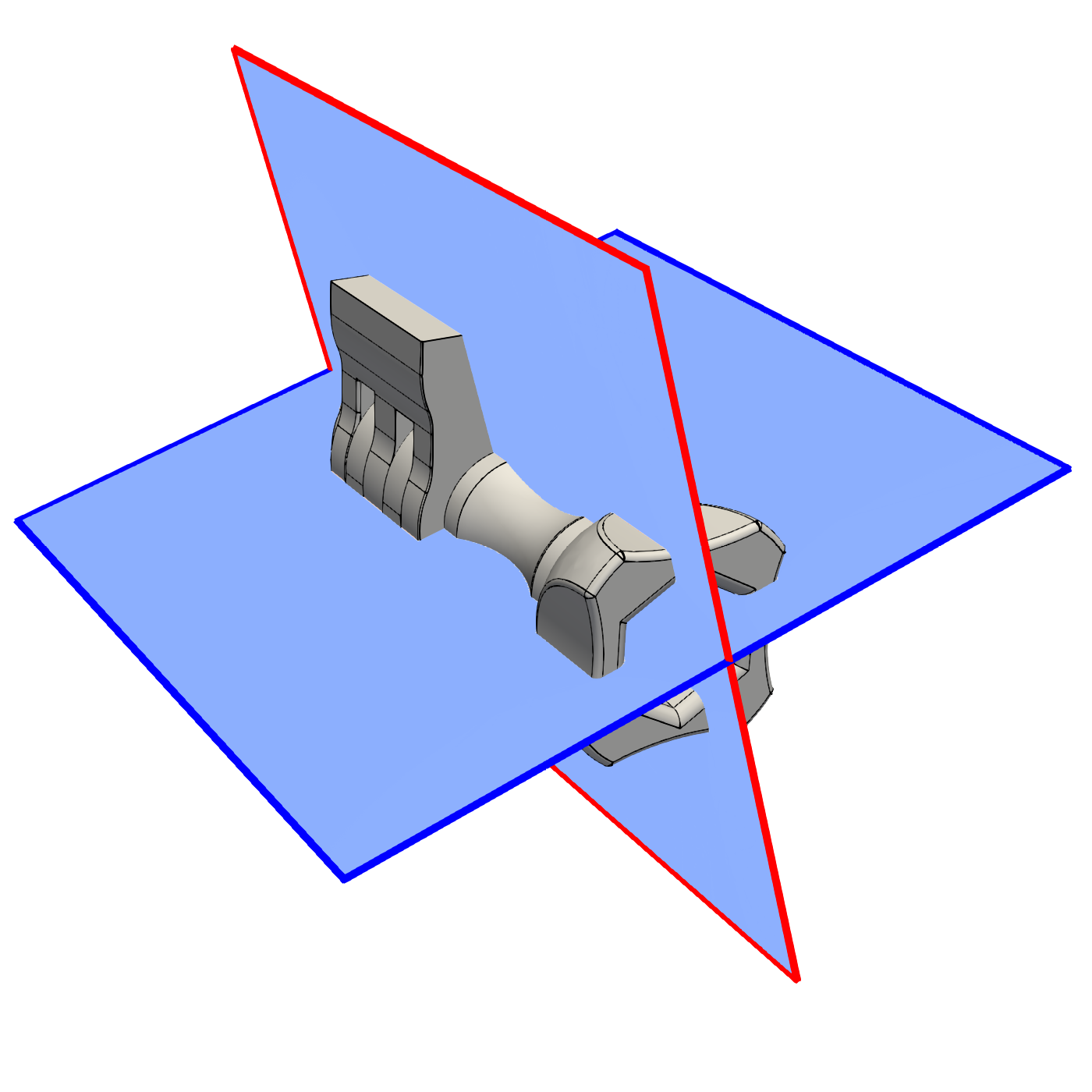}
\end{tabular}
\begin{tabular}{lr}
    Shape                          & Bolt                        \\\cmidrule{1-2}
    Number of NURBS Patches        & 248                 \\
    Number of Trimming Curves      & 1161                \\\cmidrule{1-2}
    \% Far-field Cases             & 99.7  \%  \\
    \% Near-field Cases            & 0.206 \%  \\
    \% Edge Cases                  & 0.016 \%  \\\cmidrule{1-2}
    Avg. Time per Query (ms)       & 5.76          \\\cmidrule{1-2}
    Avg. Far-field Case Time (ms)  & 0.0152           \\
    Avg. Near-field Case Time (ms) & 0.402          \\
    Avg. Edge Case Time (ms)       & 45.2          \\
    \end{tabular}
\end{minipage}
\caption{This ``Bolt'' assembly is derived from the ABC dataset model with index 3450. Note that there is a small section where adjacent watertight parts within the assembly overlap, increasing the integer winding number by one.}
\end{figure}

\vspace{1cm}

\begin{figure}
\begin{minipage}{0.5\textwidth}
\centering
\begin{tabular}{lr}
    \includegraphics[width=0.45\linewidth]{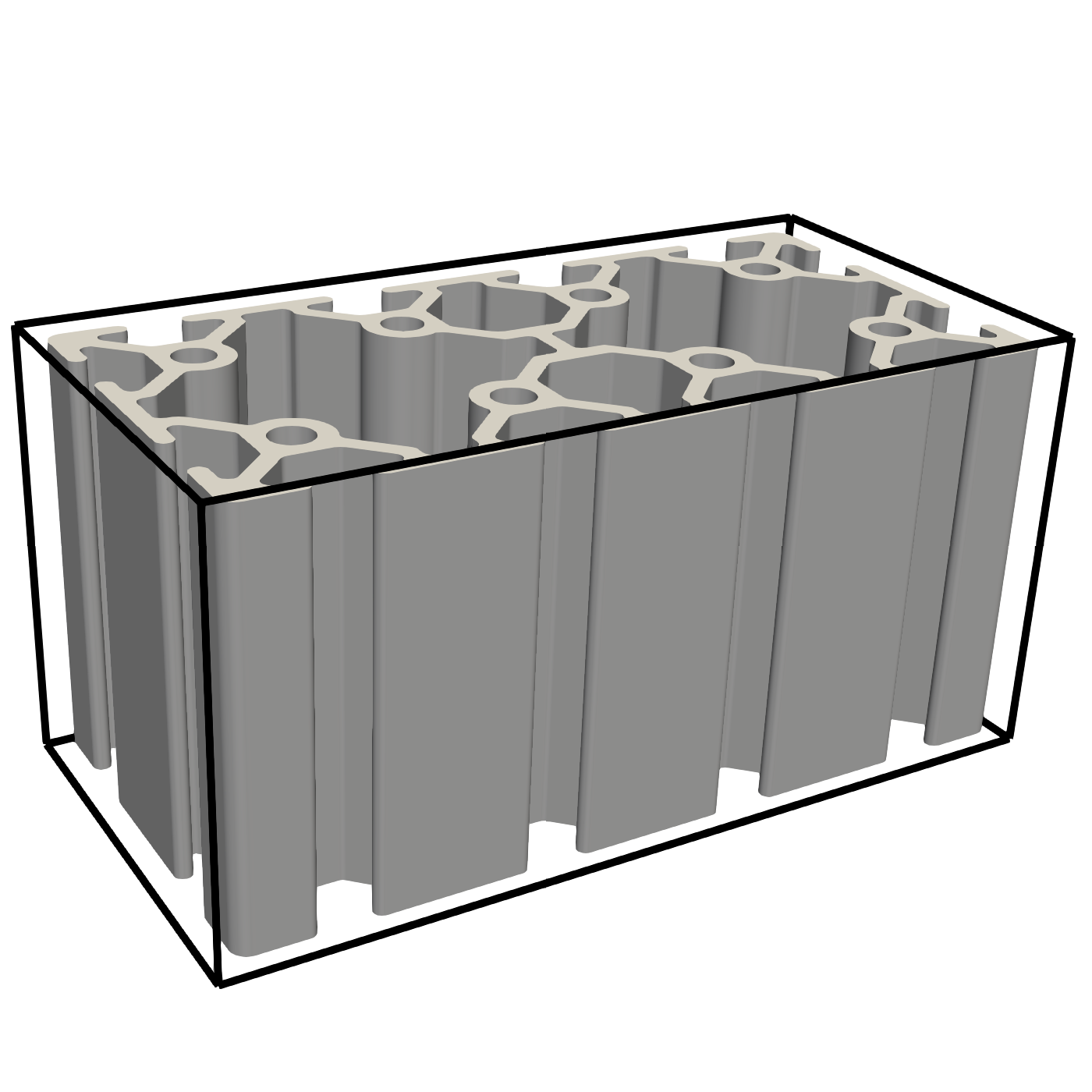}
    &
    \includegraphics[width=0.45\linewidth]{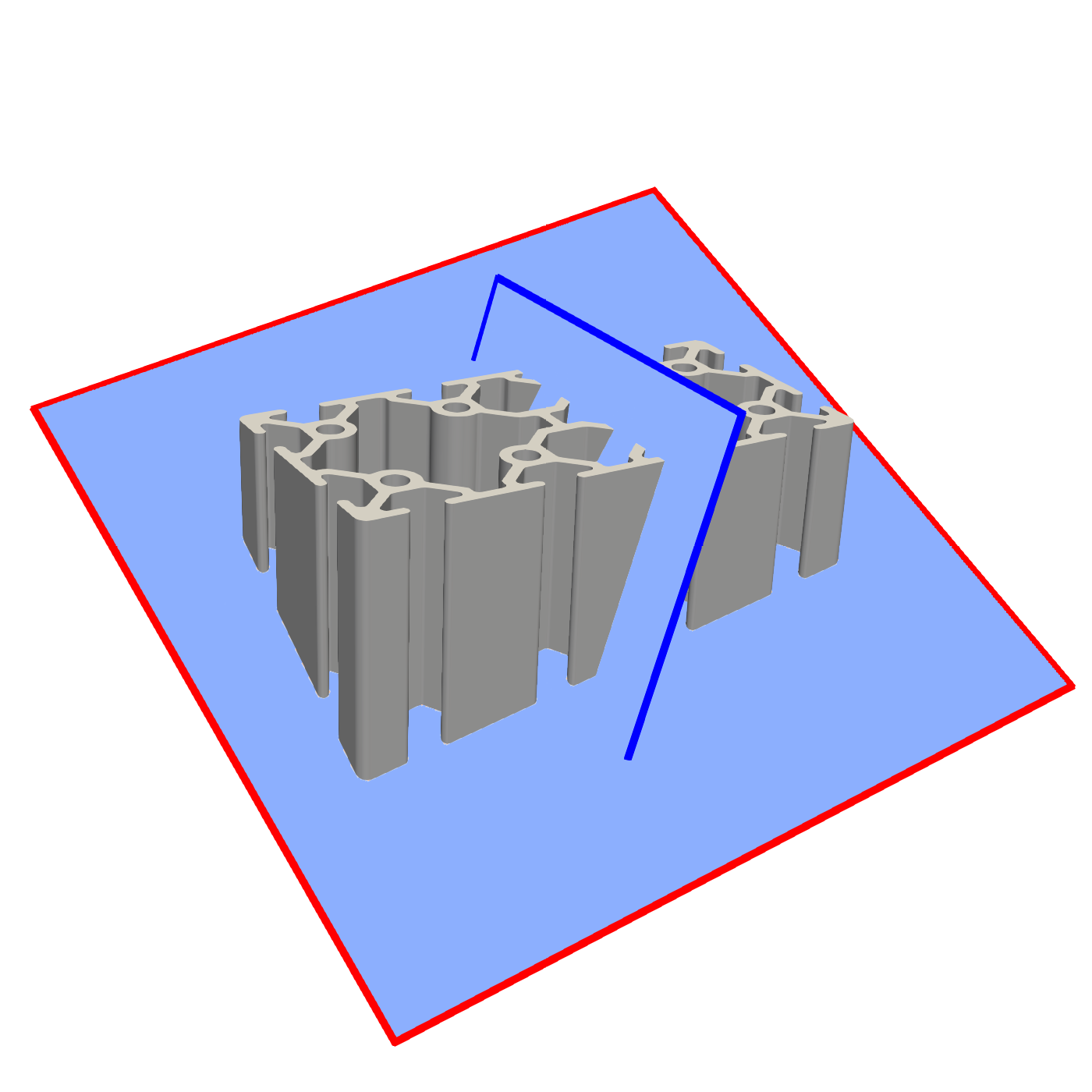}
\end{tabular}
\begin{tabular}{lr}
    Shape                          & Slide                        \\\cmidrule{1-2}
    Number of NURBS Patches        & 298                 \\
    Number of Trimming Curves      & 1776                \\\cmidrule{1-2}
    \% Far-field Cases             & 99.8  \%  \\
    \% Near-field Cases            & 0.144 \%  \\
    \% Edge Cases                  & 0.0019 \%  \\\cmidrule{1-2}
    Avg. Time per Query (ms)       & 3.71          \\\cmidrule{1-2}
    Avg. Far-field Case Time (ms)  & 0.0063           \\
    Avg. Near-field Case Time (ms) & 2.37          \\
    Avg. Edge Case Time (ms)       & 30.6          \\
    \end{tabular}
\end{minipage}%
\begin{minipage}{0.5\textwidth}
\centering
\begin{tabular}{lr}
    \includegraphics[width=0.45\linewidth]{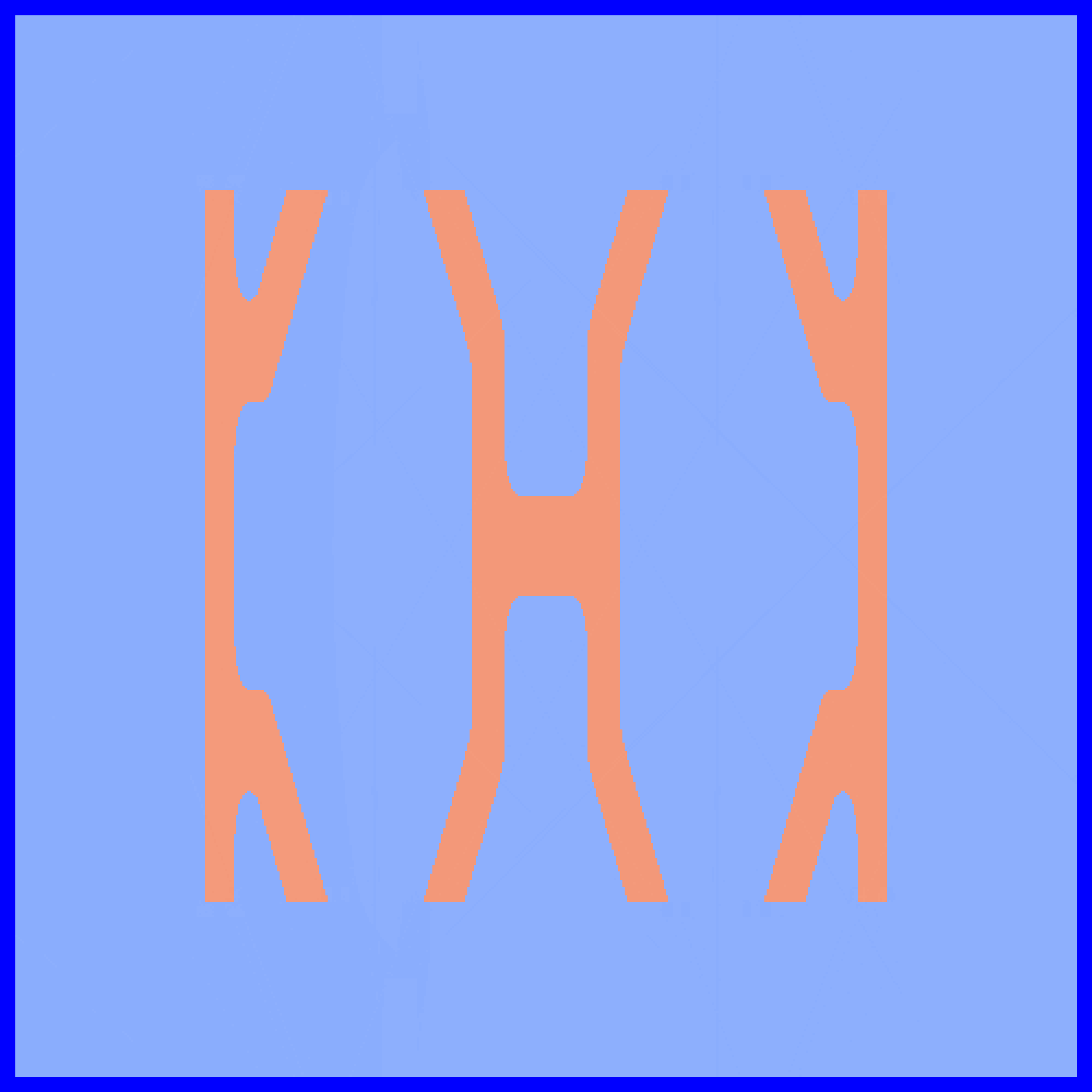} &
    \includegraphics[width=0.45\linewidth]{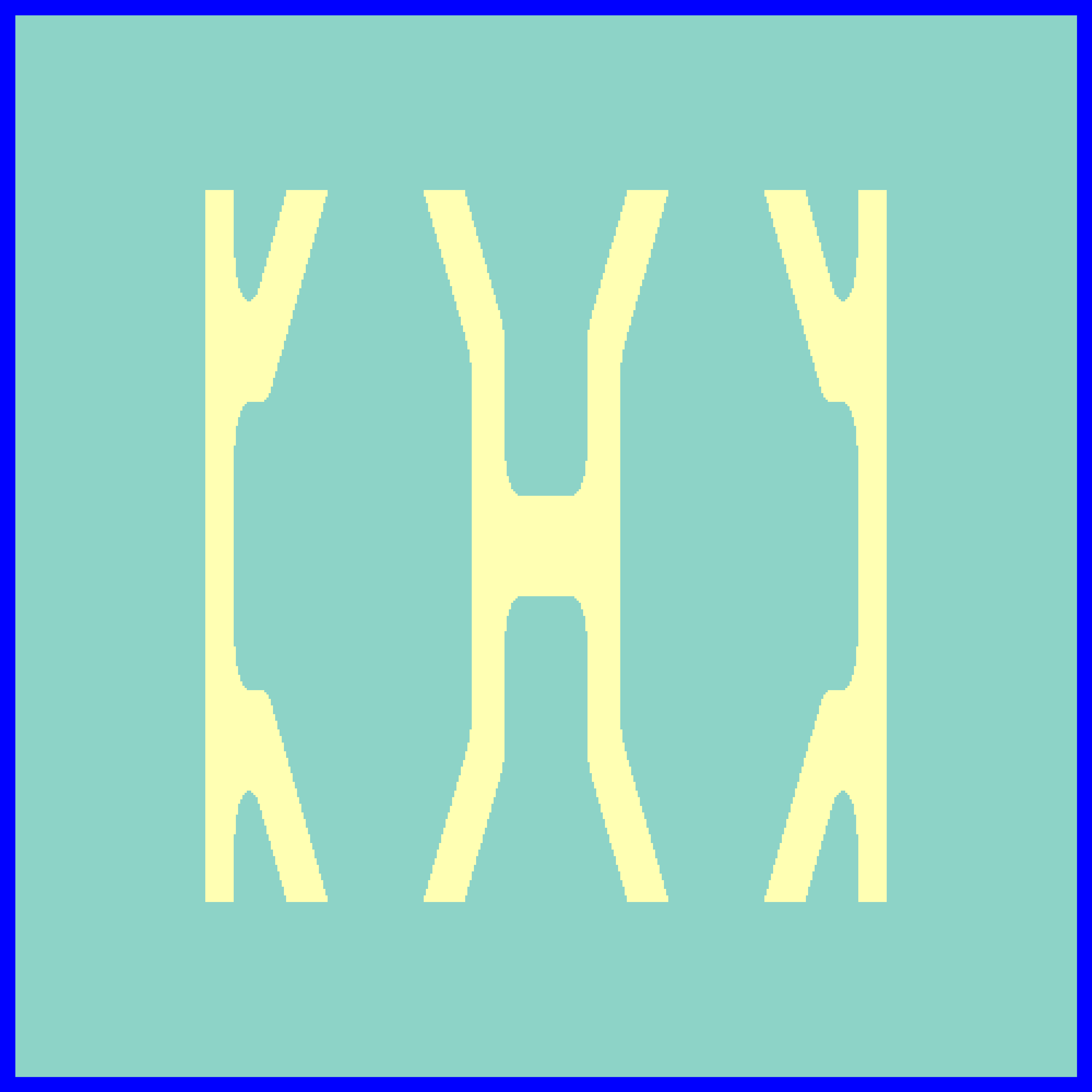} \\
    \includegraphics[width=0.45\linewidth]{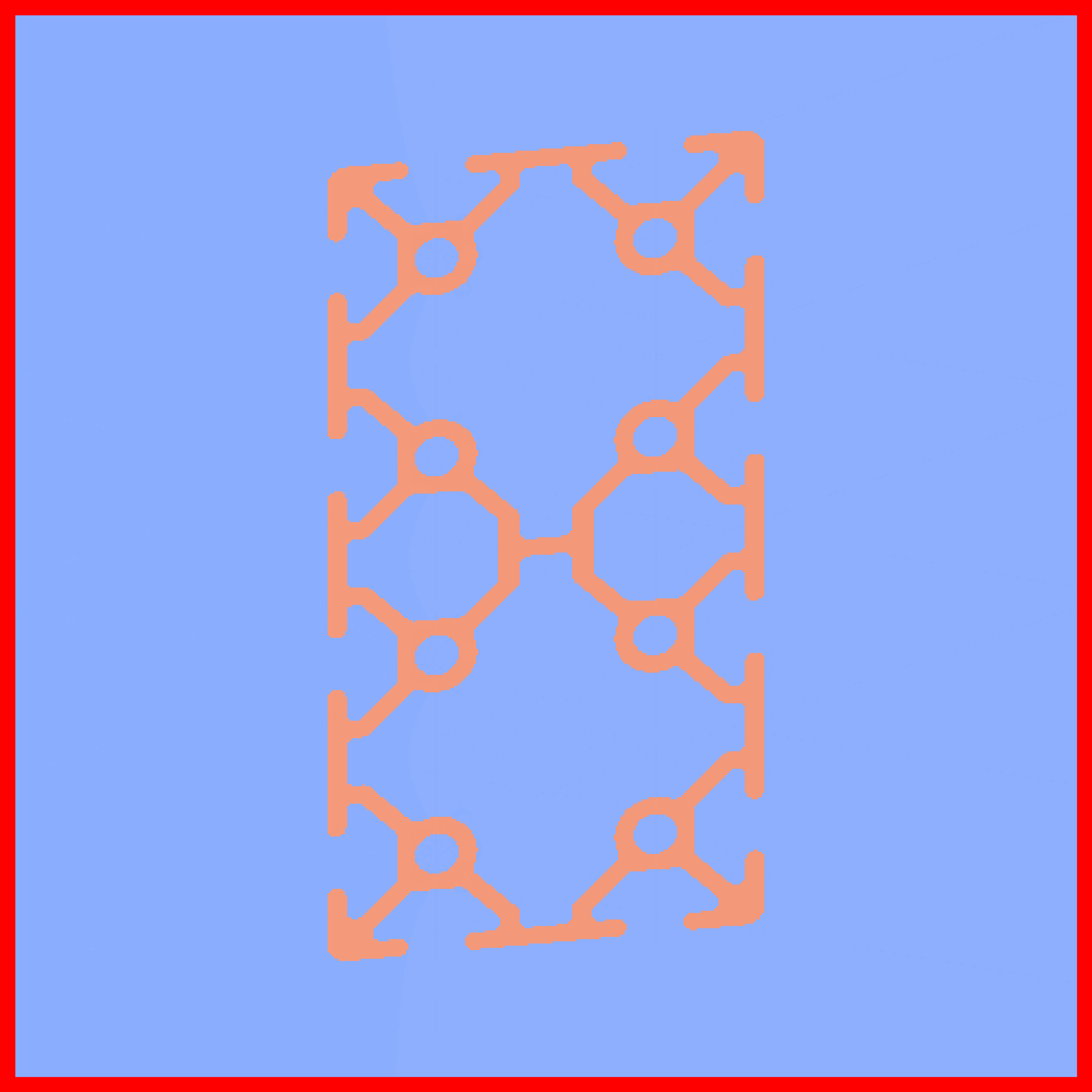} &
    \includegraphics[width=0.45\linewidth]{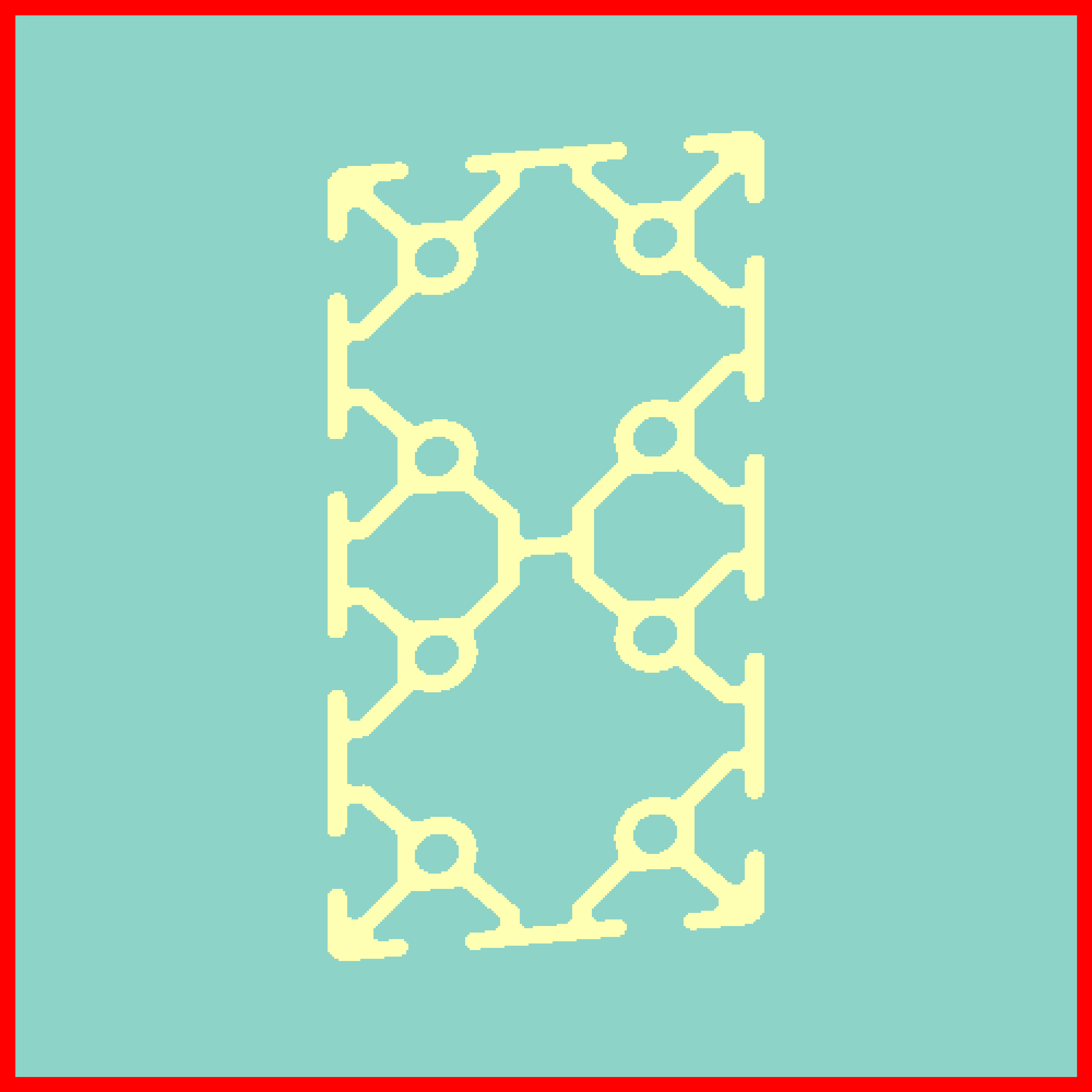} \\
    \includegraphics[width=0.45\linewidth]{summary_figures/gwn_colorbar_15.png} &
    \includegraphics[width=0.45\linewidth]{summary_figures/categories_colorbar_2.png} \\
\end{tabular}
\end{minipage}
\caption{This ``Slide'' shape is derived from the ABC dataset model with index 4237.}
\end{figure}

\end{document}